\documentclass{aa}  

\usepackage{natbib}
\usepackage{graphicx}
\usepackage{txfonts}
\usepackage[dvipsnames]{xcolor}
\usepackage{hyperref}
\usepackage{amsmath}
\newcommand{\modif}[1]{#1}
\newcommand{\modiff}[1]{#1}

\begin{document} 
  \title{VLTI/PIONIER survey of disks around post-AGB binaries\thanks{Based on VLTI observations 093.D-0573 and 094.D-0865}}
  \subtitle{Dust sublimation physics rules!}
  \author{J. Kluska\inst{1}
     \and
           H. Van Winckel\inst{1}
         \and
           M. Hillen\inst{1}
        \and
        J.-P. Berger\inst{2}
        \and
        D. Kamath\inst{3,4}
        \and 
        J.-B. Le Bouquin\inst{2,5}
        \and
        M. Min\inst{6}
          }

  \institute{
  Instituut voor Sterrenkunde (IvS), KU Leuven, Celestijnenlaan 200D, 3001 Leuven, Belgium\\ \email{jacques.kluska@kuleuven.be}
         \and
             Univ. Grenoble Alpes, CNRS, IPAG, F-38000 Grenoble, France
        \and
            Research School of Astronomy and Astrophysics, Australian National University, Cotter Road, Weston Creek ACT 2611, Australia
        \and
            Department of Physics and Astronomy, Macquarie University, Sydney, NSW 2109, Australia
        \and
           Department of Astronomy, University of Michigan, 1085 S. University, Ann Arbor, MI 48109, USA
        \and
            Netherlands Institute for Space Research (SRON), Sorbonnelaan 2, 3584 CA Utrecht, The Netherlands
             }

  \date{Received XXX; accepted XXX}

  \abstract
   {Post-AGB binaries are surrounded by circumbinary disks of gas and dust that are similar to protoplanetary disks found around young stars.}
   {We aim to understand the structure of these disks and identify the physical phenomena at play in their very inner regions. 
   We want to understand the disk-binary interaction and to further investigate the comparison with protoplanetary disks.}
   {We have conducted an interferometric snapshot survey of 23 post-AGB binaries in the near-infrared ($H$-band) using VLTI/PIONIER.
   We have fitted the multiwavelength visibilities and closure phases with purely geometrical models with an increasing complexity (including two point-sources, an azimuthally modulated ring and an over-resolved flux) in order to retrieve the sizes, temperatures and flux ratios of the different components.}
   {All sources are resolved and the different components contributing to the $H$-band flux are dissected.
   The environment of these targets is very complex: 13/23 targets need models with thirteen or more parameters to fit the data. 
   We find that the inner disk rims follow and extend the size-luminosity relation established for disks around young stars with an offset toward larger sizes. The measured temperature of the near-infrared circumstellar emission of post-AGB binaries is lower ($T_\mathrm{sub}\sim$1\modif{2}00\,K) than for young stars, probably due to a different dust mineralogy and/or gas density in the dust sublimation region. }
   {The dusty inner rims of the circumbinary disks around post-AGB binaries are ruled by dust sublimation physics.
   Additionally a significant amount of the circumstellar $H$-band flux is over-resolved (14 targets have more than 10\% of their non-stellar flux over-resolved) hinting for more structure from a yet unknown origin (disk structure or outflow). The amount of over-resolved flux is larger than around young stars.
   Due to the complexity of these targets, interferometric imaging is a necessary tool to reveal the interacting inner regions in a model-independent way. }


\maketitle


\section{Introduction}

Binarity is widely present in all kinds of stars \citep[25\% of low-mass stars and more than 80\% of high-mass stars have at least one companion;][]{Duchene2013} and binary research constitutes a main domain of stellar astrophysics.
Binary evolution gives birth to diverse phenomena such as thermonuclear novae, supernovae type Ia, sub-luminous supernovae, merger events generating detectable gravitational waves, and objects with lower initial mass such as sub-dwarf B-stars, barium stars, cataclysmic variables, and asymmetric planetary nebulae (PNe).
Understanding the diverse impact of binarity in stellar evolution is therefore crucial but is, also, still poorly understood.

In this paper we focus on observations of post-asymptotic giant branch (pAGB) binaries, objects in fast transition ($\sim$10$^{5}$ years) between the AGB and the PNe stages, that are surrounded by a circumbinary disk \citep[][]{VanWinckel2018}.

Binarity is playing a central role in the formation of the dusty disks, which were first postulated from the infrared excess in the spectral energy distribution (SED). 
These excesses cannot be attributed to expanding detached shells \citep[e.g.,][]{deRuyter2006}.
Most of the disk sources were then discovered to be binaries through radial velocity measurements \citep[][]{VanWinckel2003,vanwinckel2009,Oomen2018}.
Those observations lead to the conclusion that pAGB disks originate from the evolved star's mass loss via a poorly understood binary interaction mechanism that happens at the end of the AGB phase for low- and intermediate-mass stars (0.8 - 8\,M$_\odot$).
Millimeter observations of CO lines with the Plateau de Bure interferometer and the Atacama Large Millimeter/submillimeter Array (ALMA) resolved the outer parts of these disks and showed them to be in Keplerian rotation, and thus stable \citep[][]{Bujarrabal2013,Bujarrabal2015,Bujarrabal2016,Bujarrabal2017,Bujarrabal2018}. 
The CO observations also revealed a disk-wind component, suggesting angular momentum transport in the disk.
Dust grains are inferred to have large sizes \modif{(ranging from a few microns to millimeter sizes)} and a high crystallinity fraction, based on analyses of the mid-infrared (mid-IR) spectral features \modif{and sub-mm spectral slopes} \citep{deRuyter2006,Gielen2008,Gielen2011,Hillen2015}.
The dust masses found in these disks are of the order of 10$^{-4}$-10$^{-3}$\,M$_\odot$ \citep{Sahai2011,Hillen2014}, but are highly model dependent.

Despite very different forming processes, pAGB disks are in many ways (infrared excess, Keplerian rotation, winds, dust mass, dust mineralogy and grain sizes) similar to protoplanetary disks (PPDs) around young stellar objects (YSOs).
Radiative transfer models of PPDs are able to successfully reproduce both the SED and infrared interferometric measurements on the few pAGB targets studied so far \citep{Hillen2014,Hillen2015,Hillen2017,Kluska2018}.
As the PPDs are well studied both observationally and theoretically, the very close similarity with the disks around pAGB binaries points toward a potential universality of physical processes in dusty circumstellar disks occupying a different parameter space (i.e. different formation process, presumably shorter lifetime, high stellar luminosity).

Also, such a similarity between those two types of disks raises the question of the planet formation efficiency in pAGB disks \citep[e.g.,][]{Schleicher2014}, especially as several planets are candidates of being formed in such second-generation disks \citep[e.g., NN Ser;][]{Volschow2014,Marsh2014,Parsons2014,Hardy2016}.

The interaction between the binary and the disk gives rise to several complex physical processes.
First\modif{ly}, these systems show indirect signs for re-accretion from the circumbinary disk onto the central system: the primary's photospheric spectrum shows depletion in elements that have the highest condensation temperature \citep[][Kamath \& Van Winckel, 2019, in press]{Maas2003,Gezer2015,Oomen2018}.
The scenario explaining this depletion is that the condensed elements are subject to radiative pressure and remain in the disk while the gas is re-accreted onto the central star(s) \citep[][]{Waters1992}.
However, this scenario needs to be confirmed by direct observations of the re-accretion.

Second\modif{ly}, spectral time series observations allowed the detection of bipolar jets linked to the secondary star \citep[e.g.,][]{Thomas2013,Gorlova2015,Bollen2017}.
These jets have their origin around the secondaries which should also be surrounded by an accretion disk. 
Interestingly, the Very Large Telescope Interferometer (VLTI) observations of one of the most studied pAGB-binaries, IRAS08544-4431, with the Precision Integrated-Optics Near-infrared Imaging ExpeRiment (PIONIER) in the near-infrared, detected a point-source emission at the position of the secondary. 
It should not be detectable if the emission was coming from a photosphere alone and it was tentatively linked with the circum-secondary accretion disk \citep[][]{Hillen2016,Kluska2018}.
The existence of this unexpected continuum emission from the secondary needs to be investigated in other systems as well.

Third, the orbits of the binaries disagree with predictions of theoretical models.
At the end of the AGB phase we expect the period distribution of the binaries to be bimodal: the systems that went through common envelope evolution should result in a shrinkage of their orbital period and wider systems should have larger orbits because of the mass loss of the primary \citep{Nie2012}.
However, radial velocity monitoring of the binary orbits revealed that the detected orbital distribution falls between these two modes and show periods that are not predicted by population studies.
Moreover, tidal circularization of the orbits is expected when the primary evolves on the giant branches, whereas observations show orbits with nonzero eccentricity values ($\sim$0.2-0.3) pointing at an eccentricity pumping mechanism \citep[][]{Oomen2018}.
Interactions between the circumbinary disk and the binary could explain some of the observed eccentricities \citep[e.g.,][]{Dermine2013,Vos2015}.
However, this mechanism is still debated as strong assumptions were made about the disk radial and vertical structure, disk viscosity and lifetime \citep[][]{Rafikov2016}.
Spatially resolved observations of the disk inner rim from which we could infer the radius, height, eccentricity, perturbation will therefore help to constrain hydrodynamic models of disk-binary interactions.

The binary eccentricity can also disturb the circumbinary disk \citep[e.g.,][]{Thun2017}.
Another possibility is that the orbital eccentricity is pumped-up by an increased mass exchange between the two stars at the periastron passage \citep[grazing envelope evolution, e.g.,][]{Kashi2018}. 
This mechanism could also delay the common envelope phase extending the final orbital period \citep{Shiber2017}.

Our previous high-spectral resolution time series and high-angular resolution interferometric data enabled us to build an archetype of a pAGB binary system.
\modif{In our current state of knowledge, i}ts building blocks are \modif{likely to be}:
\begin{itemize}
    \item a pAGB primary
    \item a main-sequence secondary surrounded by an accretion disk which launches a wide bipolar jet
    \item a binary orbit that is eccentric and not predicted by population synthesis models
    \item a circumbinary disk with a dust inner rim at a radius of several astronomical units, likely ruled by dust sublimation physics and that is azimuthally perturbed \citep[e.g.,][]{Kluska2018}
    \item a near-infrared extended flux from a yet unknown origin.
\end{itemize}
Here, we present an interferometric snapshot survey in the near-infrared of 23 systems in order to test this archetype.
We focus on the general properties of the different components contributing to the $H$-band flux in those systems.
The paper is organized as follows.
We describe the photometric and interferometric observations in Sect.\,\ref{sec:obs} and the geometric models in Sect.\,\ref{sec:models}.
We show the results in Sect.\,\ref{sec:results}, discuss them in Sect.\,\ref{sec:discussion} before concluding in Sect.\,\ref{sec:conclusions}.


\section{Observations}
\label{sec:obs}

\subsection{Photometry}

\modiff{The best photosphere fit to the visible part of the photometry of the sources was used to extrapolate the stellar spectrum to the $H$-band (1.65$\mu$m).}
We used the targets photometry of \citet[][]{Hillen2017,Oomen2018}.
We have compiled archival photometry \modif{on all our targets except V494\,Vel (Figs.\,\ref{fig:SED1}, \ref{fig:SED2} and \ref{fig:SED3})}.
The photometric data is coming from Johnson-Cousins bands \citep[][]{1966CoLPL...4...99J,1975RMxAA...1..299J,1978A&AS...34..477M,1997A&AS..124..349M,2001KFNT...17..409K,2002yCat.2237....0D,2006yCat.2168....0M,2007PASP..119.1083R,2008AJ....136..735L,2008PASP..120.1128O,2012yCat.5137....0A,2016MNRAS.463.4210N,2016yCat.2336....0H}, Geneva photometry \citep[][]{1997A&AS..124..349M} and Str\"omgren photometry \citep[][]{1998A&AS..129..431H,2001A&A...373..625P}.
For some targets, we also used photometry from TYCHO \citep[][]{2000A&A...355L..27H,1997A&A...323L..57H,1998AJ....115.1212U} and GAIA \citep[][]{2016yCat.1337....0G}.
For near-infrared photometry we used 2MASS \citep[][]{2003yCat.2246....0C} while for mid and far-infrared we used AKARI \citep[][]{2007PASJ...59S.369M,2010A&A...514A...1I}, WISE \citep[][]{2012yCat.2311....0C}, IRAC \citep[][]{2009yCat.2293....0S}, IRAS \citep[][]{1988SSSC..C......0H} and MSX \citep[][]{2003yCat.5114....0E}.

\modif{To fit the SED we have first derived stellar parameters (such as effective temperature, $T_\mathrm{eff}$, surface gravity, $\log g$; see Table\,\ref{table:1}) from existing spectra of the stars \citep[][]{Waelkens1991,Giridhar1994,VanWinckel1997,VanWinckel1998,Giridhar2000,Dominik2003,Maas2002,Maas2003,Maas2005,deRuyter2006} using Kurucz models \citep[][]{Kurucz1993}.}
\modif{We then defined allowed ranges around those values that the fitting algorithm can explore ($T_\mathrm{eff}\pm250\,K$; $\log g \pm0.3$ or between 0 and 2.5 if not constrained by the spectrum). 
We then minimized the $\chi^2$ via a Levenberg-Marquardt algorithm between the photometric data and the reddened model (the results are displayed in Table\,\ref{tab:SEDfit}).}

\subsection{Interferometry}

The interferometric observations were obtained with PIONIER located at the VLTI at Mount Paranal in Chile.
PIONIER recombines light from four telescopes in the near-infrared $H$-band (\modif{between 1.5 and 1.85}$\mu$m).
The interferometric observables are the squared visibility amplitude (V$^2$), that is a measure of the \modif{degree of} spatial resolution of the source \modif{by a given baseline at a given wavelength}, and closure phase (CP), that is a measure of the departure from point-symmetry of the target.

The target selection (see Table\,\ref{table:1}) was based on 1) the identification of the object as a post-AGB star, 2) the presence of an H-band excess \citep[][]{deRuyter2006,Gezer2015} and 3) observability with PIONIER on the VLTI.

\begin{table*}
\caption{Binary post-AGB stars in our sample.}             
\label{table:1}      
\centering                          
\begin{tabular}{c l l c c c c c c c c}        
\hline\hline                 
\# & Target & IRAS & RA  & DEC & dist & Teff &  $\log g$ & P &Ref. \\ 
   &        &      &     &     & [pc] &  [K]  &  & [days] &  \\
\hline                        
1	&\object{AC\,Her}	&18281+2149	&	18 30 16.2	&	+21 52 00	&	$ 1231^{+   46}_{  -43}$	&	$ 5500^{+  250}_{-  250}$	&	$  0.5^{+  2.0}_{-  0.5}$	&	1189$\pm$1.2 & (4)\\[1pt]
2	&\object{AI\,Sco}	&17530-3348	&	17 56 18.5	&	-33 48 43	&	$11886^{+ 4870}_{-3412}$	&	$ 5000^{+   250}_{-  250}$&	$  1.8^{+  0.7}_{-  1.8}$	&	977&(2)\\[1pt]
3	&\object{EN\,TrA}	&14524-6838	&	14 57 00.6	&	-68 50 22	&	$ 2751^{+  386}_{ -303}$	&	$ 6000^{+  250}_{-  250}$	&	$  1.0^{+  0.0}_{-  0.5}$	&	1493$\pm$7&(1)\\[1pt]
4	&\object{HD\,93662}	&10456-5712	&	10 47 38.3	&	-57 28 02	&	$ 1045^{+   94}_{  -80}$	&	$ 4250$	&	$  0.5$	&	572$\pm$6&(3)\\[1pt]
5	&\object{HD\,95767}	&11000-6153	&	11 02 04.3	&	-62 09 42	&	$ 3820^{+  503}_{ -401}$	&	$ 7600^{+  250}_{-   250}$	&	$  2.0^{+  0.5}_{-  1.0}$	&	1989$\pm$61&(4)\\[1pt]
6	&\object{HD\,108015}	&12222-4652	&	12 24 53.5	&	-47 09 07	&	$ 3867^{+  795}_{ -582}$	&	$ 7000^{+  250}_{-    250}$	&	$  1.5^{+  1.0}_{-  0.5}$	&	906.3$\pm$5.9&(4)\\[1pt]
7	&\object{HD\,213985}	&22327-1731	&	22 35 27.5	&	-17 15 26	&	$  644^{+   27}_{  -25}$ &	$ 8250^{+  250}_{-  250}$	&	$  1.5^{+  0.5}_{-  0.5}$	&	259.6$\pm$0.7&(4)\\[1pt]
8	&\object{HR\,4049}	&10158-2844	&	10 18 07.5	&	-28 59 31	&	$ 1574^{+  487}_{ -315}$	&	$ 7500^{+    250}_{-  250}$	&	$  1.0^{+  1.5}_{-  0.5}$	&	430.6$\pm$0.1&(4)\\[1pt]
9	&	&05208-2035	&	05 22 59.4	&	-20 32 53	&	$ 1480^{+   78}_{  -71}$	&	$ 4000^{+  100}_{-  170}$	&	$  0.5^{+  2.0}_{-  0.5}$	&	234.38$\pm$0.04&(4)\\[1pt]
10	&	&08544-4431	&	08 56 14.1	&	-44 43 10	&	$ 1470^{+  193}_{ -154}$	&	$ 7250^{+  250}_{-  250}$	&	$  1.5^{+  1.0}_{-  1.0}$	&	501.1$\pm$0.1&(4)\\[1pt]
11	&	&10174-5704	&	10 19 16.8	&	-57 19 25	&	$ 2613^{+  668}_{ -452}$	&	$ 6000^{+  250}_{-   250}$	&	$  1.0^{+  1.5}_{-  0.5}$	&	323$\pm$50&(3)\\[1pt]
12	&	&15469-5311	&	15 50 43.8	&	-53 20 43	&	$ 3179^{+  613}_{ -449}$	&	$ 7500^{+  250}_{-  250}$	&	$  1.5^{+  1.0}_{-  1.0}$	&	390.2$\pm$0.7&(4)\\[1pt]
13	&	&17038-4815	&	17 07 36.6	&	-48 19 08	&	$ 4330^{+ 1254}_{ -823}$	&	$ 4750^{+  250}_{-  250}$	&	$  1.0^{+  1.5}_{-  0.5}$	&	1394$\pm$12&(4)\\[1pt]
14	&	&18123+0511	&	18 14 49.3	&	+05 12 55	&	$ 6196^{+ 2256}_{-1443}$	&	$ 5000^{+  250}_{-  250}$	&	$  1.0^{+  1.5}_{-  0.5}$	&	-&-\\[1pt]
15	&	&19125+0343	&	19 15 01.1	&	+03 48 42	&	$ 4131^{+  905}_{ -645}$	&	$ 7750^{+  250}_{-  250}$	&	$  1.5^{+  1.0}_{-  0.5}$	&	519.7$\pm$0.7&(4)\\[1pt]
16	&\object{IW\,Car}	&09256-6324	&	09 26 53.3	&	-63 37 48	&	$ 1811^{+  107}_{  -96}$	&	$ 6700^{+  500}_{-  500}$	&	$  2.0^{+  1.0}_{-  1.5}$	&	1449&(2)\\[1pt]
17	&\object{LR\,Sco}	&17243-4348	&	17 27 53.6	&	-43 50 46	&	$ 7325^{+ 3025}_{-1946}$	&	$ 6250^{+  250}_{-  250}$	&	$  1.0^{+  1.5}_{-  0.5}$	&	$\sim475$&(3)\\[1pt]
18	&\object{PS\,Gem}	&07008+1050	&	07 03 39.6	&	+10 46 13	&	$ 2835^{+  385}_{ -306}$	&	$ 6000^{+  250}_{-  100}$	&	$  1.0^{+  1.5}_{-  0.5}$	&	-&-\\[1pt]
19	&\object{R\,Sct}	&18448-0545	&	18 47 28.9	&	-05 42 18	&	$ 1273^{+ 1079}_{ -410}$	&	$ 4500^{+  500}_{-  500}$	&	$  0.9^{+  1.6}_{-  0.9}$	&	-&-\\[1pt]
20	&\object{RU\,Cen}	&12067-4508	&	12 09 23.8	&	-45 25 34	&	$ 1822^{+  190}_{ -158}$	&	$ 6000^{+  250}_{-  250}$	&	$  1.5^{+  1.0}_{-  1.5}$	&	1489$\pm$10&(4)\\[1pt]
21	&\object{SX\,Cen}	&12185-4856	&	12 21 12.5	&	-49 12 41	&	$ 3870^{+  721}_{ -538}$	&	$ 6000^{+  250}_{-  250}$	&	$  1.0^{+  0.5}_{-  1.0}$	&	564.3$\pm$7.6&(4)\\[1pt]
22	&\object{U\,Mon}	&07284-0940	&	07 30 47.4	&	-09 46 36	&	$ 1067^{+  120}_{  -98}$	&	$ 5000^{+  250}_{-  250}$	&	$  0.0^{+  2.5}_{-  0.0}$	&	2550$\pm$143&(4)\\[1pt]
23	& \object{V 494 Vel}	& 09400 -4733	&	09 41 51.9	&	-47 46 57.8	&	$ 2018^{+  147}_{ -129}$ & - & - & - & -\\[1pt] 
\hline                                   
\end{tabular}
\tablefoot{
The distances are from \citet[][]{BJ2018}
}
\tablebib{
(1)~\citet{vanwinckel2009}; (2) \citet{KissBodi2017}; (3) \citet{Hillen2017}; (4) \citet{Oomen2018}.
}
\end{table*}

Most targets were observed as part of a dedicated observing programme (European Southern Observatory (ESO) program 093.D-0573, PI: Hillen), but some were also observed as a supplement to the imaging campaign on \object{IRAS\,08544-4431} (ESO program 094.D-0865; PI: Hillen). 
This explains, to some extent, the diversity in uv-coverage among the sample (see Table\,\ref{tab:log}).
The observations of IRAS08544-4431 are described in \citet[][]{Hillen2016}.
Each observation of the science star was bracketed with two calibrator stars to well interpolate the transfer function and calibrate the squared visibilities (V$^2$) and closure phases (CP).
The observations were taken using the small (A1-B2-C1-D0), the intermediate (D0-G1-H0-I1) and the extended (A1-G1-J3-K0) configurations depending on the expected size and luminosity of the object.
Therefore, not all the targets have observations on the three configurations (see Fig.\,\ref{fig:uv1} and \ref{fig:uv2}) causing an in-homogeneous (u, v)-coverage throughout the whole dataset.
All the targets were observed with a grism, leading to spectrally dispersed data with a low resolution ($R\sim$30). 
We are therefore sensitive to the continuum only.
The data was reduced using the \texttt{pndrs} software \citep{lebouquin2011}.
The entire dataset is available on the Optical interferometry DataBase (OiDB)\footnote{accessible at oidb.jmmc.fr}

All the targets have squared visibilities significantly below unity (see Figs.\,\ref{fig:data1}, \ref{fig:data2} and \ref{fig:data3}) meaning that at least a fraction of the near-infrared emission is spatially resolved in all of them.
For several targets the closure phases are showing a nonzero signal indicating departure from point symmetry.

\section{Model fitting}
\label{sec:models}

In order to analyze the dataset we performed a fitting with geometrical models with increasing complexity.
We stress that the dataset is diverse in both the sizes of our targets and the obtained (u, v)-coverage.
The challenge is that the observational constraints differ from object to object.
Because of the sparsity of the (u, v)-coverages, our models do not take into account any intrinsic variability of the inner source as there is orbital motion and/or large amplitude pulsations.
We define several classes of models in Sect.\,\ref{sec:s}, \ref{sec:sr}, \ref{sec:br} and \ref{sec:b}.
We also describe our fitting strategy in Sect.\,\ref{sec:strategy}.
We start by describing in Sect.\,\ref{sec:modspec} the way these models attribute spectral dependence between the different components.

\subsection{Model definition and spectral dependence}
\label{sec:modspec}

Thanks to the linearity property of the Fourier transform, the analytic models are defined in the Fourier plane as a linear combination of different geometrical components.
The weights of this combination are the relative fluxes of the components.
Those flux contributions are defined as:
\begin{eqnarray}
\sum_\mathrm{i} f^\mathrm{i}_0 &=& 1 \label{eqn:norm}\\
f^\mathrm{i}_0 &\geqslant& 0 \\
f^\mathrm{i}_0 &\leqslant& 1 ,
\end{eqnarray}
where $f^\mathrm{i}_0$ is the flux ratio of the $i$-th component at 1.65$\mu$m.

In the models there are four possible components: the primary star, the secondary star, the \modiff{circumbinary} ring and the background.
Not all components are present in all the models.
To extrapolate the flux ratios over the observed wavelength range the components are assigned with a spectral dependence law ($f^\mathrm{i}$).

The spectral dependence of the primary star is taken from the photospheric flux in the $H$-band from the best-fit from the SED\footnote{\modif{The used parameters to model for the photospheric flux are shown Table\,\ref{tab:SEDfit}.}} and is normalized to unity at 1.65$\mu$m.
\modiff{This is possible as the contribution from the secondary and the ring are negligible in the visible because of their lower temperature and high contrast with the primary.}
The fluxes of the secondary star and the background are defined as a power-law with wavelength and a spectral index ($d_\mathrm{i} = \frac{d \log F_\lambda}{d \log \lambda}$) such that:
\begin{equation}
f^\mathrm{i} = f^\mathrm{i}_0 \Bigg(\frac{\lambda}{1.65\mu\mathrm{m}}\Bigg)^\mathrm{d_\mathrm{i}},
\end{equation}
where $\lambda$ is the wavelength of an observation and $i$ is either $sec$ or $bg$ if it is the secondary star or the background respectively.
Finally, the ring spectral dependence is defined as a black body function at a given temperature $T_\mathrm{ring}$ that is normalized to unity at 1.65$\mu$m:
\begin{equation}
f^\mathrm{ring} = f^\mathrm{ring}_0 \Bigg(\frac{BB(\lambda,T_\mathrm{ring})}{BB(1.65\mu\mathrm{m},T_\mathrm{ring})}\Bigg),
\end{equation}
where $BB$ is the black body function.

In the following sections the model geometrical descriptions are presented as well as the full model equations.

\subsection{Single star and background flux: s\texttt{0-1}}
\label{sec:s}

This first set of models includes two components: the star and the background flux.
\begin{itemize}
\item \textbf{The star:} 
The star is geometrically defined by the diameter of its uniform disk ($UD_\mathrm{prim}$). 
The stellar visibility $V^*(u,v)$ is therefore:
\begin{equation}
    V^*(u,v) = 2\frac{J_1(\pi UD_\mathrm{prim} \sqrt{u^2+v^2})}{\pi UD_\mathrm{prim} \sqrt{u^2+v^2}}
    \label{eqn:Vdisk},
\end{equation}
where $u$ and $v$ are the coordinates in the Fourier domain and $J_1$ is the first order Bessel function.

\item \textbf{The background:}
The background is the over-resolved flux which means that this component is fully resolved even for the smallest baseline.
Its visibility ($V^\mathrm{bg}$) equals 0 for all baselines.

\item \textbf{The final model} is a linear combination of the visibilities of the star and the background with as factors their flux contribution that depend on the observed wavelength (see Sect.\,\ref{sec:modspec}).
The spectral dependence of the background is either assumed to be a black-body (model \texttt{s0}) or a power-law (model \texttt{s1}).
We made this choice as the \texttt{s0} model is the starting point to all the models and, in the absence of the ring component, it is already giving a good indication for the temperature of the environment whereas in all the other models the background is modeled as a power-law.
The final visibility is expressed as:
\begin{eqnarray}
V^\mathrm{tot,s}(u,v) &=& \frac{f^\mathrm{prim} V^*(u,v) }{f^\mathrm{prim} + f^\mathrm{bg}}.
\end{eqnarray}
Thanks to Eqn.\,\ref{eqn:norm} we have:
\begin{equation}
f^\mathrm{bg}_0 = 1 - f^\mathrm{prim}_0.
\end{equation}

\end{itemize}

\subsection{Single star and a ring: sr\texttt{0-6}}
\label{sec:sr}

In this set of models there are three components: the primary star, the ring and the background.
The star is modeled as a point source and the circumstellar matter is modeled by a Gaussian ring that can be inclined and modulated, depending on the complexity of the model.
\begin{itemize}
\item \textbf{The star:}
The visibility of a point source is $V^*=1$. The position of the star with respect to the cent\modiff{e}r of the ring is defined by $x_0$ and $y_0$. Because of this shift, the complex visibility of the star is:
\begin{equation}
    V^*(u,v) = \exp-2 i \pi (x_0 u + y_0 v),
    \label{eqn:Vstar}
\end{equation}
where $u$ and $v$ are the spatial frequencies in the West-East and South-North directions.
\item \textbf{The ring:}
The ring is first defined as an infinitesimal ring distribution. Its visibility ($V^\mathrm{ring0}(u,v)$) equals to:
\begin{equation}
    V^\mathrm{ring0}(u,v) = J_\mathrm{0}(\pi\rho'\theta),
\end{equation}
where $J_\mathrm{0}$ is the Bessel function of the 0$^\mathrm{th}$ order, $\theta$ the diameter of the ring and $\rho'$ is the spatial frequency of a data point corrected for inclination ($inc$) and position angle ($PA$) of the ring such as:
\begin{eqnarray}
    \rho' &=& \sqrt{u'^2 + v'^2}\\
    u' &=& u \cos PA + v \sin PA \\
    v' &=& (-u \sin PA + v \cos PA) \cos inc.
\end{eqnarray}
This ring is then azimuthally modulated using a set of sinusoidal functions having a period 2$\pi$ and $\pi$ called first order and second order modulations respectively.
The ring visibility with first order modulation ($V^\mathrm{ring1}(u,v)$) and with first and second order modulations ($V^\mathrm{ring2}(u,v)$) can be written as:
\begin{eqnarray}
    V^\mathrm{ring1}(u,v) &=& V^\mathrm{ring0} -i (c_\mathrm{1} \cos\alpha + s_\mathrm{1} \sin\alpha)  J_\mathrm{1}(\pi\rho'\theta) \\
    V^\mathrm{ring2}(u,v) &=& V^\mathrm{ring1} - (c_\mathrm{2} \cos2\alpha + s_\mathrm{2} \sin2\alpha)  J_\mathrm{2}(\pi\rho'\theta),
\end{eqnarray}
where $c_1$ and $s_1$ are the first order modulation coefficients, $c_2$ and $s_2$ are the second order modulation coefficients, $J_\mathrm{1}$ and $J_\mathrm{2}$ are the first and second order Bessel functions and $\alpha$ is the azimuthal angle of the ring, starting at the major-axis (see Fig.\,\ref{fig:modpedago}).

\begin{figure}
\centering
\includegraphics[width=8.5cm]{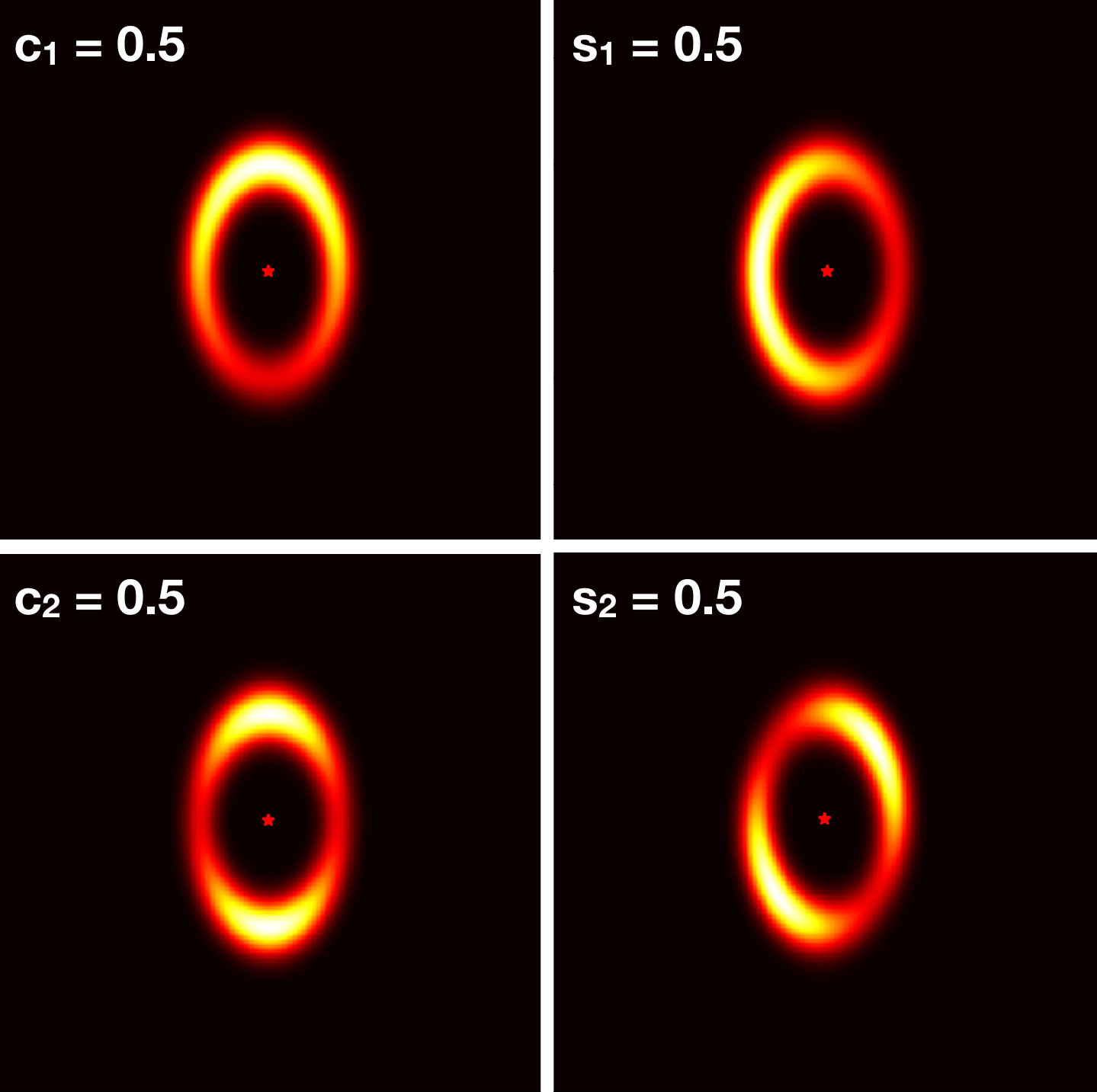}
\caption{Illustration of the four azimuthal modulation coefficients on the geometry of the ring. For each quadrant all modulation parameters are kept to zero apart from the one displayed.}
\label{fig:modpedago}
\end{figure}

Finally, to have a Gaussian ring, one needs to convolve the infinitesimal ring in the image space by a Gaussian, which is equivalent to a multiplication in the Fourier domain:
\begin{equation}
    V^\mathrm{ring}(u,v) = V^\mathrm{ring2}(u,v) \exp \frac{\pi \frac{\theta}{2} \delta\theta \sqrt{u^2 + v^2} }{4 \ln 2}, \label{eqn:ring}
\end{equation}
where $\delta\theta$ is the ratio \modiff{of the ring full width at half maximum to its radius}.

\item \textbf{The background:}
The extended flux is modeled by an over-resolved emission that has a null visibility. 

\item \textbf{The final model:}

As the flux ratios of the three components are normalized to 1 at 1.65$\mu$m (Eqn.\ref{eqn:norm}), $f^\mathrm{ring}_0$ is defined as:
\begin{equation}
    f^\mathrm{ring}_0 = 1 - f^\mathrm{prim}_0 - f^\mathrm{bg}_0.
\end{equation}
The final visibility can therefore be written as a linear combination of the three components of the model such as:
\begin{equation}
    V^\mathrm{tot}(u,v) = \frac{f^\mathrm{prim} V^*(u,v) + f^\mathrm{ring} V^\mathrm{ring}(u,v) }{ f^\mathrm{prim} + f^\mathrm{ring} + f^\mathrm{bg} }.
\end{equation}

\end{itemize}

\subsection{Binary: b\texttt{1-2}}
\label{sec:b}
This set of models is made of two stars and a background flux.
\begin{itemize}
\item \textbf{The binary:}
The two stars are defined as uniform disks (see Equation\,\ref{eqn:Vdisk}).
The position of the primary star is defined by the two position parameters ($x_0,y_0$) as in Eqn.\,\ref{eqn:Vstar}.
The visibility of the primary star is therefore defined as:
\begin{equation}
V^\mathrm{prim}(u,v) = 2\frac{J_1(\pi UD_\mathrm{prim} \sqrt{u^2+v^2})}{\pi UD_\mathrm{prim} \sqrt{u^2+v^2}} \exp-2 i \pi (x_0 u + y_0 v). 
\end{equation}
The secondary star is centered at position (0,0) and its visibility equation is identical to Eqn.\,\ref{eqn:Vdisk}:
\begin{equation}
V^\mathrm{sec}(u,v) = 2\frac{J_1(\pi UD_\mathrm{sec} \sqrt{u^2+v^2})}{\pi UD_\mathrm{sec} \sqrt{u^2+v^2}}. 
\end{equation}

\item \textbf{The final model:}
The normalization of the fluxes (Eqn.\,\ref{eqn:norm}) gives:
\begin{equation}
f^\mathrm{prim}_0  = 1 - f^\mathrm{sec}_0 - f^\mathrm{back}_0.
\end{equation}
In this set of models $f^\mathrm{prim}_0$ is therefore not fitted and is computed from the two other flux ratios.
The final visibility is therefore:
\begin{equation}
V^\mathrm{tot,b\#}(u,v) = \frac{f^\mathrm{prim} V^\mathrm{prim}(u,v) + f^\mathrm{sec} V^\mathrm{sec}(u,v) }{ f^\mathrm{prim} + f^\mathrm{sec} + f^\mathrm{back} }.
\end{equation}

\end{itemize}

\subsection{Binary and a ring: br\texttt{1-5}}
\label{sec:br}

The last set of models have four components: the primary, the secondary, a ring and a background.
\begin{itemize}
\item \textbf{The binary:}
The primary star is defined as a uniform disk which is shifted w.r.t. to the cent\modiff{e}r of the disk by ($x_0,y_0$) such as:
\begin{equation}
V^\mathrm{prim}(u,v) = 2\frac{J_1(\pi UD_\mathrm{prim} \sqrt{u^2+v^2})}{\pi UD_\mathrm{prim} \sqrt{u^2+v^2}} \exp-2 i \pi (x_0 u + y_0 v). \label{eqn:prim}
\end{equation}
The coordinates of the secondary ($x_\mathrm{sec},y_\mathrm{sec}$) is defined in reference to the coordinates of the primary such as the coordinates of the secondary are:
\begin{eqnarray}
x_\mathrm{sec} &=& -rM x_0 \\
y_\mathrm{sec} &=& -rM y_0,
\end{eqnarray}
where $rM$ is the mass ratio between the primary and the secondary.
The visibility of the secondary is:
\begin{equation}
V^\mathrm{sec}(u,v) = \exp-2 i \pi (x_\mathrm{sec} u + y_\mathrm{sec} v). \label{eqn:sec}
\end{equation}
The binary separation is limited to be lower or equal to one third of the ring diameter projected along the binary separation.

\item \textbf{The full model:}
The flux ratio normalization (Eqn.\,\ref{eqn:norm}) is as follows:
\begin{equation}
f^\mathrm{ring}_0 = 1 - f^\mathrm{prim}_0 - f^\mathrm{sec}_0 - f^\mathrm{back}_0.
\end{equation}
The ring-to-total flux ratio is therefore not fitted and is recovered from this equation.

The total visibility is therefore:
\begin{align}
V^\mathrm{tot,br\#}(u,v,\lambda) = &\frac{ f^\mathrm{prim}V^\mathrm{prim}(u,v) }
{ f^\mathrm{prim} + f^\mathrm{sec} + f^\mathrm{ring} + f^\mathrm{back}  }\\[8pt]
+&\frac{  f^\mathrm{sec} V^\mathrm{sec}(u,v) + f^\mathrm{ring} V^\mathrm{ring}(u,v) }
{ f^\mathrm{prim} + f^\mathrm{sec}+ f^\mathrm{ring} + f^\mathrm{back}  }\nonumber,
\end{align}

\end{itemize}
where $V^\mathrm{ring}(u,v)$, $V^\mathrm{prim}(u,v)$ and $V^\mathrm{sec}(u,v)$ are defined from Eqn.\,\ref{eqn:ring}, \ref{eqn:prim} and \ref{eqn:sec} respectively.

\subsection{Strategy}
\label{sec:strategy}

\begin{figure}
\centering
\includegraphics[width=8.5cm]{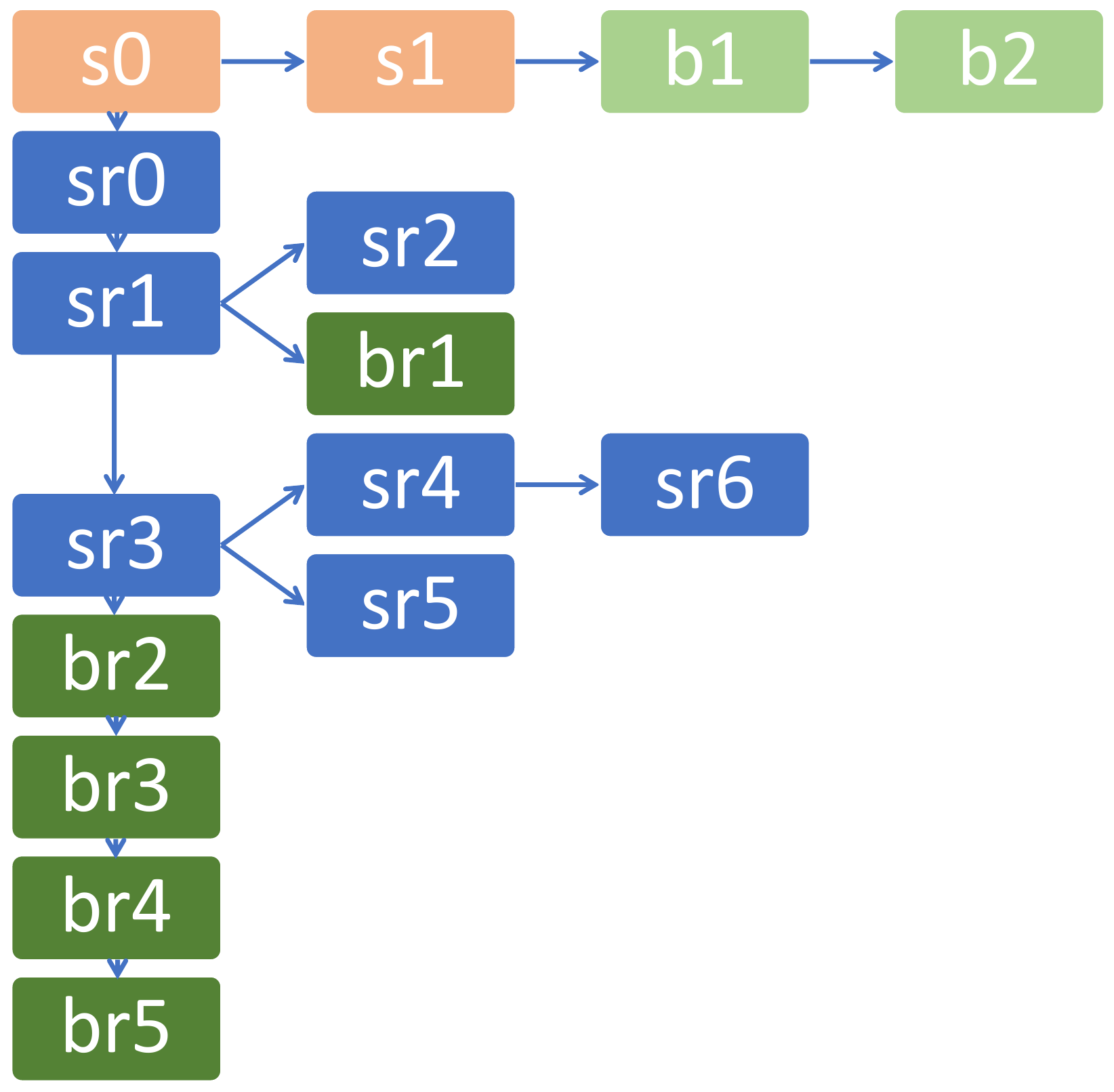}
\caption{The tree of models. The single star and background models are in orange, the single star and a ring models are in blue, the binary models are in light green and the binary and ring models are in dark green.}
\label{fig:tree}
\end{figure}

\begin{table*}
\caption{Matrix of parameters used in the different models. s indicates that the background and the secondary have the same spectral behavior \modiff{as} the primary as fitted from the SED.}
\label{tab:modvar}
\centering
\begin{tabular}{c c c c c c c c c c c c c c c c c c c c c c}
\hline
\hline
Model & \modif{n$_{par}$} & $f_{prim,0}$ & $f_{sec,0}$ & $f_{bg0}$ & $d_{sec}$ & $d_{bg}$ & $T_\mathrm{ring}$ & $\theta$ & $\delta \theta$ & $inc$ & $PA$ & $c_1$ & $s_1$ & $c_2$ & $s_2$ & $x_0$ & $y_0$ & $rM$ & UD$_\mathrm{prim}$ & UD$_\mathrm{sec}$ \\
\hline
s0 & \modif{2} & \checkmark & - & - & - & - & \checkmark & - & - & - & - & - & - & - & - & - & - & - & 0 & - \\
s1 & \modif{3} & \checkmark & - & - & - & \checkmark & - & - & - & - & - & - & - & - & - & - & - & - & \checkmark & - \\
\hline
sr0 & \modif{5} & \checkmark & - & \checkmark & - & $s$ & \checkmark & \checkmark & \checkmark & 0 & - & 0 & 0 & 0 & 0 & 0 & 0 & - & 0 & - \\
sr1 & \modif{7} & \checkmark & - & \checkmark & - & $s$ & \checkmark & \checkmark & \checkmark & \checkmark & \checkmark & 0 & 0 & 0 & 0 & 0 & 0 & - & 0 & - \\
sr2 & \modif{8} & \checkmark & - & \checkmark & - & \checkmark & \checkmark & \checkmark & \checkmark & \checkmark & \checkmark & 0 & 0 & 0 & 0 & 0 & 0 & - & 0 & - \\
sr3 & \modif{9} & \checkmark & - & \checkmark & - & s & \checkmark & \checkmark & \checkmark & \checkmark & \checkmark & \checkmark & \checkmark & 0 & 0 & 0 & 0 & - & 0 & - \\
sr4 & \modif{11} & \checkmark & - & \checkmark & - & s & \checkmark & \checkmark & \checkmark & \checkmark & \checkmark & \checkmark & \checkmark & 0 & 0 & \checkmark & \checkmark & - & 0 & - \\
sr5 & \modif{11} & \checkmark & - & \checkmark & - & s & \checkmark & \checkmark & \checkmark & \checkmark & \checkmark & \checkmark & \checkmark & \checkmark & \checkmark & 0 & 0 & - & 0 & - \\
sr6 & \modif{13} & \checkmark & - & \checkmark & - & s & \checkmark & \checkmark & \checkmark & \checkmark & \checkmark & \checkmark & \checkmark & \checkmark & \checkmark & \checkmark & \checkmark & - & 0 & - \\
\hline
b1 & \modif{7} & - & \checkmark & \checkmark & \checkmark & \checkmark & - & - & - & - & - & - & - & - & - & \checkmark & \checkmark & - & \checkmark & 0 \\
b2 & \modif{8} & - & \checkmark & \checkmark & \checkmark & \checkmark & - & - & - & - & - & - & - & - & - & \checkmark & \checkmark & - & \checkmark & \checkmark \\
\hline
br1 & \modif{11} & \checkmark & \checkmark & \checkmark & $s$ & $s$ & \checkmark & \checkmark & \checkmark & \checkmark & \checkmark & 0 & 0 & 0 & 0 & \checkmark & \checkmark & \checkmark & 0 & 0 \\
br2 & \modif{13} & \checkmark & \checkmark & \checkmark & $s$ & $s$ & \checkmark & \checkmark & \checkmark & \checkmark & \checkmark & \checkmark & \checkmark & 0 & 0 & \checkmark & \checkmark & \checkmark & 0 & 0 \\
br3 & \modif{15} & \checkmark & \checkmark & \checkmark & $s$ & $s$ & \checkmark & \checkmark & \checkmark & \checkmark & \checkmark & \checkmark & \checkmark & \checkmark & \checkmark & \checkmark & \checkmark & \checkmark & 0 & 0 \\
br4 & \modif{16} & \checkmark & \checkmark & \checkmark & $s$ & $s$ & \checkmark & \checkmark & \checkmark & \checkmark & \checkmark & \checkmark & \checkmark & \checkmark & \checkmark & \checkmark & \checkmark & \checkmark & \checkmark & 0 \\
br5 & \modif{17} & \checkmark & \checkmark & \checkmark & \checkmark & $s$ & \checkmark & \checkmark & \checkmark & \checkmark & \checkmark & \checkmark & \checkmark & \checkmark & \checkmark & \checkmark & \checkmark & \checkmark & \checkmark & 0 \\
\hline
\end{tabular}
\end{table*}

\begin{table}
\caption{Table of allowed parameter ranges}
\label{tab:paramranges}
\centering
\begin{tabular}{c c c c }
\hline
\hline
\modif{Parameter} & \modif{unit} & \modif{minimal value} & \modif{maximal value} \\
 $f_{prim,0}$ & - &  0 & 1 \\
 $f_{sec,0}$ & - & 0 & 1\\
 $f_{bg0}$ & - & 0 & 1 \\ 
 $d_{sec}$ & - & 0 & 1 \\ 
 $d_{bg}$ & - & 0 & 1 \\ 
 $T_\mathrm{ring}$ & [K] & 500 & 10000 \\ 
 $\theta$ & [mas] & 0.01 & 500 \\ 
 $\delta \theta$ & - & 0 & 20 \\ 
 $inc$ & [$^\circ$] & 0 & 90 \\ 
 $PA$ & [$^\circ$] & 0 & 360 \\ 
 $c_1$ & - & -1 & 1 \\ 
 $s_1$ & - & -1 & 1 \\ 
 $c_2$ & - & -1 & 1 \\ 
 $s_2$ & - & -1 & 1 \\ 
 $x_0$ & [mas] & -30 & 30 \\ 
 $y_0$ & [mas] & -30 & 30 \\ 
 $rM$ & - & 0 & 20 \\ 
 UD$_\mathrm{prim}$ & [mas] & 0.01 & $+\infty$ \\ 
 UD$_\mathrm{sec}$ & [mas] & 0.01 & $+\infty$ \\
\hline
\end{tabular}
\end{table}

\modif{Our models are fitted to both} V$^2$ and CP.
\modif{However, there is a large variety of data in our sample as s}ome targets have a signal with low complexity, such as \object{AC Her} (\#1), but others have a high complexity, such as \object{U\,Mon} (\#22).
Recent studies of IRAS\,08544-4431 (\#10) showed that model \texttt{br5} with 17 parameters is needed to reproduce the interferometric dataset \citep{Hillen2016}.
We therefore fit a tree of models, starting with our simplest model (\texttt{s0}, see Sect.\,\ref{sec:s}), up to the most complex model (\texttt{br5}, see Sect.\,\ref{sec:br}), inspired by \citet{Hillen2016} with 17 parameters.
We start by fitting the simplest model to all the targets. 
We then fit models with increasing complexity using the best parameters from the previous model as the starting point for the next model (adding the new parameters).
Some targets are best fitted by models that do not include a ring but just a binary.
We therefore included several forks in the model tree (see Fig.\,\ref{fig:tree}).

For each model a first fit is performed using a genetic algorithm implemented with \texttt{DEAP} \citep[][]{DEAP}.
The initial population is defined with a random flat distribution over the initial parameters.
For the parameters that are not used in the previous model the distribution spans over all the \modif{allowed} values of a parameter \modif{that we defined in Table\,\ref{tab:modvar}}.
For the parameters already used in the previous model in the tree, we span over 30\% of its best-fit value for the previous model.
The mutation is Gaussian over the parameter allowed distribution and has a probability of 5\%.
The individual selection is made through a three-rounds tournament.
Then, using the best fit from the genetic algorithm as a starting point, a MCMC minimization is performed with the \texttt{emcee} package \citep[][]{MCMC} to determine the error bars on the parameters.
When a model has a larger $\chi^2$ than the previous model in the model tree we redo the minimization as this indicates that a local minimum was reached.

As it is not possible to directly compare two models with a different number of parameters, 
\modif{we wanted to use a criterion that is applicable to all our sample.}
\modif{Such criteria exists from information theory and}
we used the Bayesian inf\modif{ormation} criterion \citep[BIC;][]{BIC} to select the best model for each target.
\modif{This criteria aims to infer the model that fits the data the best without over fitting and is used in several studies of stellar physics to infer the most likely model that reproduce the data from a set of different models \citep[e.g.,][]{Degroote2009,Aerts2018,Matra2019}. 
The BIC criteria was developped for model inference and is more conservative in the choice of the most likely model (higher penalty for models with more parameters to fit) than the Akaike information criterion (AIC) that was developed for prediction purposes.
The BIC can be written as follows:}
\begin{equation}
\mathrm{BIC} = -2\mathcal{\hat{L}} + n_\mathrm{par} \log{n_\mathrm{data}},
\end{equation}
\modif{where $\mathcal{\hat{L}}$ is the maximum likelihood values found for the optimal model parameters, $n_\mathrm{par}$ is the number of optimized parameters and $n_\mathrm{data}$ is the number of data points.}

\modif{Under the assumptions we make, i.e. data points are independent and the error distribution is Gaussian, the BIC can be written as:}
\begin{equation}
\mathrm{BIC} = \chi^2 + n_\mathrm{par} \log{n_\mathrm{data} },
\end{equation}
\modif{where} $\chi^2 = \sum_{i=1}^{n_\mathrm{data}} \Big(\frac{y_\mathrm{i}-m_\mathrm{i}}{\sigma_\mathrm{i}}\Big)^2$ \modif{with} $y_\mathrm{i}$ \modif{is $i$-th data point}, $m_\mathrm{i}$ \modif{the $i$-th model point and} $\sigma_\mathrm{i}$ \modif{the error bar of th $i$-th data point.}
The values of the BIC \modif{and }$\chi^2$ for each target are displayed on Figs.\,\ref{fig:BIC1} and \ref{fig:BIC2}.
\modif{To compare the models, we have selected the model with the lowest BIC. 
However, the differences between the BIC values for a given dataset can be small and several models can be considered. 
Usually, models with a difference of more than 10 with the BIC of the best model can be ruled out whereas a difference between 10 and 6 can be considered as moderately strong evidence for the best model, between 6 and 2 as positive evidence in favour of the best model and less than 2 as weak evidence \citep[e.g.,][]{Aerts2018}.}

\section{Results}
\label{sec:results}

In this section we present a first analysis of the fit results by discussing the important and most reliable parameters related to the ring morphology such as the near-infrared sizes, the temperatures or the fluxes of the different components.
We focus on the circumbinary environments in our comparison with YSOs.

\subsection{Results of the model selection}

\begin{figure}
\centering
\includegraphics[width=8.5cm]{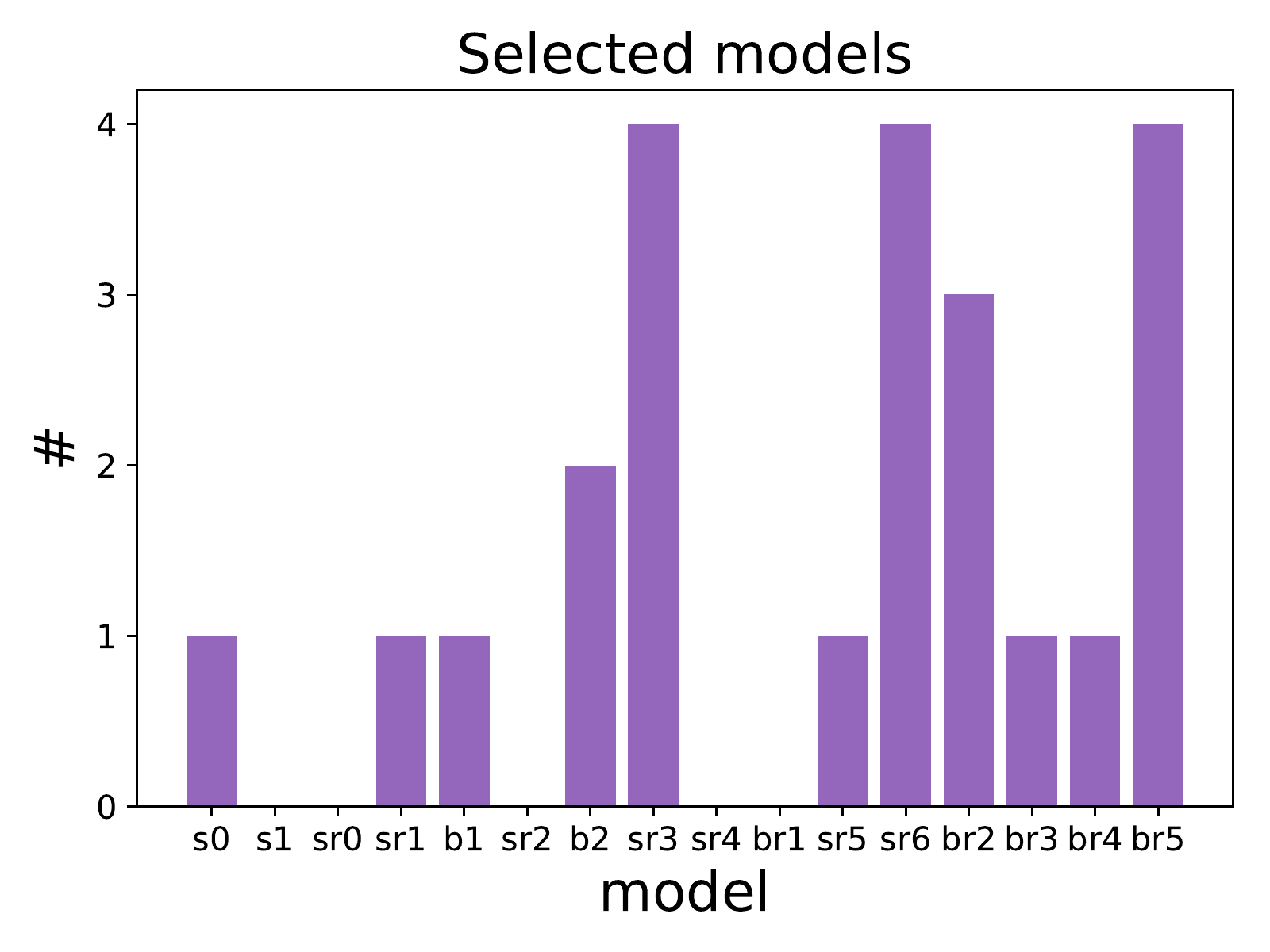}
\includegraphics[width=8.5cm]{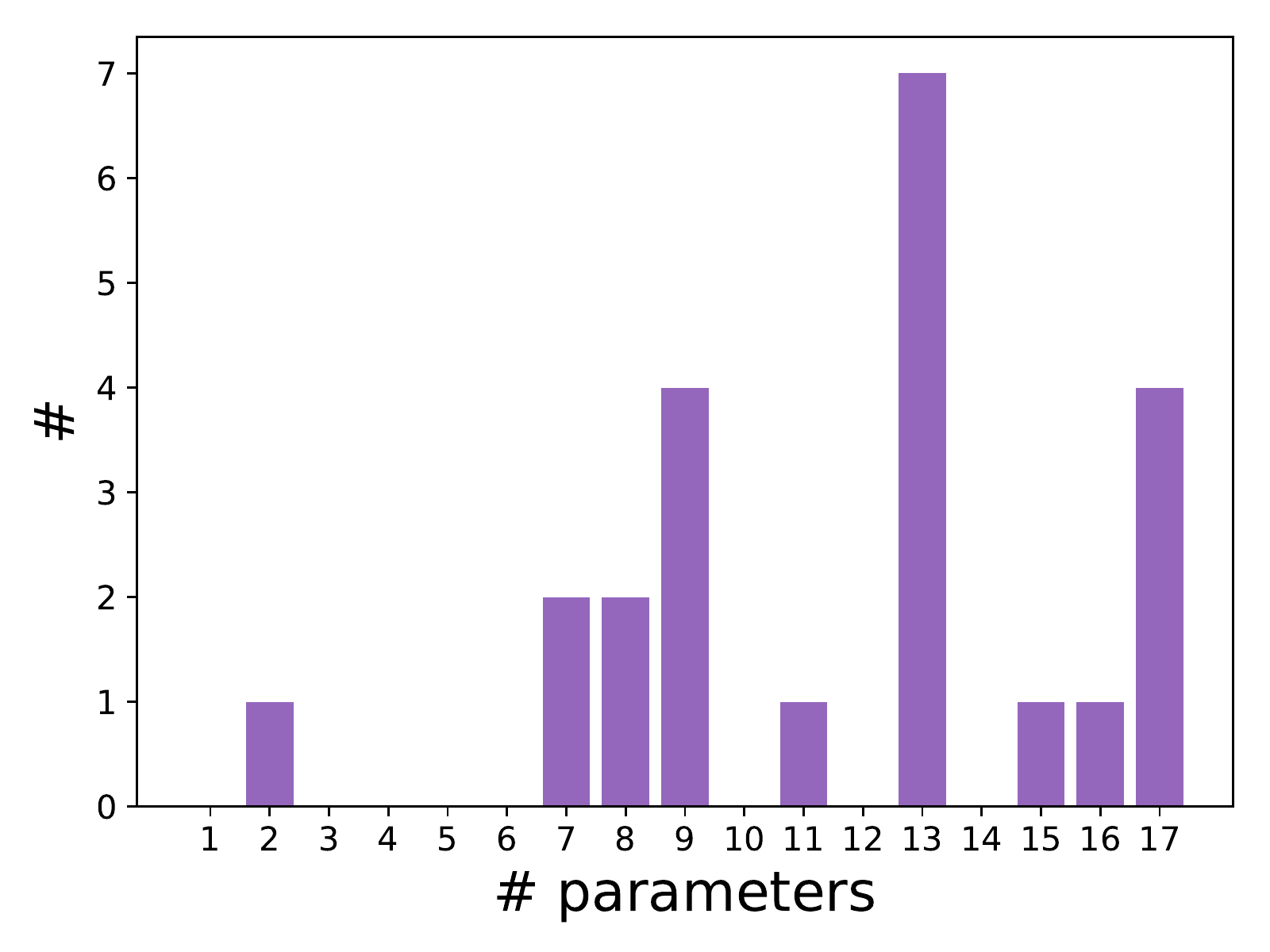}
\caption{Statistics of model selection. Top: number of targets per model. Bottom: number of targets per number of parameter of the selected model (complexity).}
\label{fig:models}
\end{figure}

\modif{
For HD\,93662, \object{IRAS\,19125+0343}, IW\,Car and R\,Sct one model has very strong evidence over all the other models ($\Delta$BIC>10 for any other model).
For all other targets, the most likely models are presented on Tables\,\ref{tab:resACHer} to \ref{tab:resV494Vel}. 
 Five targets have the most likely model to have at least strong evidence (6<$\Delta$BIC<10) over other models (AI\,Sco, HD\,108015, HD\,213985, \object{IRAS\,05208-2035}, \object{IRAS\,17038-4815}).
Most of the likely models have similar parameters (i.e. within the error bars).
In the rest of our analysis we use the most likely model (i.e. the model with the lowest BIC value).}

The best-fit parameters are presented in Table\,\ref{tab:res1} and \ref{tab:res2}.
\modif{As an estimation of the fit quality, the} \modif{reduced $\chi^2$ (} $\chi^2_\mathrm{red}$ \modif{=} $\chi^2/n_\mathrm{data}$\modif{)} value\modif{s} ranges from 0.5 to 6.6.
The number of targets per model is displayed in the top panel of Fig.\,\ref{fig:models}.
Out of the sixteen models \modif{eleven} were selected.
For one target the simplest model was preferred (AC Her, \#1).
For three targets a binary model was preferred with no ring (HD93662 (\#4), PS Gem (\#18) and RU Cen (\#20)).
For \modif{ten} targets models of a single star with a ring was selected while for ten targets a model with a binary surrounded by a ring was preferred.
For \modif{twelve} targets models with a binary were preferred and for nineteen targets models with a ring were preferred (their images are displayed in Figs.\,\ref{fig:modImages1} and \ref{fig:modImages2}).

On the bottom panel of Fig.\,\ref{fig:models} we can see how many models there are per number of parameters.
Models with six or less parameters are not fitting the data well enough (apart for AC Her, \#1).
Thirteen targets ($\sim$55\%) were fitted by models with thirteen parameters or more.
This shows the complexity of the resolved structures. 

\modif{The case of AC\,Her (\#1) is interesting as it was previously observed in mid-infrared with the MIDI instrument at the VLTI \citep[][]{Hillen2015}. A disk with an inner rim at 68\,au was fitted to the data, which corresponds to 42\,mas. A rim of this size would be over-resolved by our observations, which is compatible with the most likely model: a star + an over-resolved flux. For the target with the richest dataset, IRAS\,08544-4431 (\#10), the most likely model corresponds to the model which was fitted in \citet[][]{Hillen2016}. The best-fit parameters are also very similar except the ring temperature $T_\mathrm{ring}$. This is due to the way the photosphereric spectrum is represented as in \citet[][]{Hillen2016} it is assumed to be in the Rayleigh-Jeans regime ($F_\lambda \propto \lambda^{-4}$) and here we use the fit to the photometry.}

\subsection{Inner rim radius ruled by dust sublimation physics}
\label{sec:sizelum}

\begin{figure}
\centering
\includegraphics[width=8.5cm]{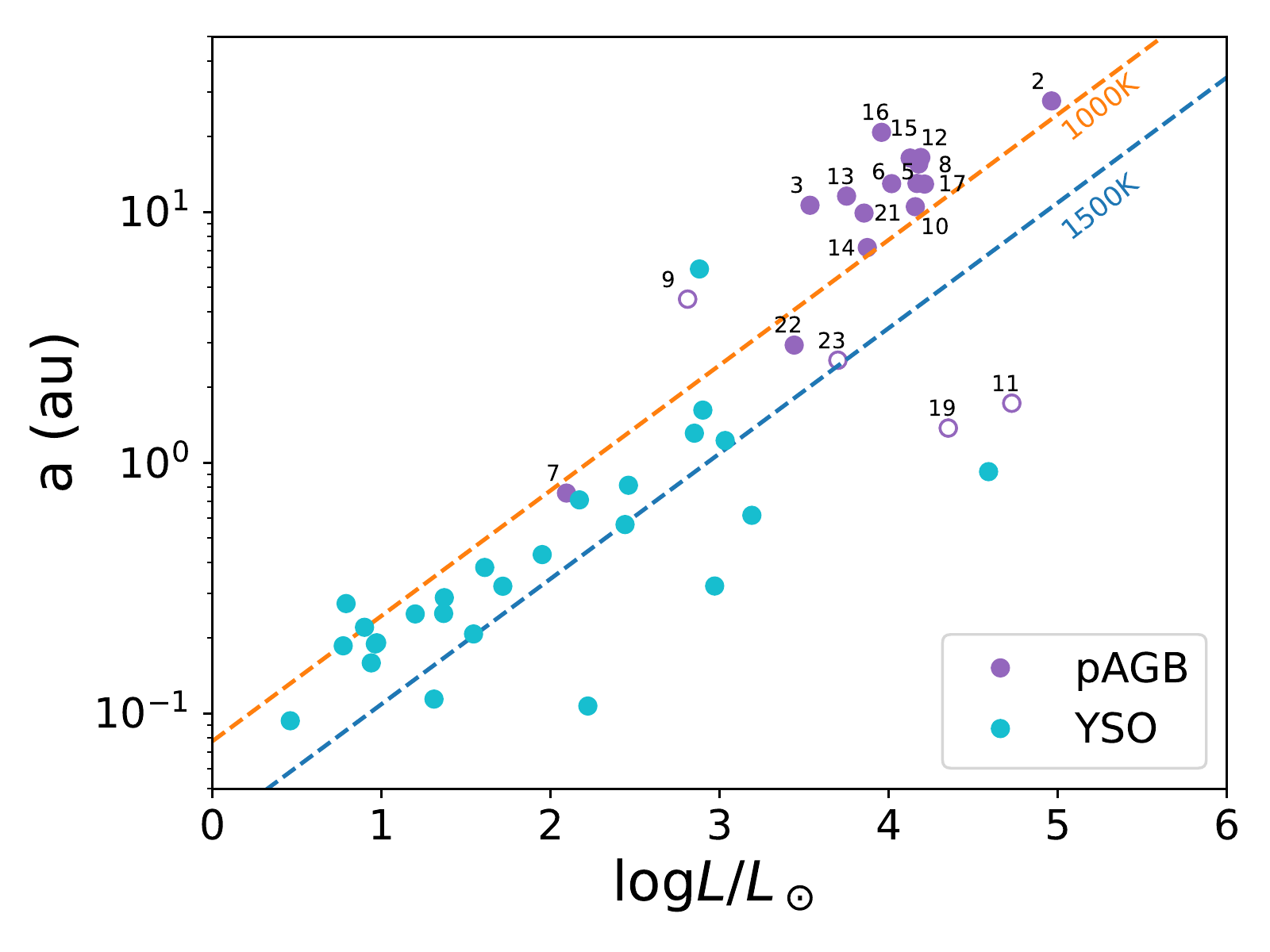}
\caption{Size-luminosity relation. The sizes are in log-scale. Purple points are pAGB binaries from this work. Purple open circles are pAGB binaries with $T_\mathrm{ring}$>7000\,K. Light blue points are Herbig Ae/Be stars from \citet{Lazareff2017}.  The blue and orange resp. dashed lines are the theoretical sublimation radius for $T_\mathrm{sub}=1500$K and $T_\mathrm{sub}=1000$K resp.}
\label{fig:sizelum1}
\end{figure}

For protoplanetary disks the size of the near-infrared extended emission (\modif{the physical radius in au:} $a$) correlates with the square root of the stellar luminosity \citep[$L_\mathrm{bol}$;][]{Monnier2002,Lazareff2017} such as:
\begin{equation}
    a = \frac{1}{2} (C_\mathrm{bw}/\epsilon)^{1/2} (L_\mathrm{bol}/4\pi\sigma T^4_\mathrm{sub})^{1/2},
\end{equation}
where $T_\mathrm{sub}$ is the sublimation temperature, $C_\mathrm{bw}$ is the backwarming coefficient \citep[][]{kama2009}, $\epsilon$\,=\,$Q_\mathrm{abs}(T_\mathrm{sub})/Q_\mathrm{abs}(T_*)$ is the dust grain cooling efficiency which is the ratio of Planck-averaged absorption cross-sections at the dust sublimation and stellar temperatures and $\sigma$ the Stefan-Boltzmann constant.
To study the size of the inner rim we use here and in the rest of the paper the half of the fitted ring diameter ($\theta$) as it was done in studies of disks around YSOs \citep[e.g.,][]{Monnier2002,Monnier2005,Lazareff2017}. The rings having a given width ($\delta\theta$) it is possible that the inner disk edge, usually defined by the location where the optical depth $\tau$ equals unity \citep[e.g.,][]{kama2009}, can be closer than the ring radius.
On Fig.\,\ref{fig:sizelum1} we plot the sizes of the ring models versus the central luminosity for pAGB binaries of our sample with, for reference, lines indicating theoretical sublimation radii for $T_\mathrm{sub}=1000$\,K and 1500\,K with $C_\mathrm{bw}=1$ and $\epsilon=1$.
In order to compute the luminosities and the physical sizes of our targets we have used the Gaia parallaxes \citep{BJ2018}.
However, as our targets are binaries with a semi-amplitude that can be of the order of the parallax, those distances are likely biased \modif{by the orbital movement of the binary}. 
Luckily, however, this does not impact the size luminosity diagram as both the physical size and the square root of the luminosity scale linearly with distance. 
An error on the distance will therefore displace a point along the size-luminosity relation.

\begin{figure}
\centering
\includegraphics[width=8.5cm]{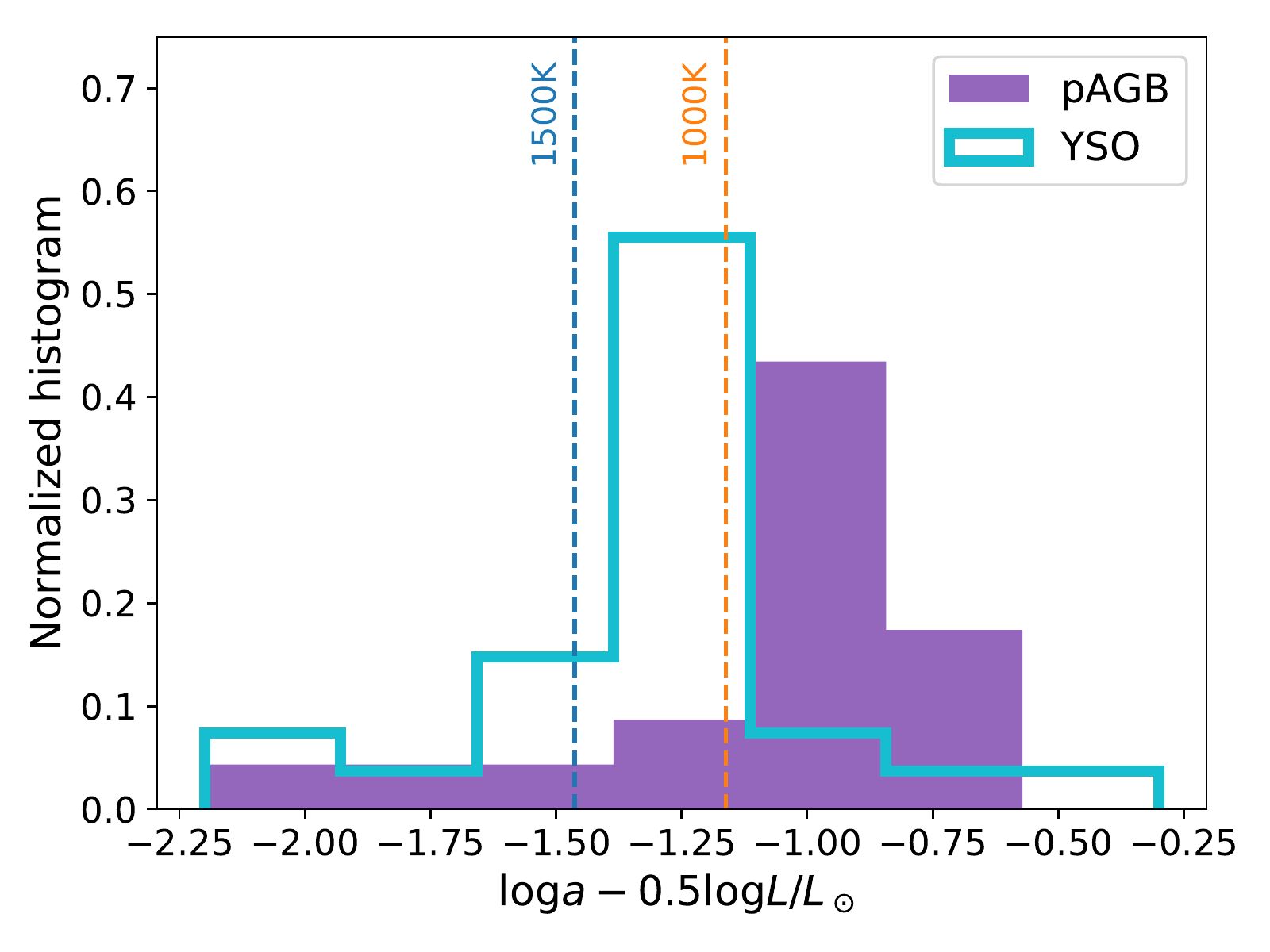}
\caption{Histogram of sizes scaled by the square root of luminosity for both pAGBs and YSOs.}
\label{fig:sizelum2}
\end{figure}

Sizes of near-infrared emission around pAGB binaries seem to scale with the stellar luminosity as it is the case for young stellar objects.
\modif{However, the sizes of pAGB circumstellar emissions are systemically always offset toward sizes larger than for circumstellar emission around YSOs.
This can be deduced more clearly from the histogram on Fig.\,\ref{fig:sizelum2}.}

There are three outliers (\object{IRAS\,10174-5704} (\#11), R Sct (\#19) and V494Vel (\#23)) with very small sizes compared to their luminosity.
Those three stars have also a large temperature for their environment (higher than 7000\,K) meaning that the traced circumstellar environment is not thermal emission from dust.

\subsection{Dust temperature}
\label{sec:T}

\begin{figure}
\centering
\includegraphics[width=8.5cm]{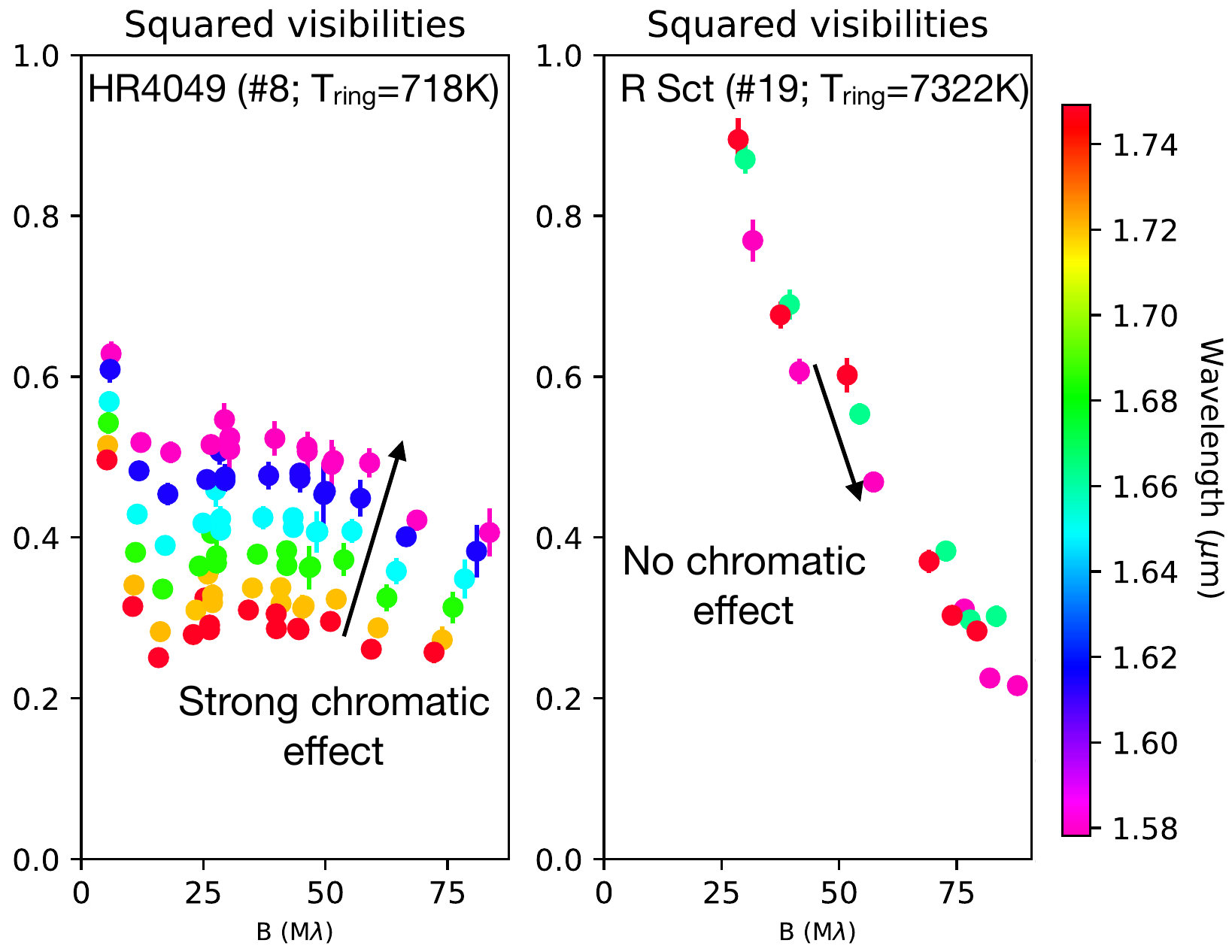}
\caption{Illustration of the chromatic effect (see the main text). Left: squared visibilities of HR4049 (\#8) showing a strong chromatic effect. Right: squared visibilities for R Sct (\#19) not showing such an effect.}
\label{fig:chrom}
\end{figure}

The spectral channels of PIONIER allow us to probe the difference in spectral index between the central star and its environment.
This is done by fitting the difference of level of the squared visibilities between the different channels.
For a target in which such a difference is present, the fraction of the total flux that is resolved by the interferometer at any given baseline depends on the wavelength, hence the observed squared visibilities as well.
This is called the chromatic effect.
When there is a large difference in temperature between a central unresolved source and its resolved environment, the squared visibility increases with s\modif{hort}er wavelength for a given baseline (see Fig.\,\ref{fig:chrom}).

We assumed the central star to have a given spectrum that we fit from photometry.
The ring models assume a black-body emission for the ring.
As we fix the primary spectrum to be what we fit from the photometry the difference in the levels of the squared visibilities per channel will be reproduced by a given ring temperature.
This temperature may not be the exact temperature at that location due to optical depth effects, but it gives a good first indication.

\begin{figure}
\centering
\includegraphics[width=8.5cm]{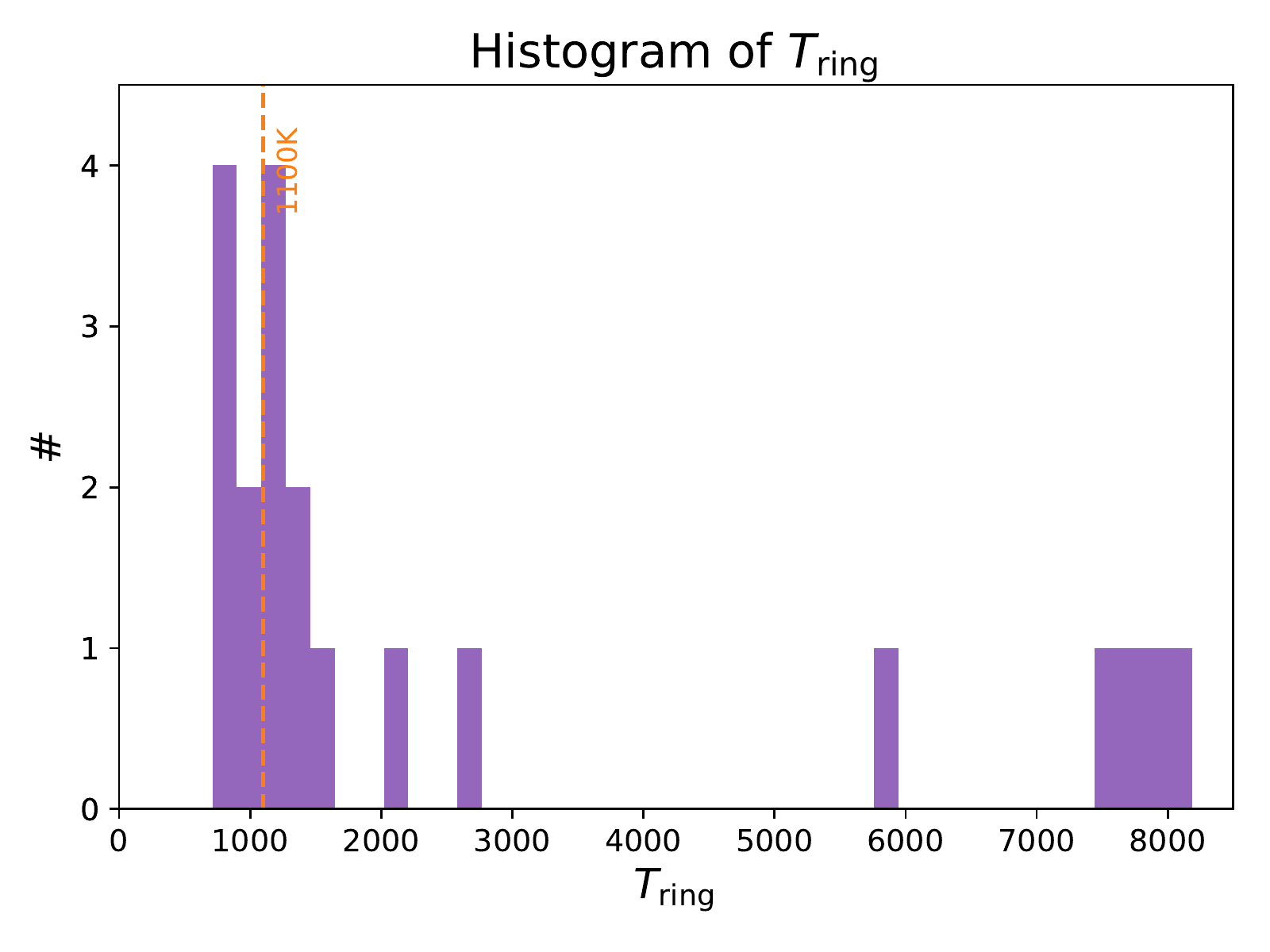}
\caption{Histogram of ring temperatures for our targets.}
\label{fig:ringT}
\end{figure}

In Fig.\,\ref{fig:ringT} we show the histogram of temperatures we found for our models.
On the one hand, two-third of the targets (13/19) have a low circumstellar emission temperature ($T_\mathrm{ring}<1600\,K$), equal or lower than the classical silicate sublimation temperatures we expect ($\sim$1500\,K) indicating that the thermal emission of the inner rim of the disk dominates.
The slightly lower temperatures agree with the shift toward larger sizes we see in the size-luminosity diagram for the pAGB with respect to YSOs.
On the other hand, four targets have a circumstellar emission with a temperature of more than 7000\,K: IRAS\,05208-2035 (\#9), IRAS\,10174-5704 (\#11), R\,Sct (\#19) and V\,494\,Vel (\#23).
Three of them are also outliers in the size-luminosity diagram (see Sect.\,\ref{sec:sizelum}) pointing toward another origin of the circumstellar flux.
For \object{IRAS\,05208-2035} (\#9), the high-stellar-to-total flux ratio (90.5$\pm$0.3\%) makes this target special.
The ring flux is perhaps stellar scattered light coming from the inner rim of the disk.
Finally, between these two categories of temperatures, two targets have a temperature around 3000-4000\,K (IRAS\,17038-4815 (\#13) and U\,Mon (\#22)). 
Interestingly, those targets are expected to have pulsation with the largest amplitude ($\Delta$mag=1.5 and 1.1 for IRAS\,17038-4815 (\#13) and U\,Mon (\#22) resp.) among our sample.
However, our data is too sparse to look for morphological changes induced by pulsations in these targets within the observations that span a large part of the pulsation cycle. 
Our models are able to reproduce the data reasonable without including any intrinsic variations.

\subsection{Disk inclinations}

\begin{figure}
\centering
\includegraphics[width=8.5cm]{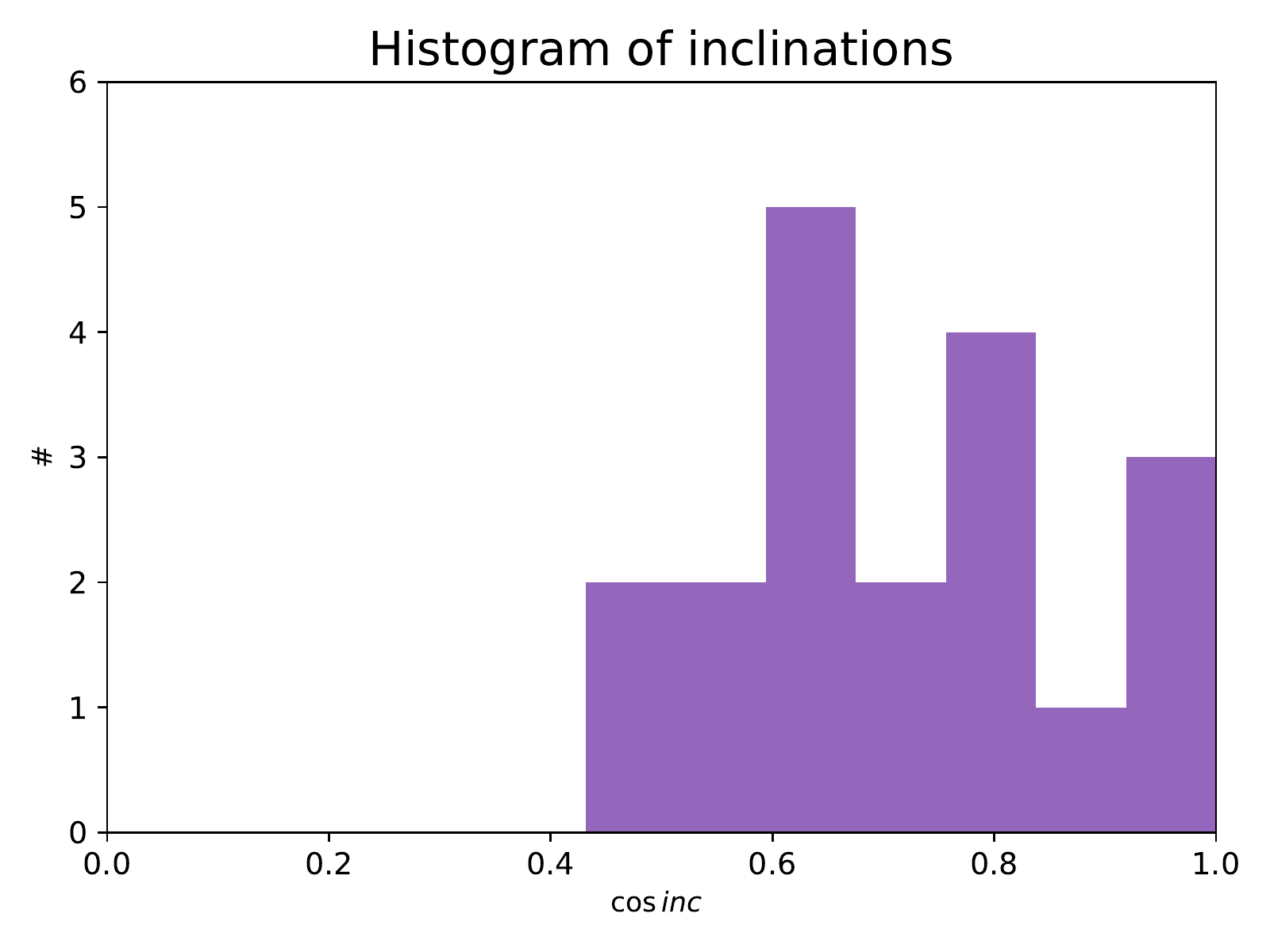}
\caption{Histogram of the cosines of fitted inclinations. 
}
\label{fig:inc}
\end{figure}

We can investigate the morphology of the circumbinary environment by looking at the ring inclinations.
As our targets are surrounded by disks, the distribution of the cosines of inclinations should be flat.
However, for spherical shells for instance, there would be a pile-up of objects at $\cos inc \sim 1$.

Fig.\,\ref{fig:inc} displays the histogram of the cosines of our ring inclinations.
There is clearly no pile-up at $\cos inc =1$.
We see a rather flat distribution with a cut-off at about $\cos inc \sim 0.5$ that corresponds to a inclination of $\sim60^\circ$.
This is likely due to observational bias as the disks that are edge-on will absorb the visible light from the central stars.
The cut-off of disk inclinations could therefore be a proxy to the characteristic thickness of the disk.
The cut-off we see would translate to a disk thickness of $h/r\sim$0.8.
However, there could be a model bias as at very high inclinations the model might not be able to reproduce the intensity distribution correctly.

\subsection{Disk width}

\begin{figure}
\centering
\includegraphics[width=8.5cm]{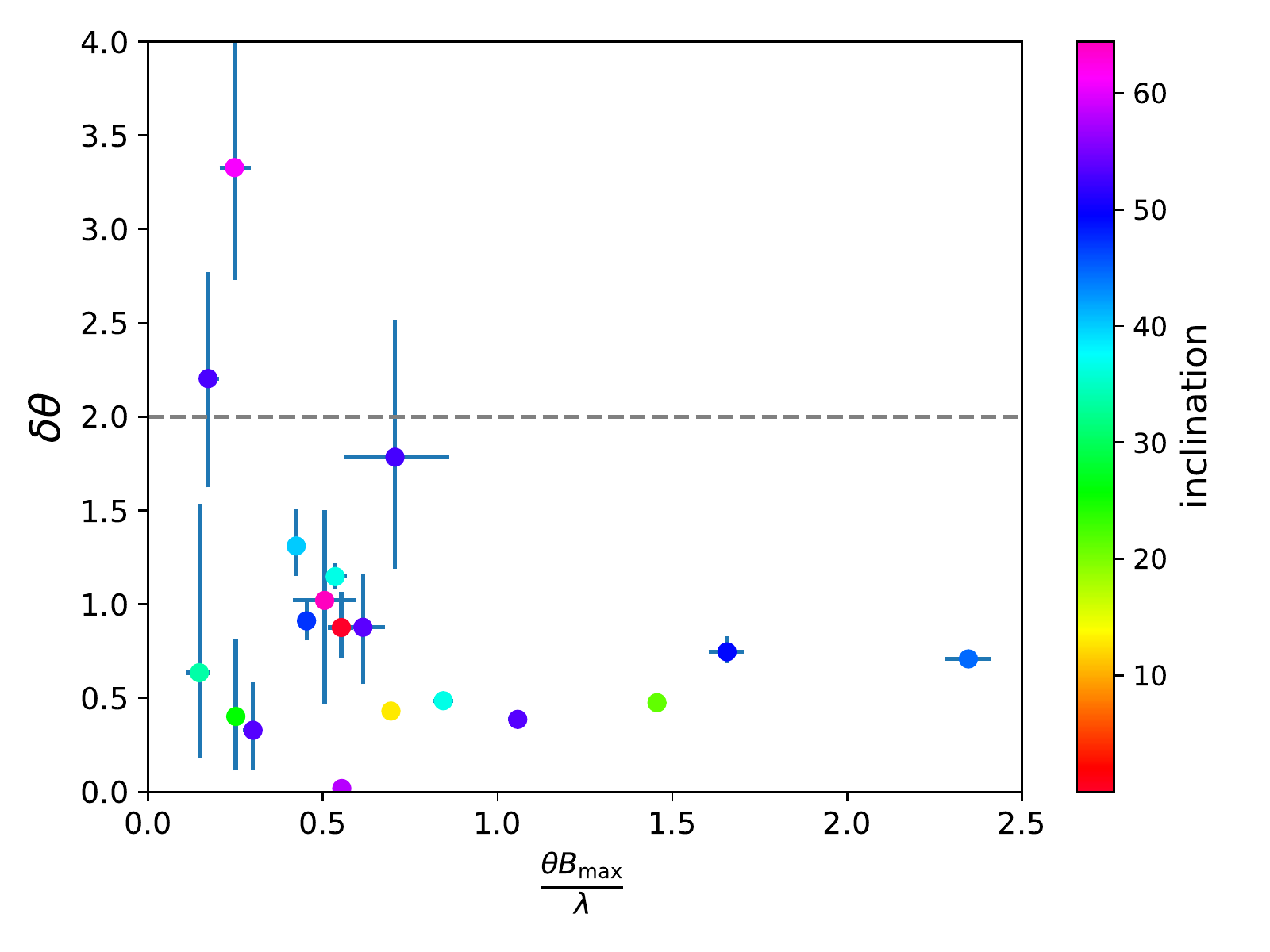}
\caption{Relative ring width (\modiff{$\delta\theta$}) against the degree of resolution of the ring. The colors indicate the ring inclination. The horizontal line indicates where the ring has no inner cavity ($\delta\theta\geq$\modiff{2}). 
}
\label{fig:ringwidths}
\end{figure}

We are sensitive to the width of the emission coming from the disk.
The parameter $\delta\theta$ measures the width of the ring in the units of the ring \modiff{radius}.
On Fig.\,\ref{fig:ringwidths} we see this ring width parameter plotted against the size of the ring divided by the angular resolution of the largest baseline.
We see that as long as the ring is resolved by the observations ($\frac{\theta B_\mathrm{max}}{\lambda}\geq0.5$) its width is better constrained and is below unity.
It means that the circumbinary dust emission is compatible with a ring and not with an emission without a cavity \modiff{($\delta\theta\geq2$)} and that the ring has a significant radial width ($\delta\theta$ between 0.5 and 1).

\subsection{Rim brightness distribution}
\begin{figure}
\centering
\includegraphics[width=8.5cm]{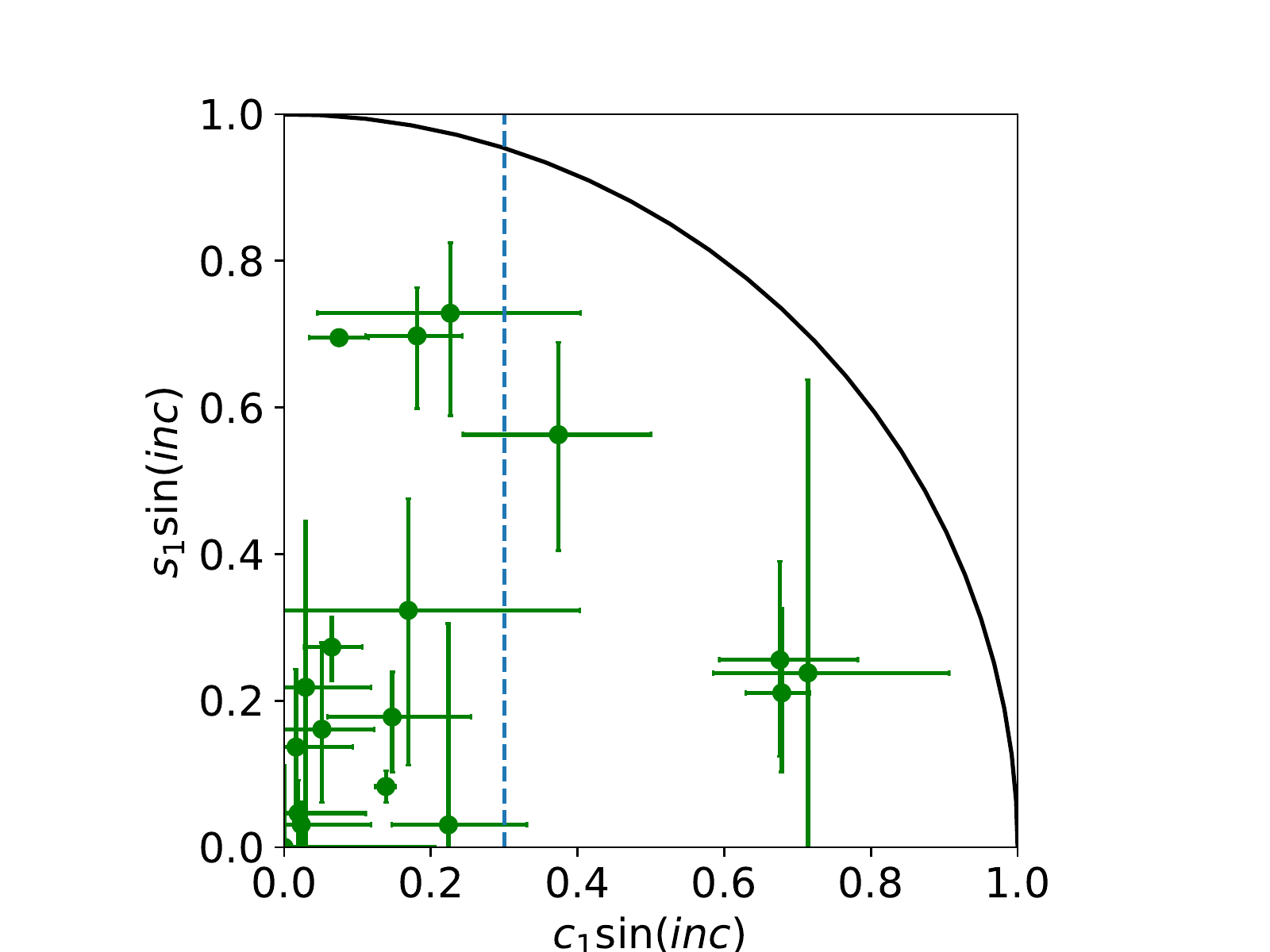}
\caption{Direction of the maximum of the first order modulation. The dashed line represent the limit for inclination-like modulations \citep[see text;][]{Lazareff2017}. The circle represent the maximum values for the modulation coefficients $c_1$ and $s_1$. 
}
\label{fig:m1}
\end{figure}

All the models with a ring require at least a first order modulation.
For a modulation due to the inclination of an optically and geometrically thick inner rim we will have a larger illumination in the direction of the minor axis \citep[e.g.,][]{Isella2005}.
If this is the case most of the targets would have a larger absolute value of the $s_1$ coefficient and an almost zero value for $c_1$.
Fig.\,\ref{fig:m1} shows that this is not the case.
The values for $c_1$ and $s_1$ are scattered.
Inclination effects would produce more points with $s_1\sin{inc}$ between 0 and 1 and $c_1\sin{inc}$ between 0 and 0.3 \citep{Lazareff2017}.
There is no clear evidence that the inclination is the main cause of the observed modulation.

Half of the targets prefer the second order modulation, also indicating that the inner rims are not ruled only by inclination effects but also by interactions with the inner binary and/or disk instabilities.

\subsection{Extended flux statistics}

\begin{figure}
\centering
\includegraphics[width=8.5cm]{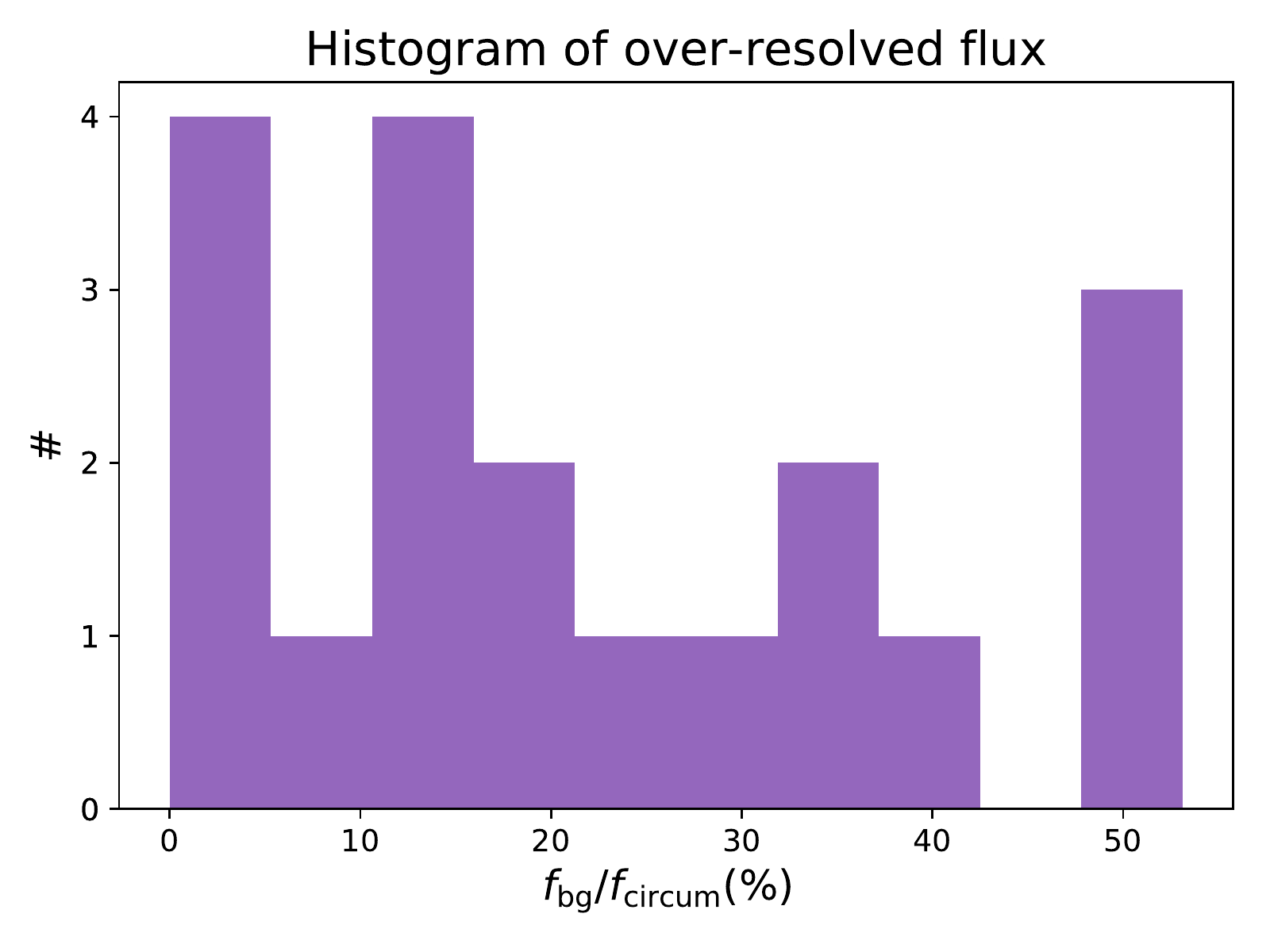}
\caption{Histogram of the fractions of the non-stellar flux that is over-resolved.
}
\label{fig:extended}
\end{figure}

In IRAS\,08544-4431 (\#10), our best studied object, $\sim$15\% of the $H$-band flux is coming from an over-resolved emission \citep{Hillen2016}.
Only half of this flux was accounted for by the radiative transfer model including a disk in hydrostatic equilibrium and scattered light \citep{Kluska2018}.
The origin of this extended flux is not clear so far.
This extended flux is easy to detect as it can be measured as the drop of visibility at short baselines.
Fig.\,\ref{fig:extended} shows the ratio of the over-resolved \modif{(f$_{bg}$)} over the total non-stellar flux \modif{(f}$_\mathrm{circum}$ \modif{= 1-f}$_\mathrm{prim}$\modif{-f}$_\mathrm{sec}$\modif{)}.
IRAS\,08544-4431 (\#10) has $\sim$42\% of its non-stellar flux to be over-resolved.
14 out of 19 targets ($\sim$75\%) have this ratio larger than 10\% and three of them have it around 50\%.







\section{Discussion}
\label{sec:discussion}

In this section we will discuss the results and interpret them in an extended context.
We will first discuss the outliers in Sect.\,\ref{sec:outliers}. 
We then discuss the shift between pAGB and YSOs in the size luminosity relation (Sect.\,\ref{sec:shift}).
Then we will discuss the relation between disk inclination and the RVb phenomenon (variable extinction/scattering in the line of sight during orbital motion; Sect.\,\ref{sec:RVb}).
We will discuss the radial structure of the disk by comparing the size of the emission in the near-infrared with the one in the mid-infrared (Sect\,\ref{sec:NIRMIR}).
Finally, in Sect.\,\ref{sec:YSOcomp}, we will discuss the differences and similarities with the disks around YSOs.

\subsection{Outliers}
\label{sec:outliers}

Four targets from our sample are outliers in the fact that the temperature of their environment is very high ($T_\mathrm{ring}$>7000\,K), significantly above any dust sublimation temperature.
Here we summarize results published in literature and give an interpretation of the origin of their pecularities.




\subsubsection{IRAS\,05208-2035 (\#9)}
This source is an outlier in \citet{deRuyter2006}: its infrared \modif{excess} starts at longer wavelengths (after $L$-band).

This visibilities show a reversed chromatic effect i.e. the squared visibility decreases with decreasing wavelengths.
It happens when the environment is bluer than the emission from the central source
Our models are not able to reproduce this effect. 
This effect can be produced by stellar light \modif{scattered on the disk surface}.

\subsubsection{IRAS\,10174-5704 (\#11)}

This source has a high temperature for its environment and a small radius.
It was included in the mid-infrared interferometric survey of post-AGB binaries with MIDI \citep{Hillen2017}.
In this sample it is standing out because of its large size.
Also in the Spitzer survey of \citet{Gielen2011} it was noticed that the spectrum of this source is dominated by amorphous silicates with no crystalline dust features.
It was postulated in the latter that IRAS\,10174-5704 is likely a luminous super-giant.
It \modiff{would} explain why this source is an outlier in our survey as well.

\subsubsection{R\,Sct (\#19)}
R\,Sct is one of the only pAGB targets for which a surface magnetic field was detected \citep[][]{Sabin2015}.
It is also classified as `uncertain' by \citet{Gezer2015}, on the basis of its WISE photometric colours.
Its binary nature is not confirmed and the SED shows a very minimal infrared excess that is more reminiscent of an outflow than of a disk.

\subsubsection{V\,494\,Vel (\#23)}
This target shows photometric fluctuations but without any periodicity \citep[][]{Kiss2007}. \modif{As for R\,Sct, the binarity nature of this source is uncertain.}

We note that the target is usually referred in previous studies as IRAS\,09400-4733 \citep[][]{Kwok1997,deRuyter2006,Kiss2007,Szczerba2007}.


\subsection{Origin of the shift between pAGBs and YSOs in the size luminosity relation}
\label{sec:shift}
There can be two explanations for the systematic offset between inner rim sizes (scaled to the squared stellar luminosity, see Fig.\,\ref{fig:sizelum2}) between pAGB and YSO sources.
\modiff{The two explanations are about factors that influence the dust sublimation radius.}

\modiff{
A first factor that could be different between pAGBs and YSOs is the dust type.
For a given gas density, different types of dust will have different sublimation temperatures.
For example silicates, that are Oxygen rich, will have lower dust sublimation temperature that Carbon-rich dust \citep[e.g.,][]{Kobayashi2011}.
Carbon-rich dust is more abundant in
PPDs than in disks around pAGB \citep[e.g.,][]{Gielen2011}.
Although amorphous C does not show significant spectral features, the absence of distinct features from other C-rich dust species in the mid-IR spectra indicates a low abundance of C in the pAGB circumbinary dust. 
Mg-rich species like Olivine on the other hand are abundantly present \citep{Gielen2008,Gielen2011,Hillen2015}.
Therefore, in pAGB the overall dust sublimation temperature will be higher and the inner disk radius will be larger as observed. 
}


\modiff{
Another factor is the local gas density.
Assuming the same dust species, the same temperature will be reached farther away from the central star in the pAGBs because of higher luminosity of the central star than in YSOs. 
Assuming the same central stellar masses, same disk masses and a similar disk structure for the two types of objects, i.e. a decreasing surface density with radius, the local density will be lower at those locations in pAGBs (also because of weaker gravity due to the central star).
As the dust sublimation temperature depends on the gas density \citep[e.g.,][]{kama2009}, it will be lower for pAGBs and hence the inner dust rim will be larger.}

\begin{figure}
\centering
\includegraphics[width=9cm]{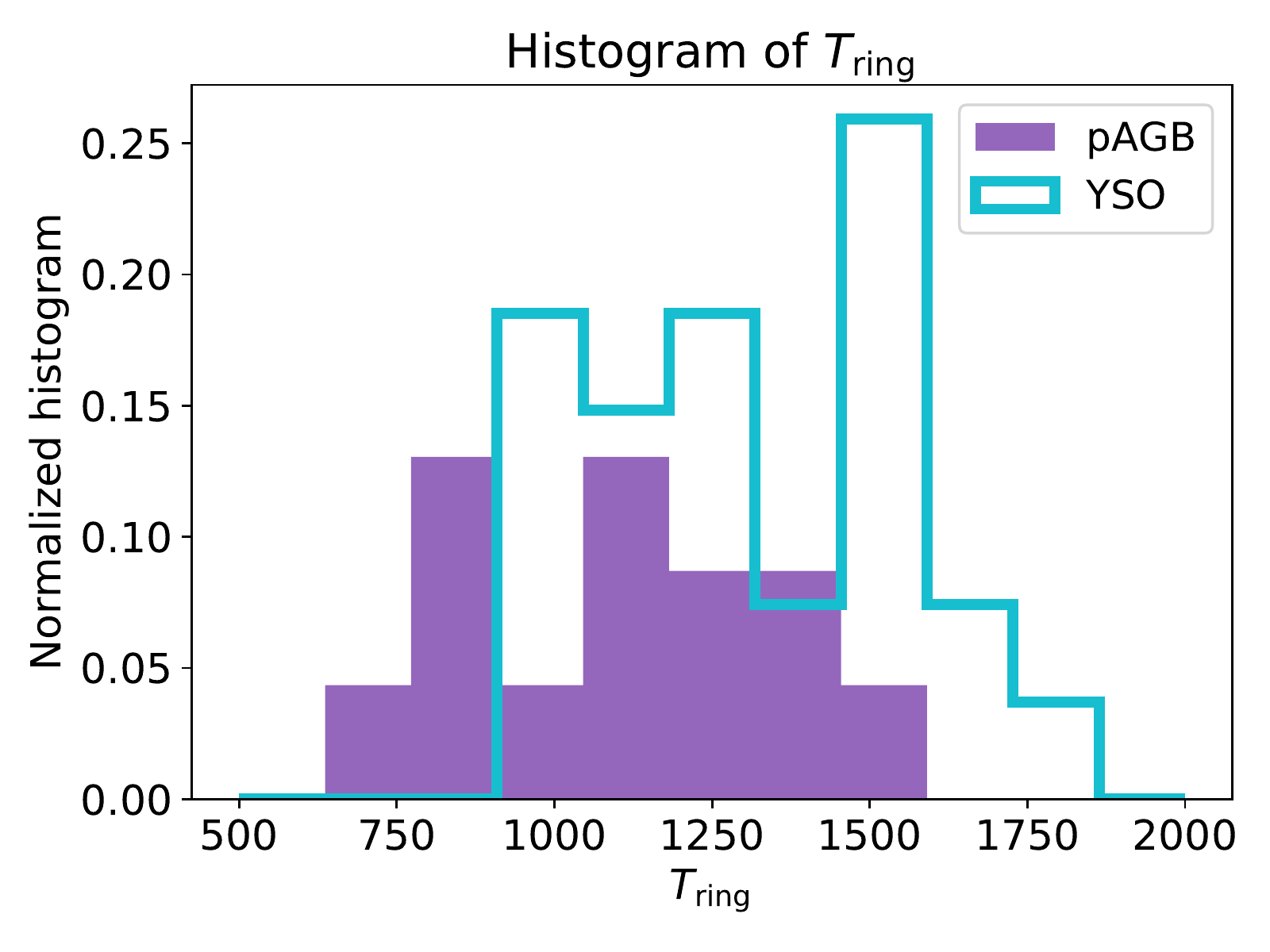}
\caption{Histogram of temperatures of the near-infrared circumbinary emission of pAGB from this work (\modif{purple}) and around YSOs from \citep{Lazareff2017} for objects having $T_\mathrm{ring}$ between 500 and 2000\,K.}
\label{fig:Tshift}
\end{figure}
We can compare the measured interferometric temperatures of the near-infrared circumbinary emission of pAGBs to those of the environment of YSOs.
Fig.\,\ref{fig:Tshift} shows that pAGBs have systemically lower measured disk rim temperatures than YSOs.

\begin{figure}
\centering
\includegraphics[width=9cm]{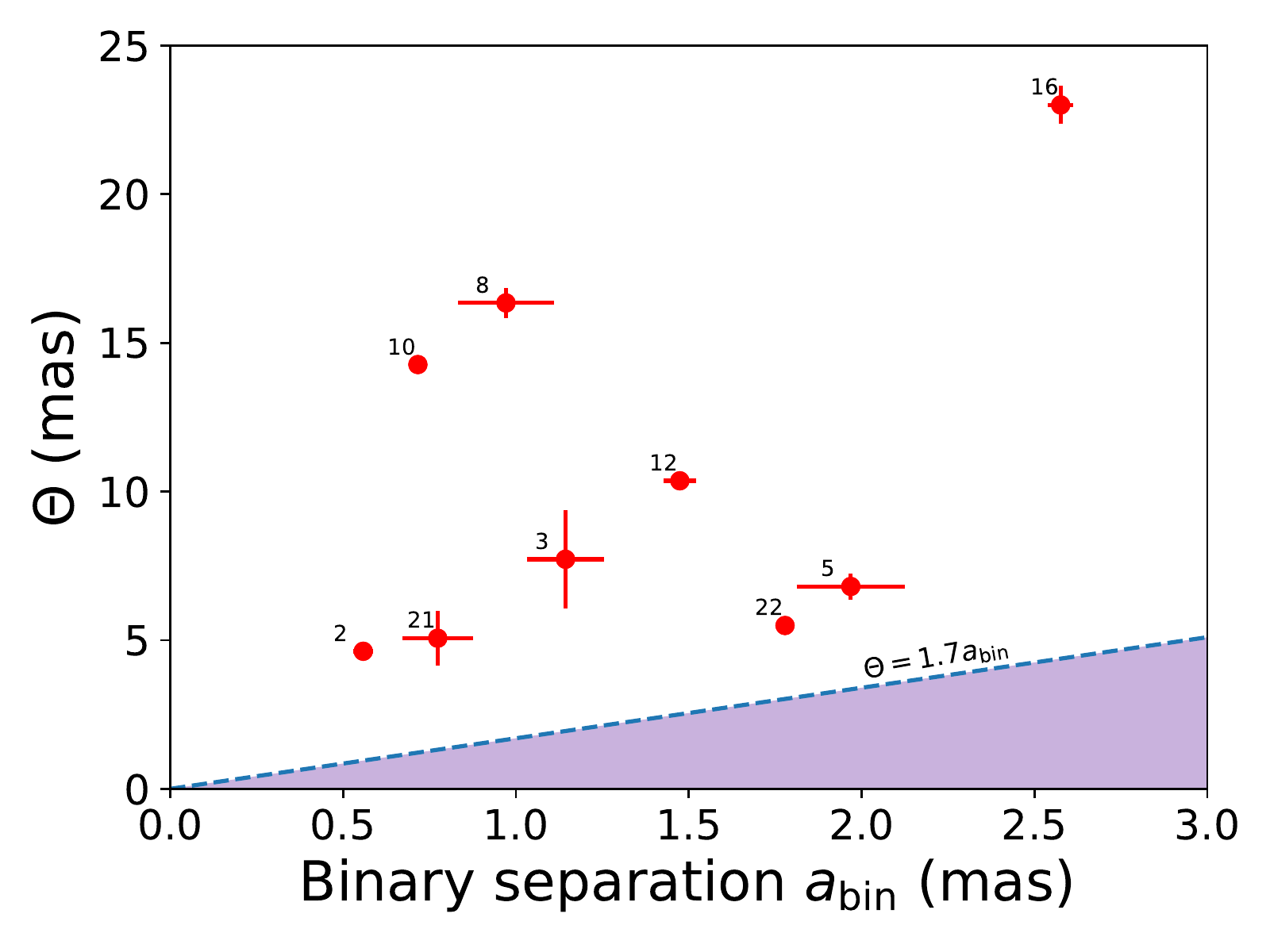}
\caption{\modif{Comparison between the ring diameter ($\Theta$) and the binary separation ($a_\mathrm{bin} = \sqrt{x_\mathrm{bin}^2 + y_\mathrm{bin}^2}$) for targets for which the most likely model contains a binary. The dashed blue line indicates the dynamical truncation diameter of the ring by the inner binary. The purple area indicates the forbidden ring diameters that are below the dynamical truncation diameter.}}
\label{fig:radbinsep}
\end{figure}

One could expect that some targets will show larger sizes that the theoretical dust sublimation radius because of the dynamical interaction between the inner binary and the disk.
This interaction would push the disk rim at a radius $\sim$1.7 times larger than the binary separation \citep{artymowicz1994}.
This would have been seen by having some targets \modif{higher up} in the size-luminosity diagram (Fig.\,\ref{fig:sizelum1}) and that some ring diameter\modiff{s} ($\Theta$) would be of the order of 1.7$\times$ the binary separation which we do not see (Fig.\,\ref{fig:radbinsep}).
Therefore the dust component is no tracing the disk dynamical truncation by the inner binary.

\subsection{Relation between the RVb phenomenon and disk inclination}
\label{sec:RVb}

\begin{figure}
\centering
\includegraphics[width=9cm]{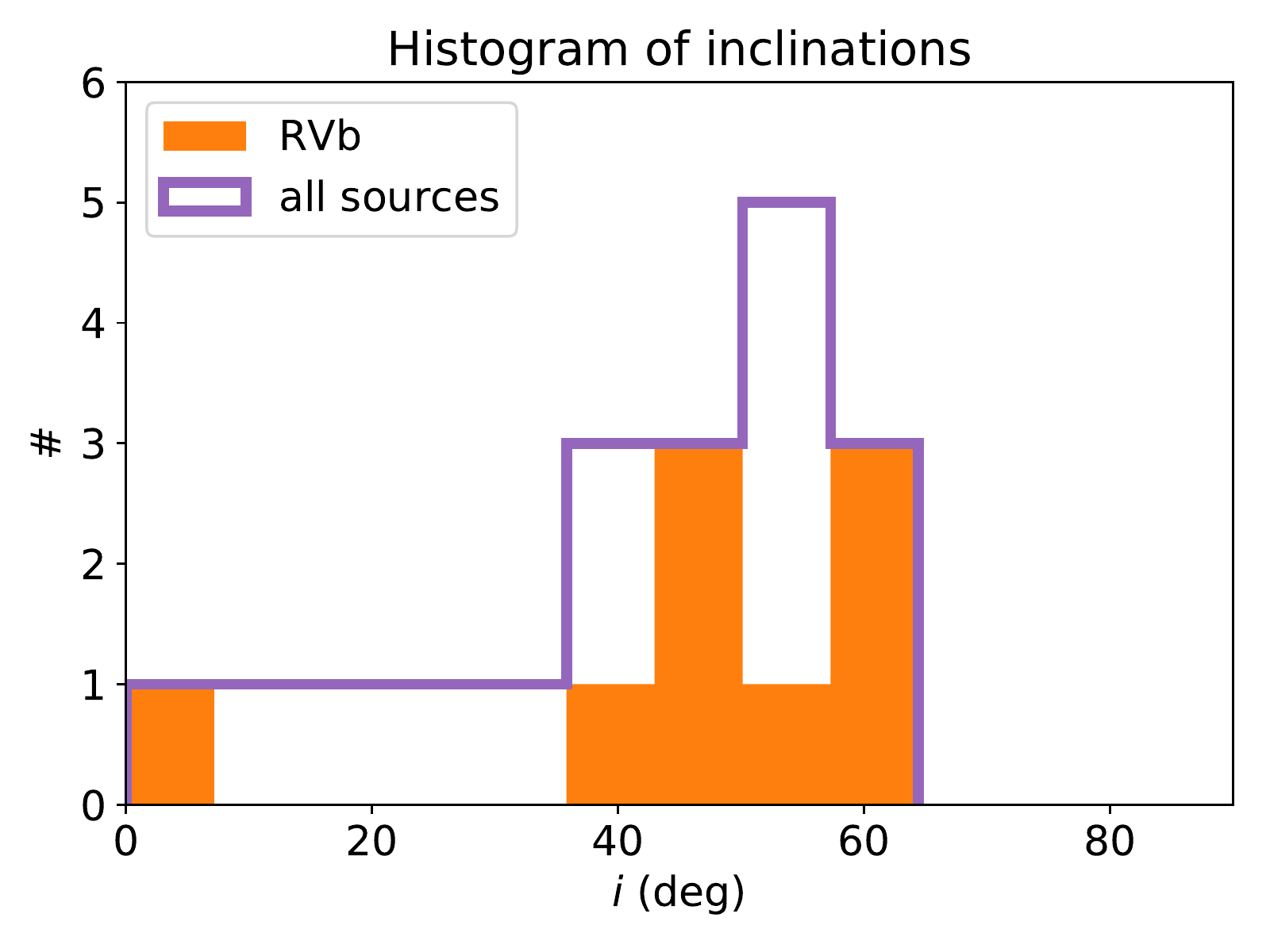}
\caption{Histogram of inclinations for objects from our sample showing the RVb phenomenon (orange) and the whole sample (purple).}
\label{fig:RVb}
\end{figure}

The RV Tauri stars are variable pulsating post-AGB stars \citep[e.g.,][]{Kiss2007,KissBodi2017,Manick2017}.
They display alternate deep and shallow minima. 
A sub-sample of these stars has a long-period variation of the mean luminosity \citep[][]{Pollard1996} and are classified as RVb.
More generally this long-period photometric variation is observed in non RV Tauri stars as well and is caused by variable extinction/scattering in the line of sight due to orbital motion of the central binary \citep[][]{Kiss2007}.
It is interpreted as caused by a highly inclined disk shadowing the primary at certain phases of the orbit \citep[][]{VanWinckel1999,Manick2017}.
As we are sensitive to the disk inclination we can test this hypothesis.

In our sample, there are nine targets displaying the RVb phenomenon: AI\,Sco (\#2), HD\,95767(\#5), HD\,213985 (\#7), HR\,4049(\#8), IRAS\,19125+0343 (\#15), IW\,Car (\#16), SX\,Cen(\#21), U\,Mon (\#22) and V\,494\,Vel (\#23) \citep[][]{Waelkens1991,Waelkens1995,Kiss2007,KissBodi2017}.
While HD\,213985 (\#7), SX\,Cen (\#21) and U\,Mon (\#22) are the most inclined disks from this survey ($inc\sim60^\circ$) confirming the inclined disk hypothesis, the other objects are only moderately inclined  implying very high disk scale-height for the disk shadowing interpretation to be true.
The histogram of inclinations for all sources and RVb sources shows that most of the RVb sources have the highest inclination (Fig.\,\ref{fig:RVb}).
One RVb source is found to have a pole-on inclination: HD\,95767 (\#5).
\modif{The other likely model for this source also point toward a pole-on orientation (Fig.\,\ref{tab:resHD95}).}
We also note that three sources have high inclinations (above 50$^\circ$) without showing the RVb phenomenon: EN\,TrA (\#3), IRAS\,05208-2035 (\#9) and \object{IRAS\,15469-5311} (\#12).
The apparent inclinations are deduced from an aspect ratio and, given the poor uv-coverage for some sources, need to be confirmed by further studies.

\subsection{Comparison between near-infrared and mid-infrared sizes}
\label{sec:NIRMIR}

\begin{figure}
\centering
\includegraphics[width=9cm]{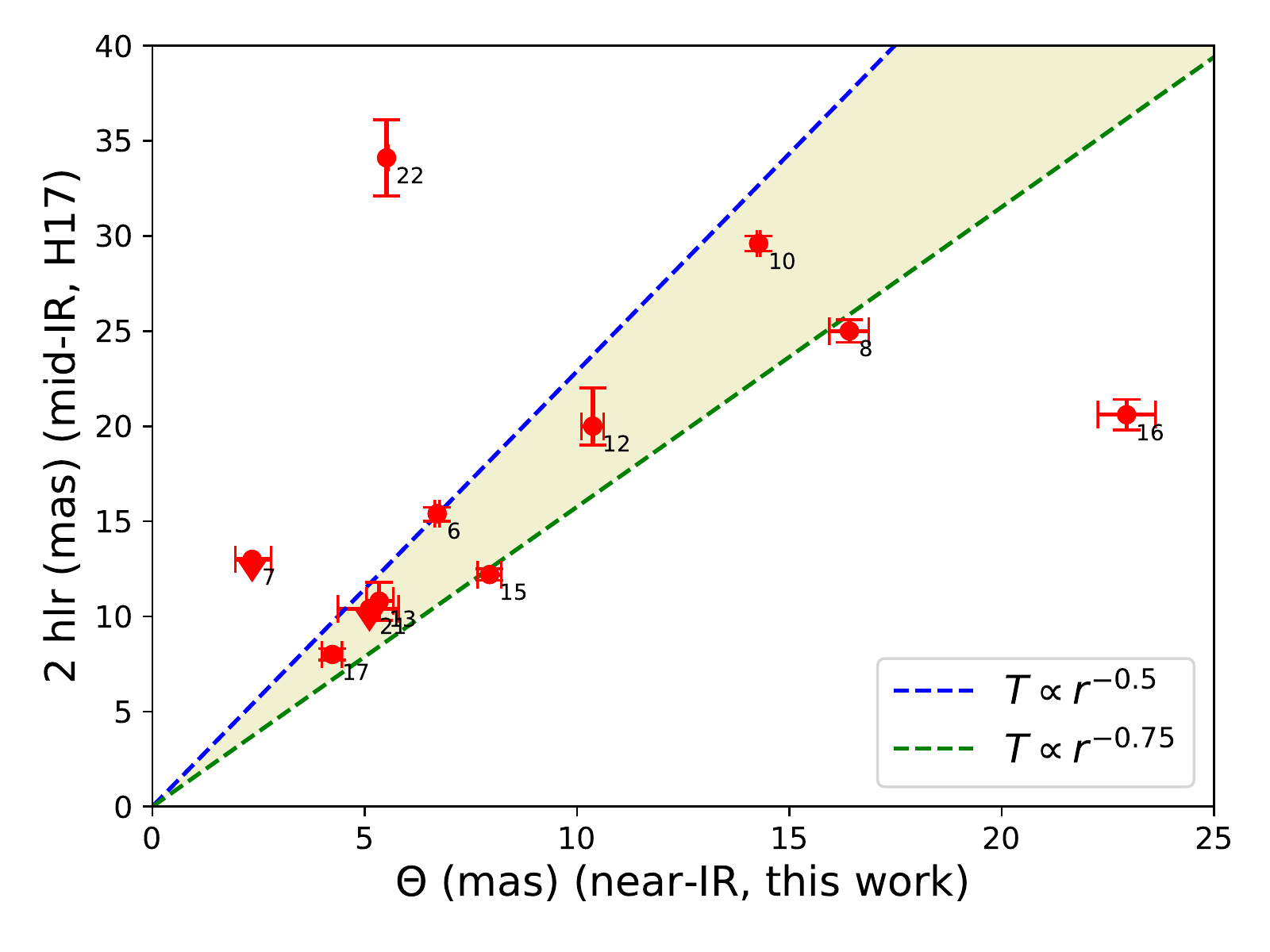}
\caption{Comparison between MIDI half light diameters and PIONIER diameters for ring like targets. The dashed lines represent the models of disks with a temperature dependence that scales with a power-law of the radius. IRAS\,10174-5704 (\#11) is outside the limits of this plot (see Sect.\,\ref{sec:NIRMIR}) and is not appearing for clarity reasons.}
\label{fig:Midi}
\end{figure}

We see in Sect.\,\ref{sec:sizelum} that the size of the near-infrared emission is ruled by dust sublimation physics as it is proportional to the square root luminosity of the central star.
However the structure of the disk can be constrained if it is observed at different wavelengths.
To do so we can compare the \modiff{diameters} of the circumstellar emission at the mid-infrared probed by MIDI at 10$\mu$m \citep{Hillen2016} with the \modiff{diameters} from this work.
The two \modiff{diameters} are plotted against each other on Fig.\,\ref{fig:Midi}.
There is a relation of proportionality where $\theta_\mathrm{MIR}$ $\approx$ 2$\theta_\mathrm{NIR}$.

Most of the models of protoplanetary disks predict a power-law for the radial temperature dependence $T\propto r^{-\alpha}$ \citep[e.g.,][]{LyndenBell1974,Kenyon1987}.
In several studies the used power-law index is either $\alpha$=0.5 or $\alpha$=0.75 \citep[e.g.,][]{Kraus2008}.
Assuming that the disk emits as a black-body with a power-law radial temperature profile and that the inner disk radius has a temperature of 1100\,K (see Sect.\,\ref{sec:T}) we can simulate the radial profile of the emission in both the near and mid-infrared.
From those profiles we can take the ratio of the half-light \modiff{diameters} \modif{($\Theta$)} between the near-infrared and the mid-infrared. 
Those ratios are reported in Fig.\,\ref{fig:Midi}.
We can see that most of our targets fall within the two limits set by the two temperature power-laws.
This points toward disks that have a similar radial temperature dependence and a smooth radial disk structure in the inner disk regions (<30\,au).

There are some sources that are standing out from this picture. 
IRAS10174-5704 (\#11) has a mid-infrared size \citep[2hlr=150$\pm$8mas\modif{;}][]{Hillen2017} $\sim$115 times larger than the near-infrared size from this work.
U\,Mon (\#22) has also a relatively large mid to near infrared size ratio.
This target is showing the RVb phenomenon and has a very complex visibility profile that could only be reproduced by the most complex model. 
The result of the fit is displaying a strong azimuthal modulation showing a complex morphology of this target. 
Finally IW\,Car (\#16) is the only outlier having a significantly small mid to near-infrared size ratio.
This target is showing an RVb phenomenon as well, however, we find a moderate inclination in our work.
It is fitted by the most complex model and has still a relatively large $\chi_\mathrm{red}^2$=3.9.
It is possible that the models are not able to reproduce all the complexity of that source and that it could have an effect on the derived size.
More observations of this target are therefore needed to confirm its status.


\subsection{Continuing the comparison between pAGB and YSOs}
\label{sec:YSOcomp}

Despite a completely different formation process, disks surrounding post-AGB binaries and those surrounding young stars are similar with respect to several criteria.
They are often compared to young intermediate mass Herbig stars.
The disks around Herbig Ae/Be stars are classified into two groups \citep[][]{Meeus2001,Maaskant2013,Menu2015}: disks with flared or gapped structure (group I) and flat disks (group II).
It was advocated that disks around post-AGB binaries are group II sources \citep[][]{deRuyter2006,Hillen2017} because of their SED and mid-infrared sizes and colors.

In this work we push the comparison further.
We recall and discuss here the main points of comparison between those two kinds of disks arising from this study.

\begin{itemize}
    \item \textbf{The near-infrared size-luminosity relation:} We show in Sect.\,\ref{sec:sizelum} that the relation extends for the near-infrared emission around post-AGB disks. As these targets have a higher luminosity than most of the young intermediate mass Herbig Ae/Be stars, we are able to probe very high luminosity regimes. This result shows that the near-infrared emission in post-AGB is also mainly ruled by dust sublimation rather than dynamical disk truncation by the inner binary. However, the shift toward larger sizes and the \modif{ring temperatures, $T_\mathrm{ring}$,} inform us \modif{on different properties of} the disks around post-AGB binaries that lowers the temperature of their dust sublimation front. We postulate that it can be due to smaller local gas density or different dust grain mineralogy or both.
    \item \textbf{Amount of extended flux:} For Herbig Ae/Be stars, the disks that have a double-peaked SED in the mid-infrared \citep[group I;][]{Meeus2001} display an over-resolved-to-non-stellar flux ratio larger than 5\% whereas the targets with a flat SED in the mid-infrared (group II) have this flux ratio lower than 5\% \citep{Lazareff2017}.
    In our sample, fourteen targets have more than 10\% of non-stellar flux to be over-resolved. 
    This ratio can reach 50\% in some cases.
    This confirms that there is a significant contribution from an over-resolved flux in the near-infrared for these targets.
    It is the first \modif{time we observe a} \modif{specific} feature \modif{of group~I Herbig disks in} pAGB disks.
    The origin of this extended flux \modif{($f_\mathrm{bg0}$)} is unknown and should be the focus of future studies.
    It can be related to the disk structure (e.g., disk flaring, presence of a gap) or to a mechanism lifting the dust from the disk (e.g., disk wind, jet from the secondary).
\end{itemize}

\section{Conclusions}
\label{sec:conclusions}
We summarize here the most important findings of our interferometric survey.
Despite the very inhomogeneous (u, v)-coverages obtained we can conclude that:
   \begin{enumerate}
      \item For \modif{most of the} sources \modif{19/23} a compact but resolved ring-like $H$-band emission component is detected which confirms the presence of a disk in pAGB binaries.
      \item Most of the targets (14/23) prefer models with more than ten parameters and several targets (\modif{6}/23) prefer the most complex models with fifteen or more, including complex azimuthal modulations, even with a limited (u, v)-plane coverage. 
      \item \modiff{There is a relation of proportionality between the size of the near-infrared circumstellar emission and the square root of the stellar luminosity as it is the case around YSOs, suggesting that the near-infrared extended emission is also linked to the dust sublimation region around pAGBs.}
      \item The measured temperature of the near-infrared circumbinary emission is lower (\modif{median ring temperature of}$\sim$1\modif{2}00\modif{-1300}\,K) for pAGB disks and the sizes \modif{of the near-infrared circumstellar emission} are systematically a bit larger than for YSOs. This can be due to different dust grain mineralogy and/or lower gas density at the sublimation front.
      \item The dust sublimation front \modiff{has} width-to-\modif{radius} ratios spanning between 0.5 and unity.
      \item A significant fraction of the near-infrared emission is over-resolved by our observations. This ratio is higher than for Herbig Ae/Be stars. The origin of this circumstellar flux is unknown.
   \end{enumerate}

Given the complexity of our targets and the limitations of geometrical modeling of the near-infrared interferometric observables, we believe that a time series of interferometric images is the only way to come to a sharp view on the physics that drives the interactions in the inner regions of these objects. The here presented survey (see Figs.\,\ref{fig:modImages1} and \ref{fig:modImages2}) and our imaging campaign of IRAS\,08544-4431 \citep[][]{Hillen2016,Kluska2018} have demonstrated the potential of near-infrared interferometric images, have shown which targets can be well resolved with the existing instrumentation and which are most interesting for follow-up campaigns.

The focus of this paper has been mostly on the circumbinary emission, as this is most reliably detected in our data. Our survey also demonstrates, however, that there is a lot of potential for detecting the companions, even if the here presented detections are likely not free of model bias. Model-independent determinations of the binary orbits will require significant investments of observing time though, as significantly better uv-coverages are required (like that of IRAS\,08544-4431), and this at various orbital phases.

To investigate the origin of the over-resolved flux (outflow, disk wind, disk structure), direct imaging in scattered light should bring strong constraints.
Finally\modif{,} a complete view of the dust \modiff{disk} structure such as millimeter observations with ALMA will allow us to investigate the possibility of second-generation planet formation in these disks.

\begin{acknowledgements}
We thank the referee for his comments that improved the clarity of the paper.
JK and HVW acknowledge support from the research council of the KU Leuven under grant number C14/17/082.
\modif{DK acknowledges support from Australian Research Council DECRA grant DE190100813.}
This research has made use of the Jean-Marie Mariotti Center \texttt{Aspro}
service\footnote{Available at http://www.jmmc.fr/aspro}.
We used the following internet-based resources: NASA Astrophysics Data System for bibliographic services; Simbad; the VizieR online catalogs operated by CDS.
\end{acknowledgements}

\bibliographystyle{aa} 
\bibliography{biblio} 

\begin{appendix}

\section{\modif{Best-fit parameters}}

\begin{sidewaystable*}
\caption{Best-fit parameters. *Note that for clarity the flux ratios $f_{prim,0}$, $f_{sec,0}$ and $f_{bg0}$ are written in \%, i.e. hundred times the value determined by the fit.}             
\label{tab:res1}      
\centering                        
\begin{tabular}{c c r c c c c c c c c c c c c c c c c c c c c c c }        
\hline\hline
\# & Star & \modif{$n_\mathrm{data}$}  & model & BIC & $\chi_\mathrm{r}^2$  & $f_{prim,0}$ & $f_{sec,0}$ & $f_{bg0}$ & $d_{sec}$ & $d_{back}$ & $T_\mathrm{ring}$ & $\theta$ & $\delta \theta$ & $inc$\\
\hline
1 &  AC\,Her &30 & s0 & 36.0 & 1.0& $91.4^{0.3}_{-0.3}$ & - & - & - & -& $5812^{2713}_{-2239}$ & - & - & - \\ [1pt] 
2 &  AI\,Sco &60 & br2 & 229.2 & 3.7& $41.1^{7.0}_{-6.7}$& $24.7^{6.5}_{-6.9}$& $1.1^{0.4}_{-0.4}$ & - & -& $1498^{63}_{-59}$& $4.6^{0.2}_{-0.2}$& $0.9^{0.1}_{-0.1}$& $47.3^{2.3}_{-2.6}$ \\ [1pt] 
3 &  EN\,TrA &51 & br2 & 88.7 & 1.0& $73.6^{1.5}_{-2.5}$& $7.8^{1.7}_{-1.2}$& $2.5^{0.9}_{-1.0}$ & - & -& $1445^{299}_{-210}$& $7.7^{1.7}_{-1.6}$& $1.8^{0.7}_{-0.6}$& $52.7^{7.8}_{-9.1}$ \\ [1pt] 
4 &  HD\,93662 &60 & b2 & 254.7 & 4.3 & -& $0.4^{0.1}_{-0.1}$& $13.8^{0.3}_{-0.3}$& $-0.7^{2.6}_{-2.2}$& $-0.6^{0.4}_{-0.4}$ & - & - & - & - \\ [1pt] 
5 &  HD\,95767 &60 & br3 & 98.0 & 0.8& $60.0^{1.5}_{-2.5}$& $12.3^{2.3}_{-1.2}$& $1.1^{0.8}_{-0.7}$ & - & -& $986^{99}_{-83}$& $6.8^{0.4}_{-0.5}$& $0.9^{0.2}_{-0.2}$& $0.002^{0.003}_{-0.001}$ \\ [1pt] 
6 & HD\,108015 &60 &  sr6 & 229.2 & 3.7& $57.8^{0.2}_{-0.2}$ & -& $8.5^{0.4}_{-0.4}$ & - & -& $1155^{28}_{-27}$& $6.7^{0.1}_{-0.1}$& $0.43^{0.02}_{-0.02}$& $12.9^{4.7}_{-5.9}$ \\ [1pt] 
7 & HD\,213985 &60 &  sr3 & 133.5 & 1.9& $57.8^{0.6}_{-0.6}$ & -& $6.4^{0.6}_{-0.6}$ & - & -& $1203^{106}_{-93}$& $2.3^{0.4}_{-0.4}$& $3.3^{0.8}_{-0.6}$& $61.0^{2.5}_{-3.1}$ \\ [1pt] 
8 &  HR\,4049 &144 & br5 & 253.9 & 1.3& $62.9^{1.0}_{-1.2}$& $2.5^{1.1}_{-0.8}$& $17.0^{0.7}_{-0.7}$& $-3.3^{1.0}_{-0.5}$ & -& $711^{24}_{-23}$& $16.4^{0.5}_{-0.5}$& $0.8^{0.1}_{-0.1}$& $49.3^{3.2}_{-3.3}$ \\ [1pt] 
9 &  IRAS\,05208-2035 &204 & sr3 & 267.8 & 1.1& $90.6^{0.3}_{-0.3}$ & -& $5.0^{0.4}_{-0.4}$ & - & -& $7871^{1484}_{-1918}$& $6.1^{0.6}_{-0.7}$& $0.9^{0.3}_{-0.3}$& $53.8^{3.7}_{-5.0}$ \\ [1pt] 
10 &  IRAS\,08544-4431 &1332 & br5 & 3343.1 & 2.4& $58.5^{0.4}_{-0.4}$& $5.2^{0.4}_{-0.4}$& $15.3^{0.2}_{-0.2}$& $-0.4^{0.8}_{-0.8}$ & -& $875^{10}_{-9}$& $14.3^{0.1}_{-0.1}$& $0.48^{0.01}_{-0.01}$& $21.3^{0.8}_{-0.8}$ \\ [1pt] 
11 &  IRAS\,10174-5704 &30 & sr1 & 45.5 & 0.9& $56.8^{8.9}_{-10.7}$ & -& $0.4^{0.5}_{-0.3}$ & - & -& $7700^{1578}_{-1877}$& $1.5^{0.3}_{-0.4}$& $0.6^{0.9}_{-0.5}$& $33.6^{2.5}_{-2.4}$ \\ [1pt] 
12 &  IRAS\,15469-5311 &90 & br4 & 190.0 & 1.6& $57.2^{0.4}_{-0.4}$& $0.1^{0.1}_{-0.1}$& $12.8^{0.5}_{-0.6}$ & - & -& $818^{17}_{-17}$& $10.4^{0.3}_{-0.3}$& $0.39^{0.03}_{-0.03}$& $53.5^{1.8}_{-2.2}$ \\ [1pt] 
13 & IRAS\,17038-4815 & 90 & sr6 & 148.6 & 1.2& $71.7^{0.2}_{-0.2}$ & -& $4.2^{0.4}_{-0.4}$ & - & -& $2132^{118}_{-104}$& $5.3^{0.3}_{-0.3}$& $1.2^{0.1}_{-0.1}$& $36.7^{5.1}_{-7.8}$ \\ [1pt] 
14 &  IRAS\,18123+0511 &60 & sr5 & 83.2 & 0.8& $61.0^{1.9}_{-2.8}$ & -& $19.2^{1.2}_{-1.4}$ & - & -& $1410^{137}_{-125}$& $2.4^{0.2}_{-0.2}$& $0.4^{0.4}_{-0.3}$& $25.9^{9.1}_{-13.4}$ \\ [1pt] 
15 &  IRAS\,19125+0343 &102 & sr6 & 233.7 & 2.0& $54.1^{0.2}_{-0.3}$ & -& $8.0^{0.5}_{-0.5}$ & - & -& $896^{23}_{-23}$& $7.9^{0.3}_{-0.3}$& $0.49^{0.02}_{-0.02}$& $36.7^{2.8}_{-3.0}$ \\ [1pt] 
16 &  IW\,Car &456 & br5 & 1798.8 & 3.9& $66.5^{0.1}_{-0.1}$& $3.5^{0.1}_{-0.1}$& $10.5^{0.3}_{-0.3}$& $1.6^{0.2}_{-0.2}$ & -& $1047^{15}_{-15}$& $23.0^{0.7}_{-0.6}$& $0.71^{0.02}_{-0.02}$& $44.6^{1.7}_{-1.8}$ \\ [1pt] 
17 & LR\,Sco &60 &  sr3 & 98.8 & 1.2& $71.2^{0.4}_{-0.4}$ & -& $2.4^{0.5}_{-0.5}$ & - & -& $1200^{54}_{-50}$& $4.2^{0.2}_{-0.3}$& $1.3^{0.2}_{-0.2}$& $40.2^{4.1}_{-4.9}$ \\ [1pt] 
18 &  PS\,Gem &24 & b1 & 38.7 & 1.0 & -& $2.4^{1.6}_{-0.3}$& $4.3^{0.7}_{-0.8}$& $-1.9^{1.4}_{-1.2}$& $1.2^{1.8}_{-2.0}$ & - & - & - & - \\ [1pt] 
19 &  R\,Sct &30 & sr6 & 139.3 & 5.6& $55.5^{3.1}_{-4.6}$ & -& $0.02^{0.03}_{-0.02}$ & - & -& $7534^{1572}_{-1648}$& $2.8^{0.3}_{-0.3}$& $0.3^{0.3}_{-0.2}$& $53.4^{3.0}_{-4.0}$ \\ [1pt] 
20 &  RU\,Cen &60 & b2 & 92.9 & 1.2 & -& $6.8^{0.2}_{-0.2}$& $4.6^{0.3}_{-0.3}$& $-1.0^{0.4}_{-0.4}$& $-1.5^{1.3}_{-1.2}$ & - & - & - & - \\ [1pt] 
21 &  SX\,Cen &60 & br2 & 97.8 & 0.9& $72.8^{3.3}_{-4.2}$& $16.8^{4.1}_{-3.2}$& $2.5^{0.6}_{-0.6}$ & - & -& $1125^{171}_{-144}$& $5.1^{0.9}_{-0.9}$& $1.0^{0.5}_{-0.6}$& $64.4^{5.4}_{-8.1}$ \\ [1pt] 
22 & U\,Mon &204 &  br5 & 483.8 & 2.1& $68.8^{1.5}_{-1.2}$& $1.5^{0.2}_{-0.2}$& $9.7^{0.3}_{-0.3}$& $-3.7^{0.4}_{-0.2}$ & -& $2619^{138}_{-132}$& $5.49^{0.03}_{-0.03}$& $0.02^{0.02}_{-0.01}$& $57.9^{1.6}_{-1.5}$ \\ [1pt] 
23 & V\,494\,Vel &84 & sr3 & 96.5 & 0.8& $74.9^{0.8}_{-0.7}$ & -& $3.2^{0.4}_{-0.4}$ & - & -& $8186^{1255}_{-1599}$& $2.6^{0.5}_{-0.4}$& $2.2^{0.6}_{-0.6}$& $53.0^{1.7}_{-1.7}$ \\ [1pt] 
\hline                                   
\end{tabular}
\end{sidewaystable*}

\begin{sidewaystable*}
\caption{Best fit parameters (continued from Table \ref{tab:res1}).}             
\label{tab:res2}      
\centering                          
\begin{tabular}{c c c c c c c c c c c c c c c c c c c c c c c c }        
\hline\hline   
\# & Star  & $PA$ & $c_1$ & $s_1$ & $c2$ & $s2$ & $x_0$ & $y_0$ & rM & UD$_\mathrm{prim,0}$ & UD$_\mathrm{sec,0}$\\ 
 \hline 
 1 & ACHer & - & - & - & - & - & - & - & - & - & - \\ [1pt] 
2 & AI\,Sco& $179^{3}_{-3}$& $0.92^{0.04}_{-0.05}$& $-0.3^{0.1}_{-0.1}$ & - & -& $-0.13^{0.03}_{-0.03}$& $-0.09^{0.02}_{-0.03}$& $2.7^{1.1}_{-0.8}$ & - & - \\ [1pt] 
3 & EN\,TrA& $13^{9}_{-7}$& $0.47^{0.13}_{-0.13}$& $0.71^{0.13}_{-0.16}$ & - & -& $-0.27^{0.09}_{-0.14}$& $-0.05^{0.06}_{-0.07}$& $3.2^{2.4}_{-1.4}$ & - & - \\ [1pt] 
4 & HD\,93662 & - & - & - & - & -& $-28.6^{0.4}_{-0.4}$& $28.8^{0.3}_{-0.3}$ & -& $1.74^{0.01}_{-0.01}$& $1.63^{0.81}_{-0.96}$ \\ [1pt] 
5 & HD\,95767& $165^{17}_{-18}$& $0.22^{0.21}_{-0.20}$& $0.34^{0.11}_{-0.09}$& $0.67^{0.17}_{-0.21}$& $0.11^{0.40}_{-0.46}$& $0.7^{0.1}_{-0.1}$& $-1.4^{0.2}_{-0.2}$& $0.2^{0.2}_{-0.2}$ & - & - \\ [1pt] 
6 & HD\,108015& $40^{16}_{-19}$& $-0.09^{0.09}_{-0.08}$& $0.21^{0.05}_{-0.05}$& $0.22^{0.05}_{-0.07}$& $-0.05^{0.14}_{-0.17}$& $0.09^{0.04}_{-0.04}$& $-0.76^{0.03}_{-0.03}$ & - & - & - \\ [1pt] 
7 & HD\,213985& $93^{3}_{-4}$& $0.26^{0.18}_{-0.18}$& $0.83^{0.10}_{-0.14}$ & - & - & - & - & - & - & - \\ [1pt] 
8 & HR\,4049& $63^{7}_{-6}$& $0.04^{0.09}_{-0.07}$& $-0.29^{0.23}_{-0.24}$& $0.30^{0.18}_{-0.19}$& $-0.76^{0.31}_{-0.15}$& $-0.6^{0.1}_{-0.2}$& $-0.1^{0.1}_{-0.1}$& $0.7^{0.6}_{-0.4}$& $0.5^{0.1}_{-0.1}$ & - \\ [1pt] 
9 & IRAS\,05208-2035& $155^{4}_{-4}$& $-0.84^{0.11}_{-0.08}$& $0.32^{0.13}_{-0.13}$ & - & - & - & - & - & - & - \\ [1pt] 
10 & IRAS\,08544-4431& $12^{3}_{-3}$& $0.38^{0.02}_{-0.02}$& $-0.23^{0.02}_{-0.02}$& $0.02^{0.03}_{-0.03}$& $-0.27^{0.01}_{-0.01}$& $0.4^{0.1}_{-0.1}$& $0.6^{0.1}_{-0.1}$& $0.02^{0.03}_{-0.02}$& $0.46^{0.02}_{-0.02}$ & - \\ [1pt] 
11 & IRAS\,10174-5704& $13^{5}_{-6}$ & - & - & - & - & - & - & - & - & - \\ [1pt] 
12 & IRAS\,15469-5311& $64^{2}_{-2}$& $0.08^{0.04}_{-0.04}$& $-0.34^{0.04}_{-0.05}$& $-0.19^{0.06}_{-0.05}$& $0.30^{0.05}_{-0.05}$& $-0.9^{0.1}_{-0.1}$& $0.72^{0.03}_{-0.03}$& $0.3^{0.3}_{-0.2}$& $0.5^{0.1}_{-0.1}$ & - \\ [1pt] 
13 & IRAS\,17038-4815& $158^{7}_{-8}$& $-0.09^{0.07}_{-0.08}$& $0.27^{0.12}_{-0.10}$& $-0.43^{0.19}_{-0.17}$& $0.29^{0.16}_{-0.15}$& $0.2^{0.1}_{-0.1}$& $0.8^{0.1}_{-0.1}$ & - & - & - \\ [1pt] 
14 & IRAS\,18123+0511& $96^{14}_{-17}$& $0.39^{0.23}_{-0.23}$& $0.74^{0.15}_{-0.21}$& $0.60^{0.21}_{-0.30}$& $0.34^{0.34}_{-0.52}$ & - & - & - & - & - \\ [1pt] 
15 & IRAS\,19125+0343& $142^{4}_{-4}$& $-0.03^{0.08}_{-0.08}$& $0.23^{0.11}_{-0.10}$& $-0.31^{0.05}_{-0.05}$& $-0.26^{0.08}_{-0.08}$& $-0.67^{0.04}_{-0.04}$& $0.16^{0.04}_{-0.04}$ & - & - & - \\ [1pt] 
16 & IW\,Car& $155^{2}_{-2}$& $0.11^{0.04}_{-0.04}$& $-0.99^{0.01}_{-0.01}$& $0.09^{0.03}_{-0.03}$& $0.04^{0.06}_{-0.06}$& $1.14^{0.03}_{-0.03}$& $-2.28^{0.04}_{-0.04}$& $0.01^{0.02}_{-0.01}$& $0.65^{0.03}_{-0.03}$ & - \\ [1pt] 
17 & LR\,Sco& $48^{8}_{-8}$& $0.23^{0.11}_{-0.09}$& $-0.28^{0.06}_{-0.08}$ & - & - & - & - & - & - & - \\ [1pt] 
18 & PS\,Gem & - & - & - & - & -& $-3.3^{35.6}_{-32.5}$& $-5.0^{21.6}_{-23.7}$ & -& $2.1^{0.5}_{-1.1}$ & - \\ [1pt] 
19 & R\,Sct& $106^{6}_{-6}$& $0.28^{0.11}_{-0.08}$& $-0.04^{0.28}_{-0.27}$& $-0.84^{0.14}_{-0.10}$& $-0.02^{0.07}_{-0.07}$& $-0.1^{0.1}_{-0.1}$& $0.07^{0.03}_{-0.03}$ & - & - & - \\ [1pt] 
20 & RU\,Cen & - & - & - & - & -& $-0.38^{0.03}_{-0.03}$& $1.8^{0.03}_{-0.03}$ & -& $0.05^{0.05}_{-0.03}$& $0.4^{0.3}_{-0.3}$ \\ [1pt] 
21 & SX\,Cen& $132^{4}_{-10}$& $-0.79^{0.19}_{-0.13}$& $-0.26^{0.40}_{-0.32}$ & - & -& $-0.14^{0.04}_{-0.05}$& $0.5^{0.1}_{-0.1}$& $0.6^{0.5}_{-0.4}$ & - & - \\ [1pt] 
22 & U\,Mon& $45^{1}_{-1}$& $-0.03^{0.10}_{-0.14}$& $0.04^{0.03}_{-0.03}$& $0.95^{0.02}_{-0.03}$& $-0.29^{0.08}_{-0.07}$& $1.42^{0.02}_{-0.02}$& $1.07^{0.04}_{-0.04}$& $0.001^{0.001}_{-0.000}$& $0.4^{0.1}_{-0.1}$ & - \\ [1pt] 
23 & V\,494\,Vel& $96^{2}_{-2}$& $-0.23^{0.06}_{-0.07}$& $0.87^{0.07}_{-0.10}$ & - & - & - & - & - & - & - \\ [1pt] 
\hline                                   
\end{tabular}
\end{sidewaystable*}

\begin{table*}
\caption{Observation log}             
\label{tab:log}      
\centering                          
\begin{tabular}{l l c c c l}        
\hline\hline                 
\# & Target & Program ID & Date & MJD & Configuration \\    
\hline                        
1 & AC\,Her & 093.D-0573(B) & 2014-06-05 & 56813.3 & A1-B2-C1-D0\\
2 & AI\,Sco & 093.D-0573(A) & 2014-05-09 & 56786.4 & A1-G1-J3-K0\\
2 & AI\,Sco & 093.D-0573(B) & 2014-06-05 & 56813.4 & A1-B2-C1-D0\\
3 & EN\,TrA & 093.D-0573(A) & 2014-05-09 & 56786.2 & A1-G1-J3-K0\\
3 & EN\,TrA & 093.D-0573(B) & 2014-06-05 & 56813.1 & A1-B2-C1-D0\\
4 & HD\,93662 & 093.D-0573(A) & 2014-05-08 & 56786.0 & A1-G1-J3-K0\\
4 & HD\,93662 & 093.D-0573(B) & 2014-06-04 & 56813.0 & A1-B2-C1-D0\\
5 & HD\,95767 & 093.D-0573(A) & 2014-05-09 & 56786.1 & A1-G1-J3-K0\\
5 & HD\,95767 & 093.D-0573(B) & 2014-06-05 & 56813.0 & A1-B2-C1-D0\\
6 & HD\,108015 & 093.D-0573(A) & 2014-05-09 & 56786.1 & A1-G1-J3-K0\\
6 & HD\,108015 & 093.D-0573(B) & 2014-06-05 & 56813.1 & A1-B2-C1-D0\\
7 & HD\,213985 & 093.D-0573(B) & 2014-06-05 & 56813.4 & A1-B2-C1-D0\\
7 & HD\,213985 & 093.D-0573(C) & 2014-06-22 & 56830.4 & A1-G1-J3-K0\\
8 & HR\,4049 & 094.D-0865(B) & 2015-01-22 & 57044.3 & D0-G1-H0-I1\\
8 & HR\,4049 & 094.D-0865(C) & 2015-01-25 & 57047.2 & A1-G1-I1-K0\\
8 & HR\,4049 & 094.D-0865(A) & 2015-02-24 & 57077.4 & B2-C1-D0\\
9 & IRAS\,05208\,-2035 & 094.D-0865(B) & 2015-01-22 & 57044.1 & D0-G1-H0-I1\\
9 & IRAS\,05208\,-2035 & 094.D-0865(C) & 2015-01-25 & 57047.1 & A1-G1-I1-K0\\
9 & IRAS\,05208\,-2035 & 094.D-0865(C) & 2015-01-25 & 57047.0 & A1-G1-I1-K0\\
9 & IRAS\,05208\,-2035 & 094.D-0865(A) & 2015-02-24 & 57077.1 & B2-C1-D0\\
11 & IRAS\,10174\,-5704 & 093.D-0573(A) & 2014-05-09 & 56786.0 & A1-G1-J3-K0\\
12 & IRAS\,15469\,-5311 & 093.D-0573(A) & 2014-05-09 & 56786.3 & A1-G1-J3-K0\\
12 & IRAS\,15469\,-5311 & 093.D-0573(A) & 2014-05-09 & 56786.2 & A1-G1-J3-K0\\
12 & IRAS\,15469\,-5311 & 093.D-0573(B) & 2014-06-05 & 56813.2 & A1-B2-C1-D0\\
13 & IRAS\,17038\,-4815 & 093.D-0573(A) & 2014-05-09 & 56786.3 & A1-G1-J3-K0\\
13 & IRAS\,17038\,-4815 & 093.D-0573(A) & 2014-05-09 & 56786.4 & A1-G1-J3-K0\\
13 & IRAS\,17038\,-4815 & 093.D-0573(B) & 2014-06-05 & 56813.2 & A1-B2-C1-D0\\
14 & IRAS\,18123\,+0511 & 093.D-0573(C) & 2014-06-22 & 56830.3 & A1-G1-J3-K0\\
14 & IRAS\,18123\,+0511 & 093.D-0573(C) & 2014-06-22 & 56830.2 & A1-G1-J3-K0\\
15 & IRAS\,19125\,+0343 & 093.D-0573(B) & 2014-06-05 & 56813.4 & A1-B2-C1-D0\\
15 & IRAS\,19125\,+0343 & 093.D-0573(C) & 2014-06-22 & 56830.3 & A1-G1-J3-K0\\
15 & IRAS\,19125\,+0343 & 093.D-0573(C) & 2014-06-22 & 56830.3 & A1-G1-J3-K0\\
15 & IRAS\,19125\,+0343 & 093.D-0573(C) & 2014-06-22 & 56830.3 & A1-G1-J3-K0\\
16 & IW\,Car & 094.D-0865(B) & 2015-01-22 & 57044.1 & D0-G1-H0-I1\\
16 & IW\,Car & 094.D-0865(B) & 2015-01-22 & 57044.1 & D0-G1-H0-I1\\
16 & IW\,Car & 094.D-0865(B) & 2015-01-22 & 57044.2 & D0-G1-H0-I1\\
16 & IW\,Car & 094.D-0865(B) & 2015-01-22 & 57044.3 & D0-G1-H0-I1\\
16 & IW\,Car & 094.D-0865(B) & 2015-01-22 & 57044.4 & D0-G1-H0-I1\\
16 & IW\,Car & 094.D-0865(C) & 2015-01-25 & 57047.1 & A1-G1-I1-K0\\
16 & IW\,Car & 094.D-0865(A) & 2015-02-24 & 57077.0 & B2-C1-D0\\
16 & IW\,Car & 094.D-0865(A) & 2015-02-24 & 57077.2 & B2-C1-D0\\
16 & IW\,Car & 094.D-0865(A) & 2015-02-24 & 57077.3 & B2-C1-D0\\
16 & IW\,Car & 094.D-0865(A) & 2015-02-24 & 57077.3 & B2-C1-D0\\
17 & LR\,Sco & 093.D-0573(A) & 2014-05-09 & 56786.3 & A1-G1-K0-J3\\
17 & LR\,Sco & 093.D-0573(B) & 2014-06-05 & 56813.3 & A1-B2-C1-D0\\
18 & PS\,Gem & 094.D-0865(A) & 2015-02-24 & 57077.1 & B2-C1-D0\\
19 & R\,Sct & 093.D-0573(A) & 2014-05-09 & 56786.4 & A1-G1-J3-K0\\
20 & RU\,Cen & 093.D-0573(A) & 2014-05-09 & 56786.0 & A1-G1-J3-K0\\
20 & RU\,Cen & 093.D-0573(B) & 2014-06-05 & 56813.0 & A1-B2-C1-D0\\
21 & SX\,Cen & 093.D-0573(A) & 2014-05-09 & 56786.1 & A1-G1-J3-K0\\
21 & SX\,Cen & 093.D-0573(B) & 2014-06-05 & 56813.1 & A1-B2-C1-D0\\
22 & U\,Mon & 094.D-0865(B) & 2015-01-22 & 57044.0 & D0-G1-H0-I1\\
22 & U\,Mon & 094.D-0865(B) & 2015-01-22 & 57044.0 & D0-G1-H0-I1\\
22 & U\,Mon & 094.D-0865(C) & 2015-01-25 & 57047.2 & A1-G1-I1-K0\\
22 & U\,Mon & 094.D-0865(A) & 2015-02-24 & 57077.2 & B2-C1-D0\\
23 & V494\,Vel & 094.D-0865(B) & 2015-01-22 & 57044.3 & D0-G1-H0-I1\\
23 & V494\,Vel & 094.D-0865(A) & 2015-02-24 & 57077.4 & B2-C1-D0\\
\hline                                   
\end{tabular}
\end{table*}

\section{\modif{Interferometric dataset}}

\begin{figure*}
\centering
\includegraphics[width=6.5cm]{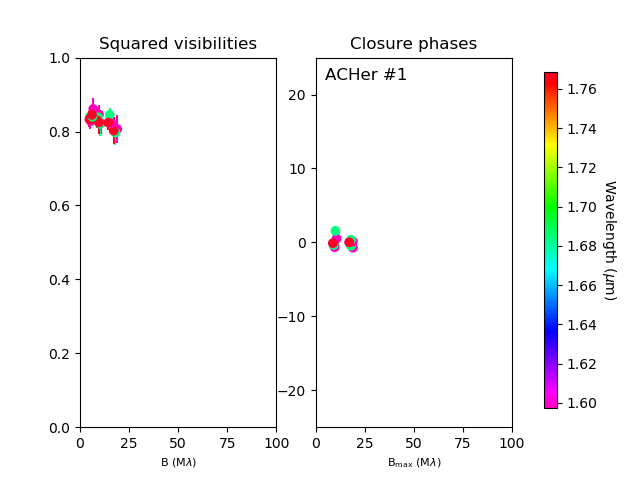} 
\includegraphics[width=6.5cm]{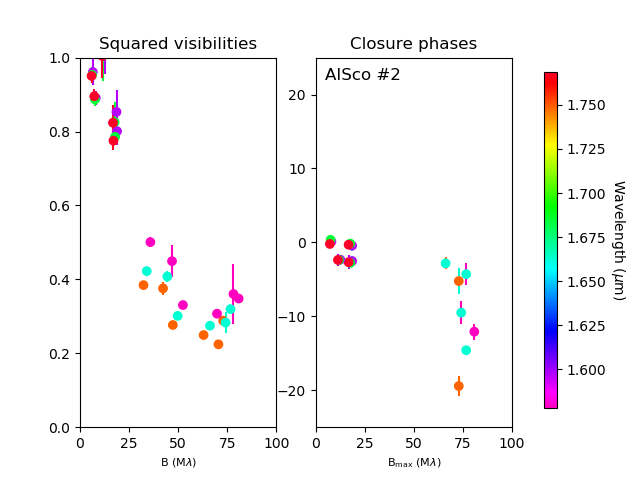}
\includegraphics[width=6.5cm]{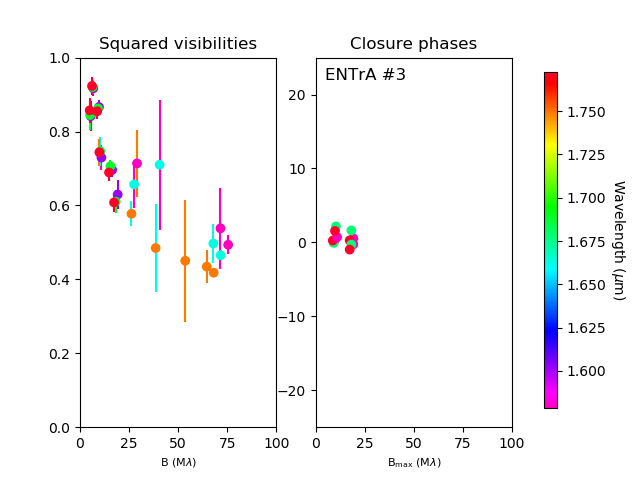}
\includegraphics[width=6.5cm]{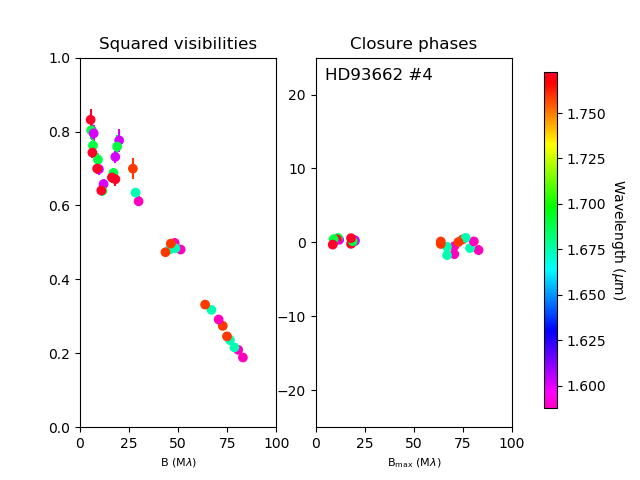}
\includegraphics[width=6.5cm]{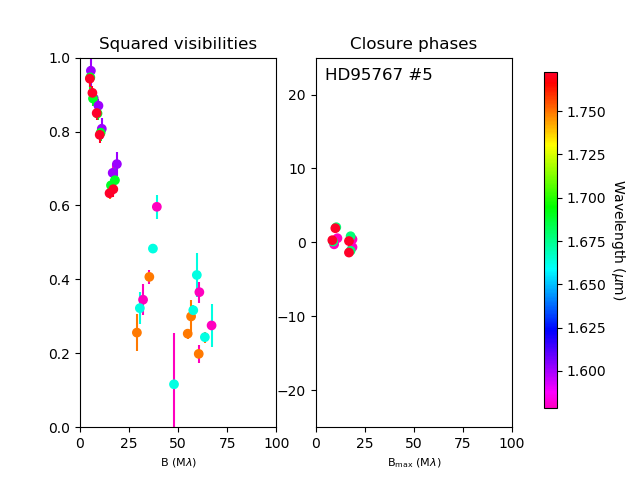}
\includegraphics[width=6.5cm]{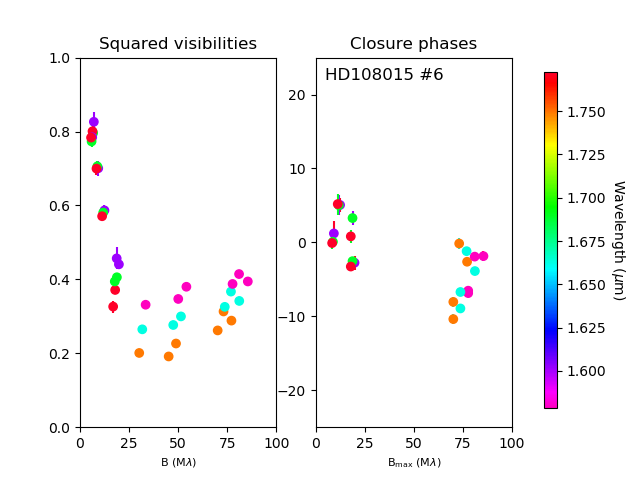}
\includegraphics[width=6.5cm]{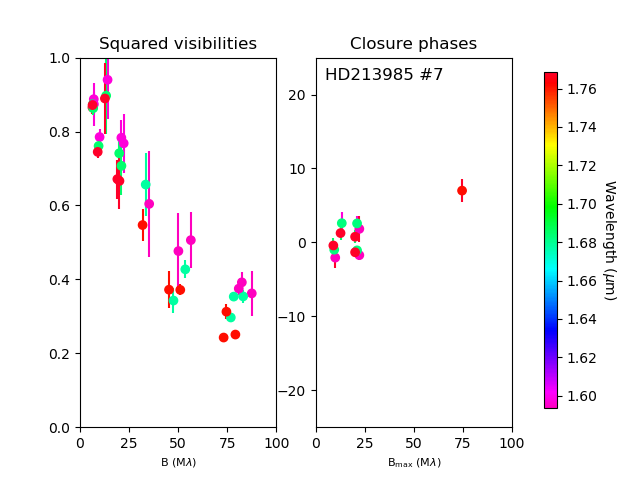}
\includegraphics[width=6.5cm]{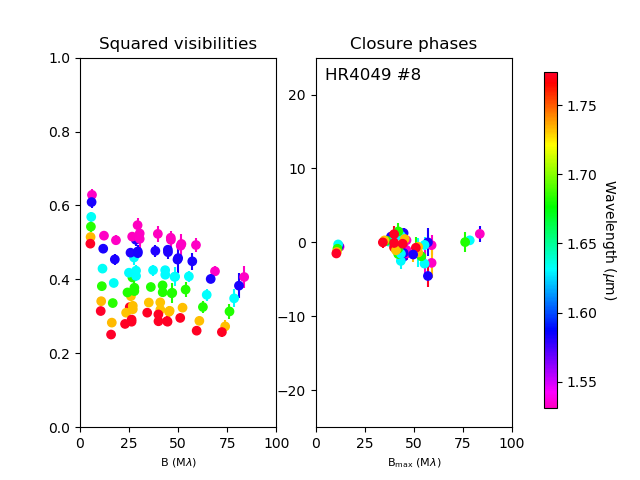}
\caption{Squared visibilities and closure phases for each target. The color represent the wavelength.}
\label{fig:data1}
\end{figure*}

\begin{figure*}
\centering
\includegraphics[width=6.5cm]{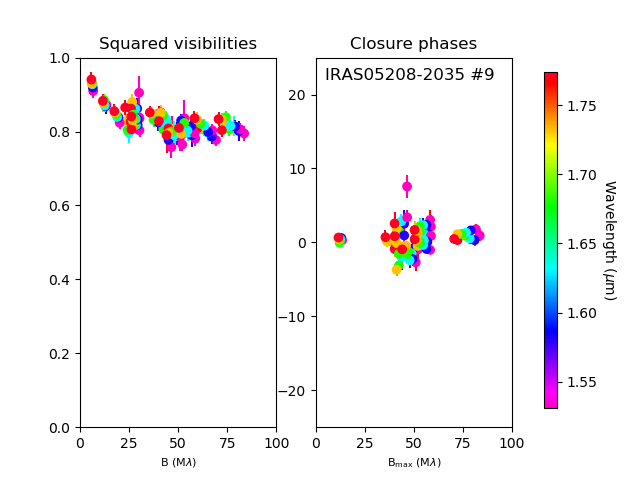}
\includegraphics[width=6.5cm]{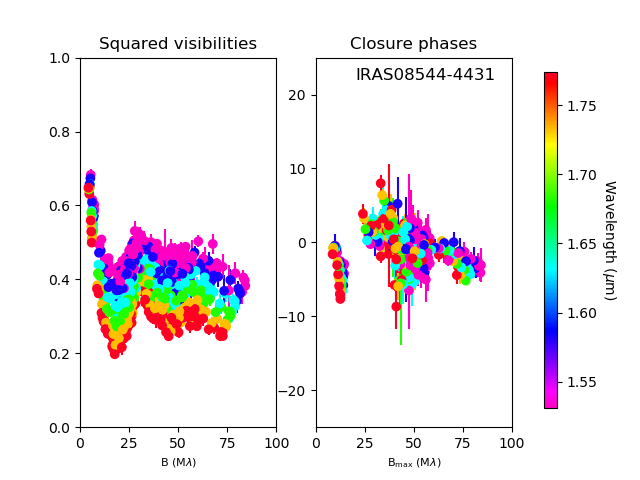}
\includegraphics[width=6.5cm]{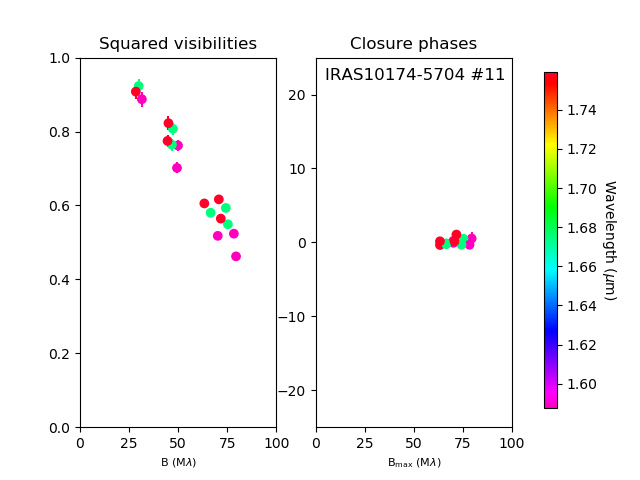}
\includegraphics[width=6.5cm]{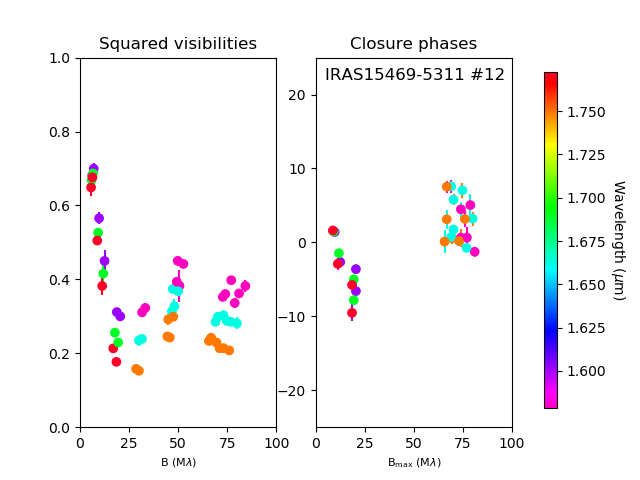}
\includegraphics[width=6.5cm]{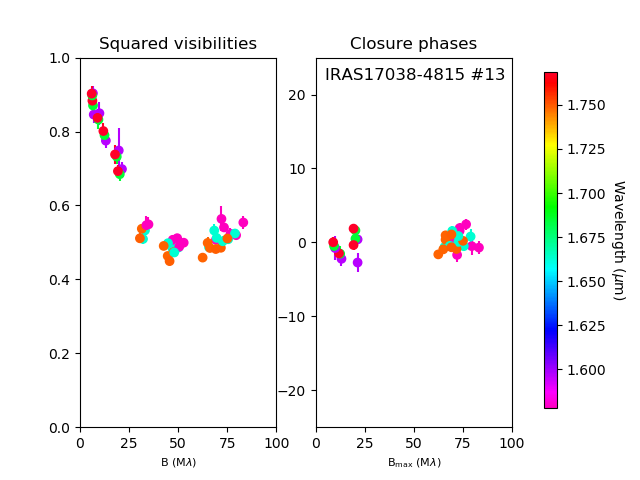}
\includegraphics[width=6.5cm]{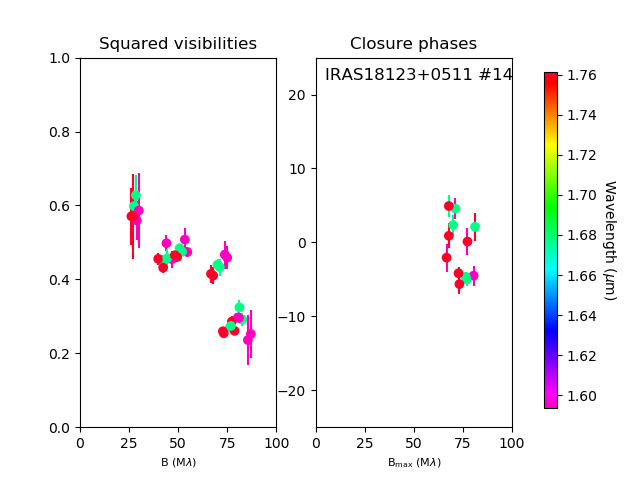}
\includegraphics[width=6.5cm]{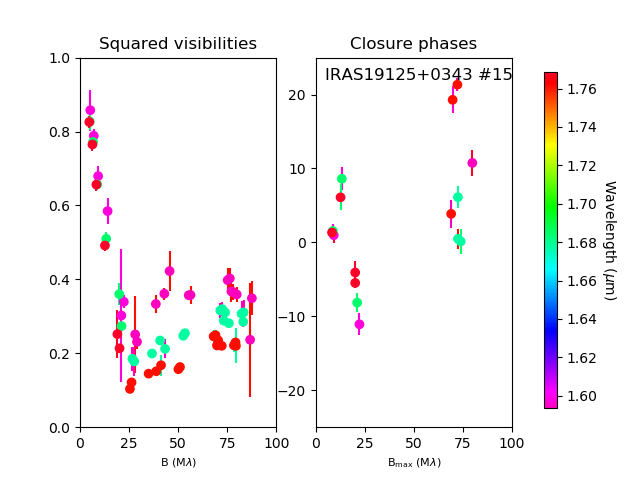}
\includegraphics[width=6.5cm]{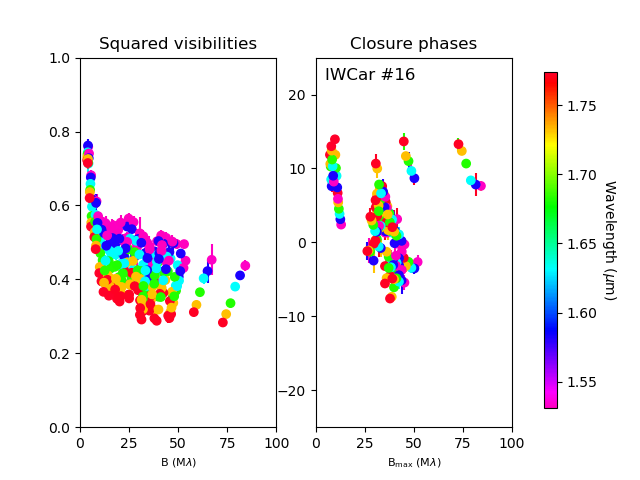}
\caption{Same as Fig.\,\ref{fig:data1}.}
\label{fig:data2}
\end{figure*}

\begin{figure*}
\centering
\includegraphics[width=6.5cm]{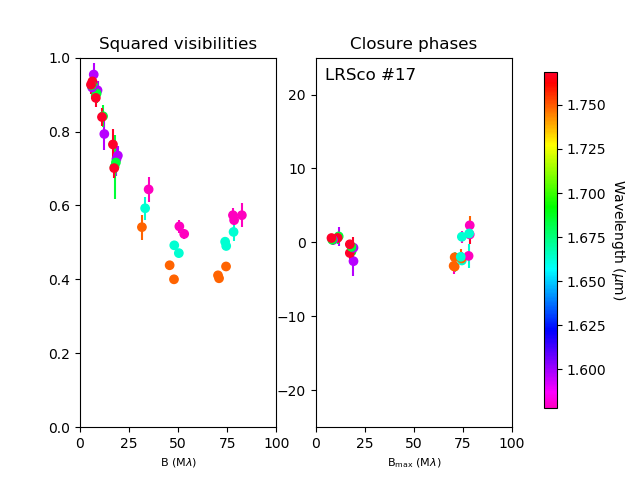}
\includegraphics[width=6.5cm]{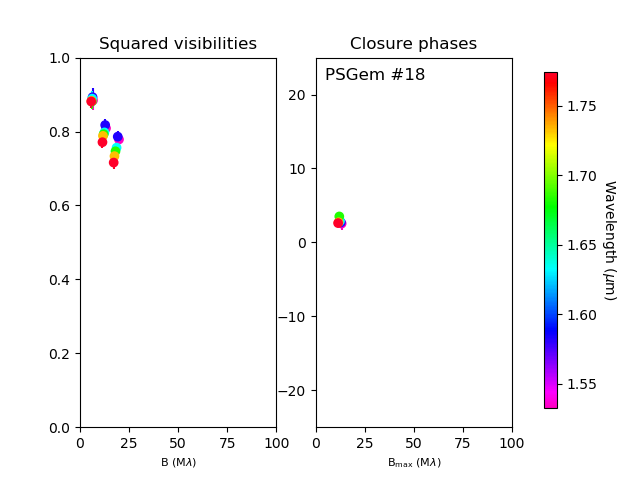}
\includegraphics[width=6.5cm]{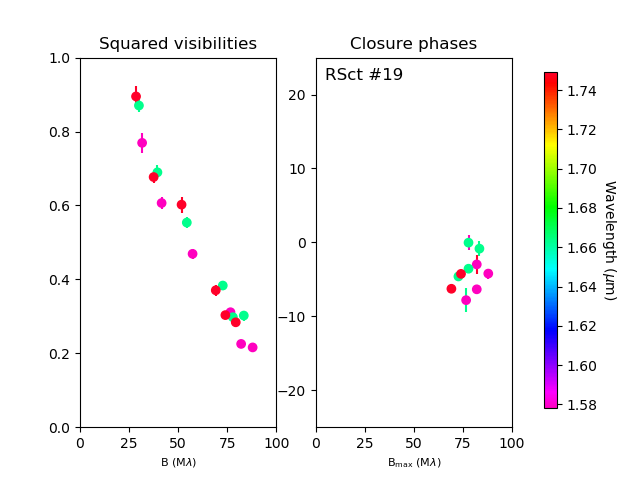}
\includegraphics[width=6.5cm]{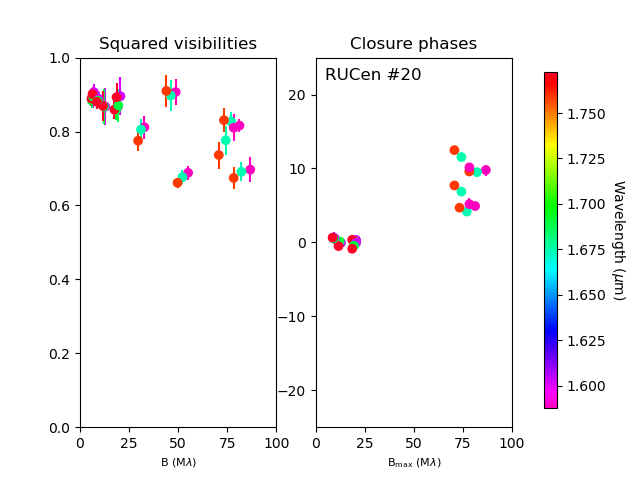}
\includegraphics[width=6.5cm]{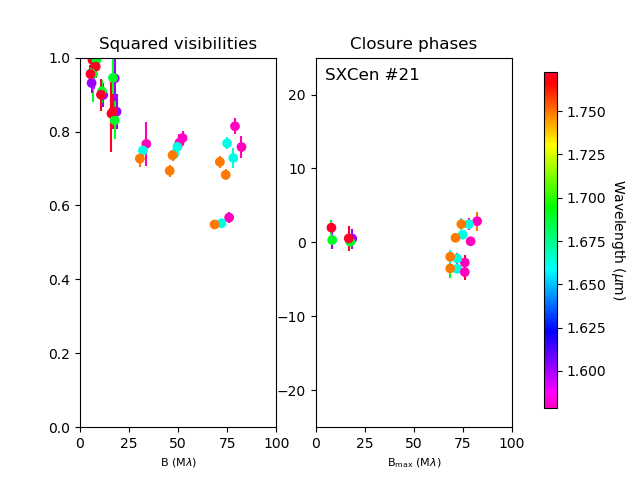}
\includegraphics[width=6.5cm]{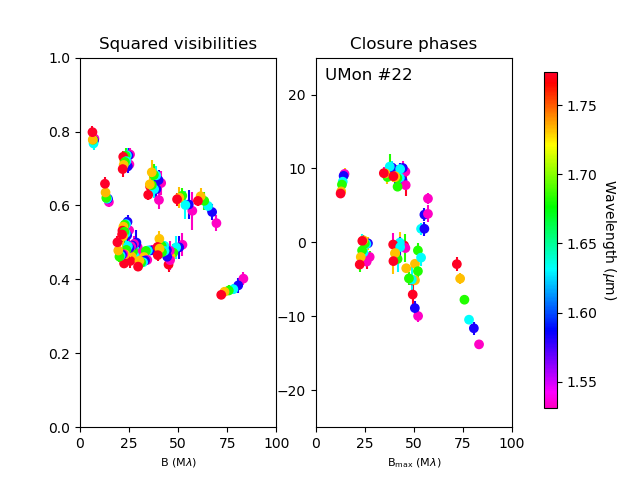}
\includegraphics[width=6.5cm]{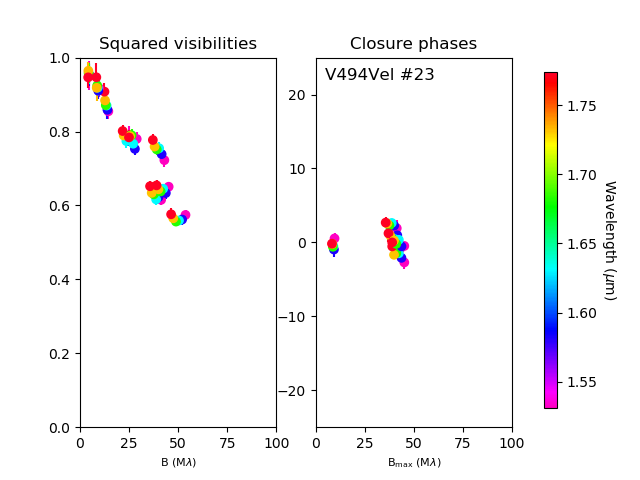}
\caption{Same as Fig.\,\ref{fig:data1}.}
\label{fig:data3}
\end{figure*}

\begin{figure*}
\centering
\includegraphics[width=4.5cm]{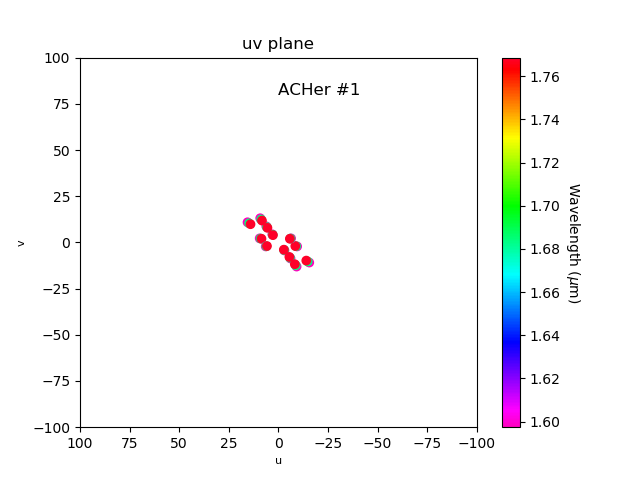}
\includegraphics[width=4.5cm]{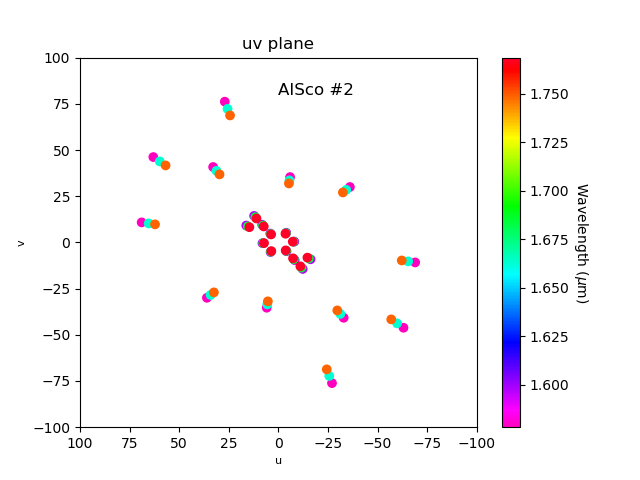}
\includegraphics[width=4.5cm]{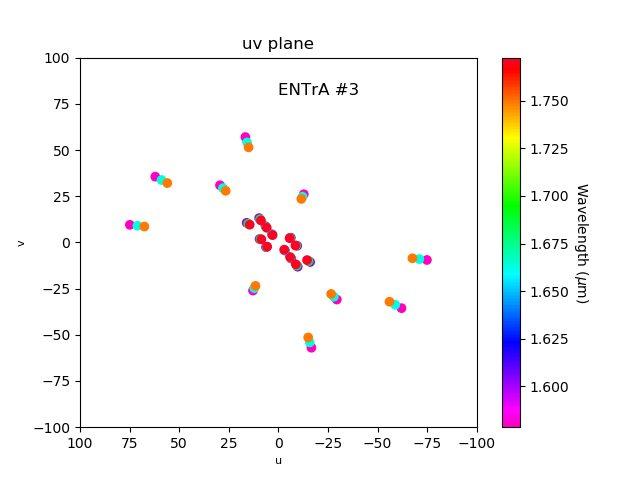}
\includegraphics[width=4.5cm]{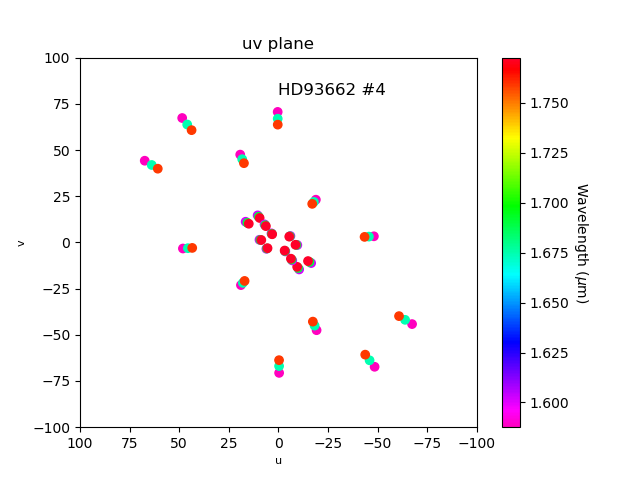}
\includegraphics[width=4.5cm]{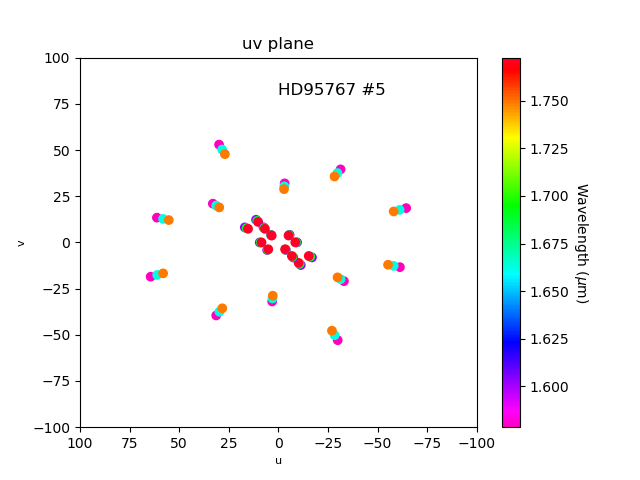}
\includegraphics[width=4.5cm]{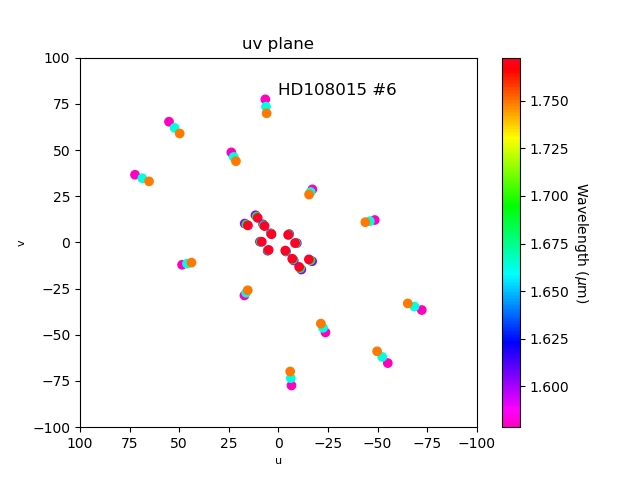}
\includegraphics[width=4.5cm]{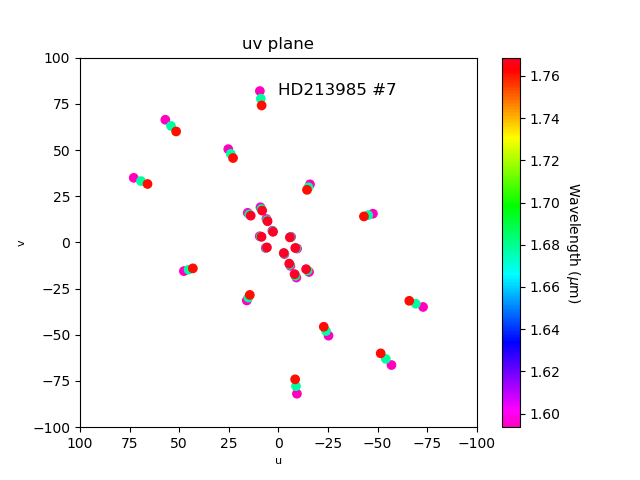}
\includegraphics[width=4.5cm]{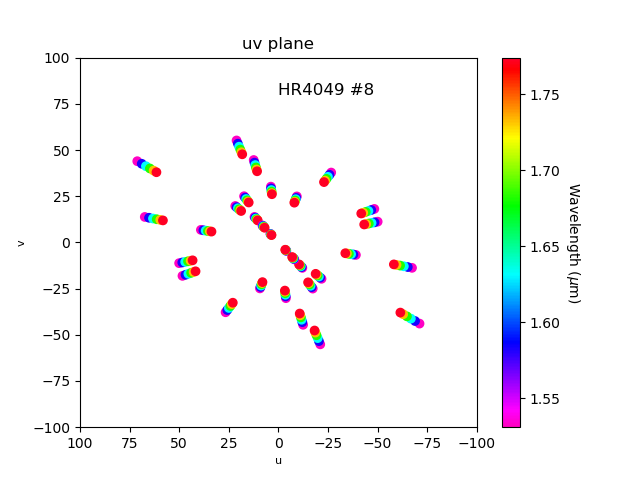}
\includegraphics[width=4.5cm]{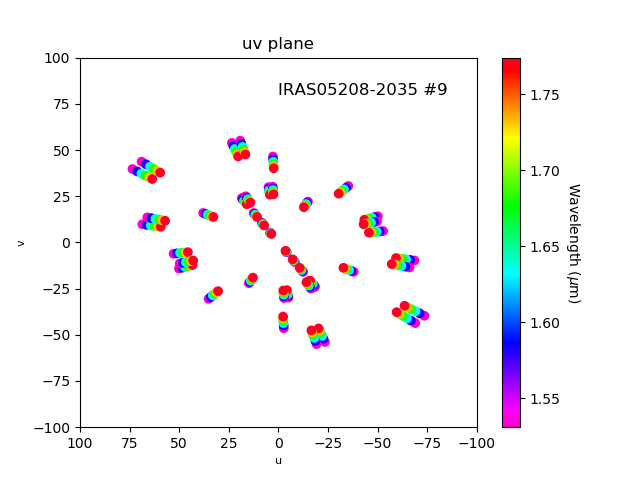}
\includegraphics[width=4.5cm]{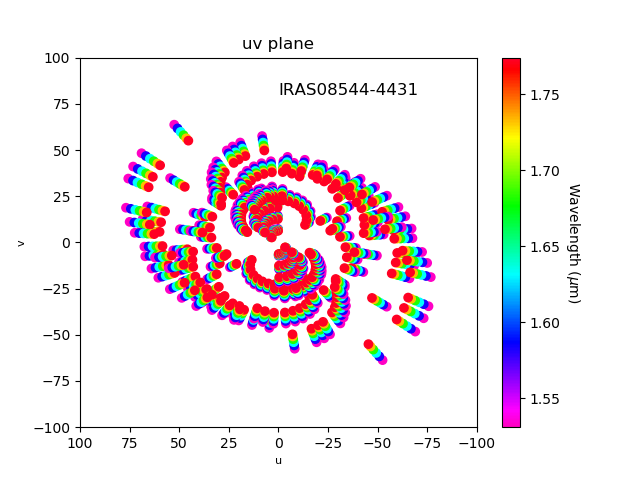}
\includegraphics[width=4.5cm]{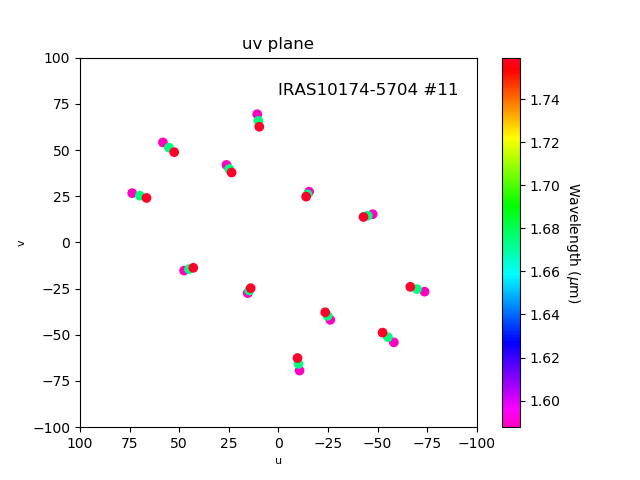}
\includegraphics[width=4.5cm]{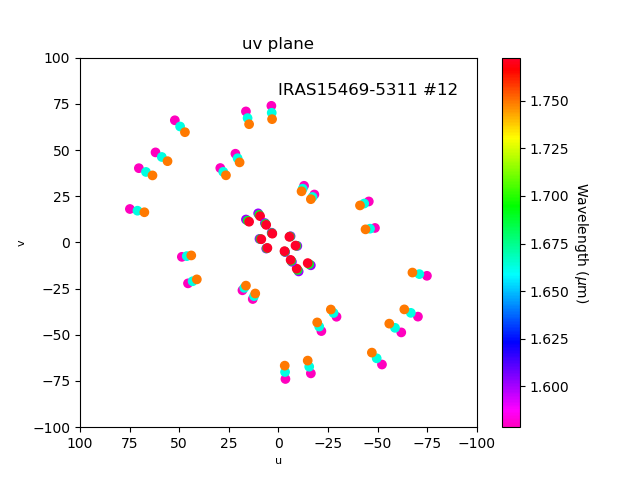}
\includegraphics[width=4.5cm]{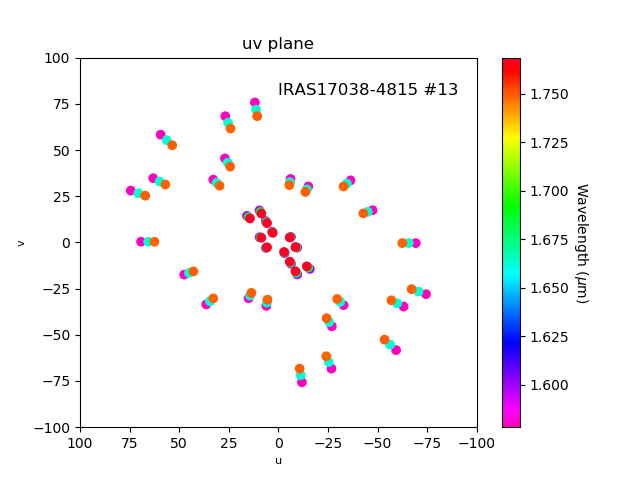}
\includegraphics[width=4.5cm]{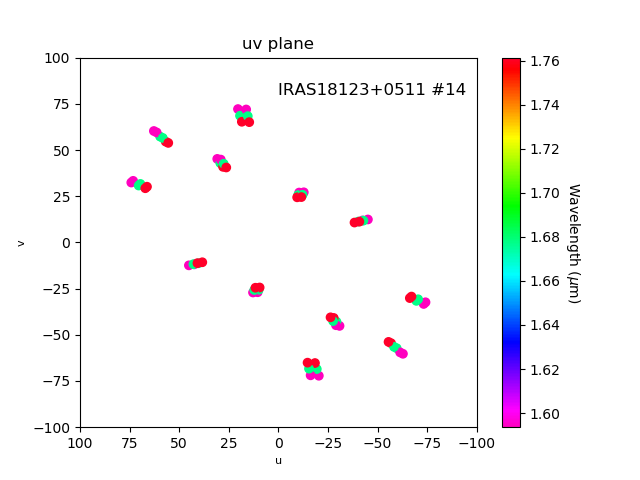}
\includegraphics[width=4.5cm]{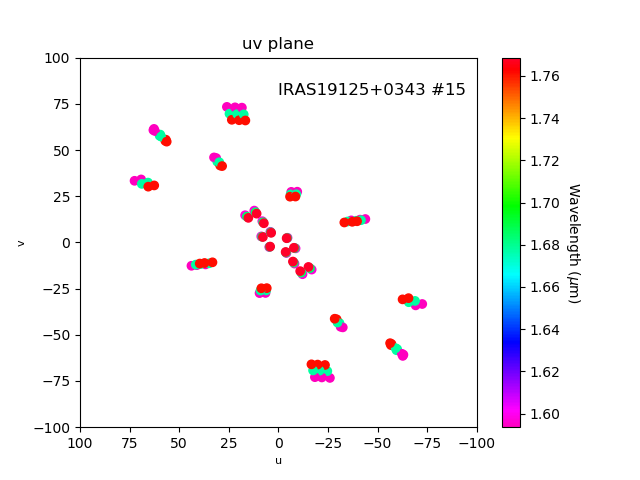}
\includegraphics[width=4.5cm]{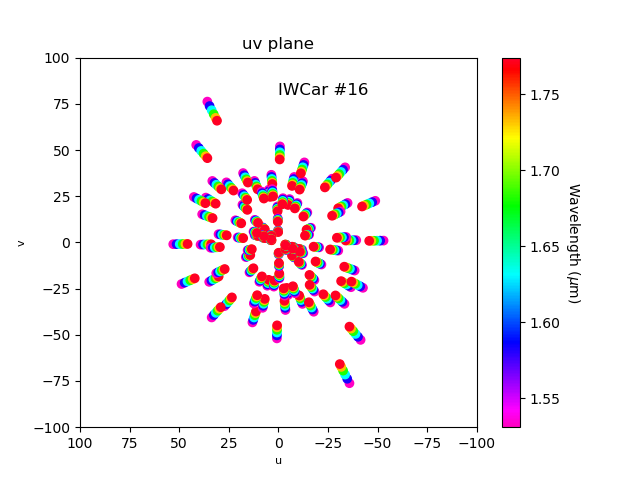}
\includegraphics[width=4.5cm]{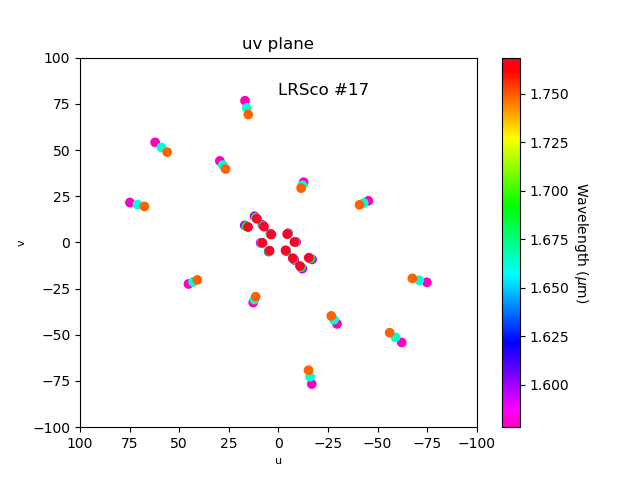}
\includegraphics[width=4.5cm]{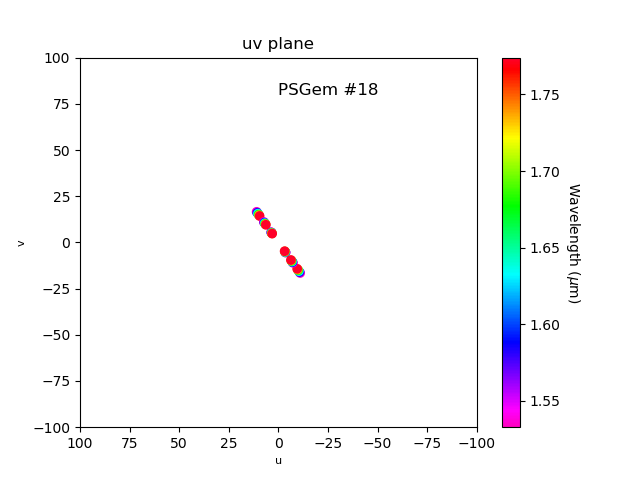}
\caption{(u, v)-plane for each target.}
\label{fig:uv1}
\end{figure*}

\begin{figure*}
\centering
\includegraphics[width=5cm]{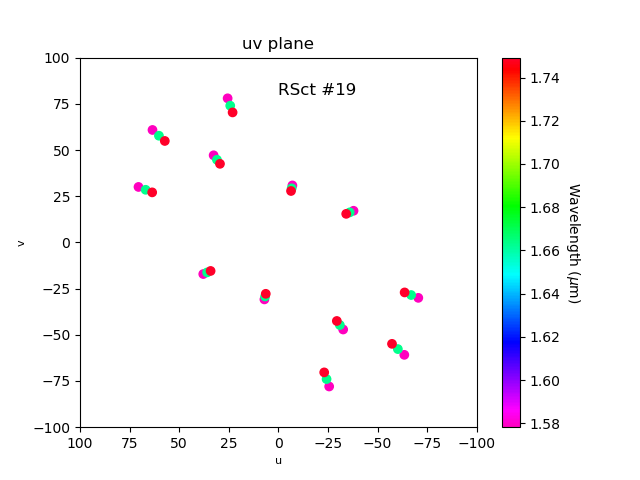}
\includegraphics[width=5cm]{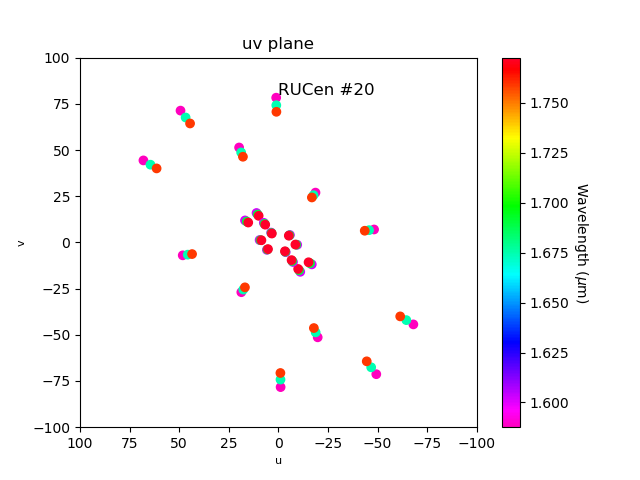}
\includegraphics[width=5cm]{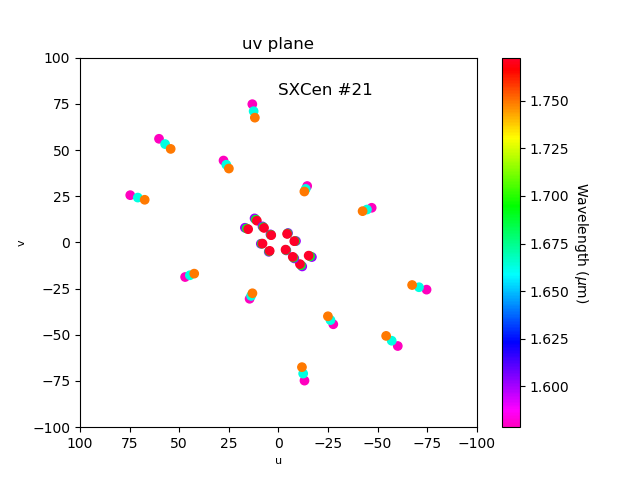}
\includegraphics[width=5cm]{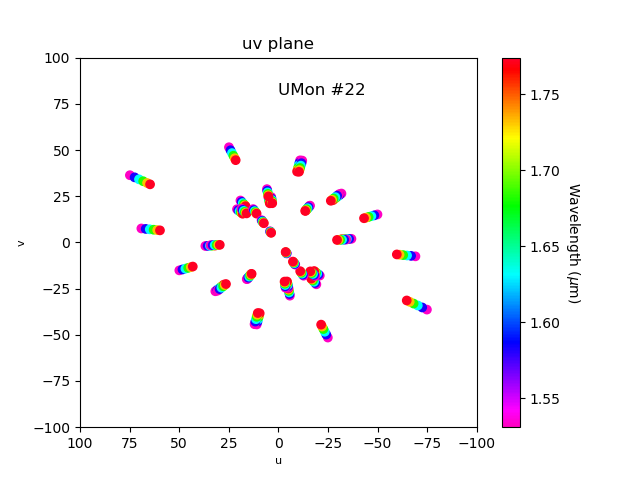}
\includegraphics[width=5cm]{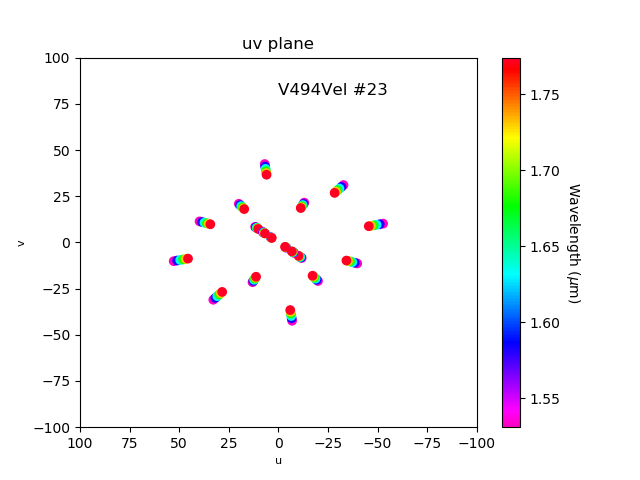}
\caption{Same as Fig.\,\ref{fig:uv1}.}
\label{fig:uv2}
\end{figure*}

\section{\modif{Images of the targets where the most likely model has a disk.}}

\begin{figure*}
\centering
\includegraphics[width=6.5cm]{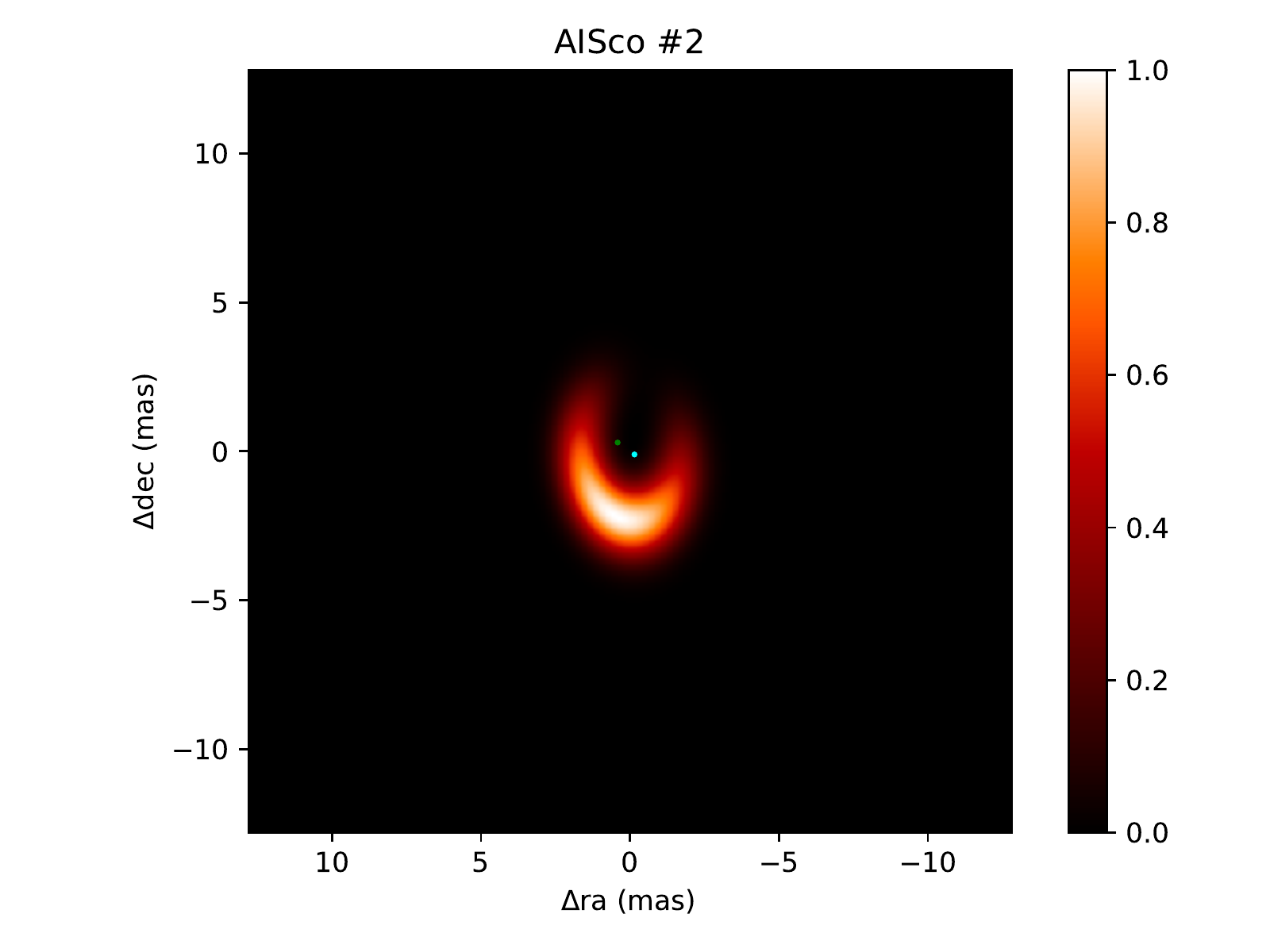}
\includegraphics[width=6.5cm]{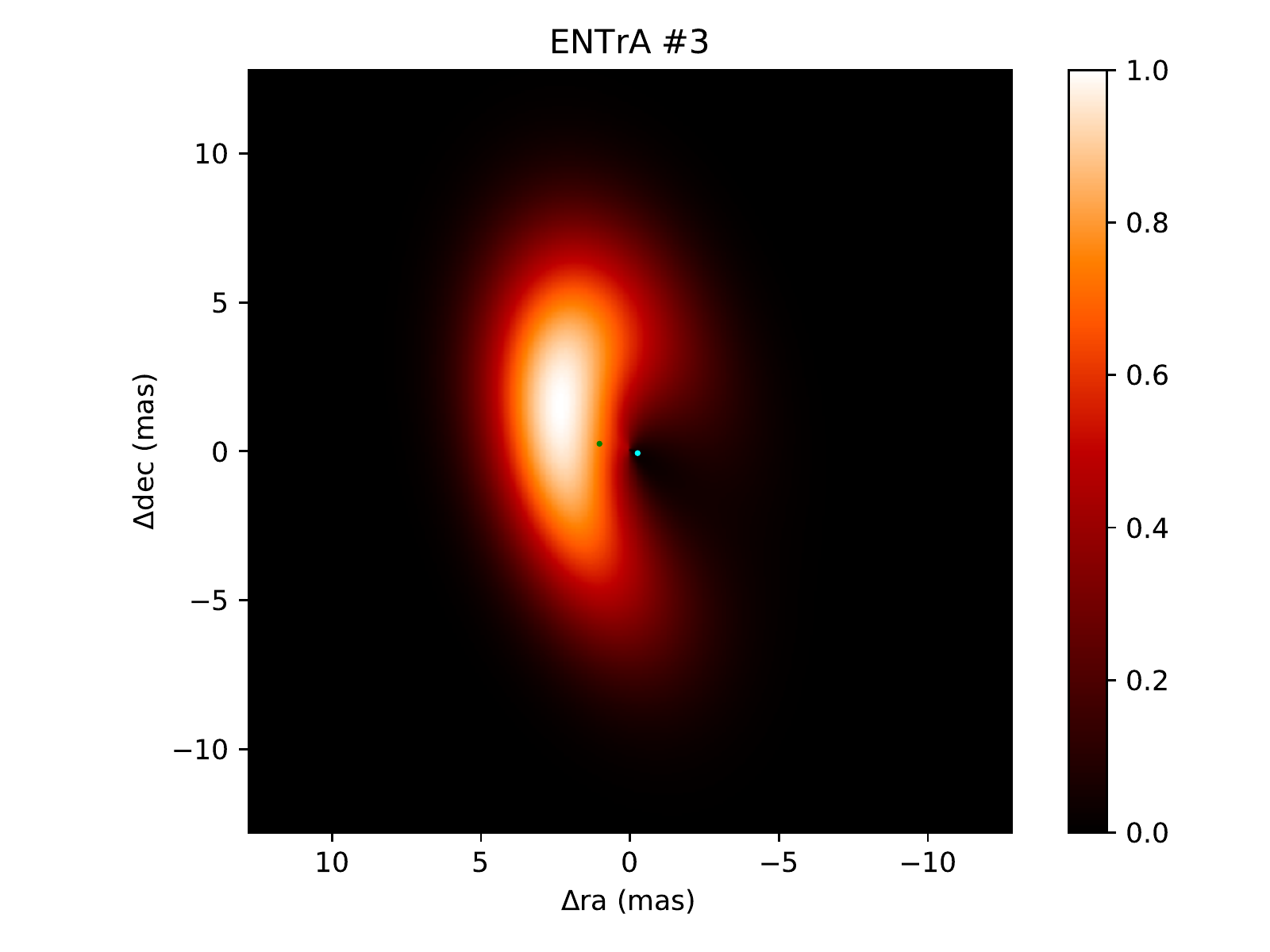}
\includegraphics[width=6.5cm]{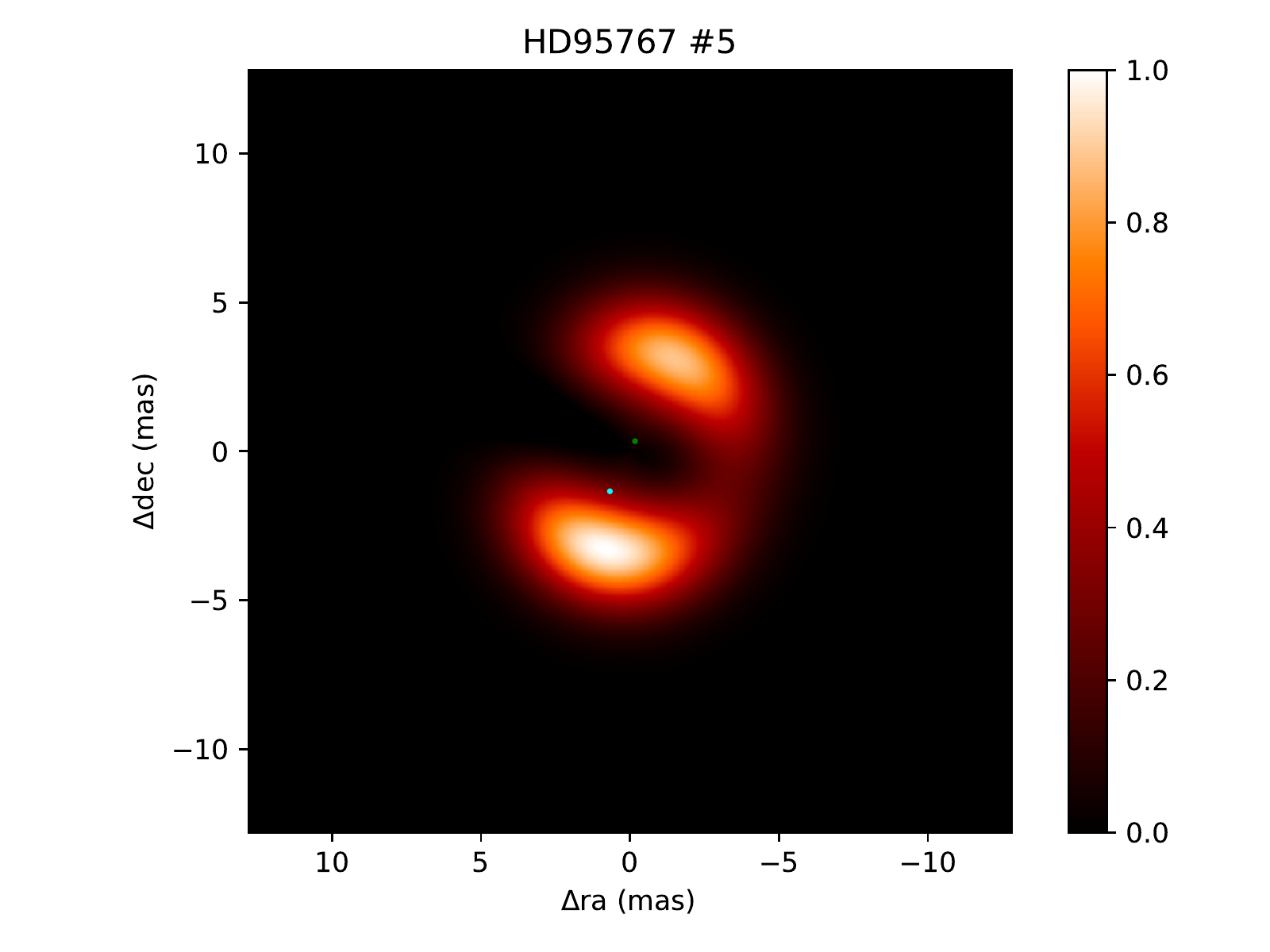}
\includegraphics[width=6.5cm]{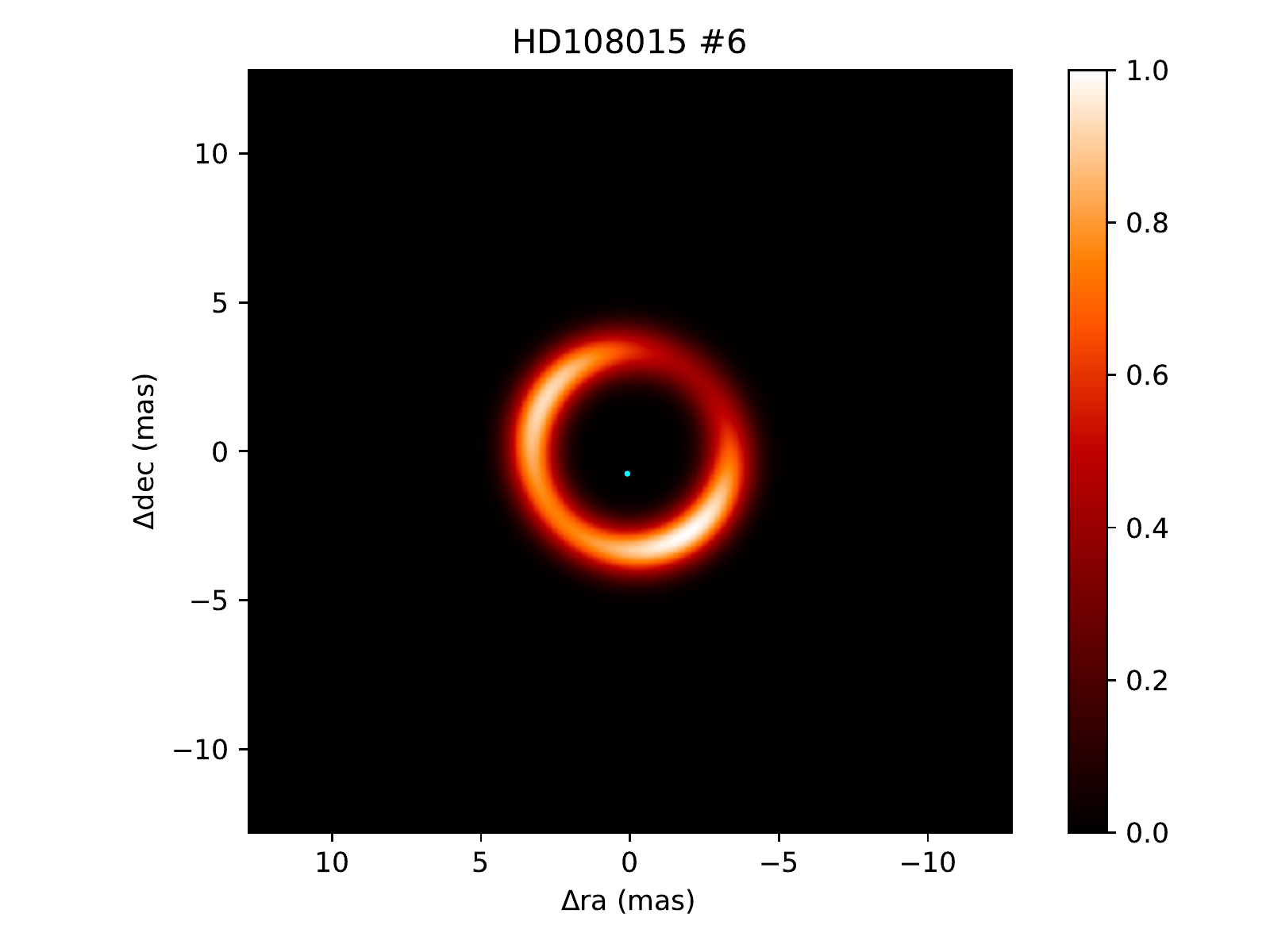}
\includegraphics[width=6.5cm]{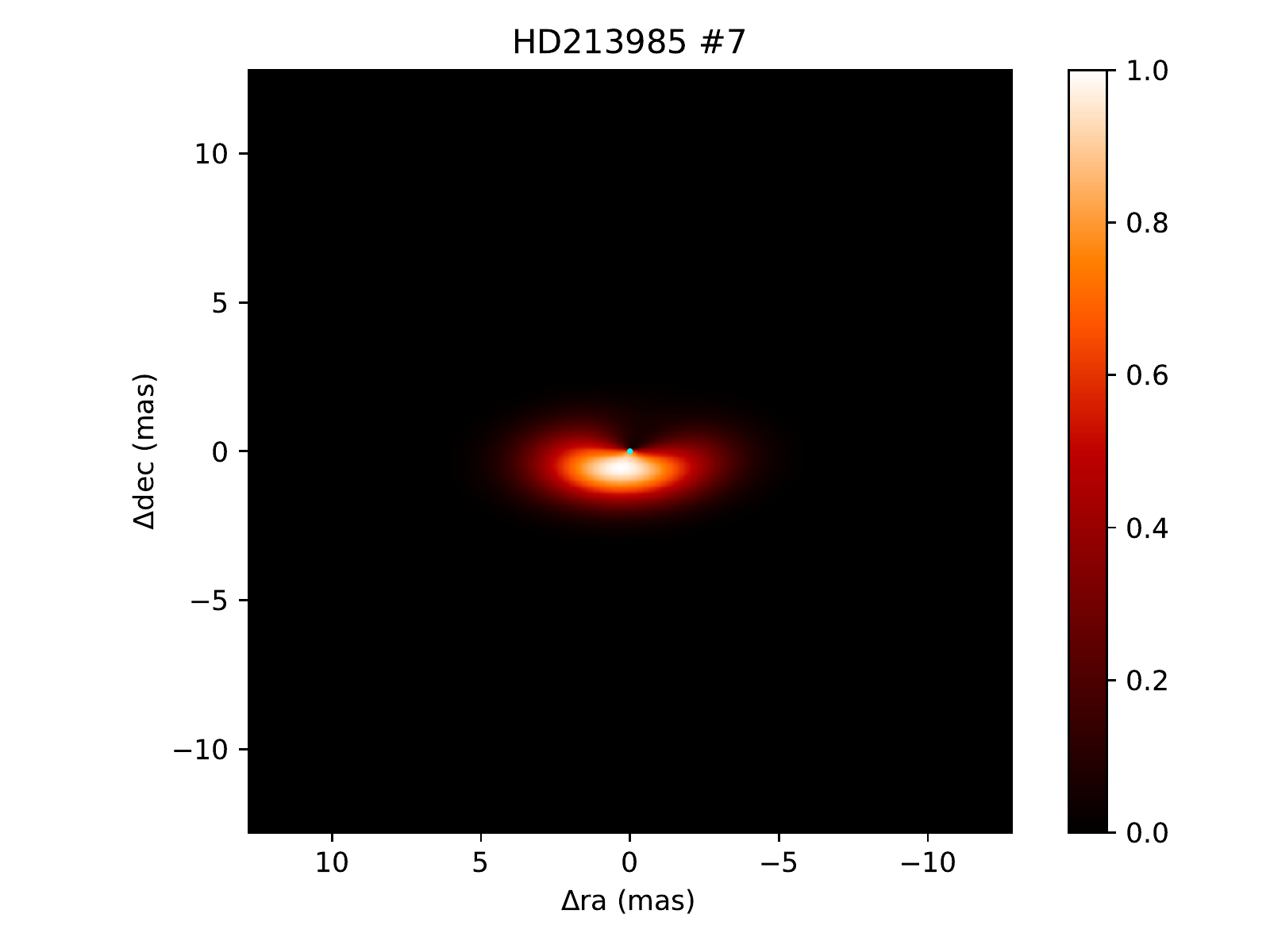}
\includegraphics[width=6.5cm]{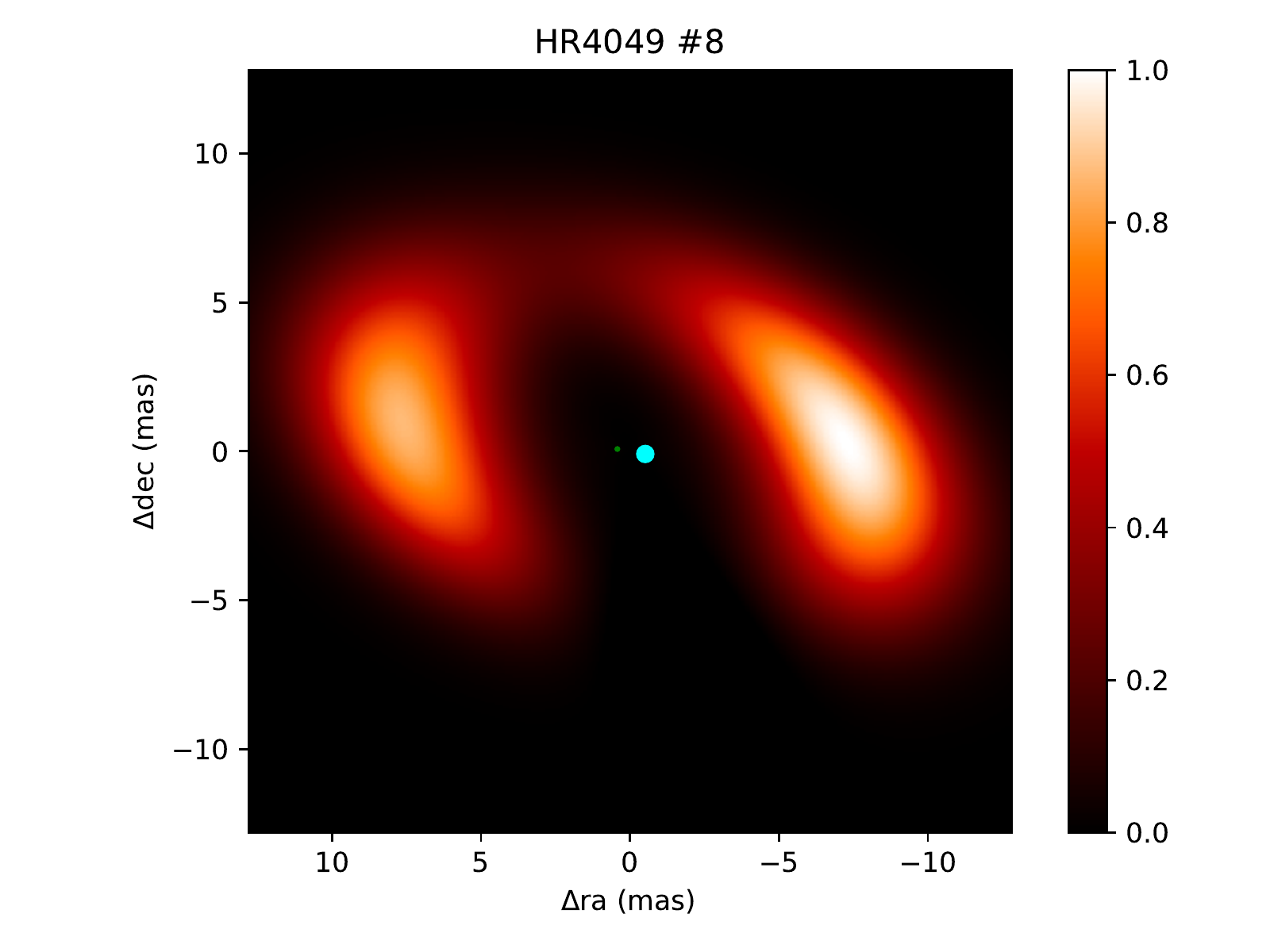}
\includegraphics[width=6.5cm]{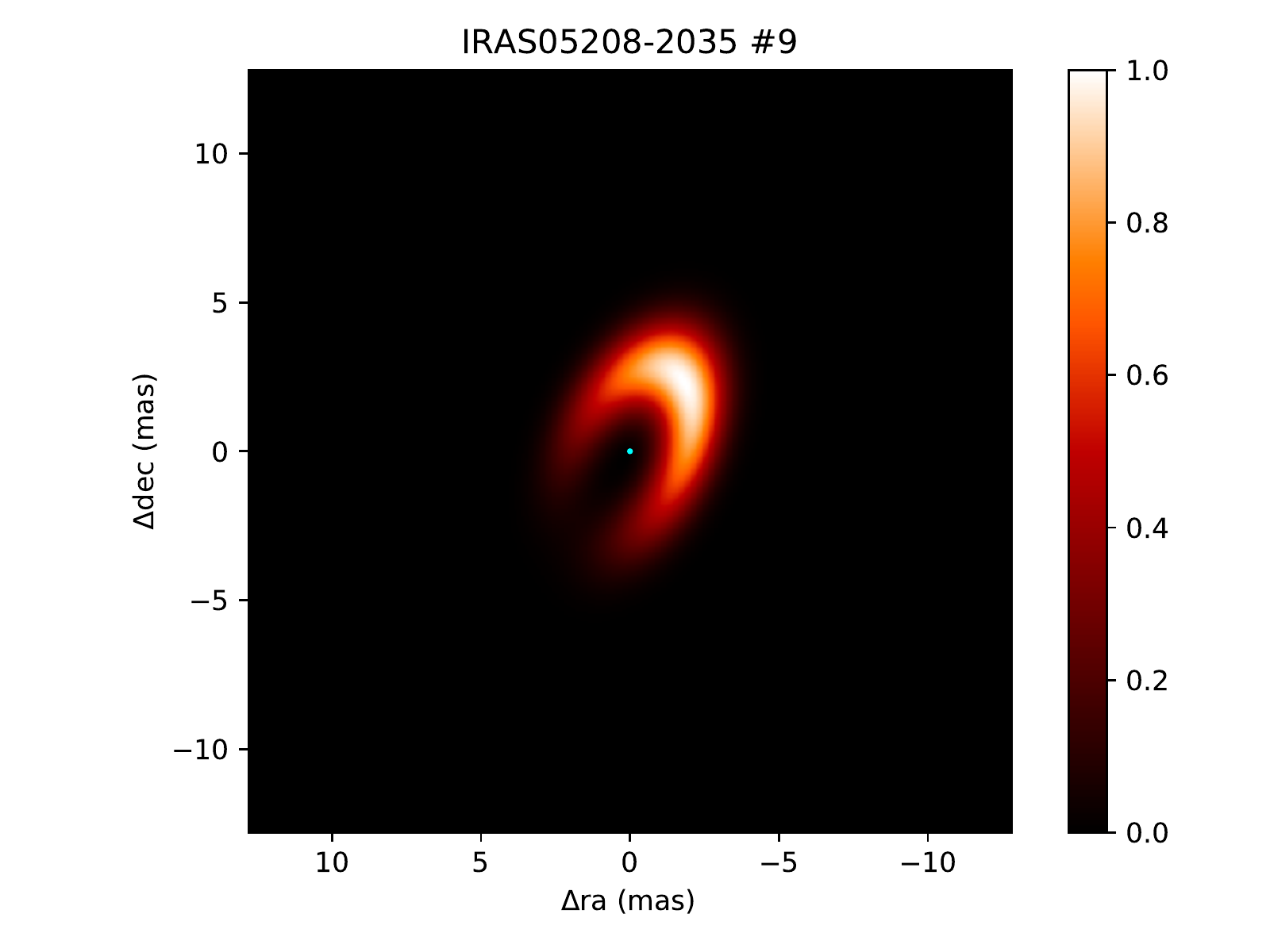}
\includegraphics[width=6.5cm]{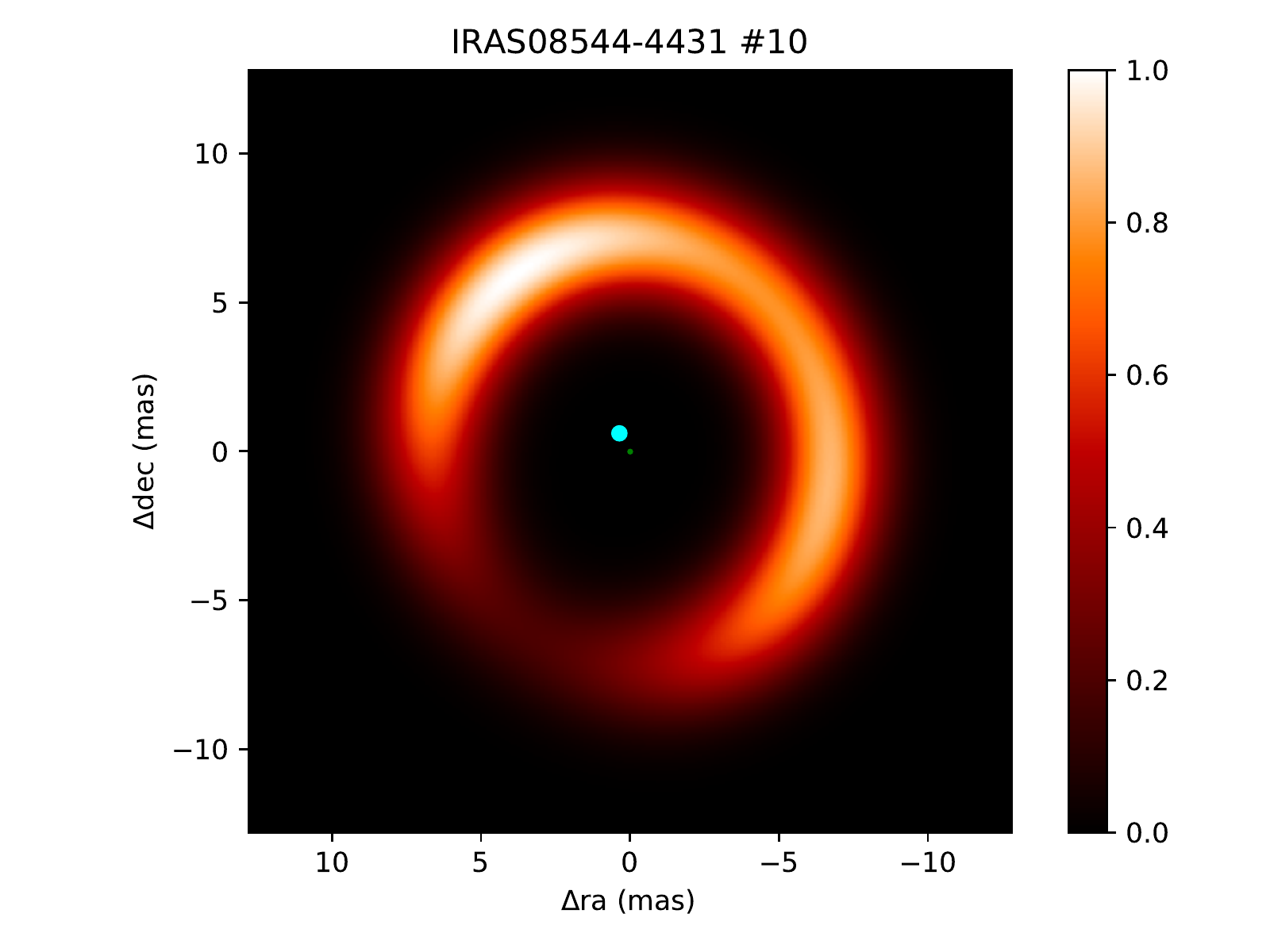}
\includegraphics[width=6.5cm]{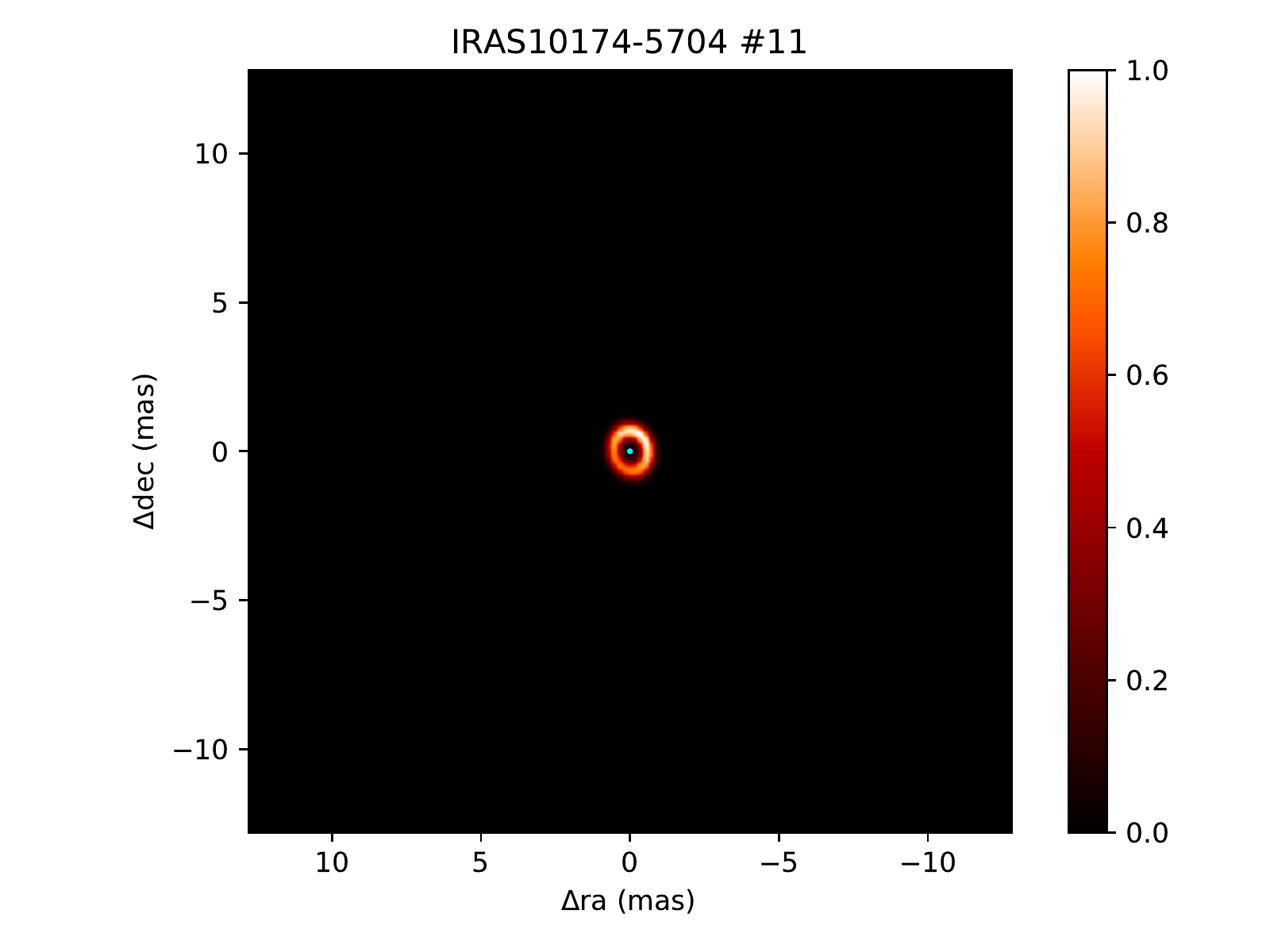}
\includegraphics[width=6.5cm]{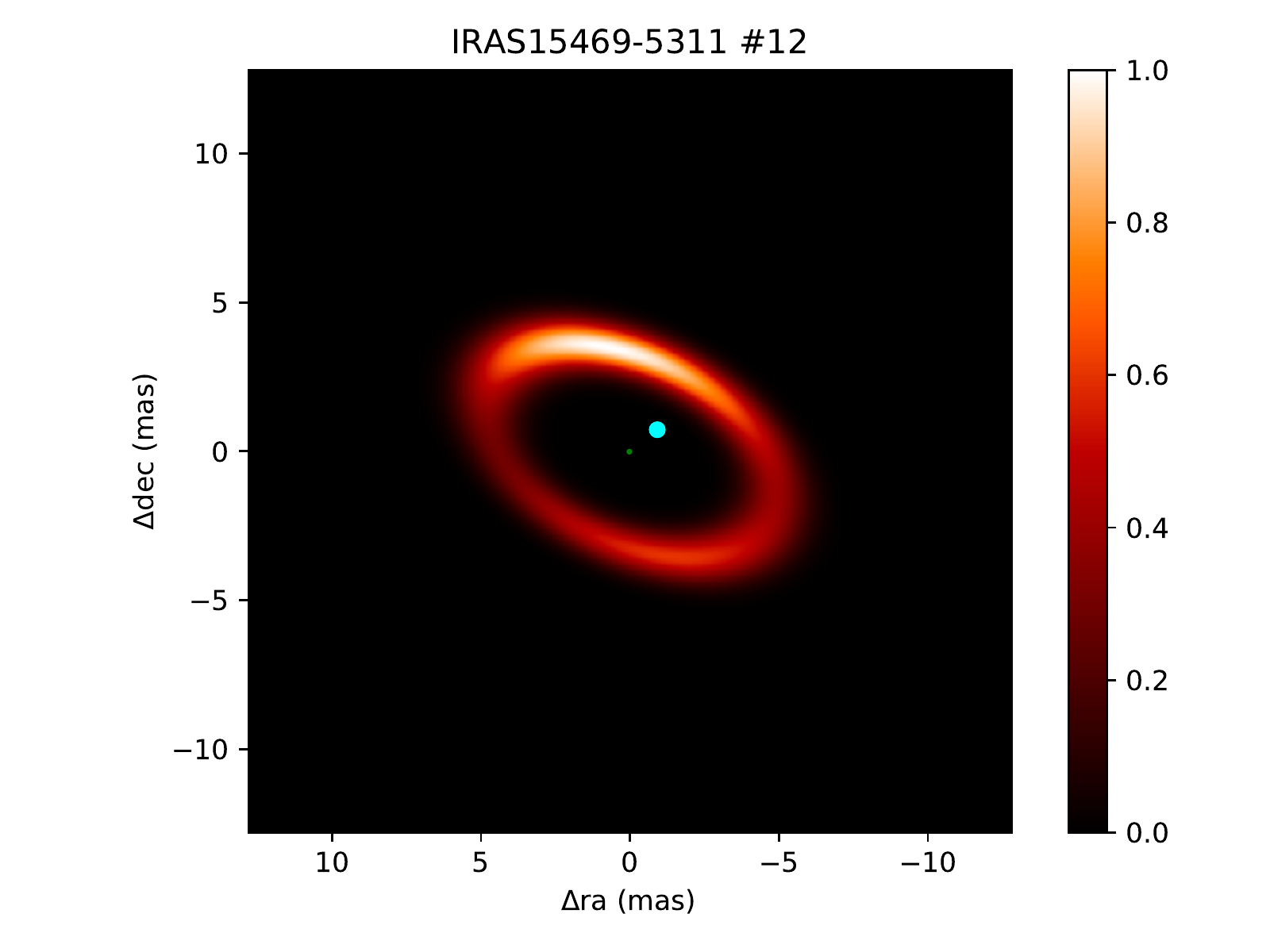}
\caption{Best fit model images. The green star represents the primary. The cyan star represents the secondary.}
\label{fig:modImages1}
\end{figure*}

\begin{figure*}
\centering
\includegraphics[width=6.5cm]{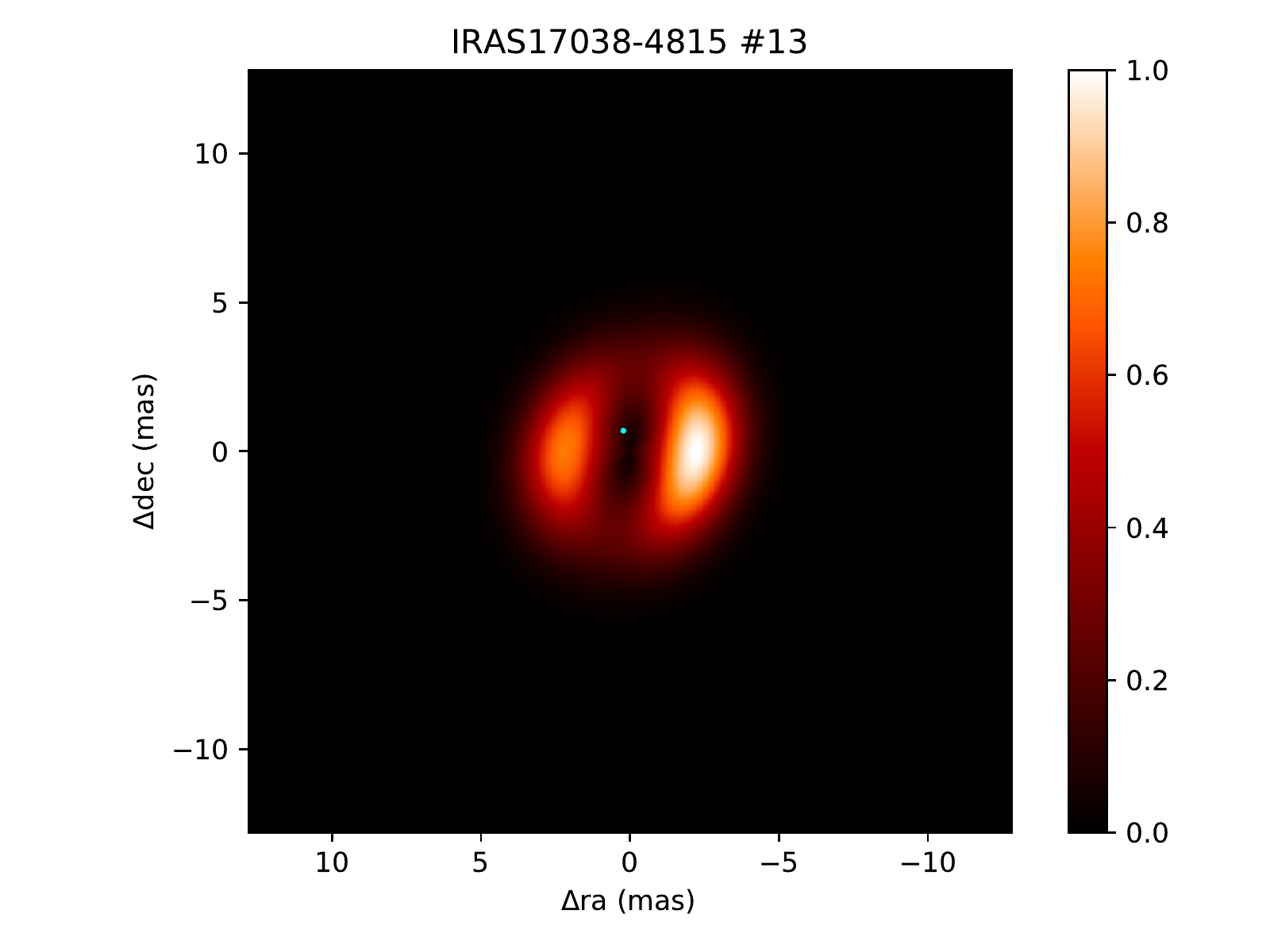}
\includegraphics[width=6.5cm]{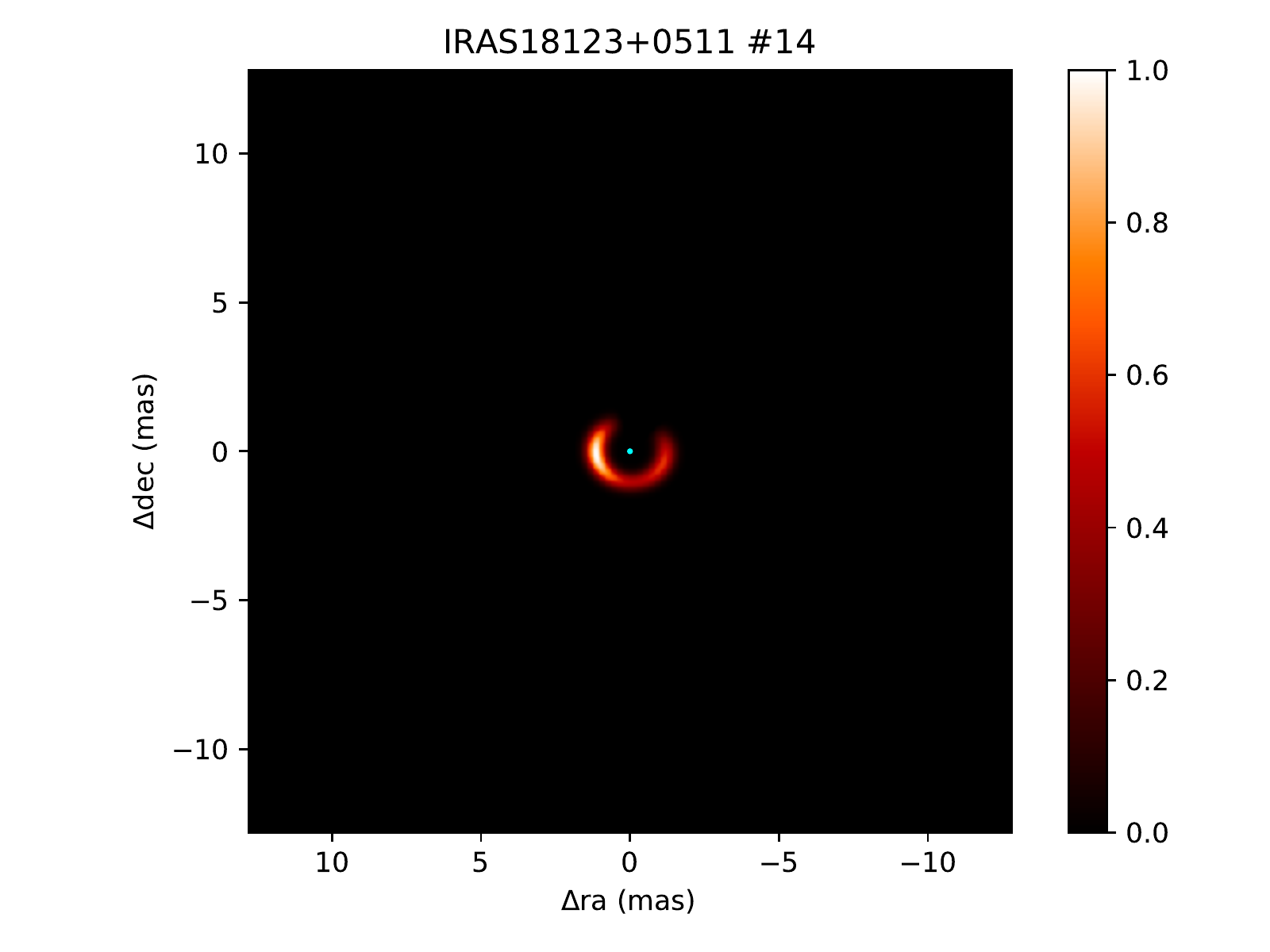}
\includegraphics[width=6.5cm]{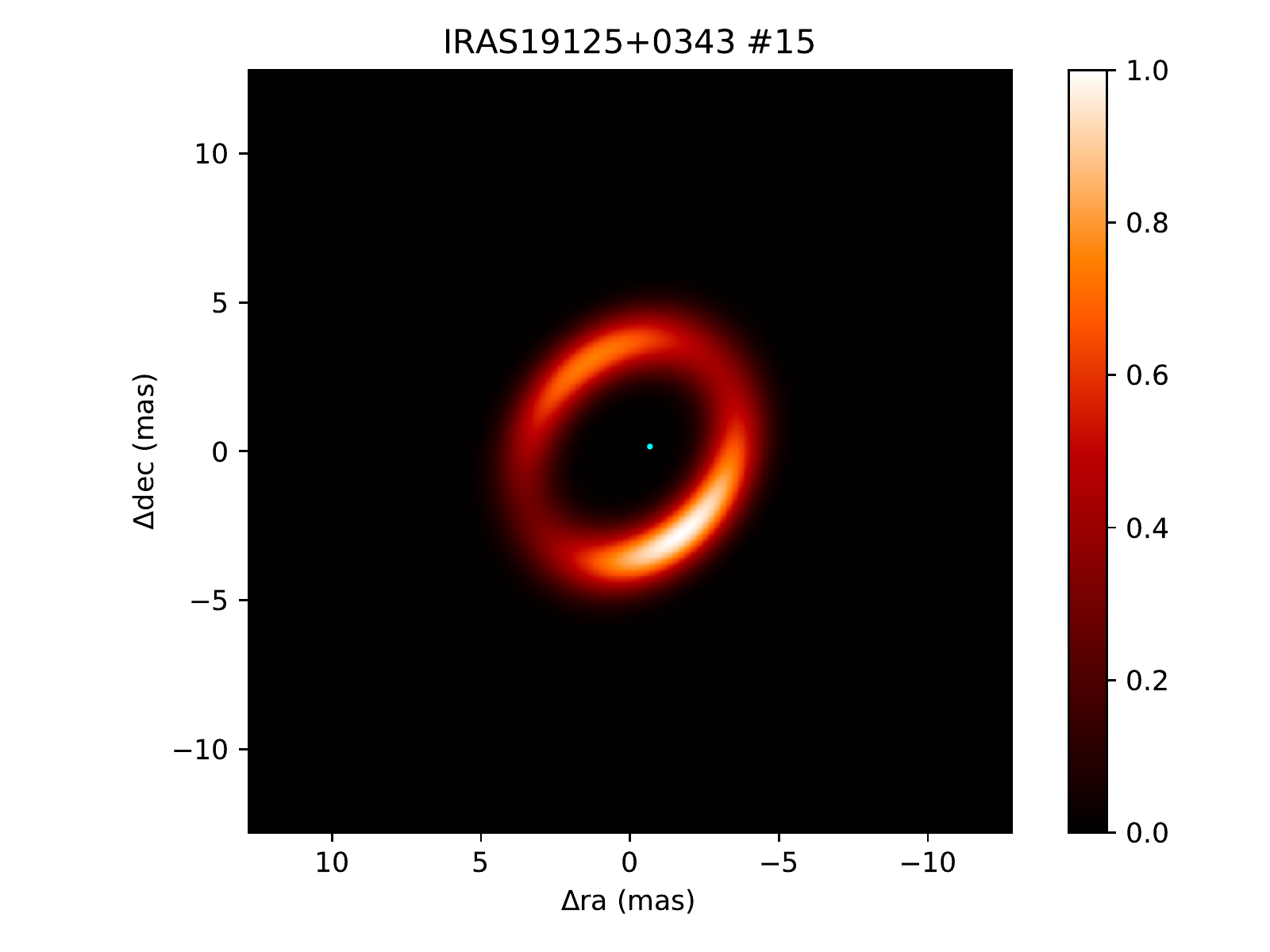}
\includegraphics[width=6.5cm]{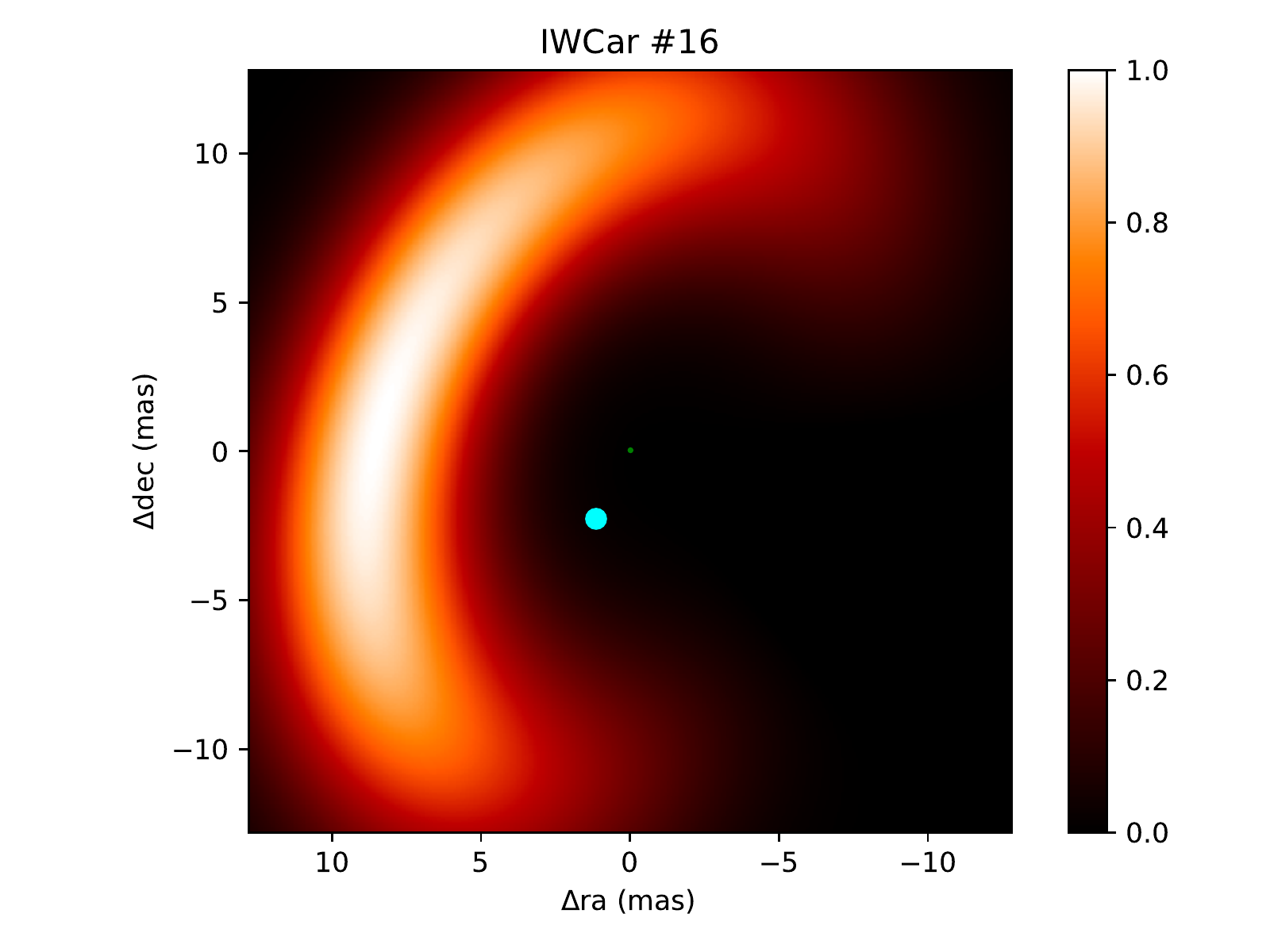}
\includegraphics[width=6.5cm]{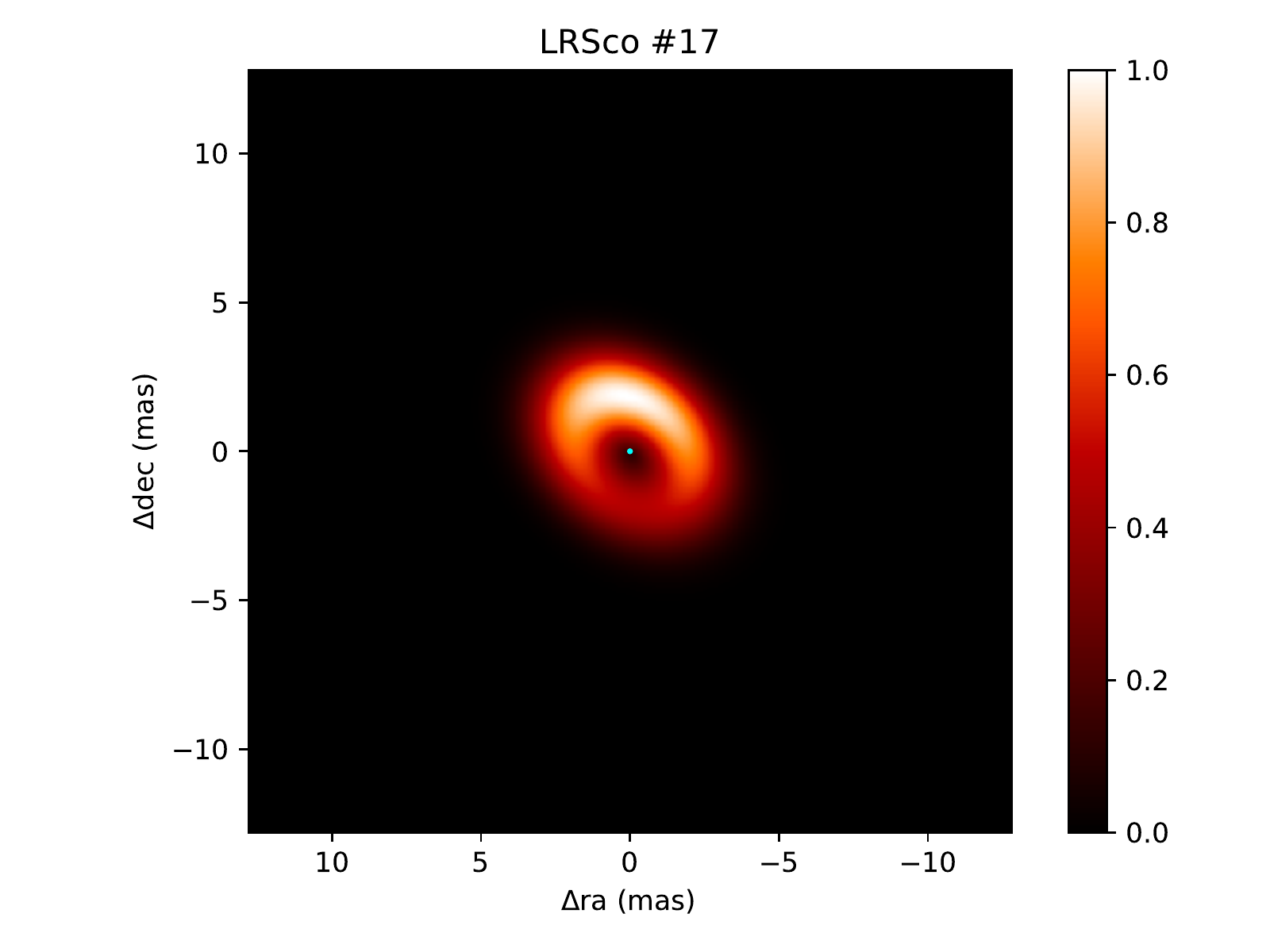}
\includegraphics[width=6.5cm]{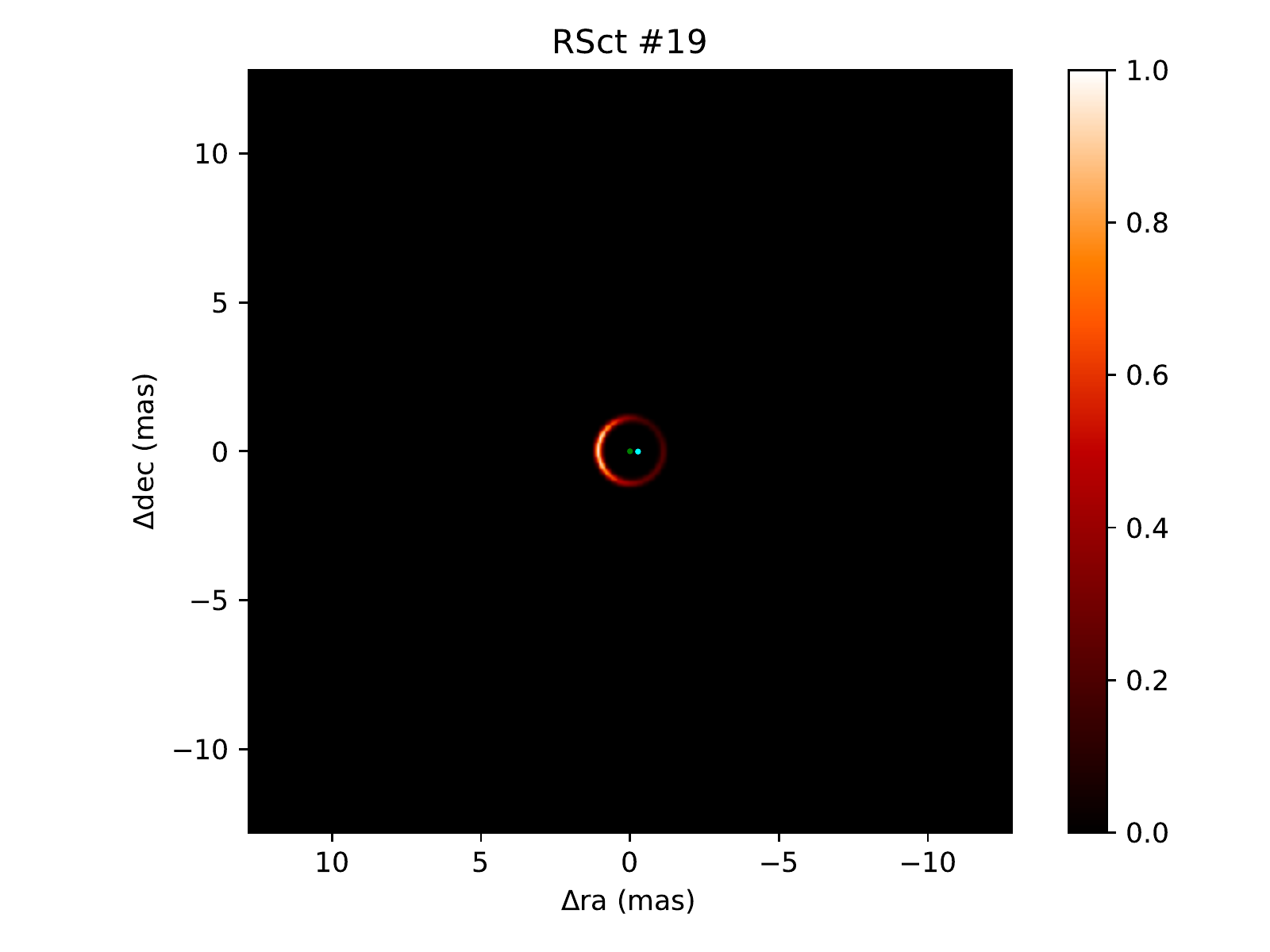}
\includegraphics[width=6.5cm]{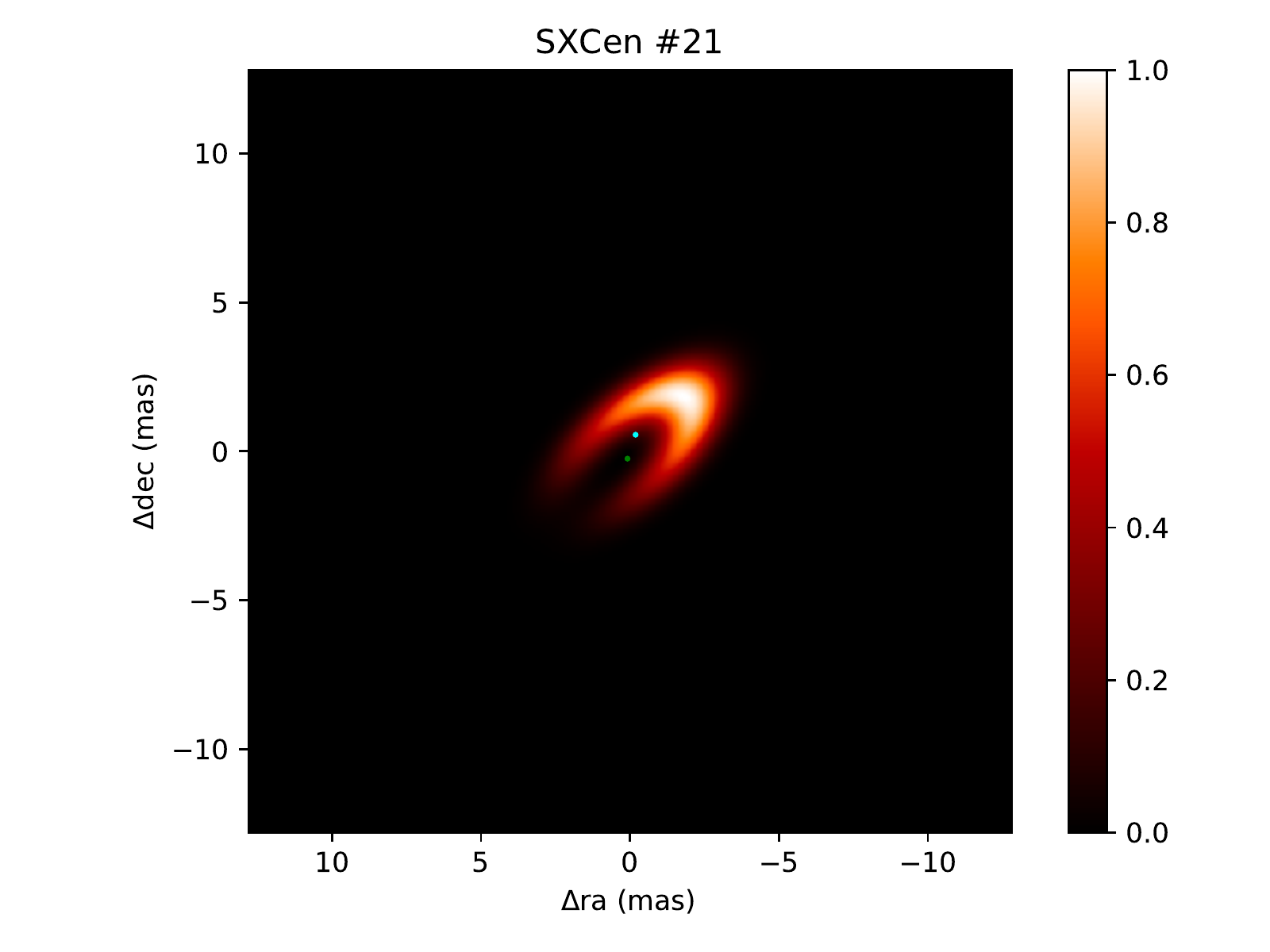}
\includegraphics[width=6.5cm]{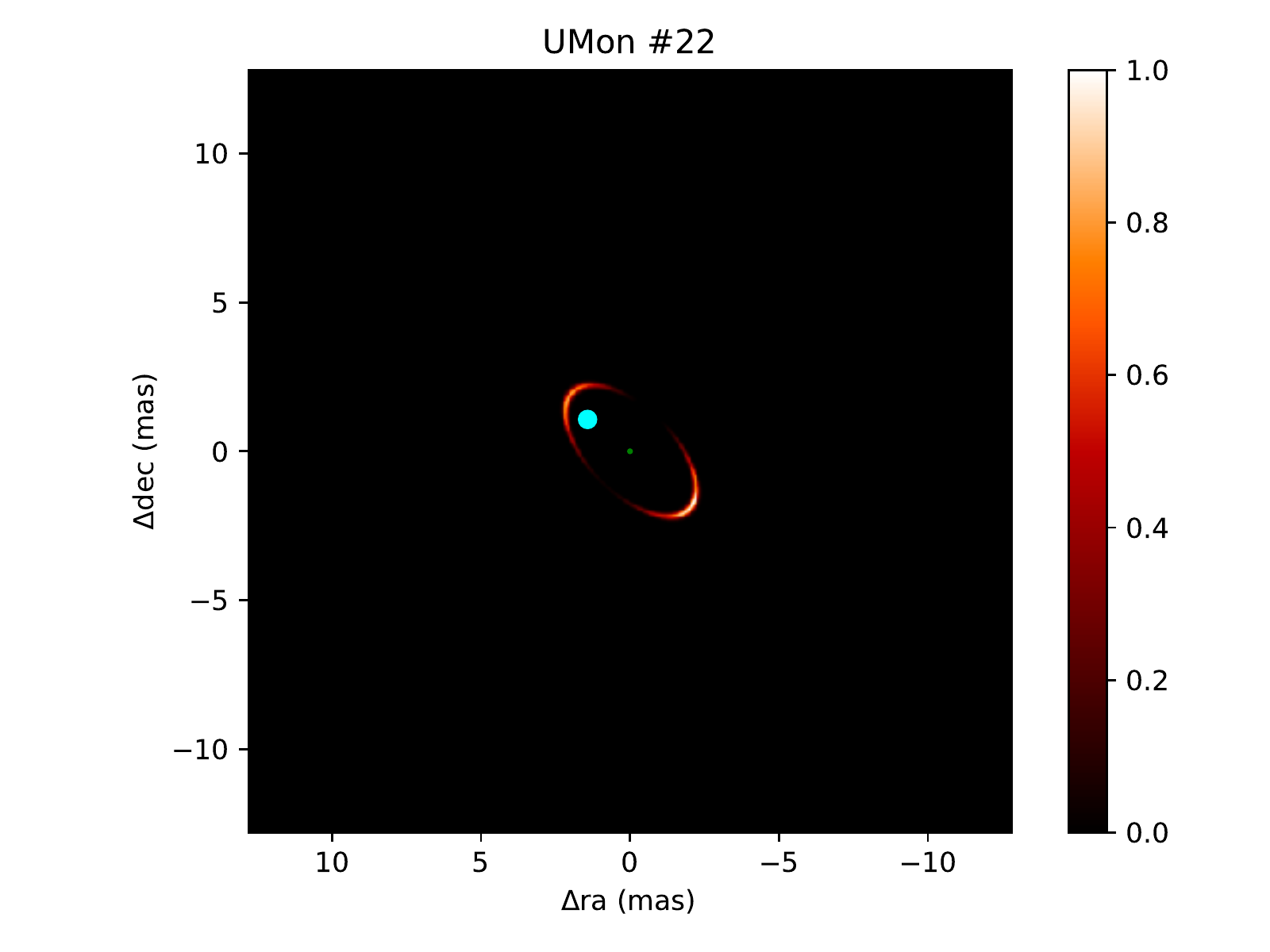}
\includegraphics[width=6.5cm]{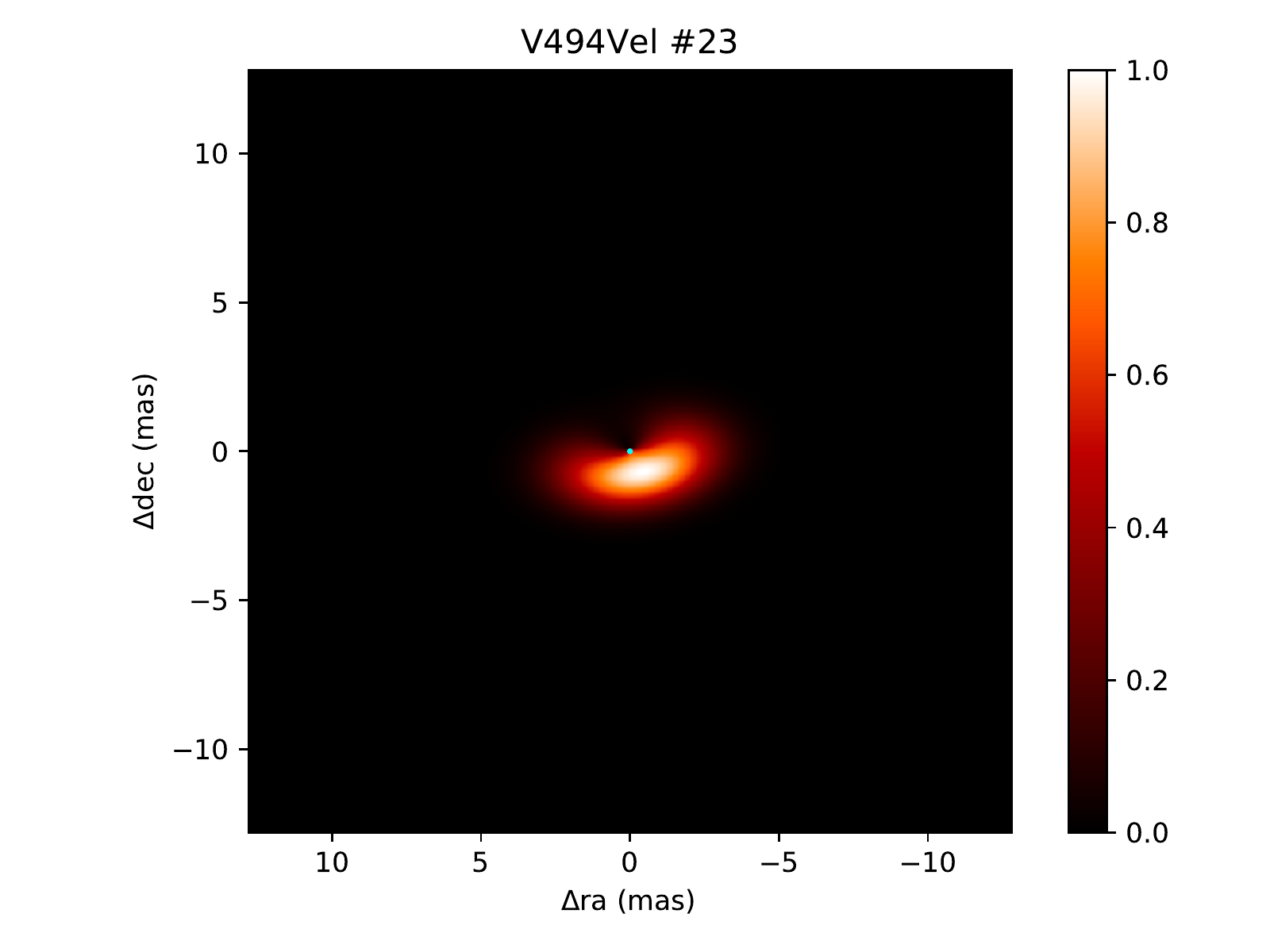}
\caption{Same as Fig.\,\ref{fig:modImages1}}
\label{fig:modImages2}
\end{figure*}

\section{\modif{Best-fit model comparison to the data.}}

\begin{figure*}
\centering
\includegraphics[width=6.5cm]{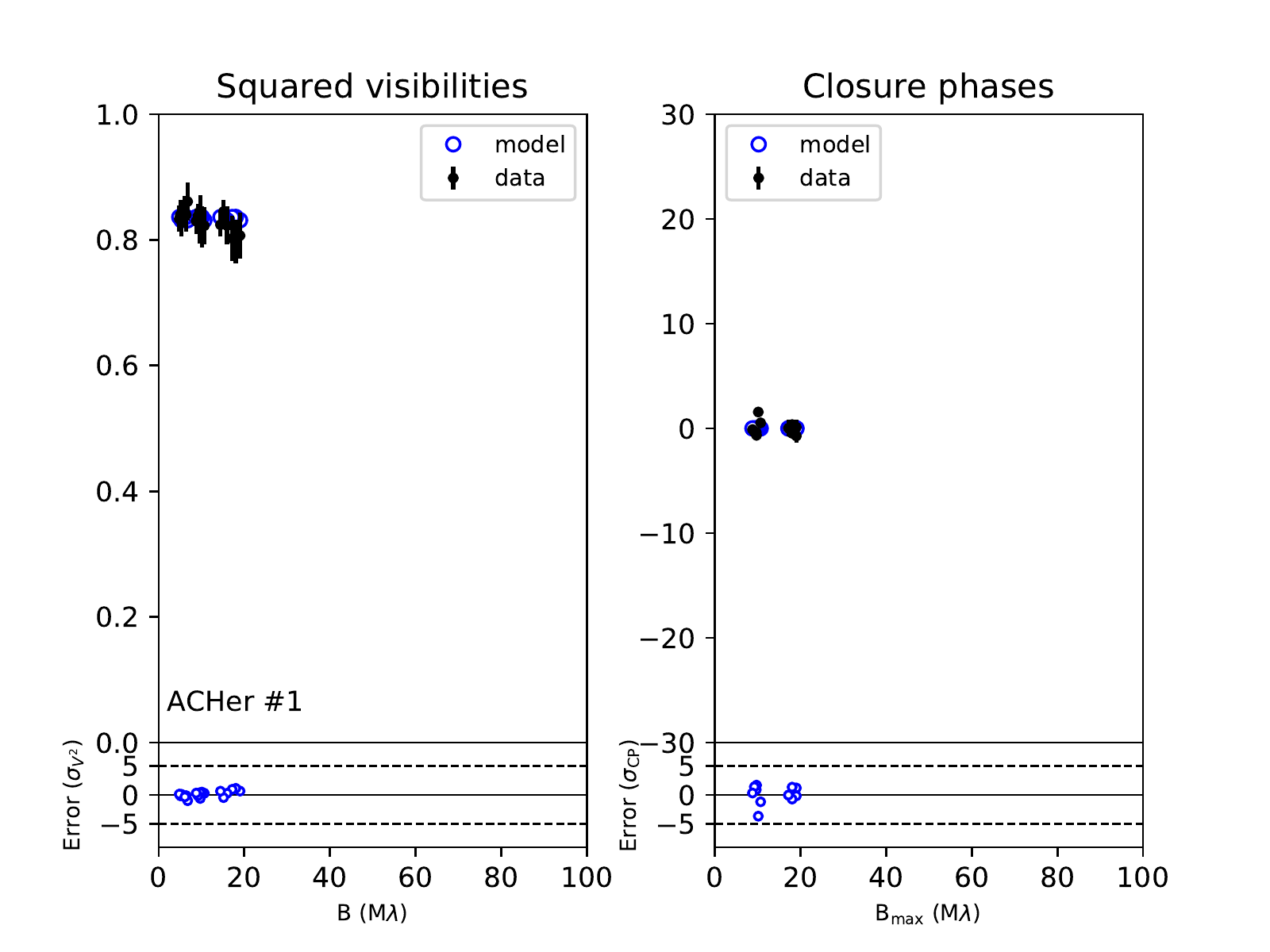} 
\includegraphics[width=6.5cm]{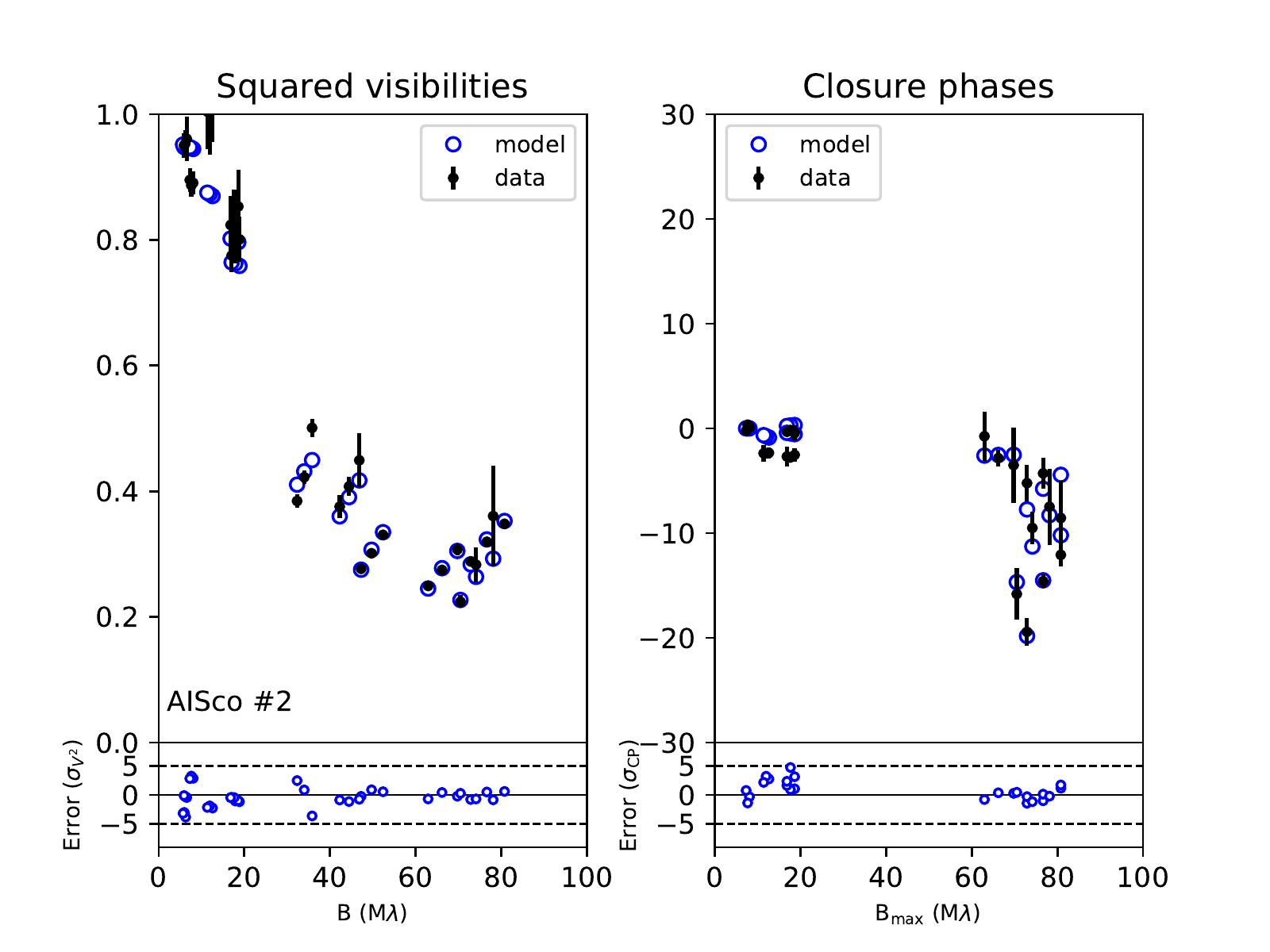}
\includegraphics[width=6.5cm]{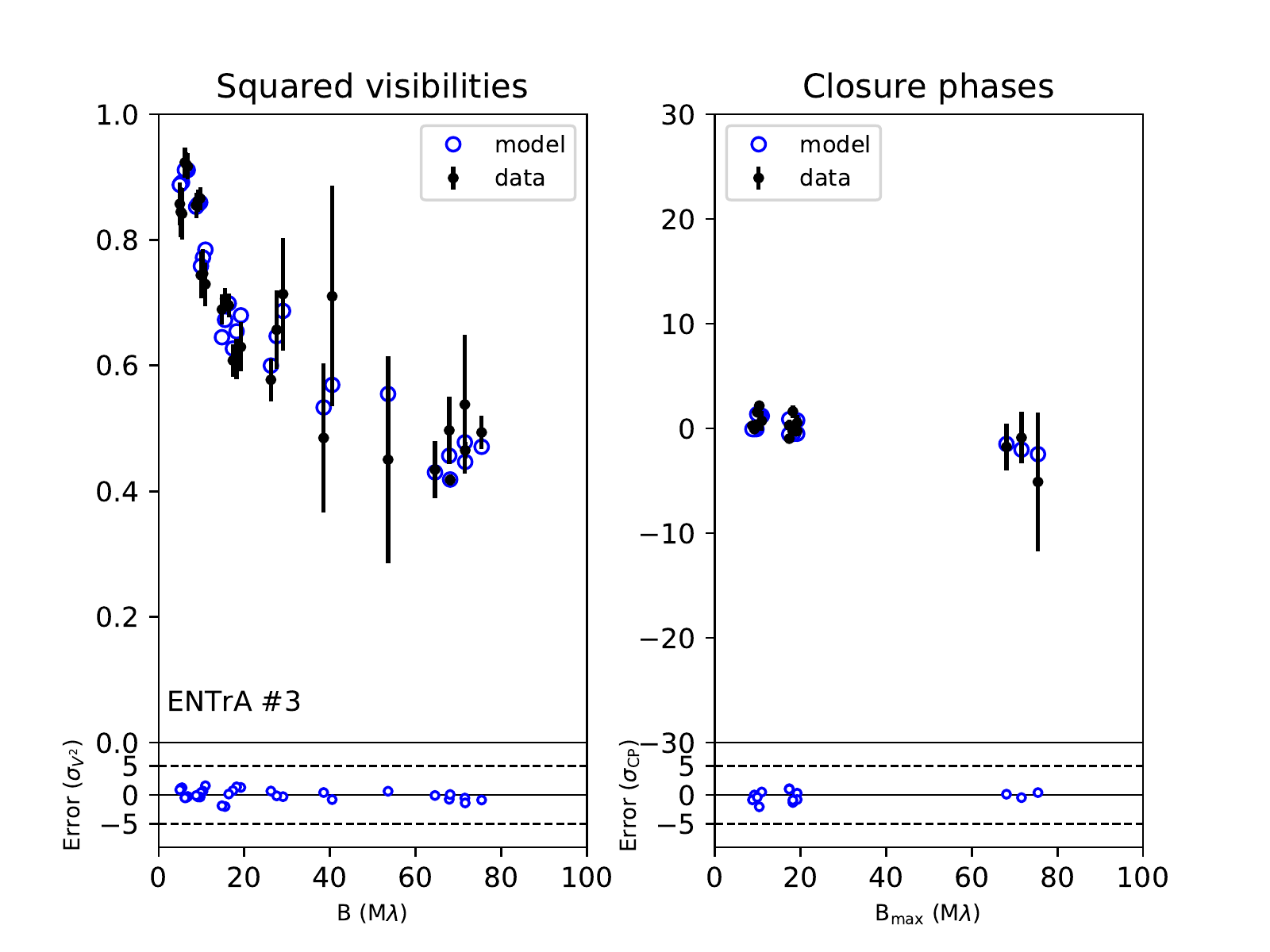}
\includegraphics[width=6.5cm]{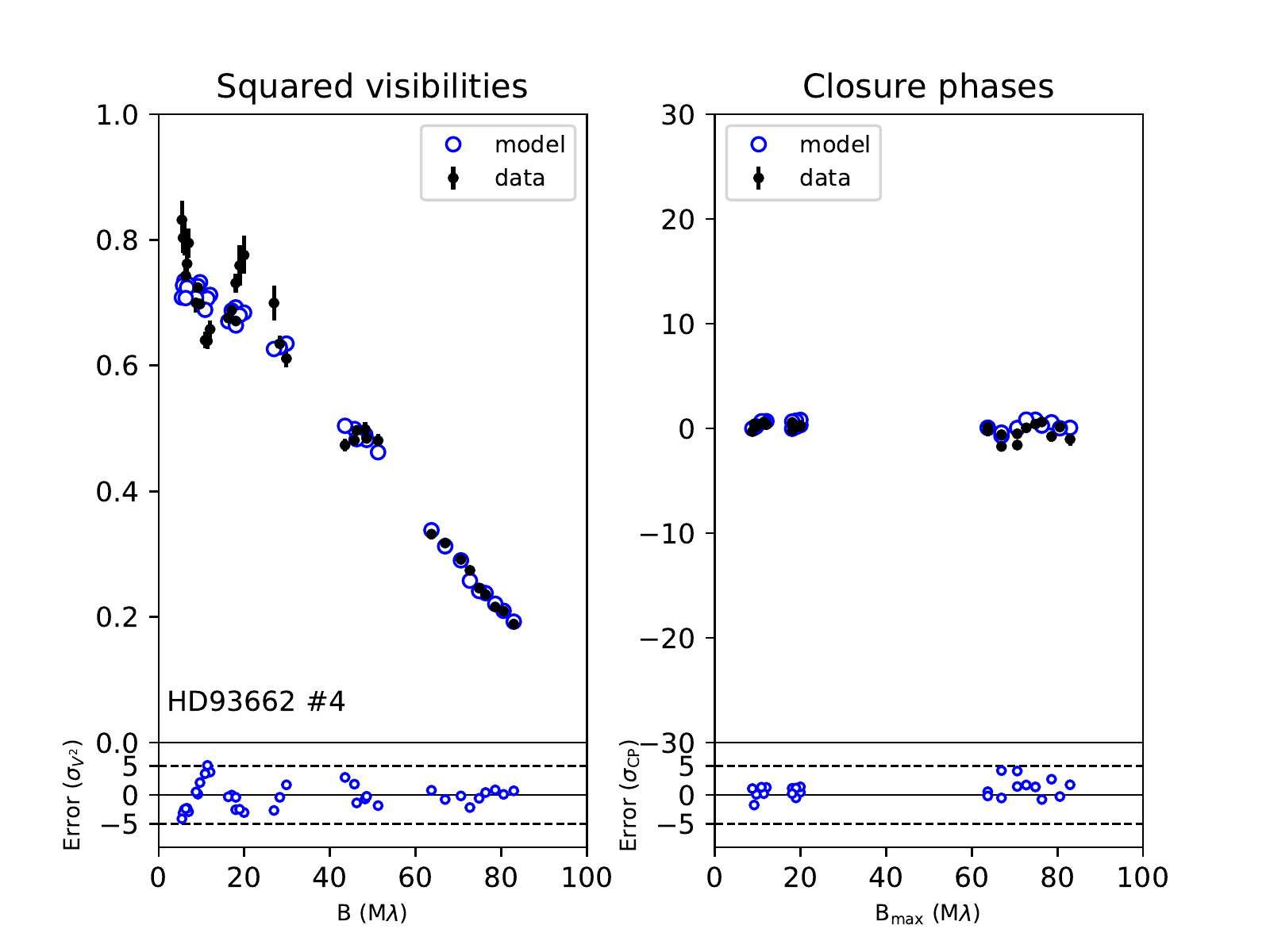}
\includegraphics[width=6.5cm]{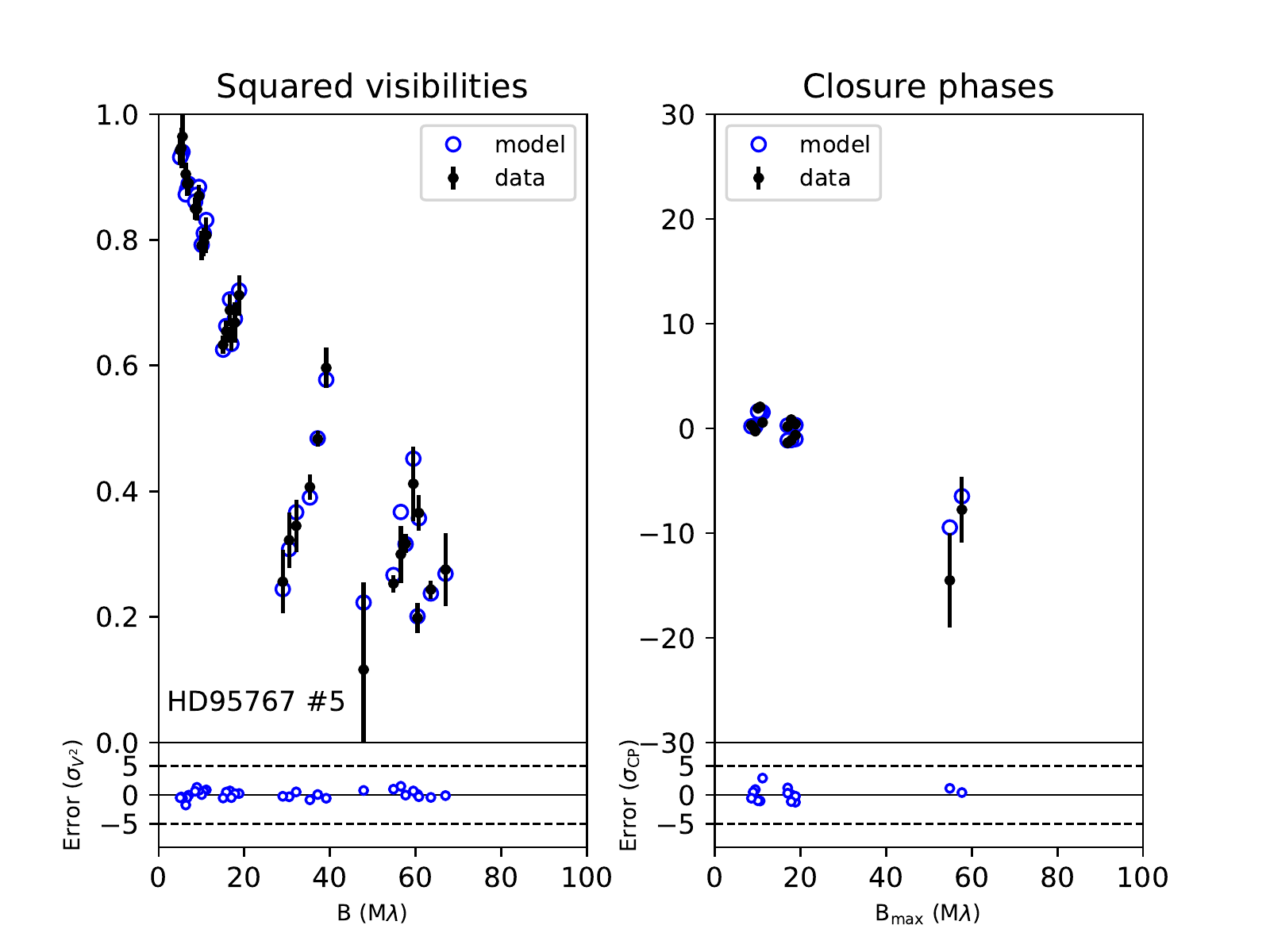}
\includegraphics[width=6.5cm]{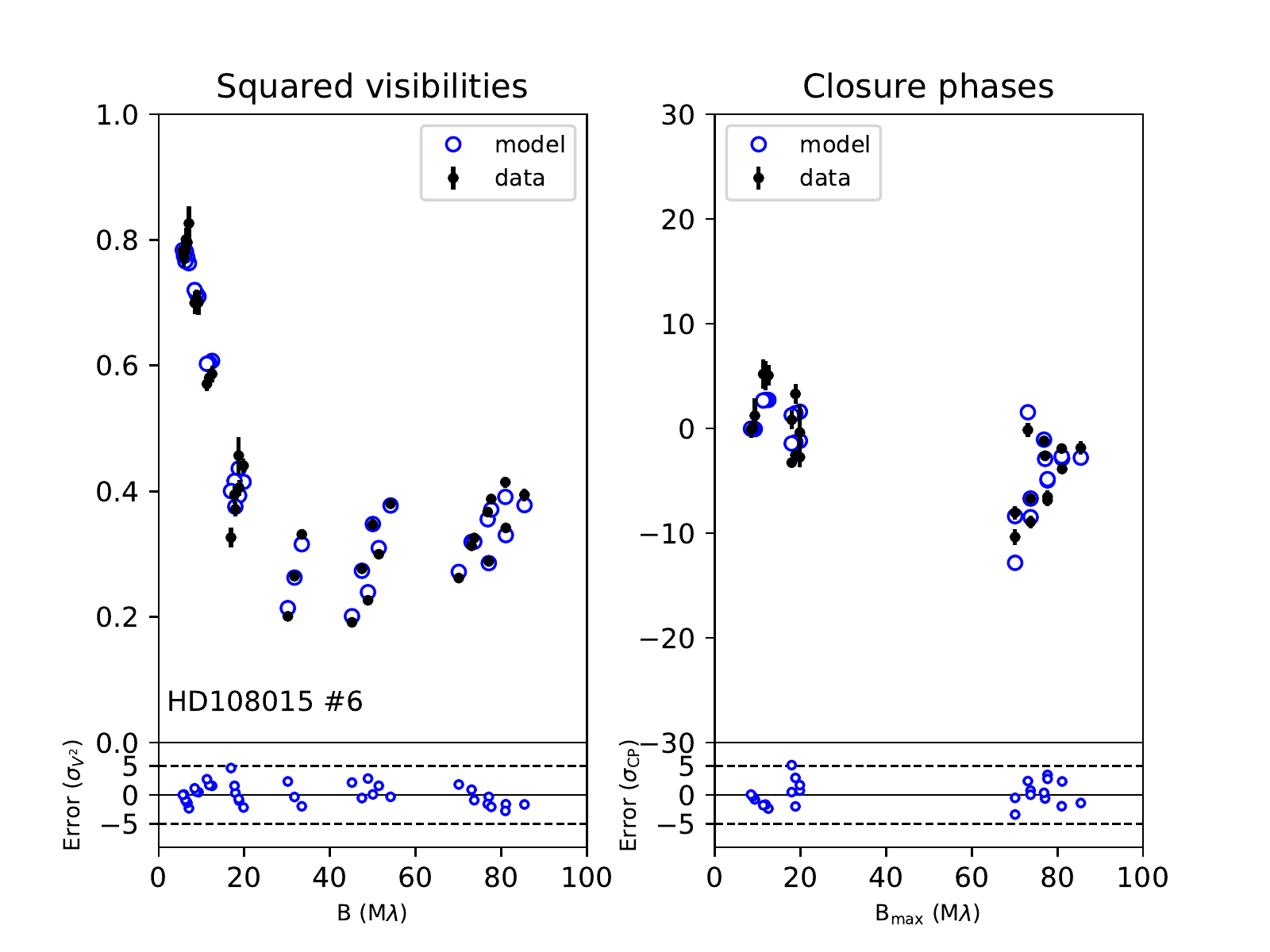}
\includegraphics[width=6.5cm]{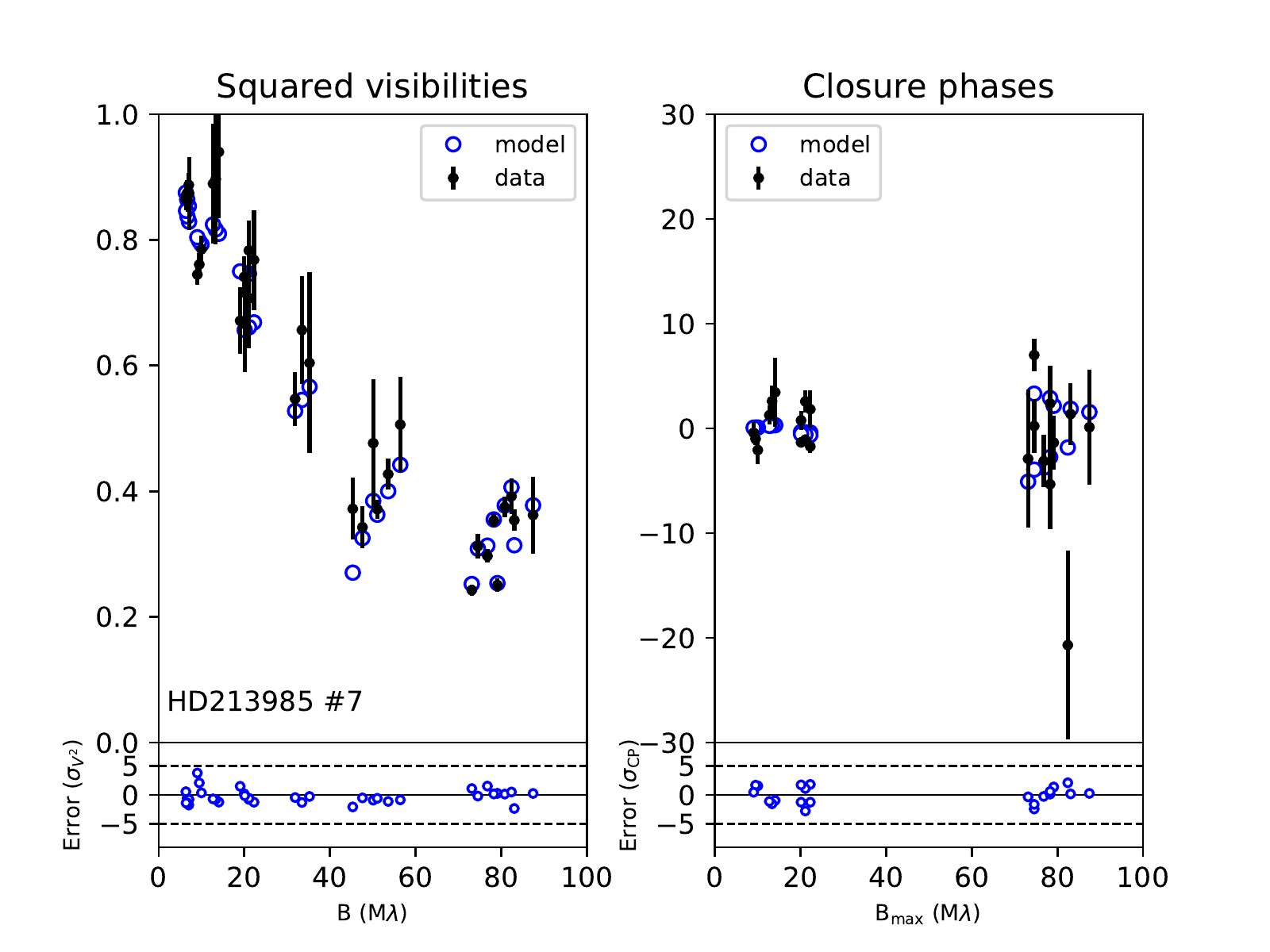}
\includegraphics[width=6.5cm]{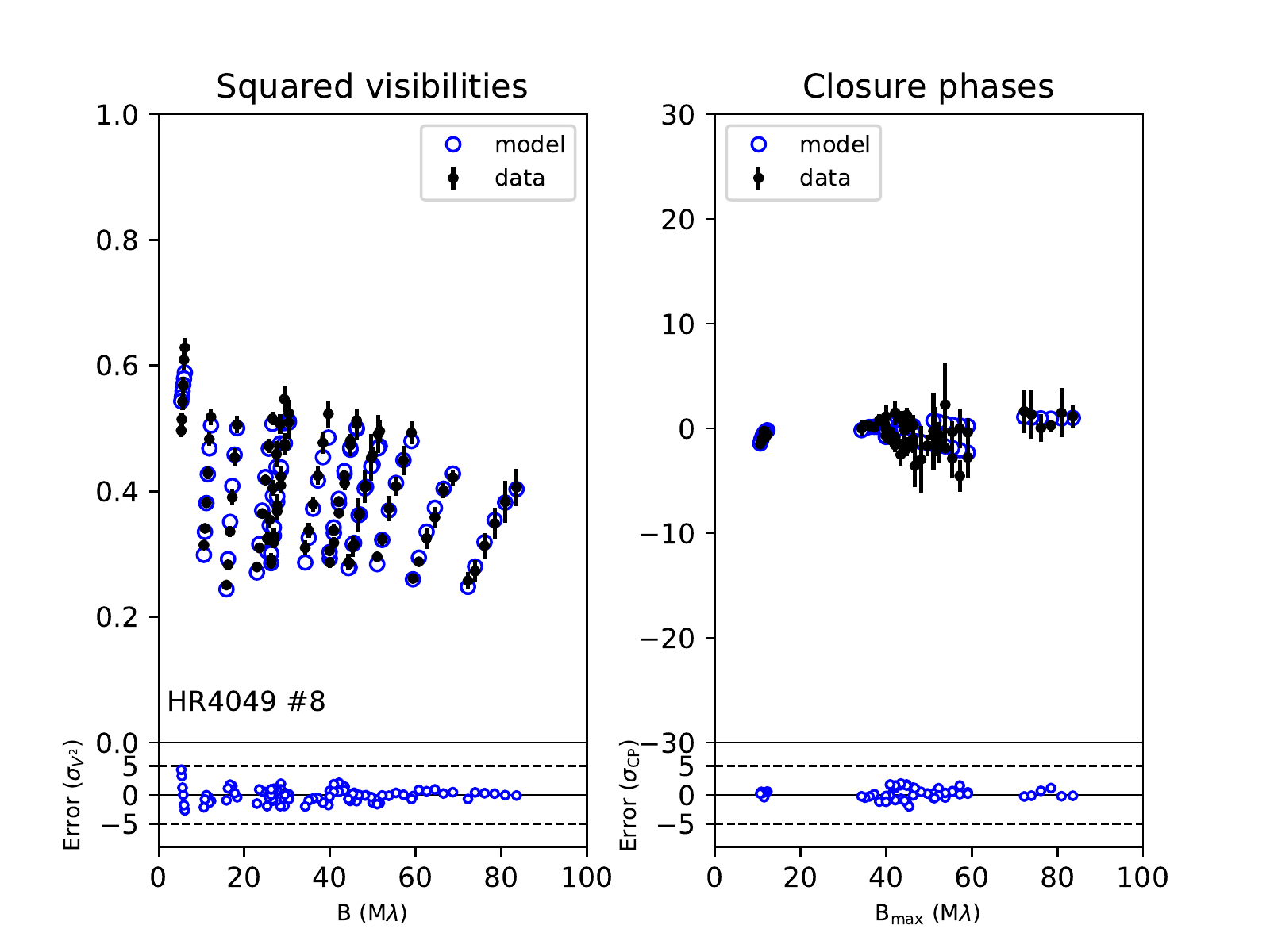}
\caption{Best fit model (blue) versus data (black) for squared visibilities and closure phases for each target. Top panels are the data and the bottom panels are the residuals in $\sigma$.}
\label{fig:Fit1}
\end{figure*}

\begin{figure*}
\centering
\includegraphics[width=6.5cm]{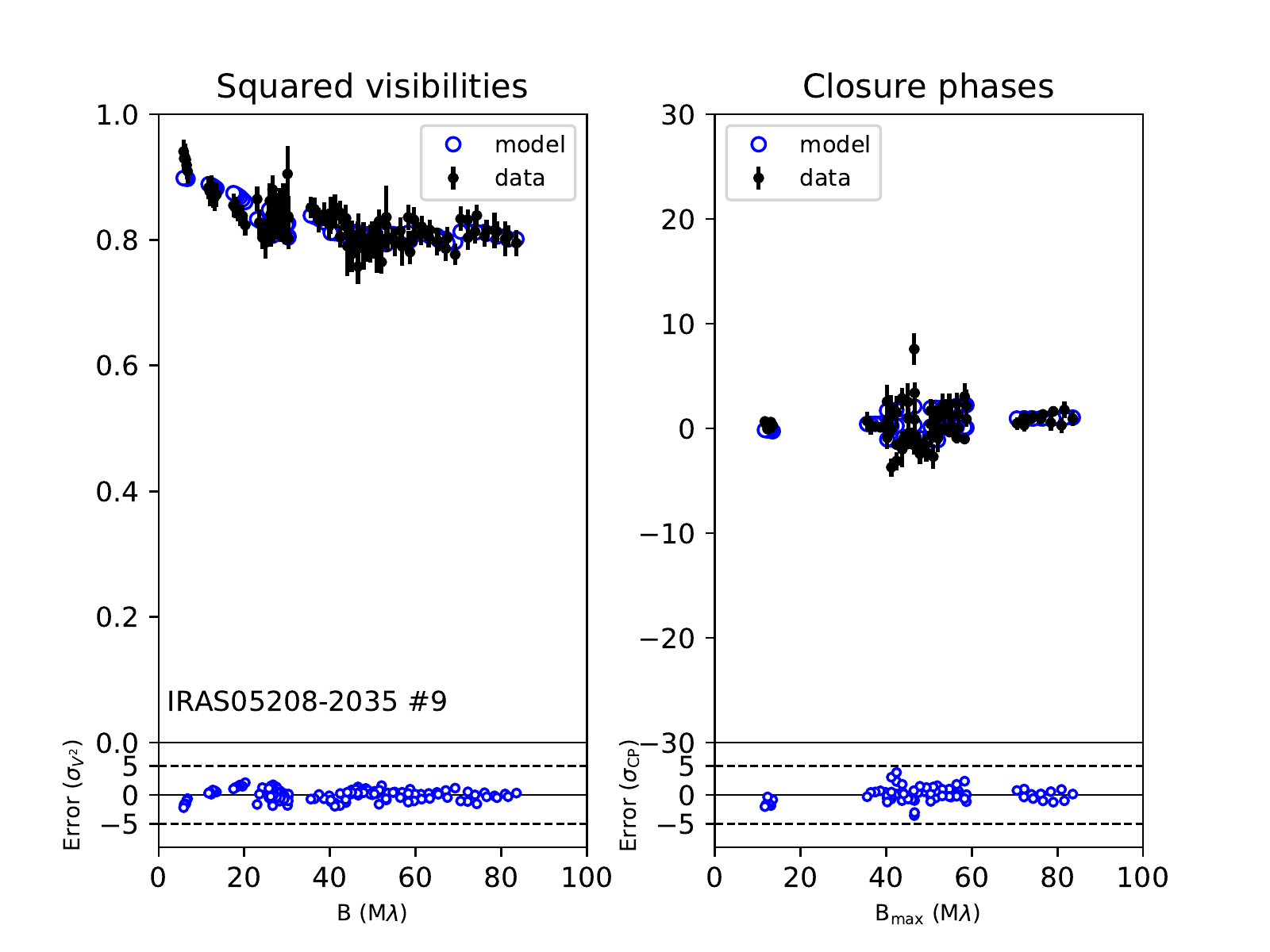}
\includegraphics[width=6.5cm]{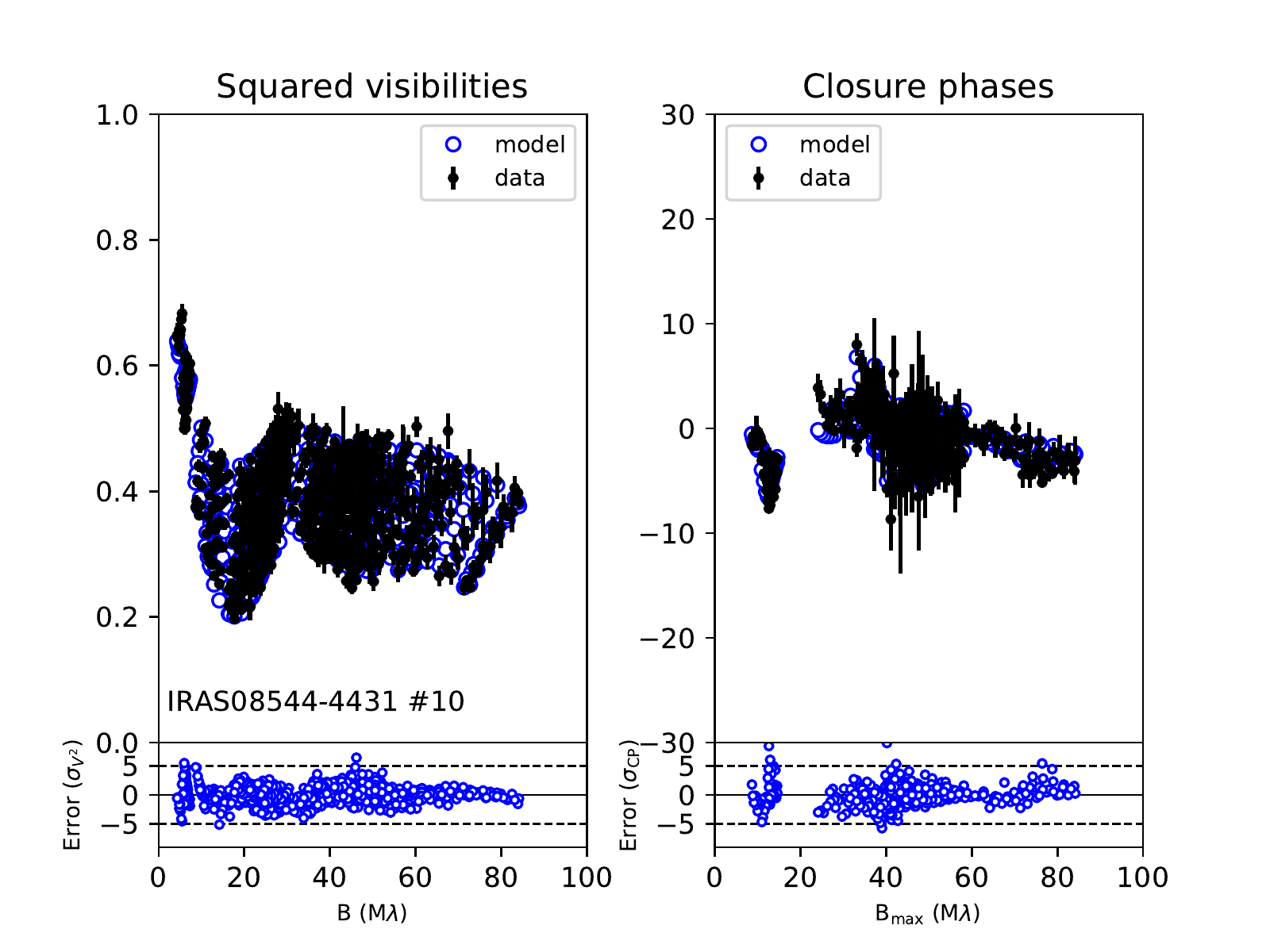}
\includegraphics[width=6.5cm]{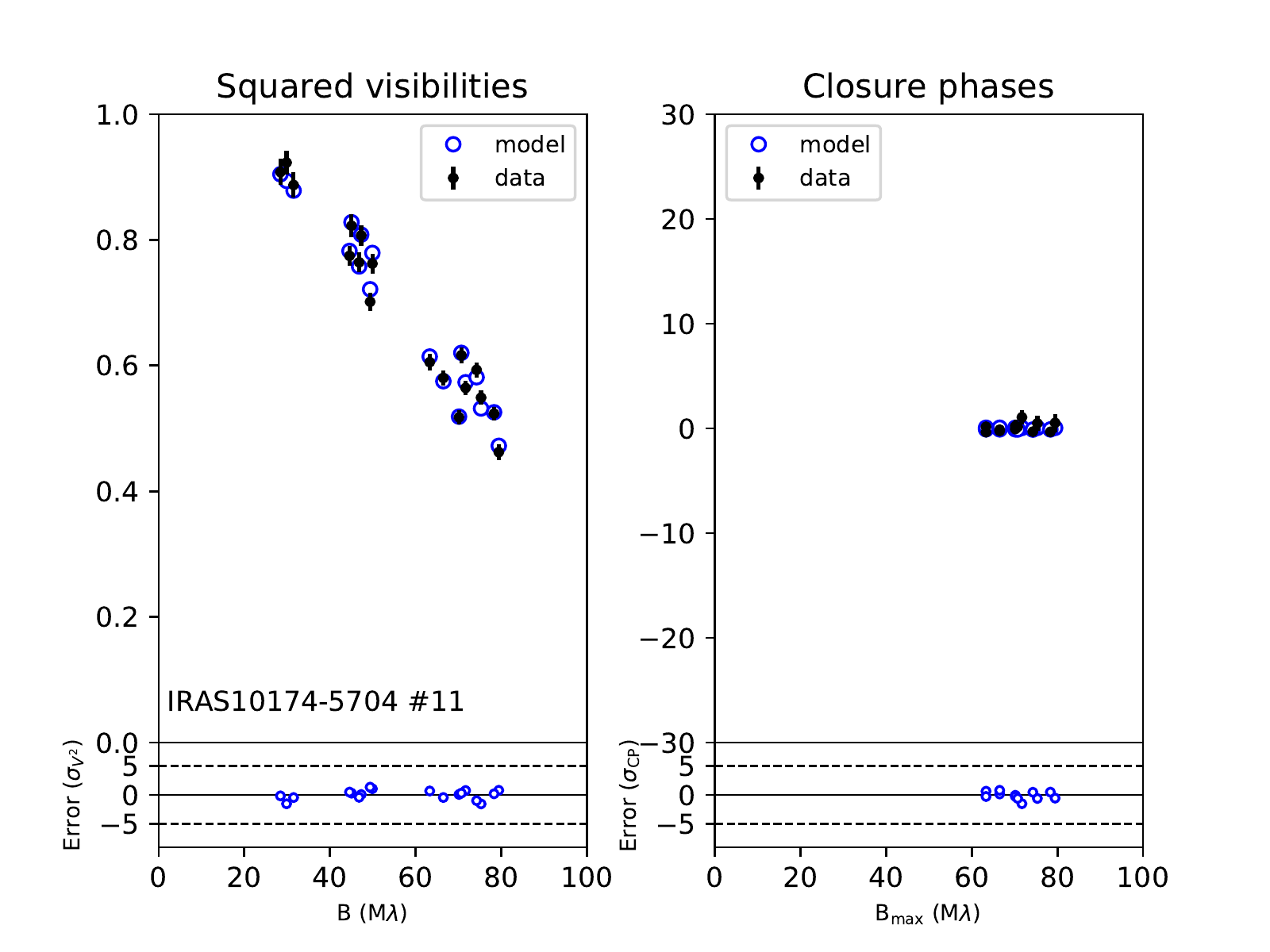}
\includegraphics[width=6.5cm]{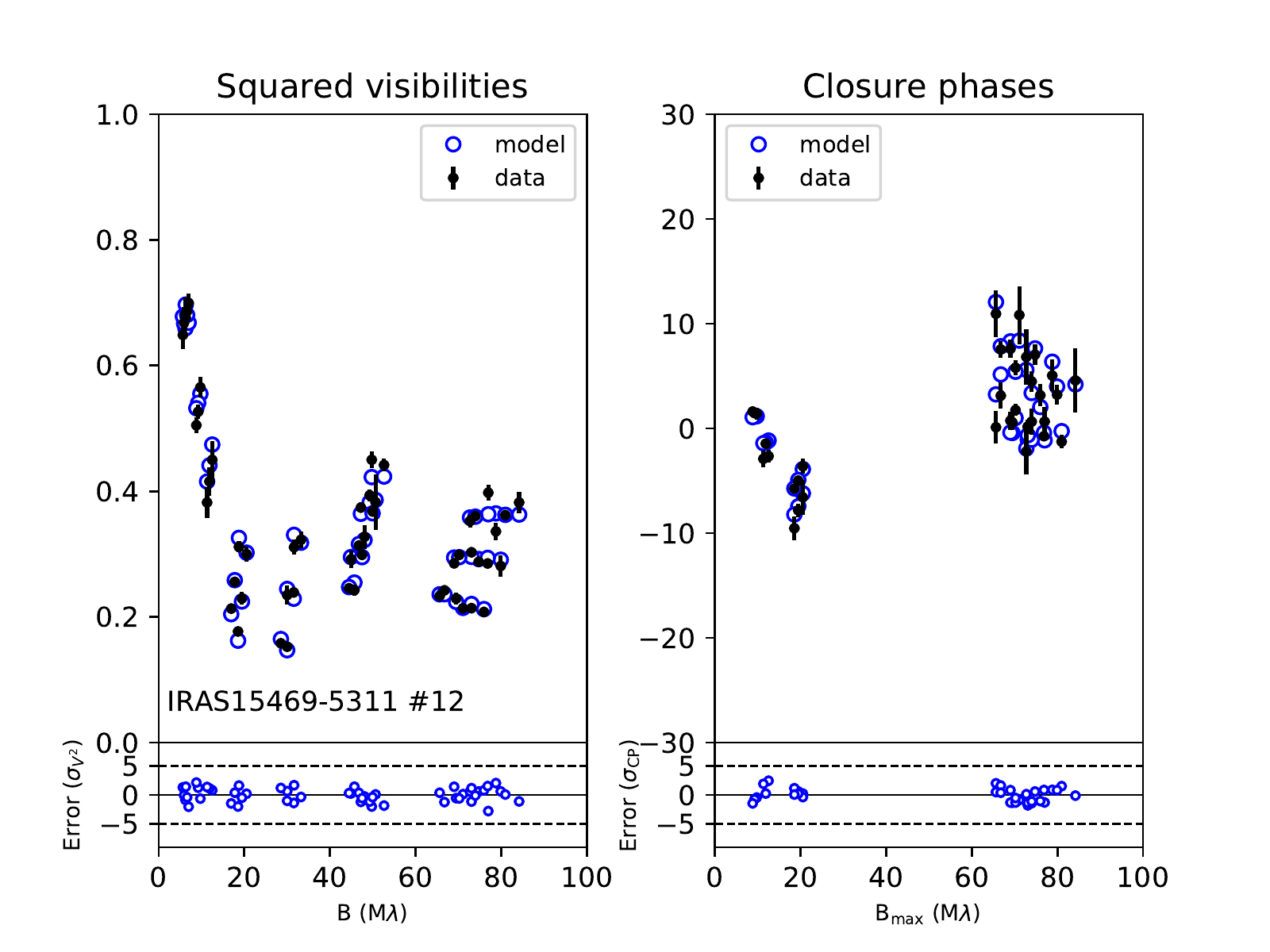}
\includegraphics[width=6.5cm]{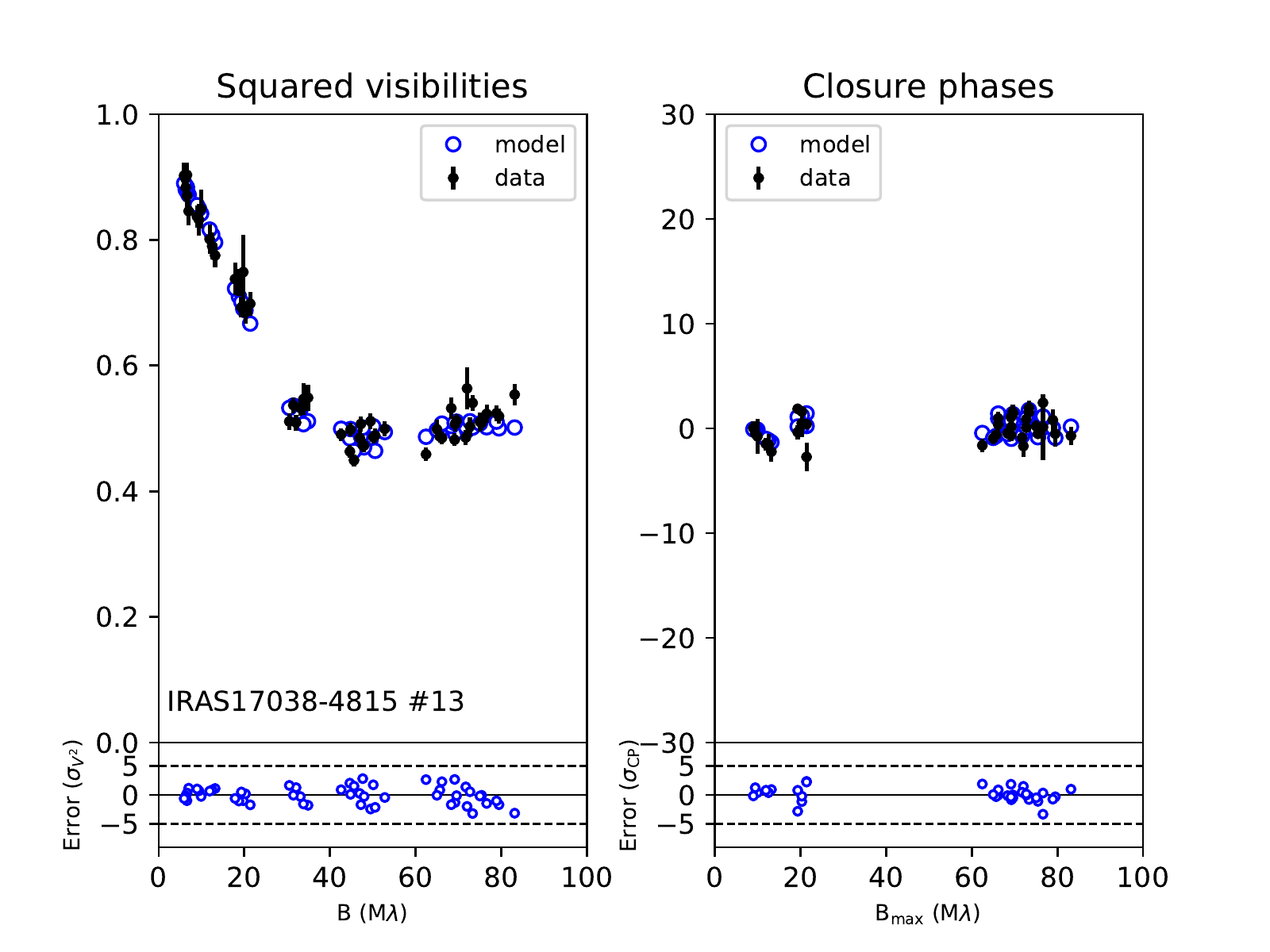}
\includegraphics[width=6.5cm]{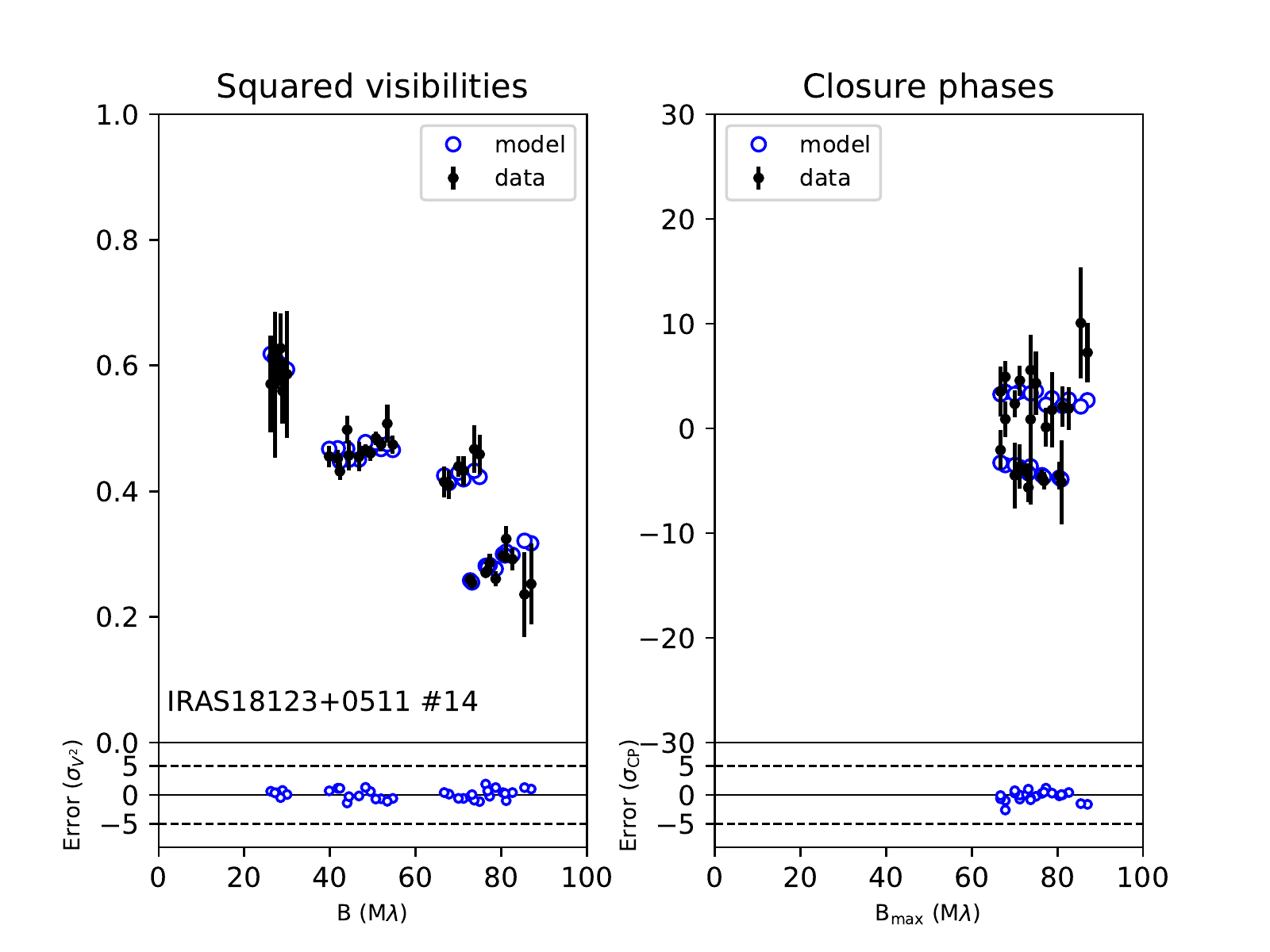}
\includegraphics[width=6.5cm]{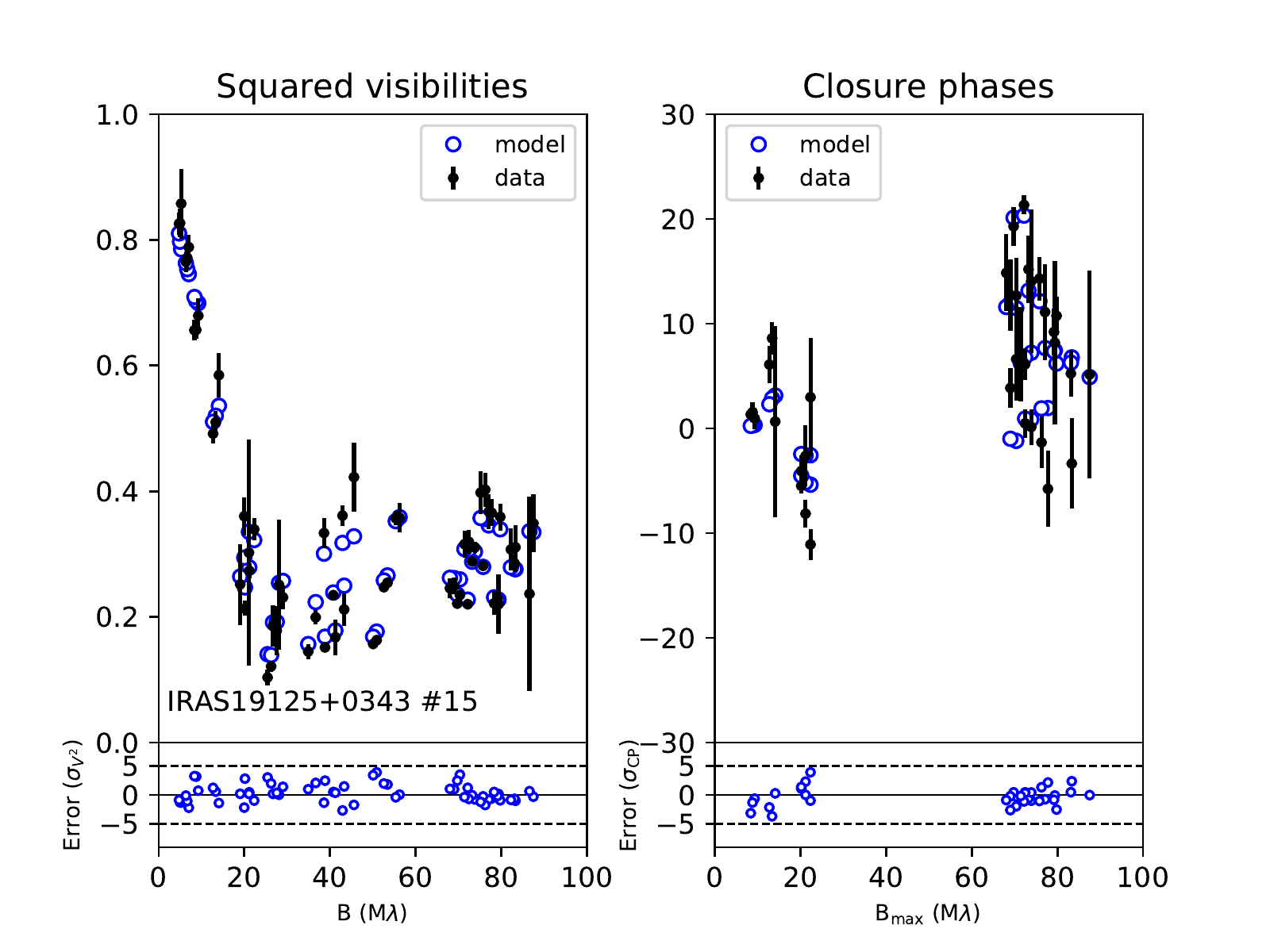}
\includegraphics[width=6.5cm]{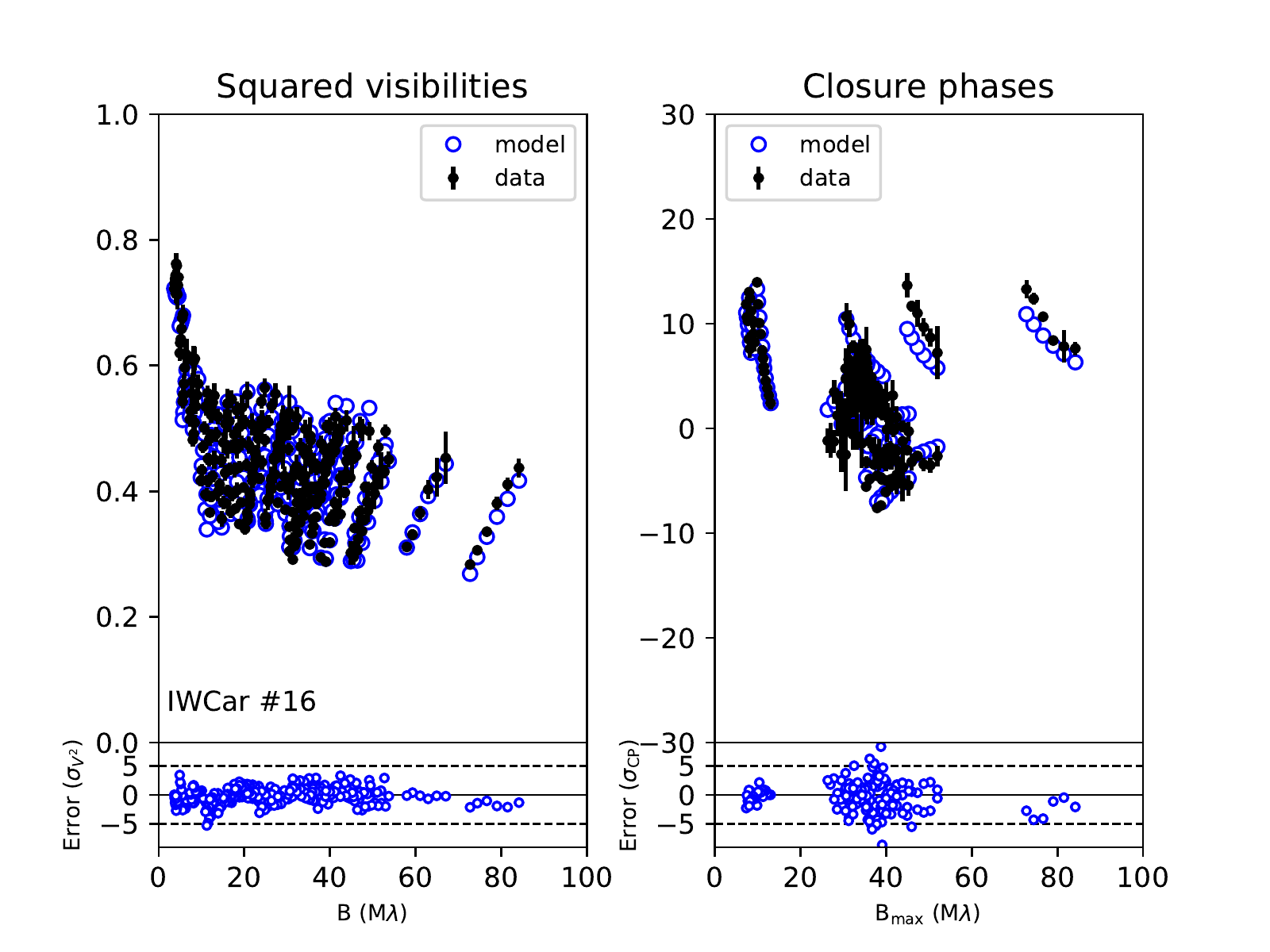}
\caption{Same as Fig.\,\ref{fig:Fit1}}
\label{fig:Fit2}
\end{figure*}

\begin{figure*}
\centering
\includegraphics[width=6.5cm]{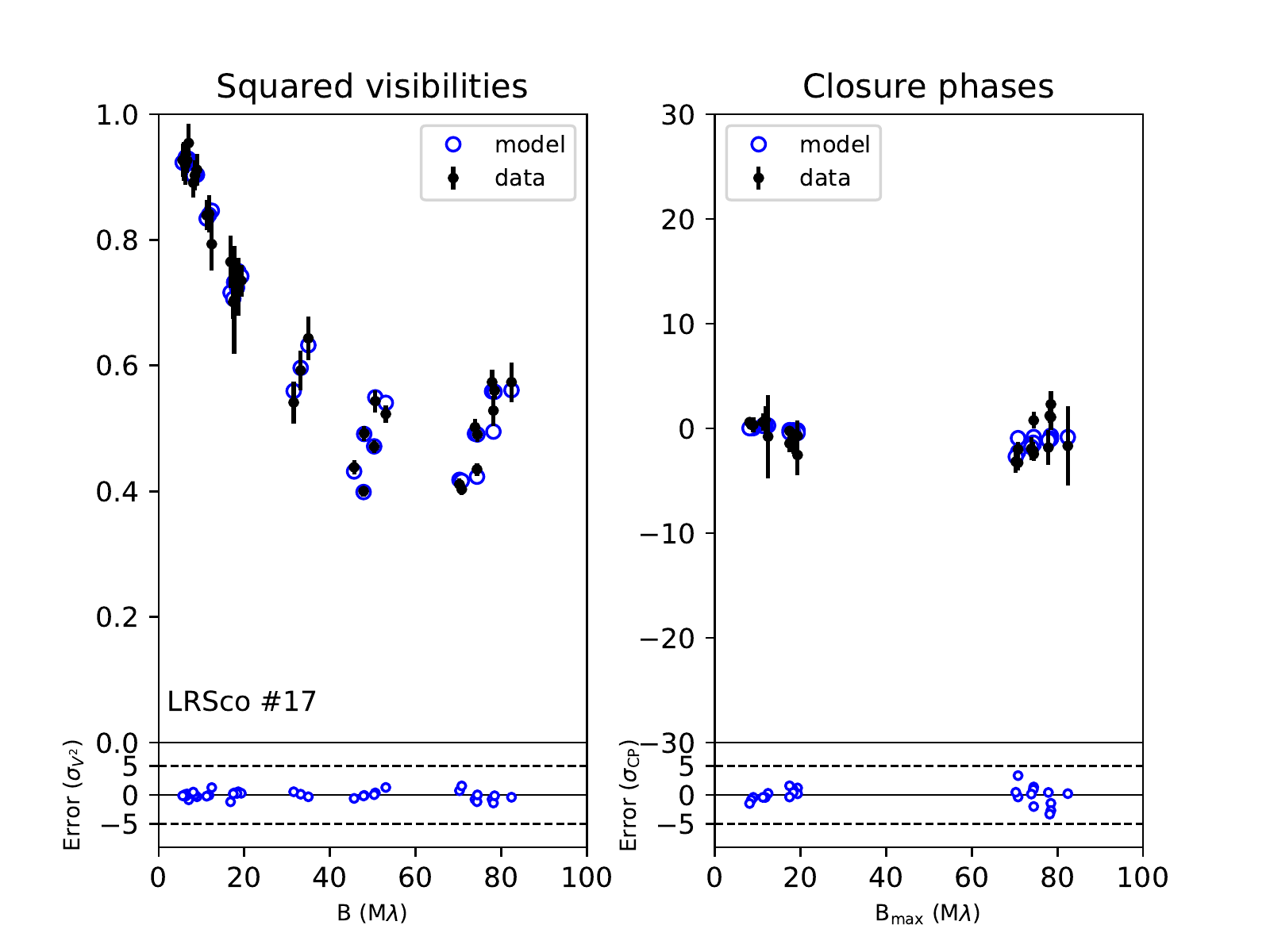}
\includegraphics[width=6.5cm]{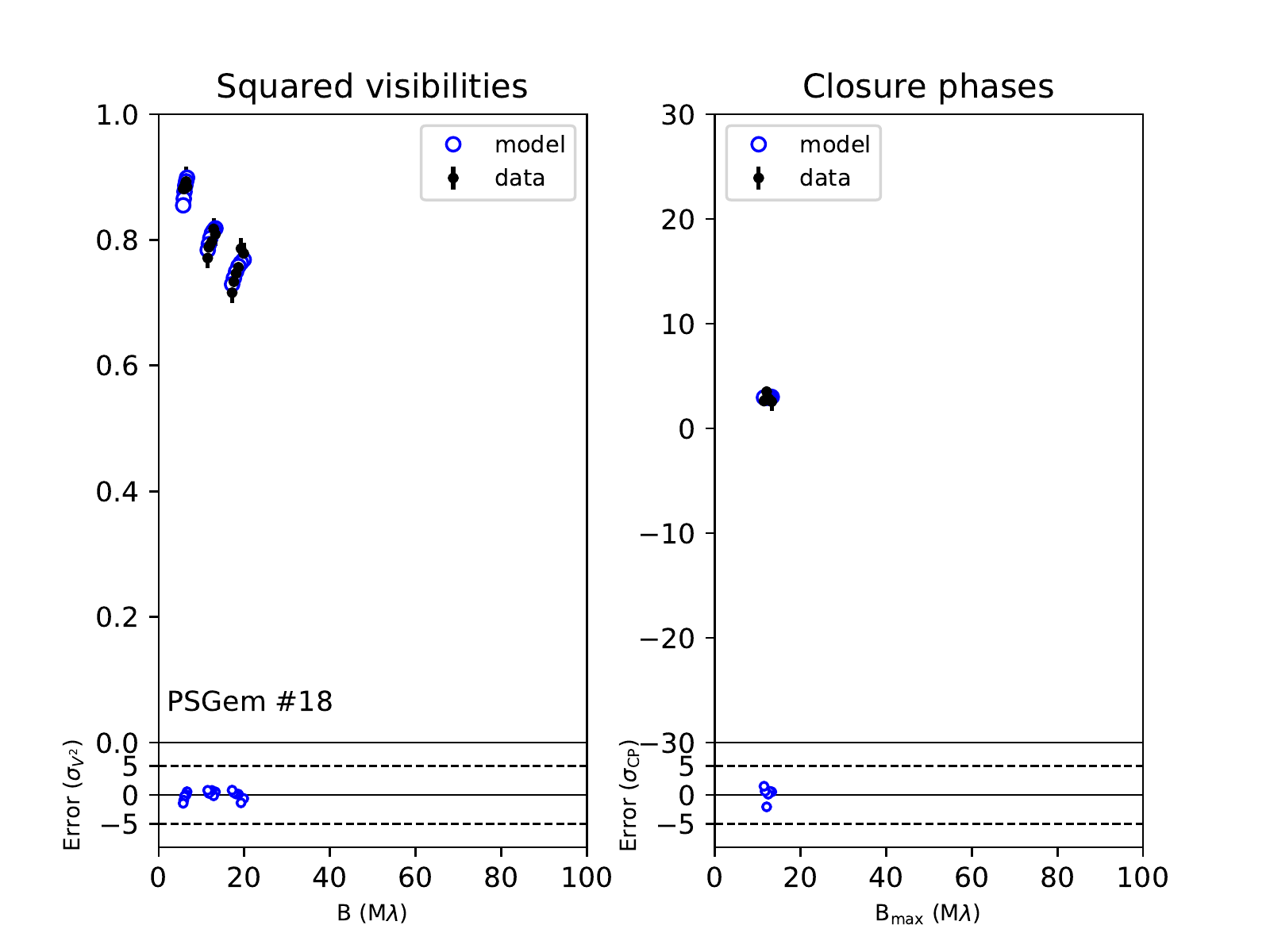}
\includegraphics[width=6.5cm]{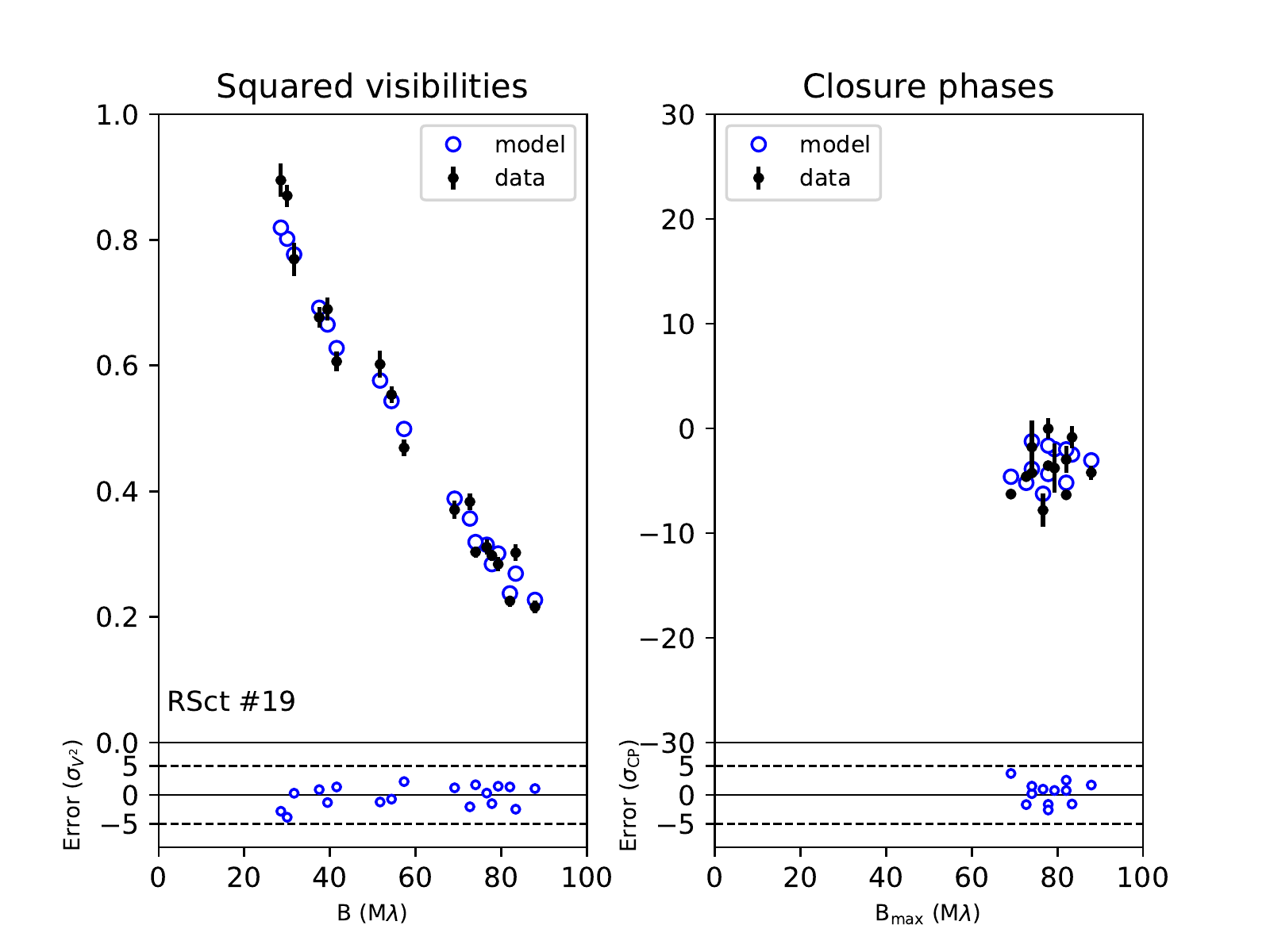}
\includegraphics[width=6.5cm]{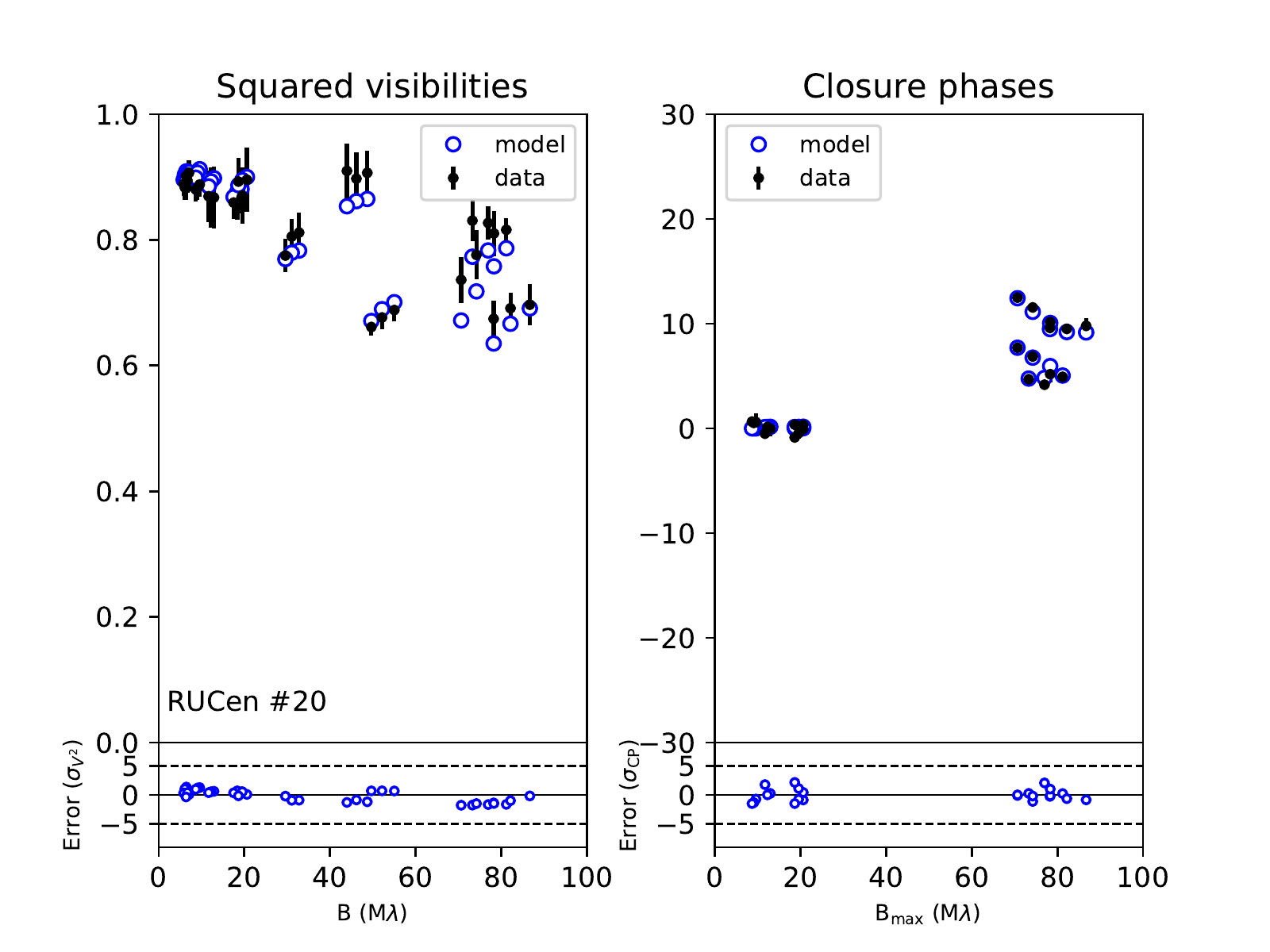}
\includegraphics[width=6.5cm]{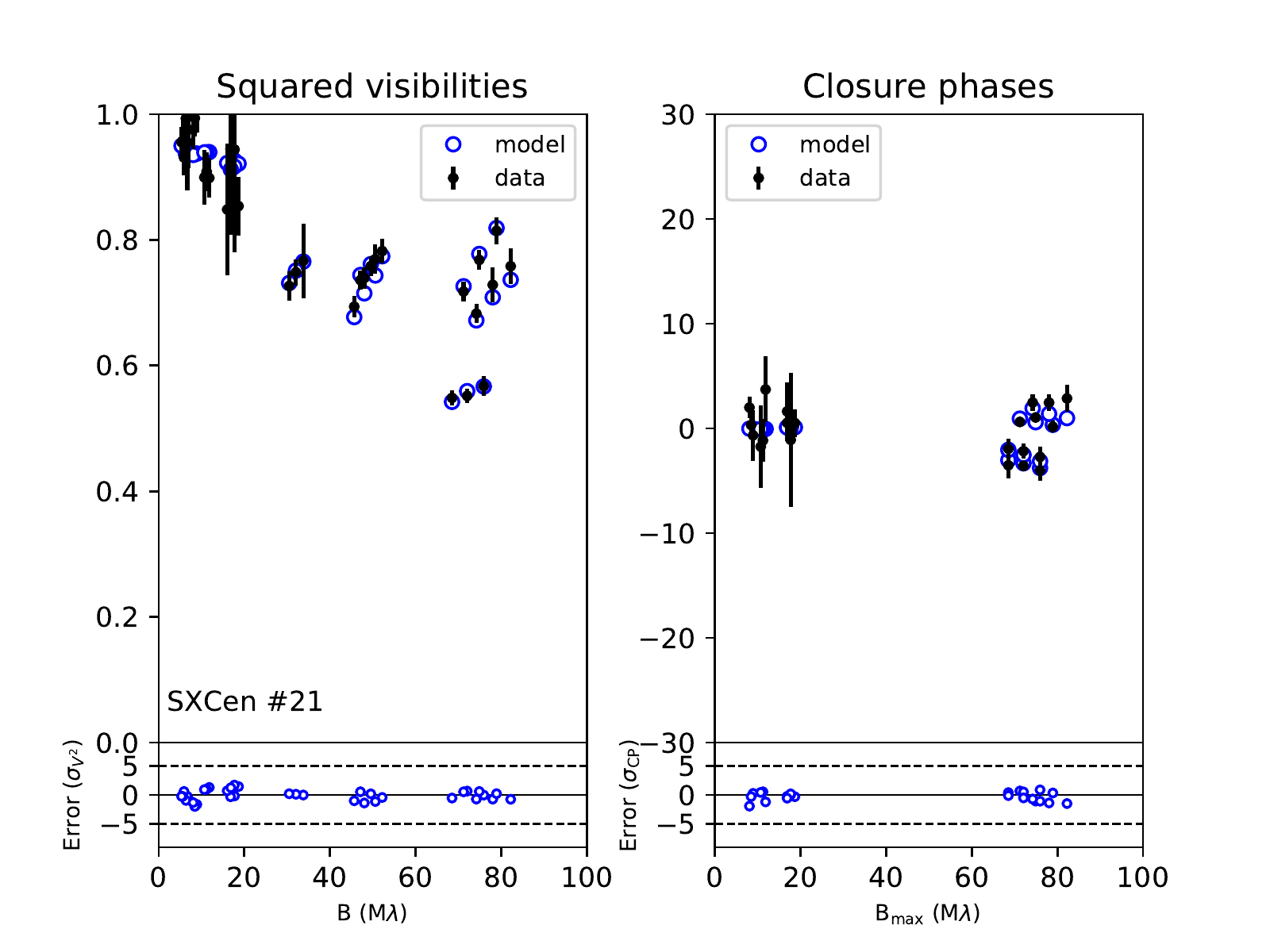}
\includegraphics[width=6.5cm]{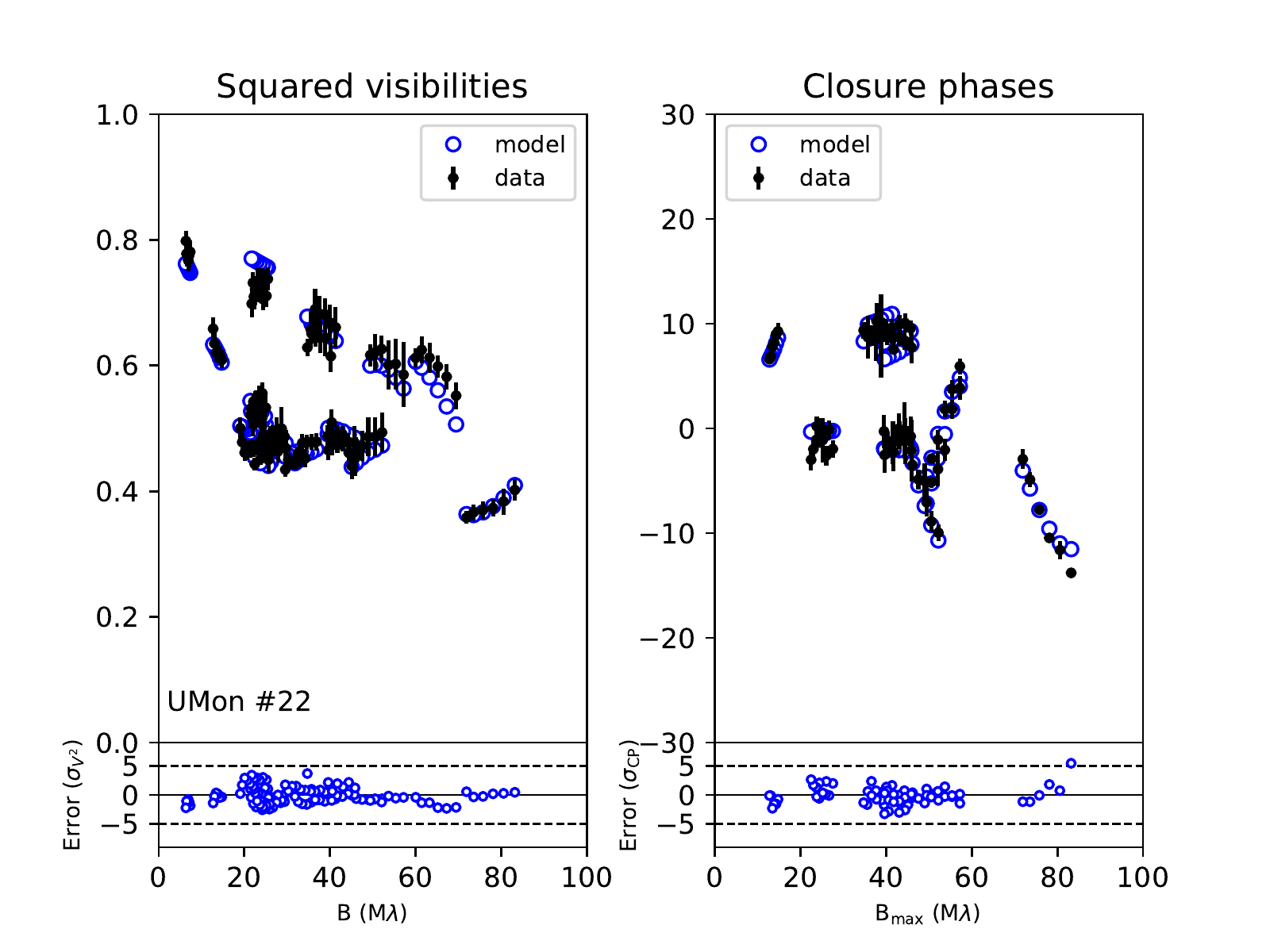}
\includegraphics[width=6.5cm]{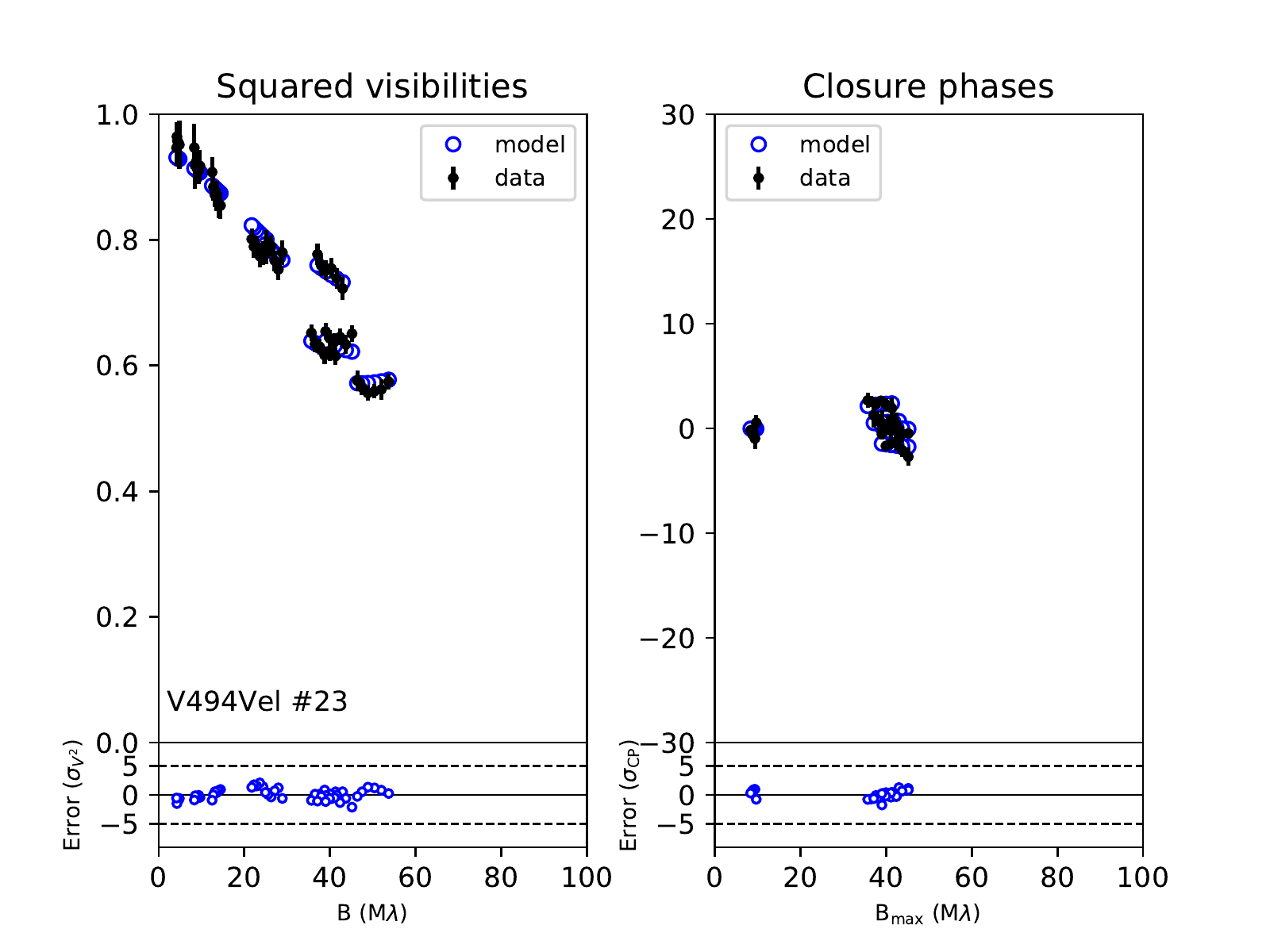}
\caption{Same as Fig.\,\ref{fig:Fit1}}
\label{fig:Fit3}
\end{figure*}

\section{\modif{The BIC and $\chi^2$ evolution per model for each target.}}

\begin{figure*}
\centering
\includegraphics[width=4.5cm]{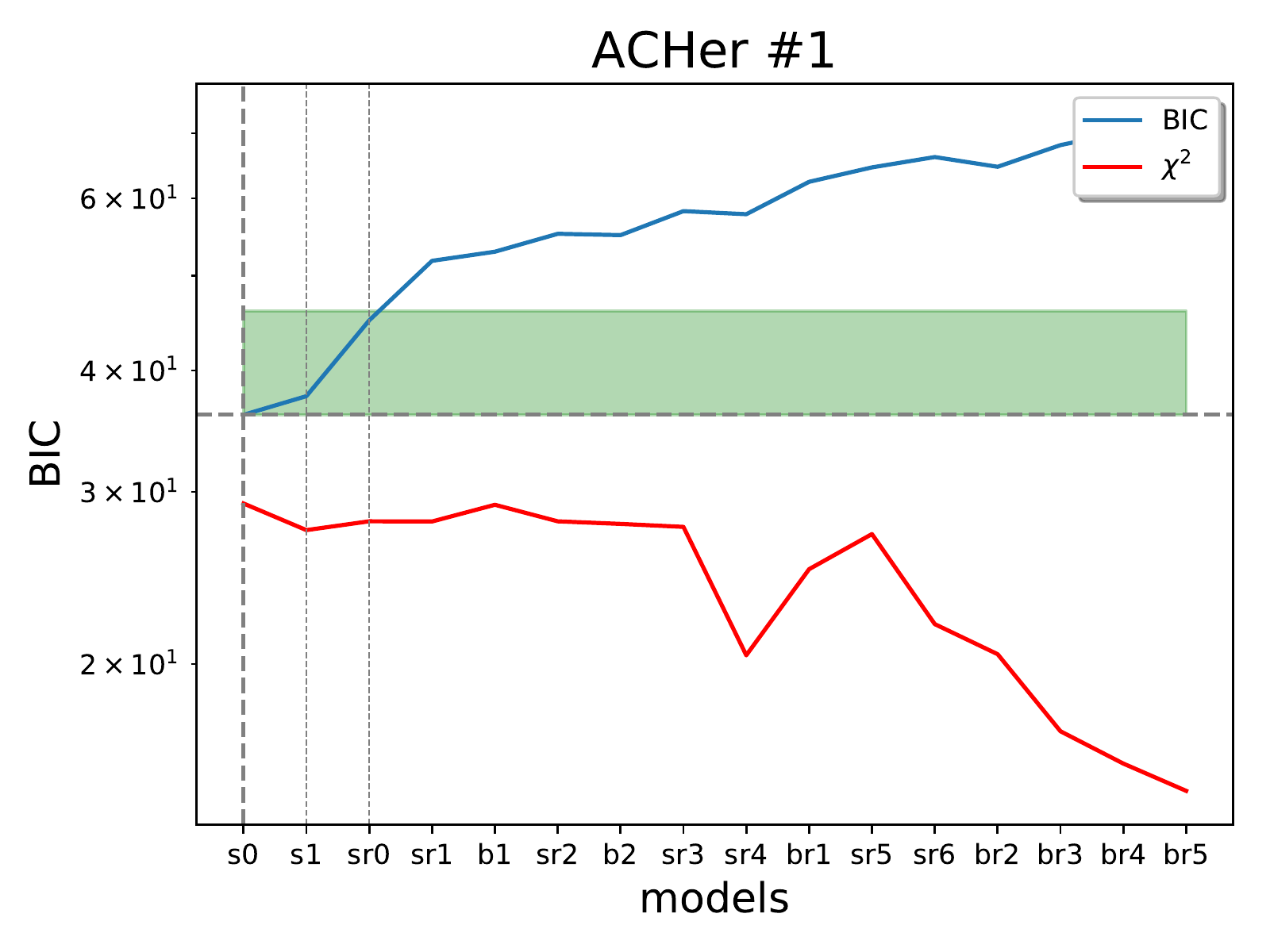} 
\includegraphics[width=4.5cm]{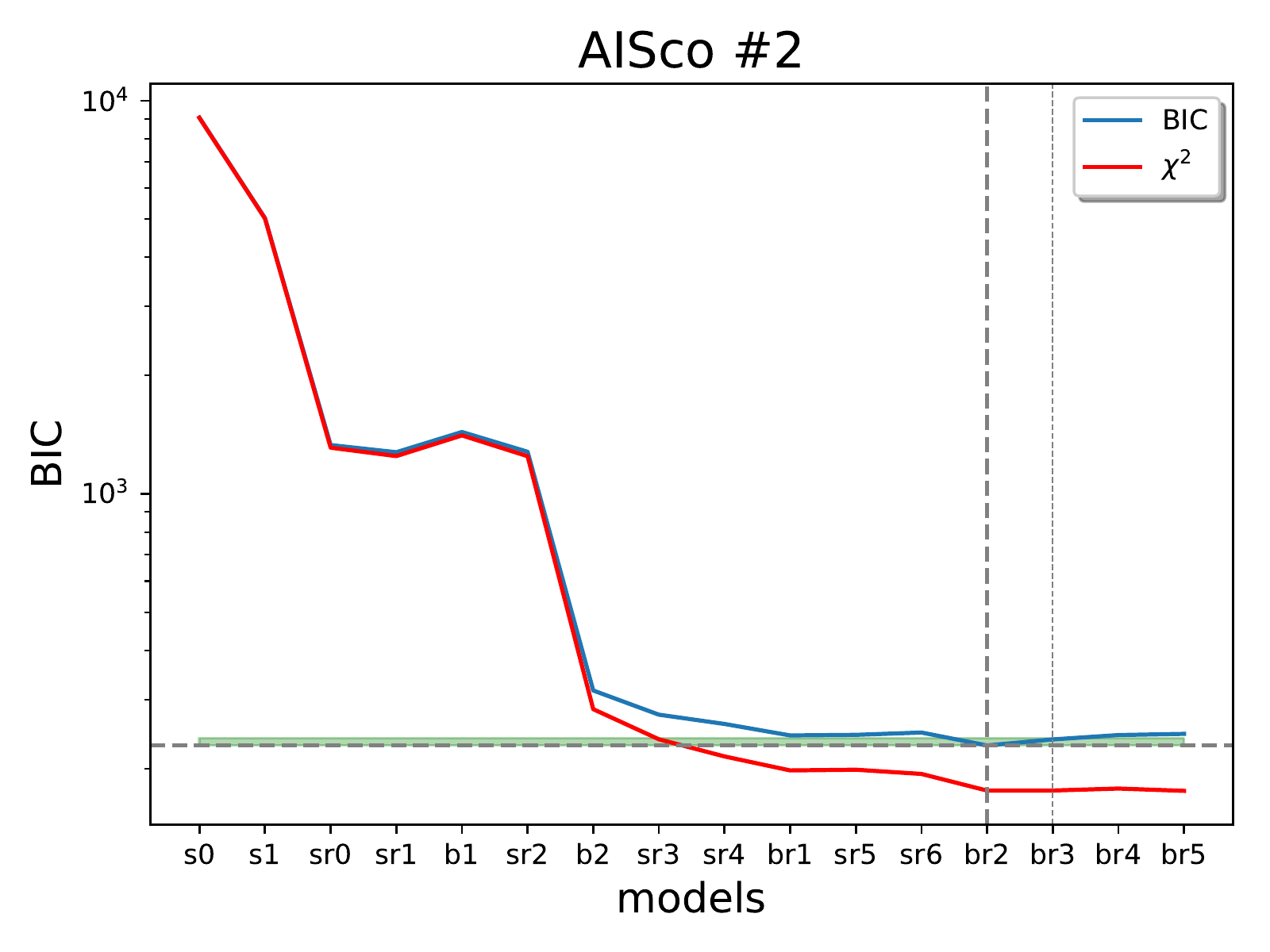}
\includegraphics[width=4.5cm]{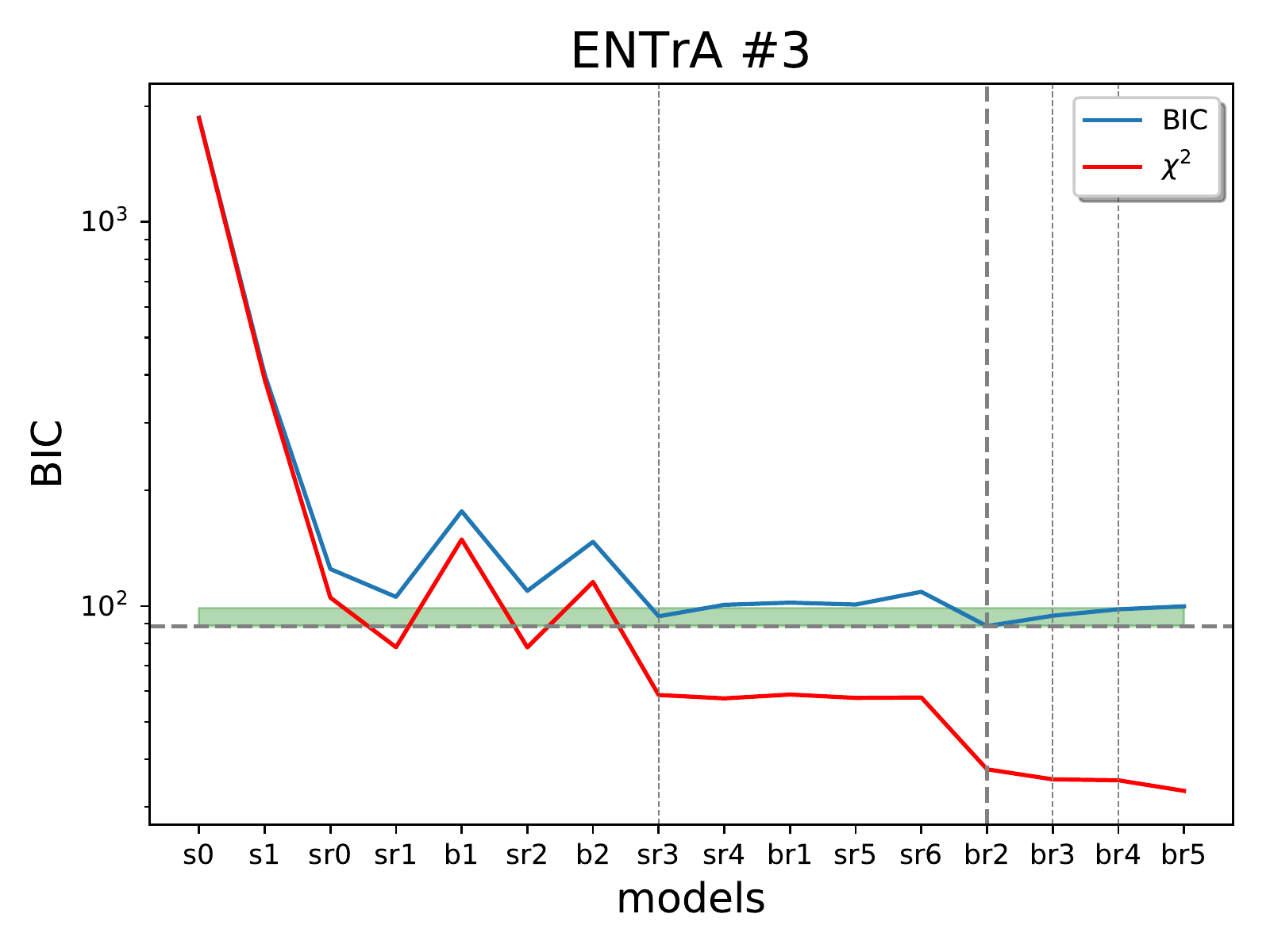}
\includegraphics[width=4.5cm]{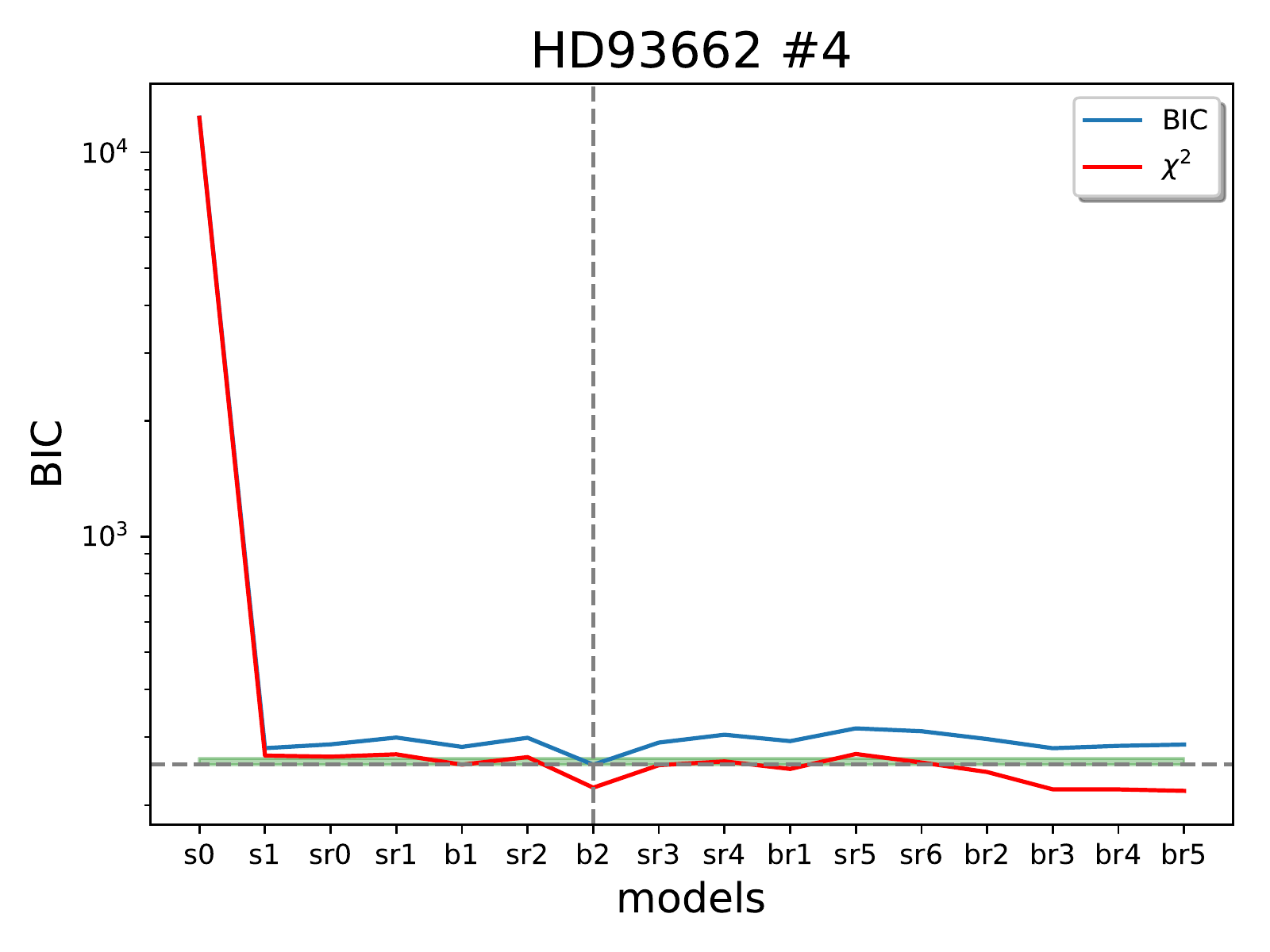}
\includegraphics[width=4.5cm]{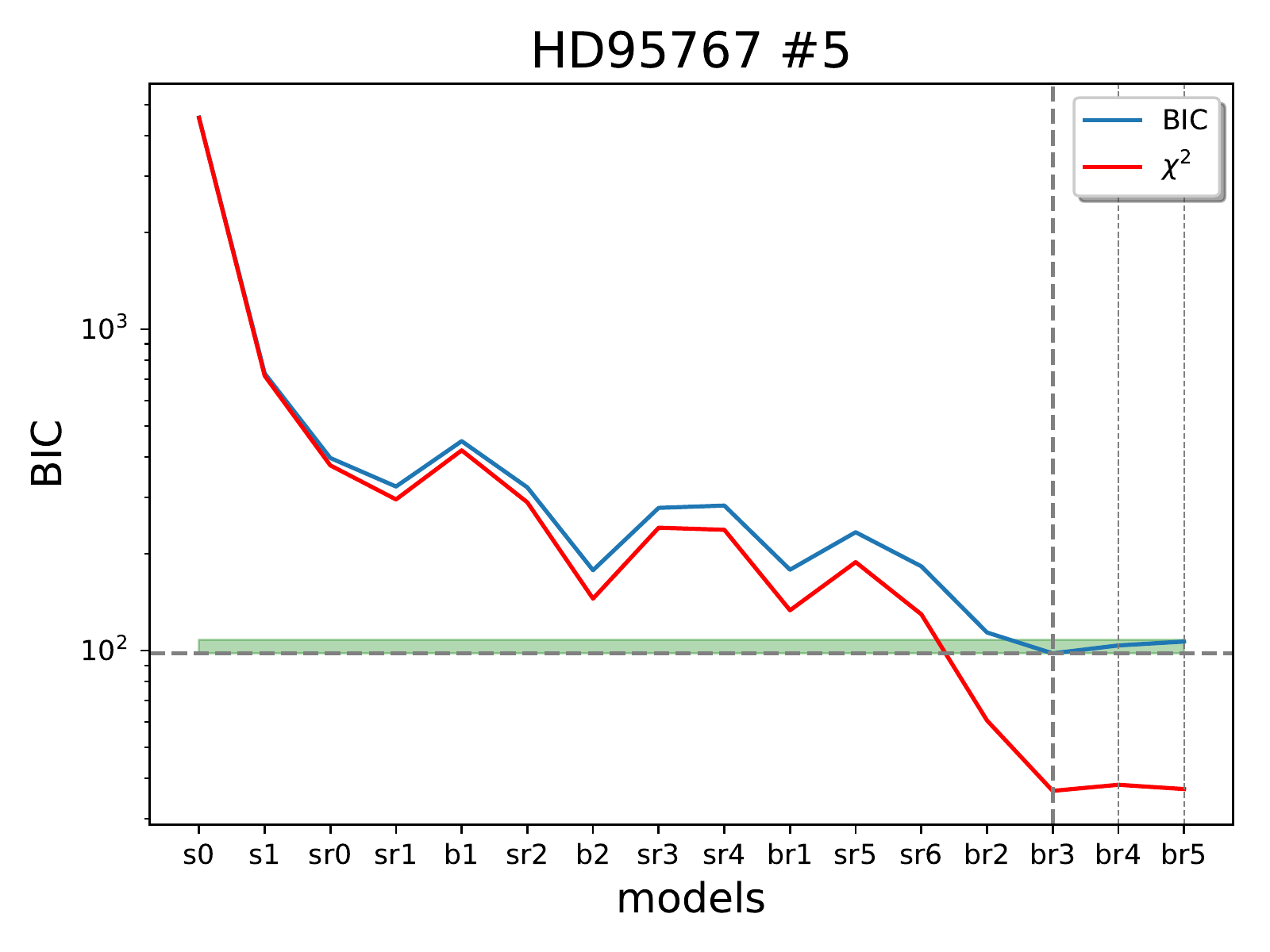}
\includegraphics[width=4.5cm]{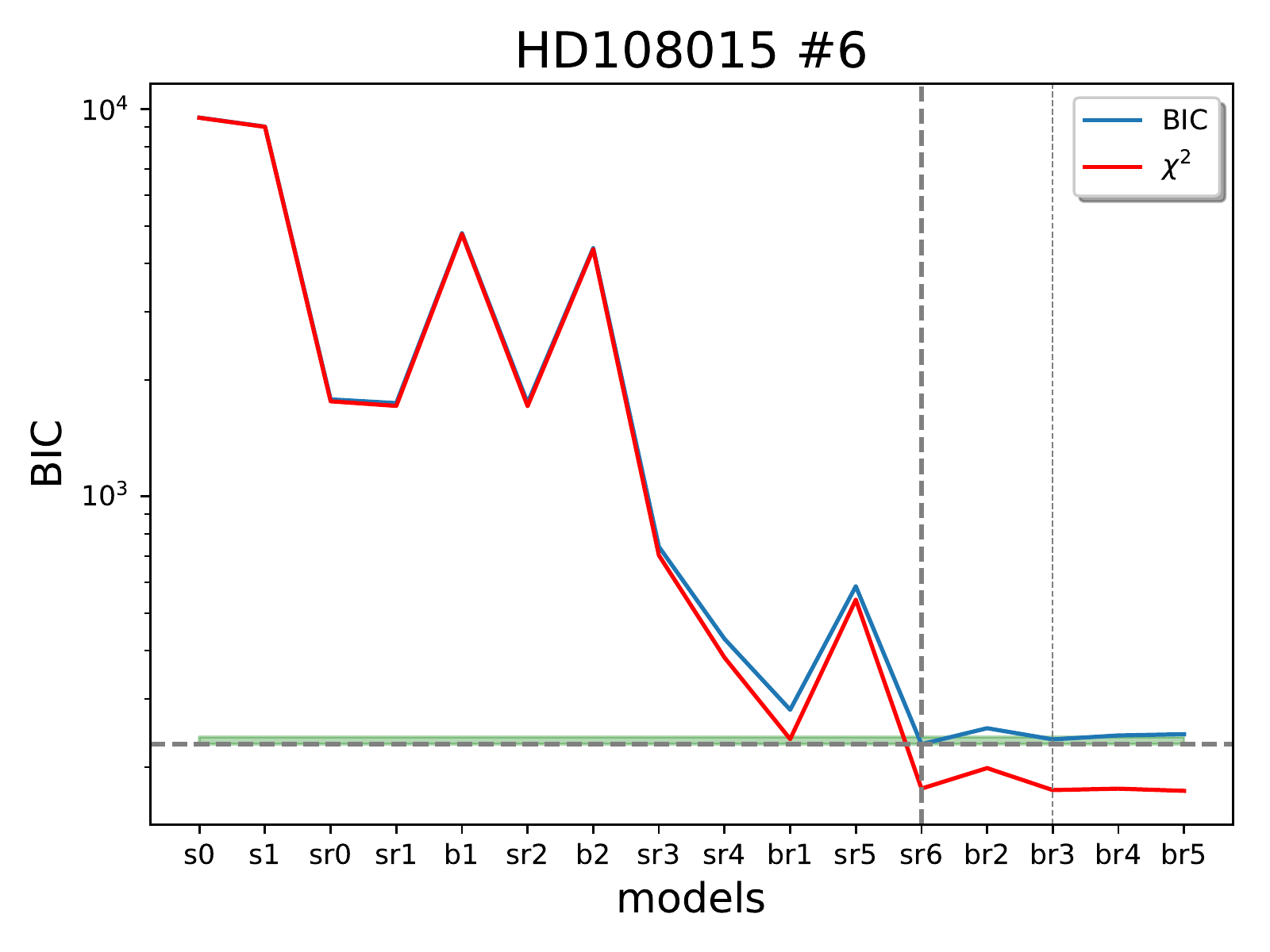}
\includegraphics[width=4.5cm]{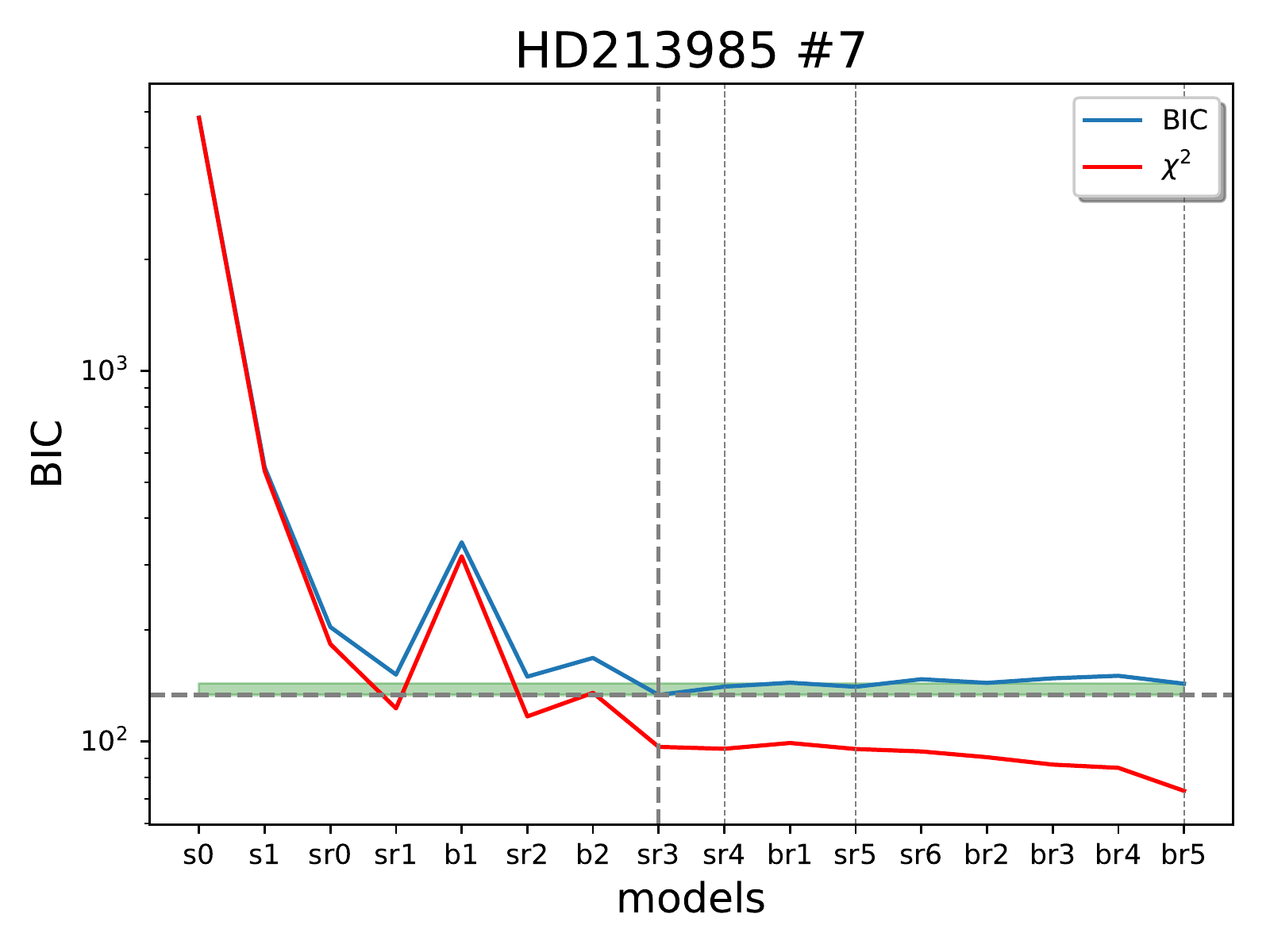}
\includegraphics[width=4.5cm]{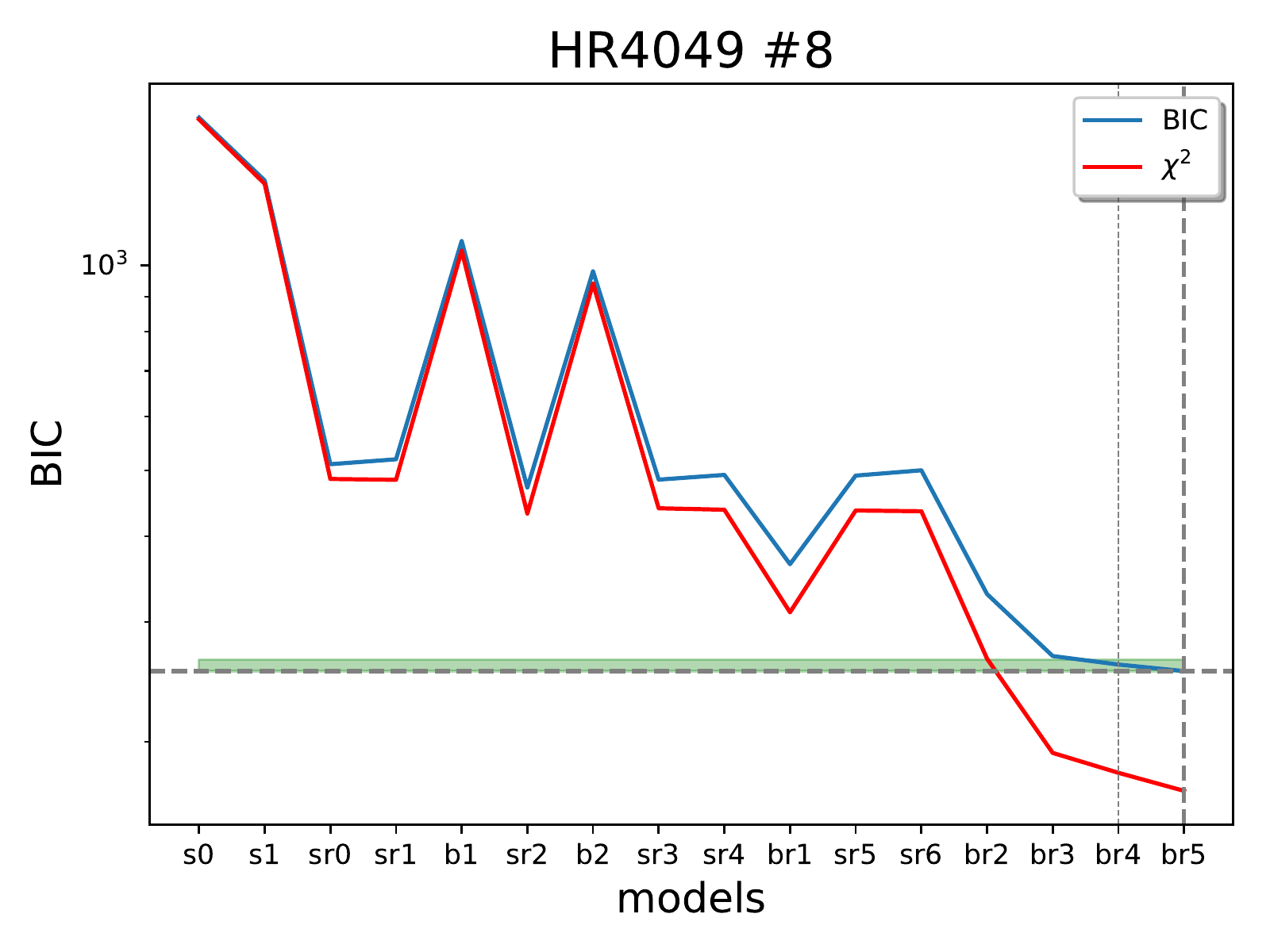}
\includegraphics[width=4.5cm]{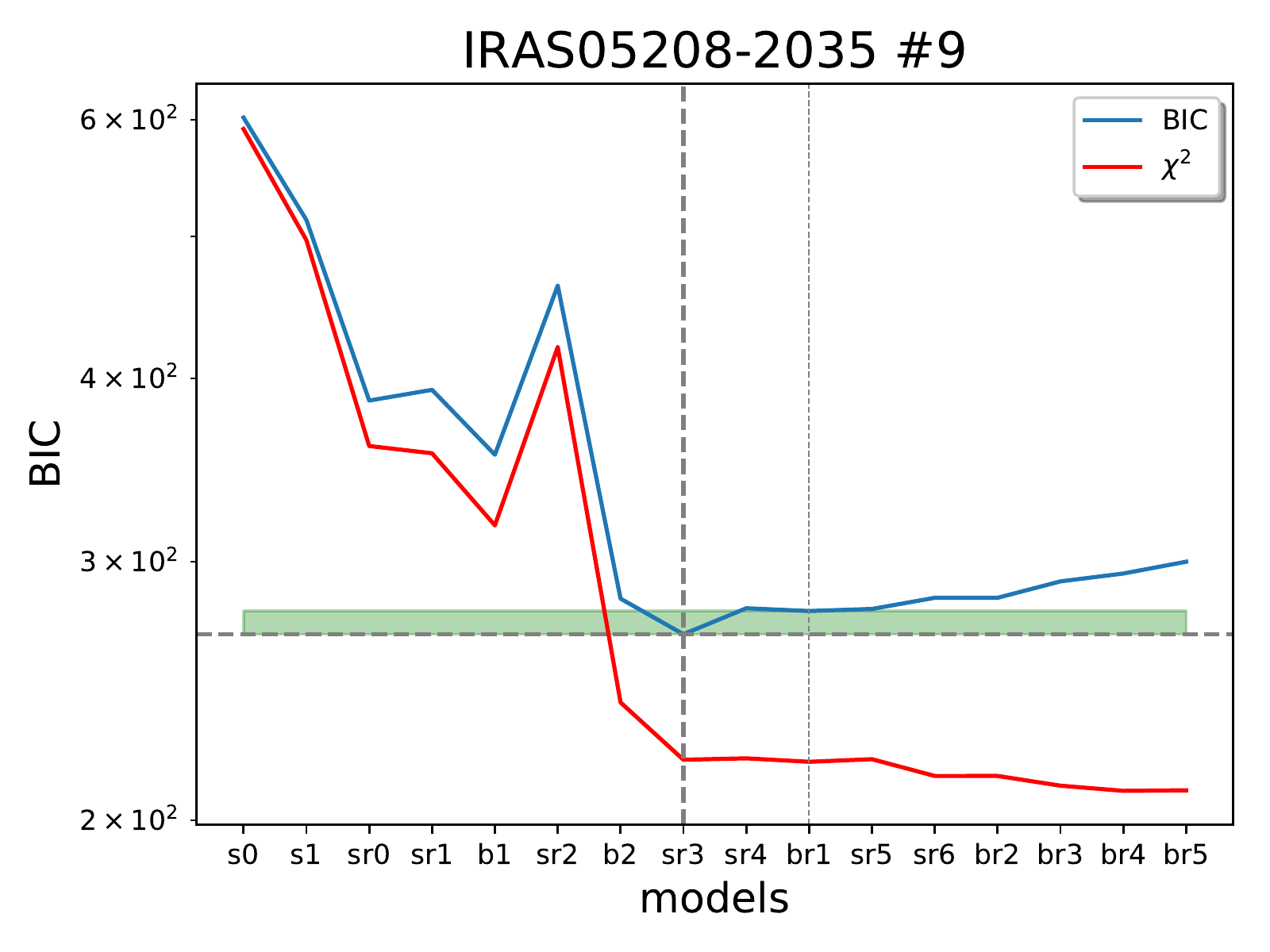}
\includegraphics[width=4.5cm]{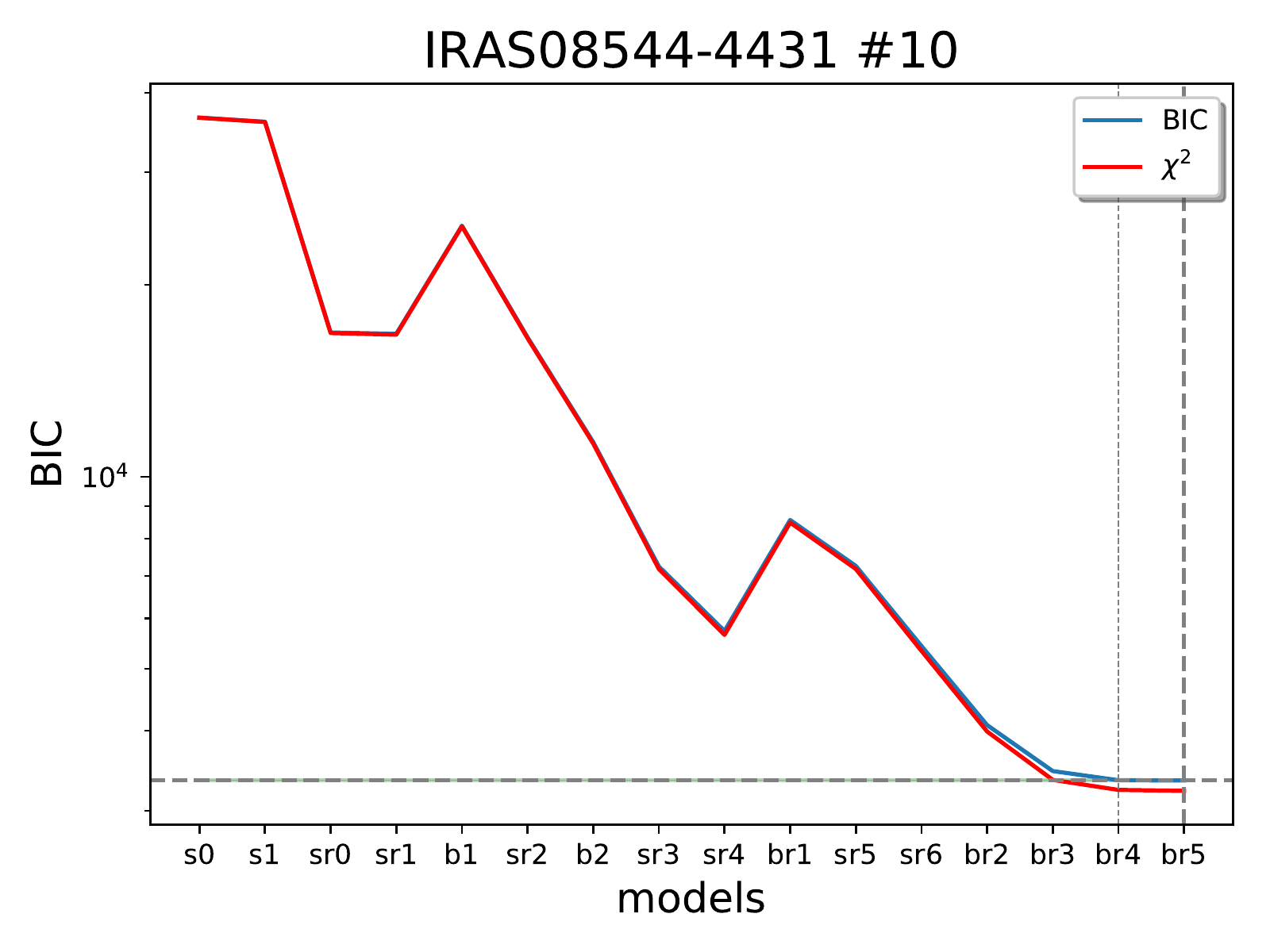}
\includegraphics[width=4.5cm]{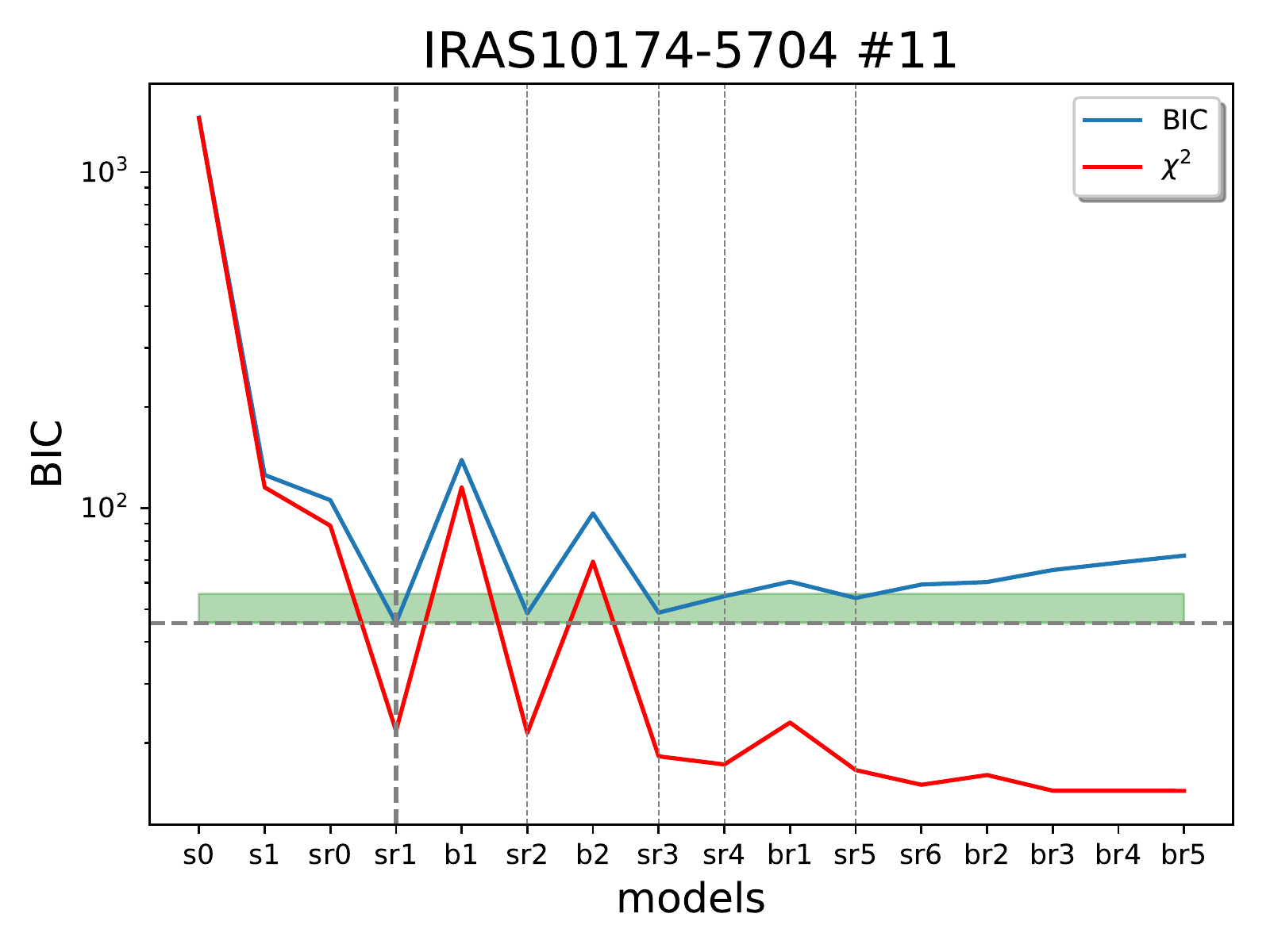}
\includegraphics[width=4.5cm]{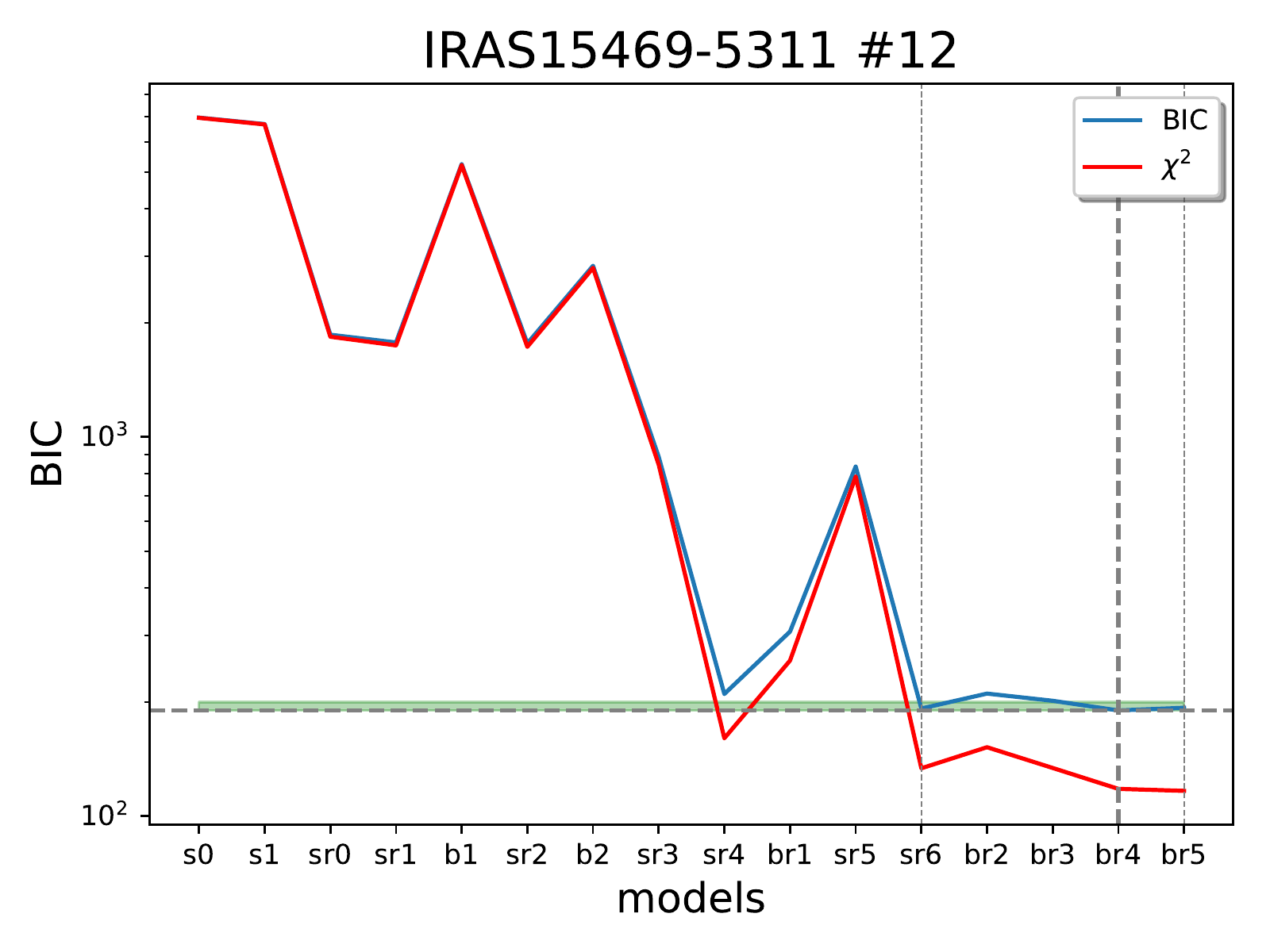}
\includegraphics[width=4.5cm]{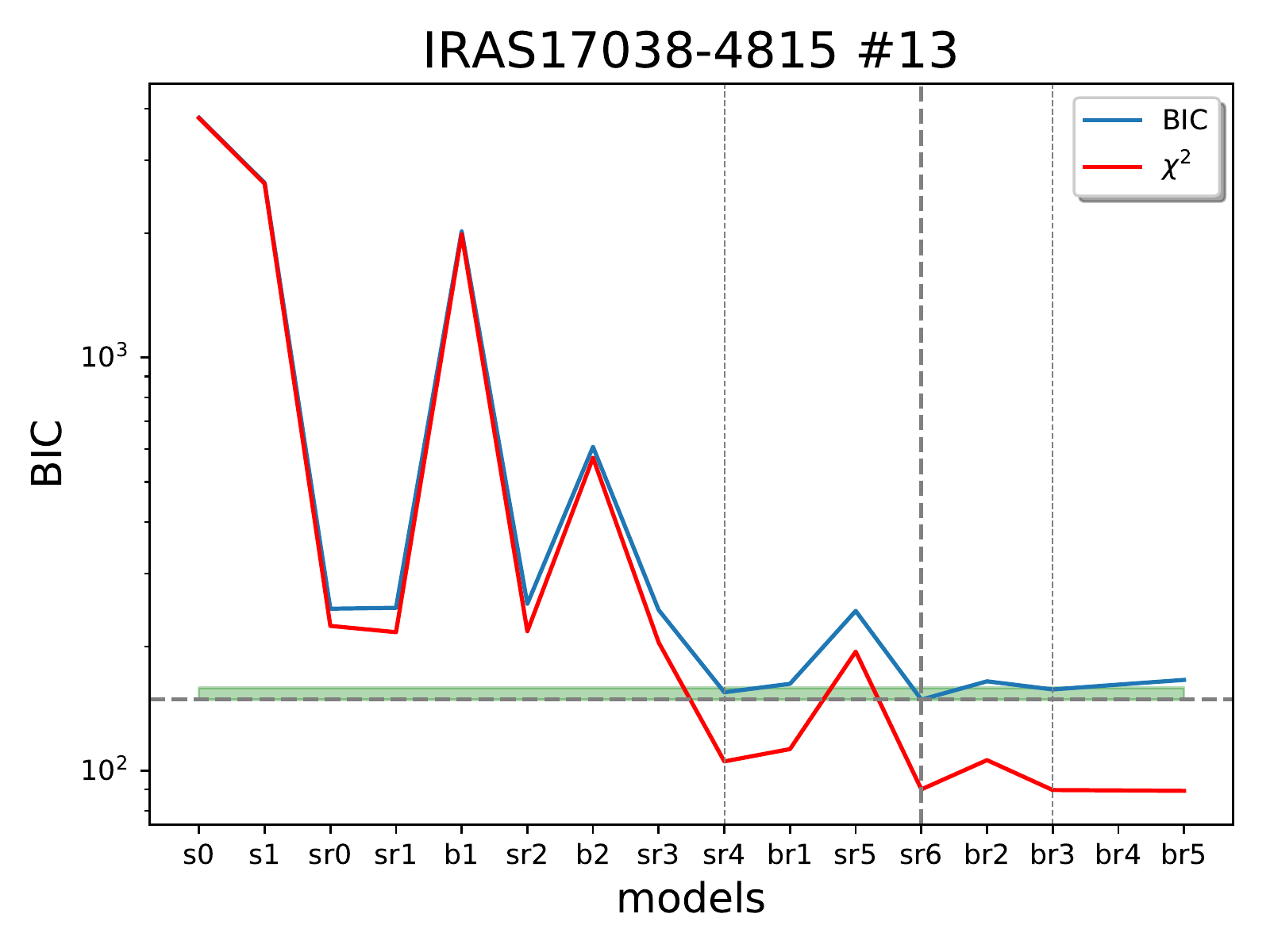}
\includegraphics[width=4.5cm]{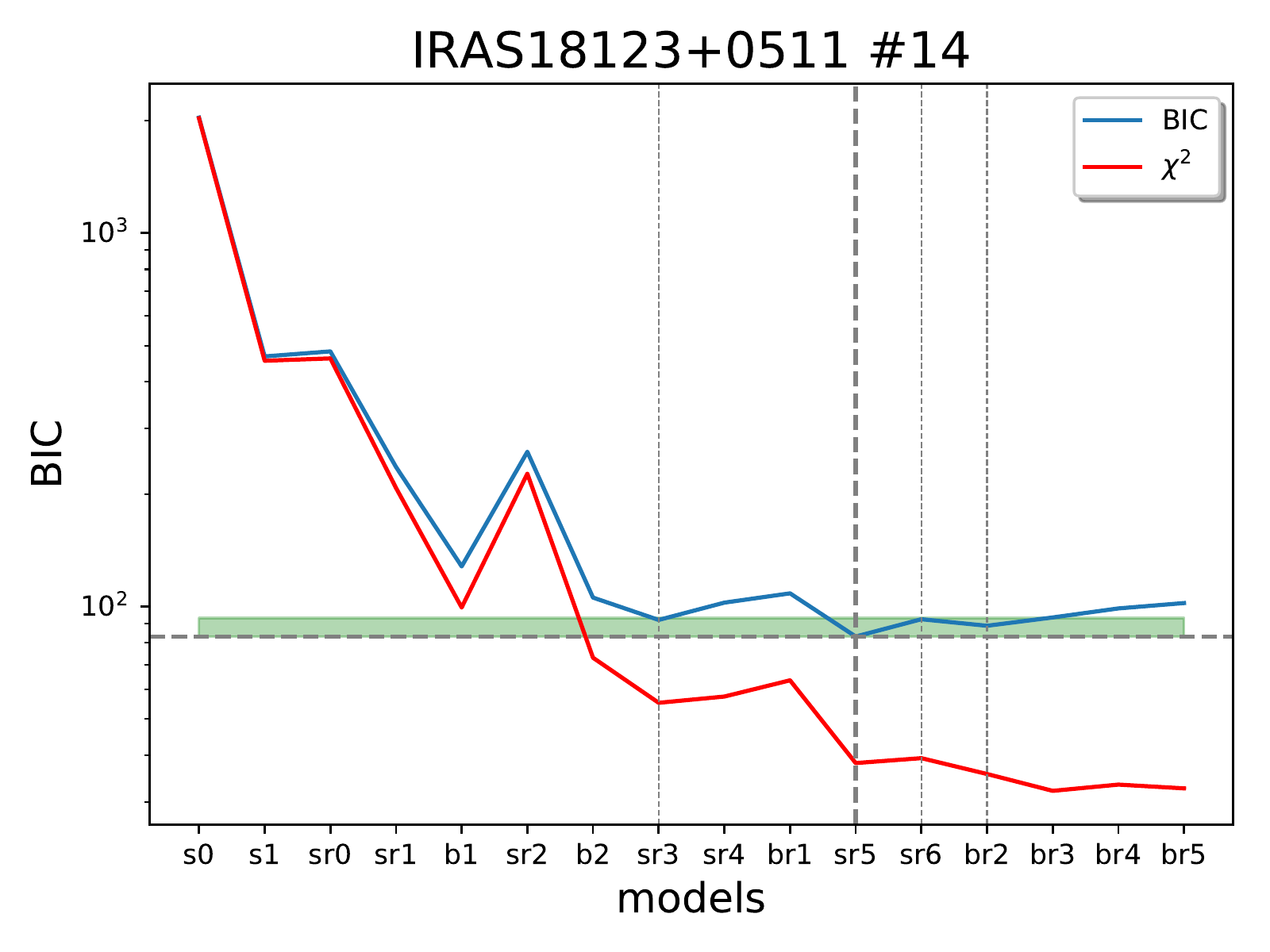}
\includegraphics[width=4.5cm]{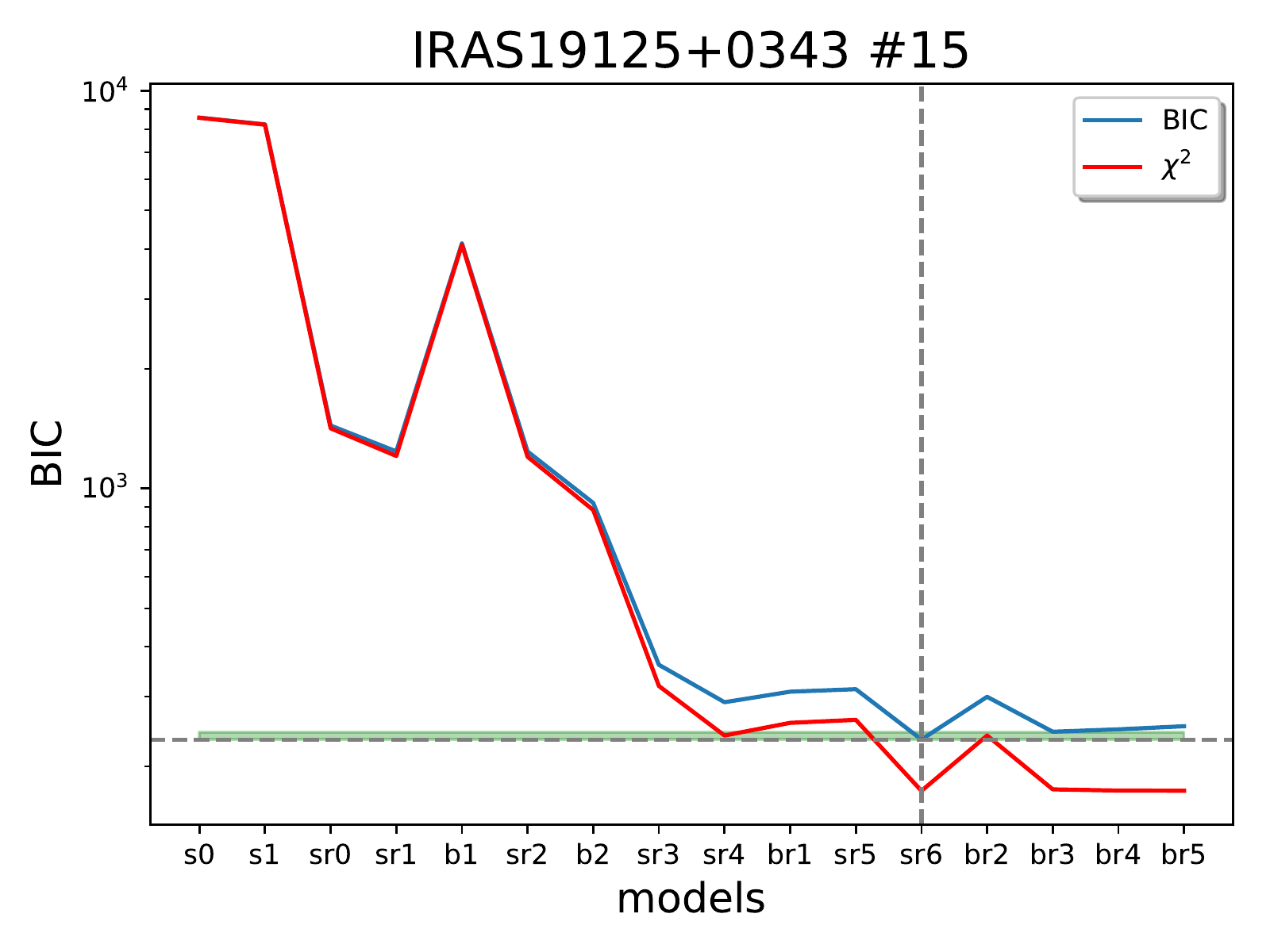}
\caption{BIC and $\chi^2$ per model for each target.}
\label{fig:BIC1}
\end{figure*}

\begin{figure*}
\centering
\includegraphics[width=4.5cm]{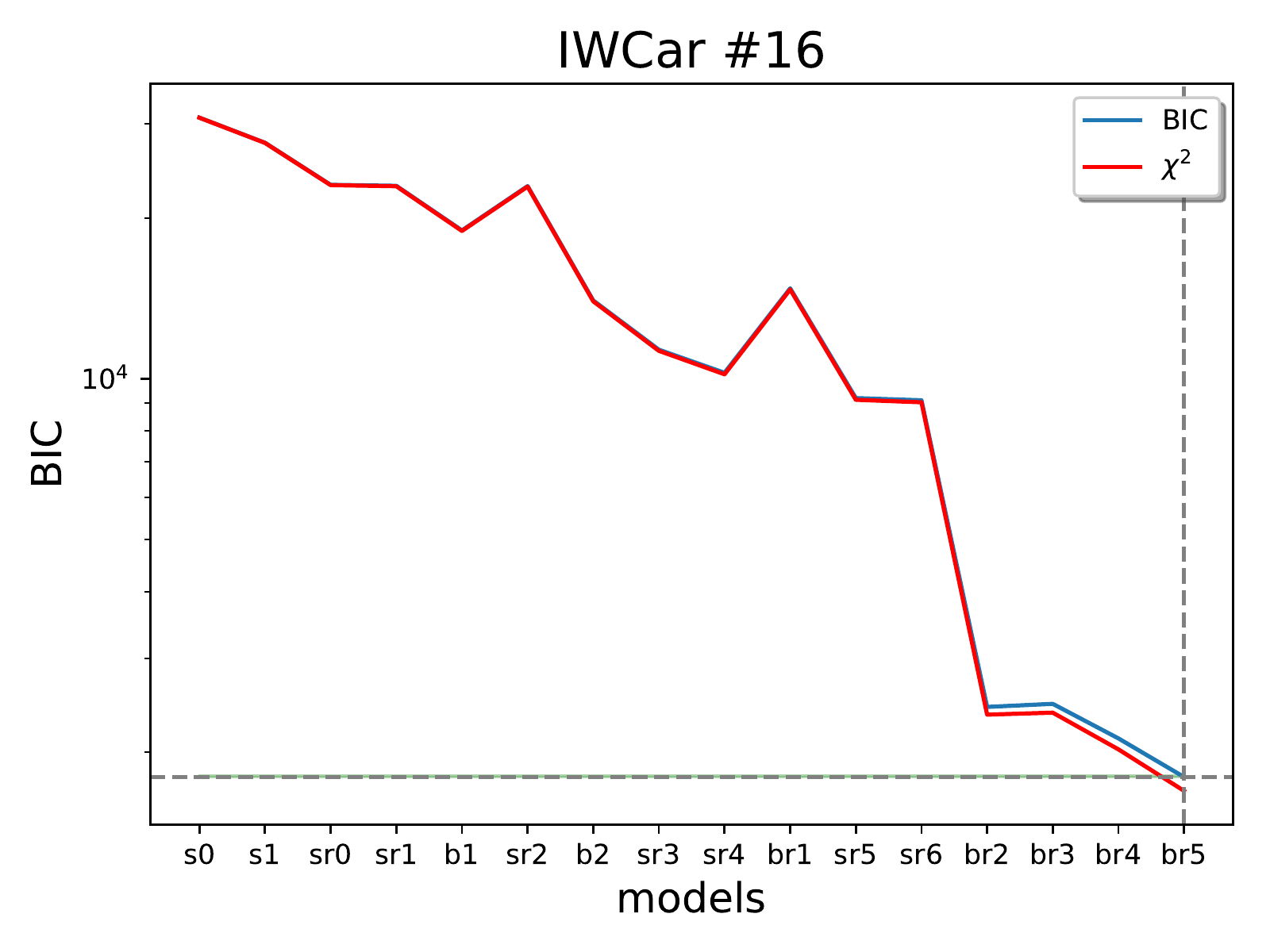}
\includegraphics[width=4.5cm]{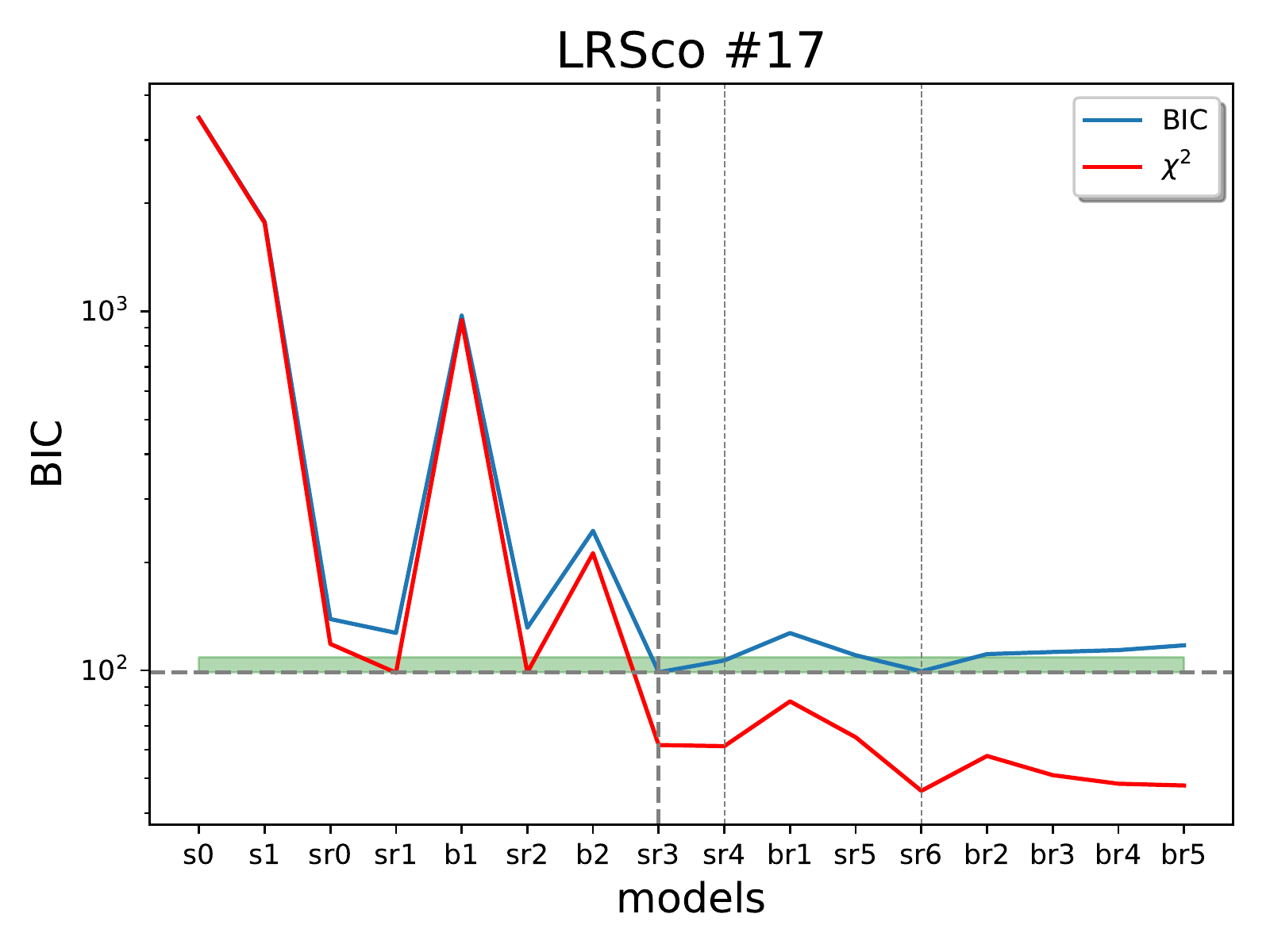}
\includegraphics[width=4.5cm]{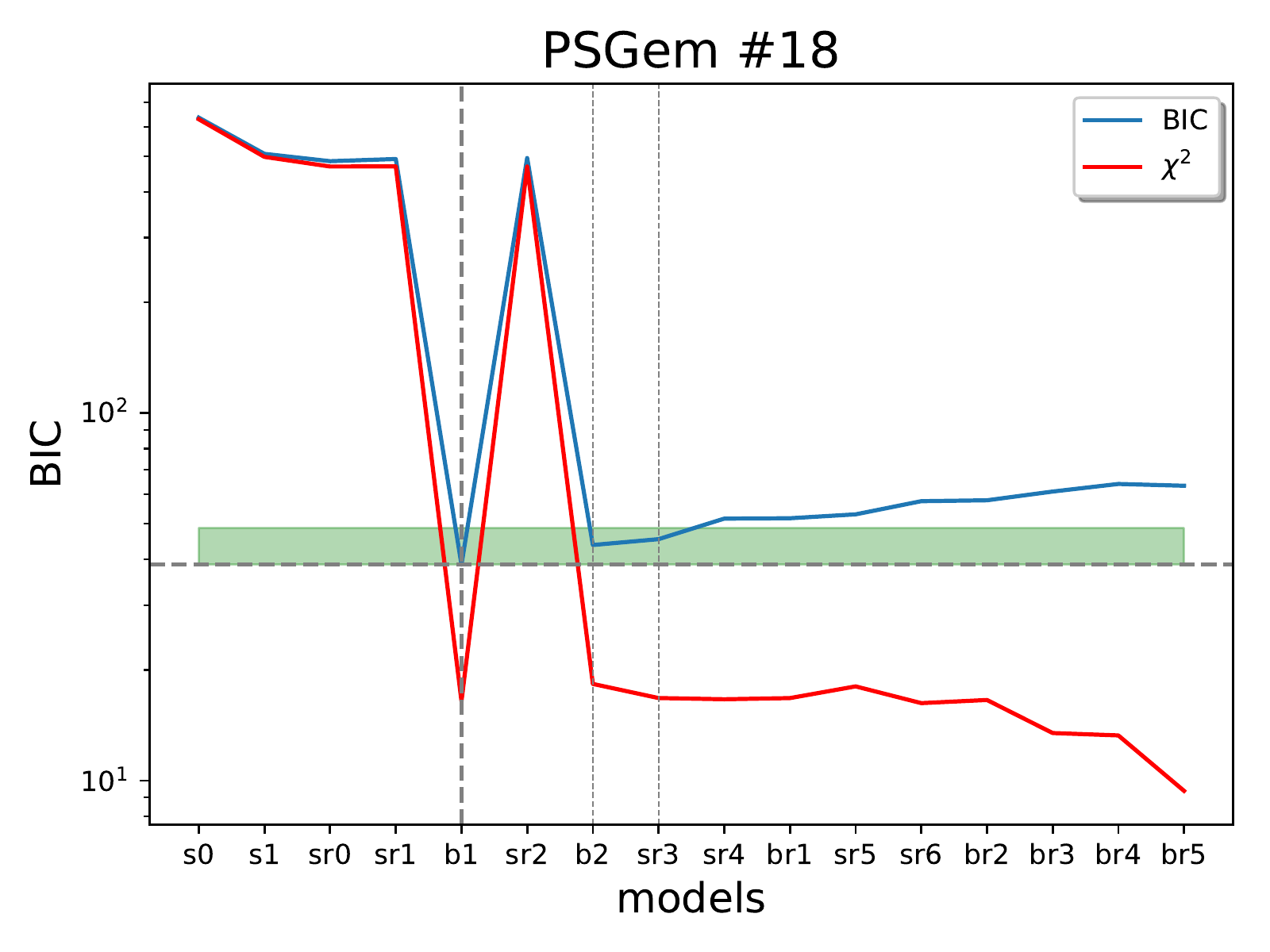}
\includegraphics[width=4.5cm]{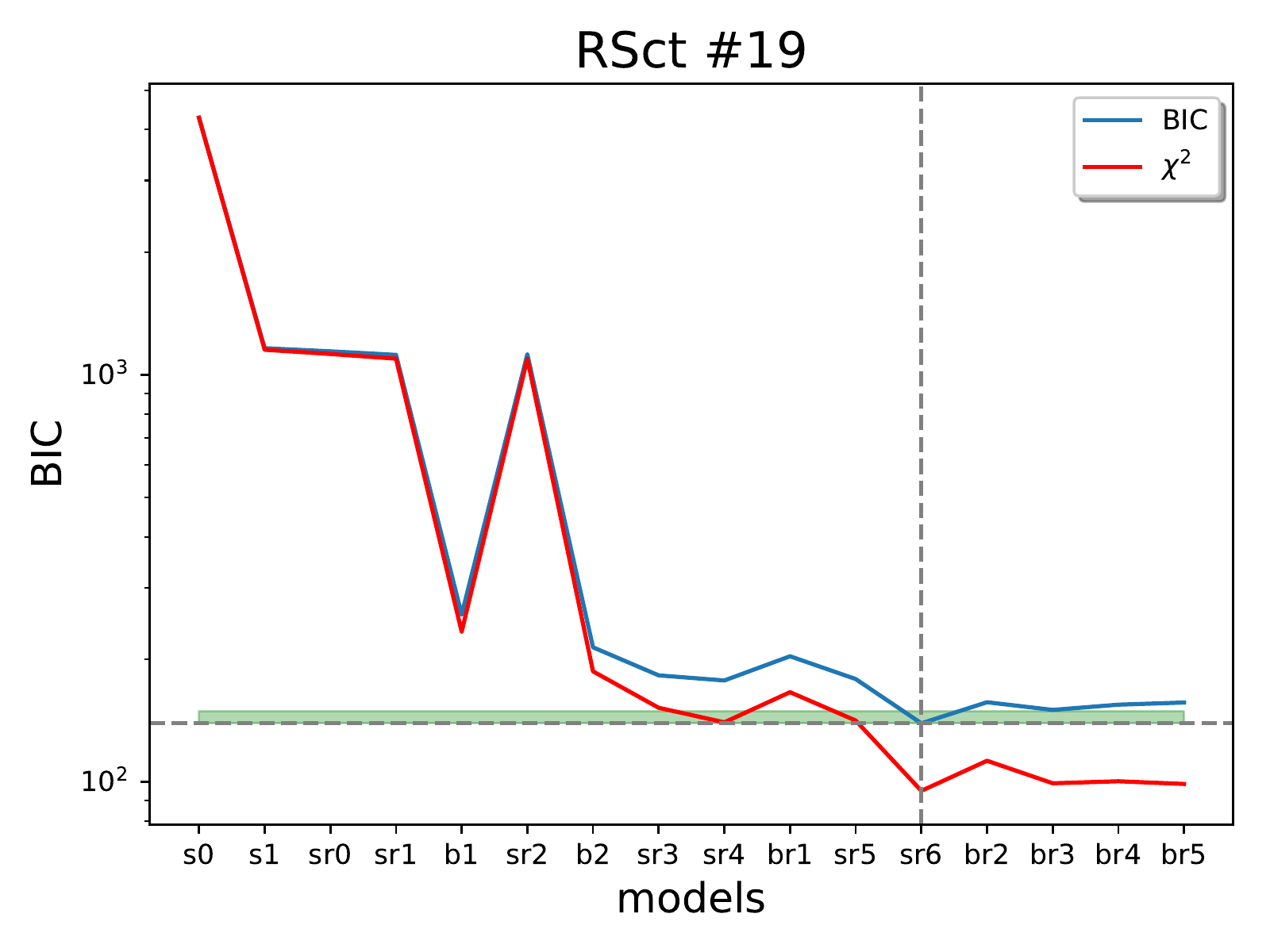}
\includegraphics[width=4.5cm]{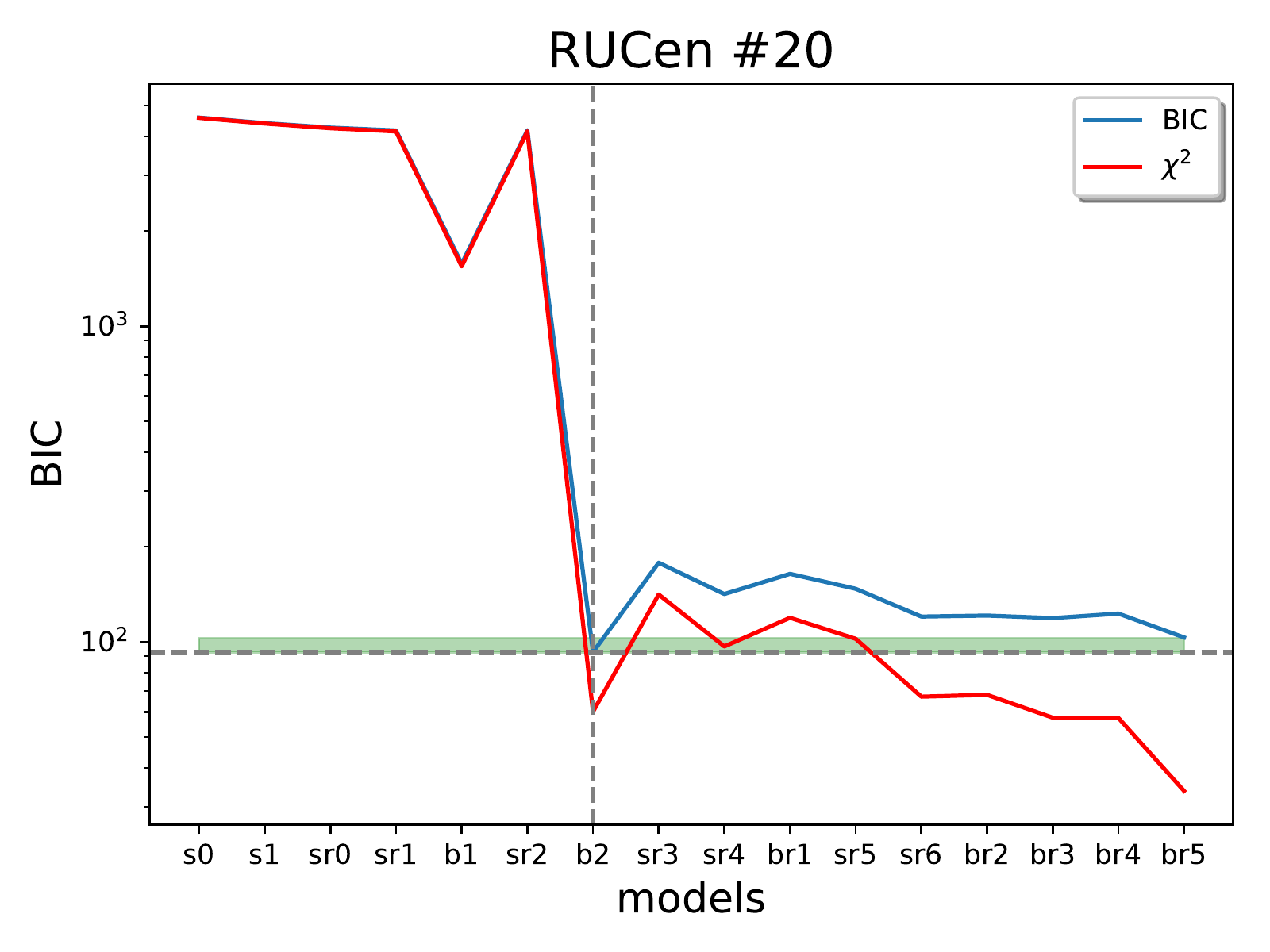}
\includegraphics[width=4.5cm]{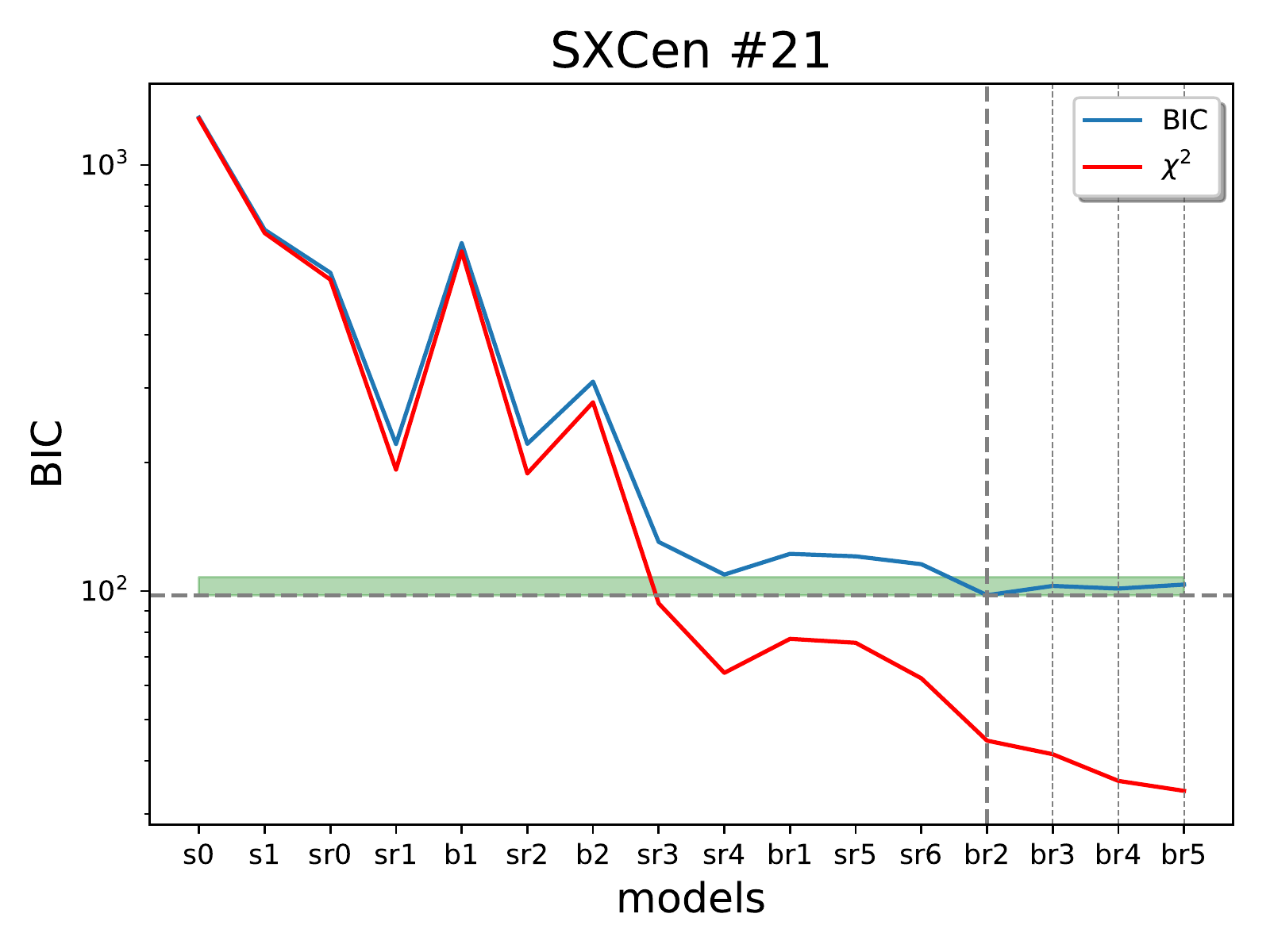}
\includegraphics[width=4.5cm]{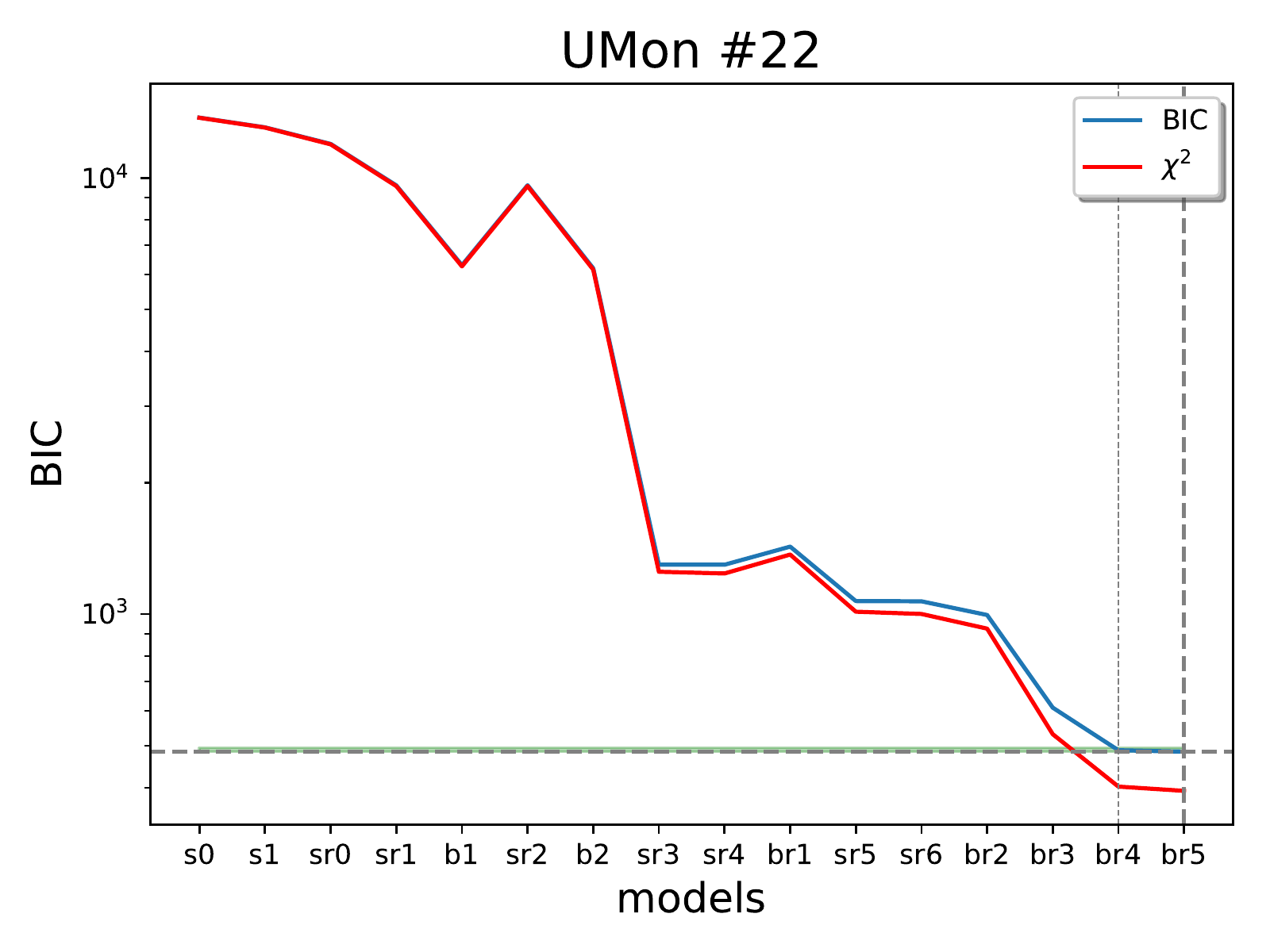}
\includegraphics[width=4.5cm]{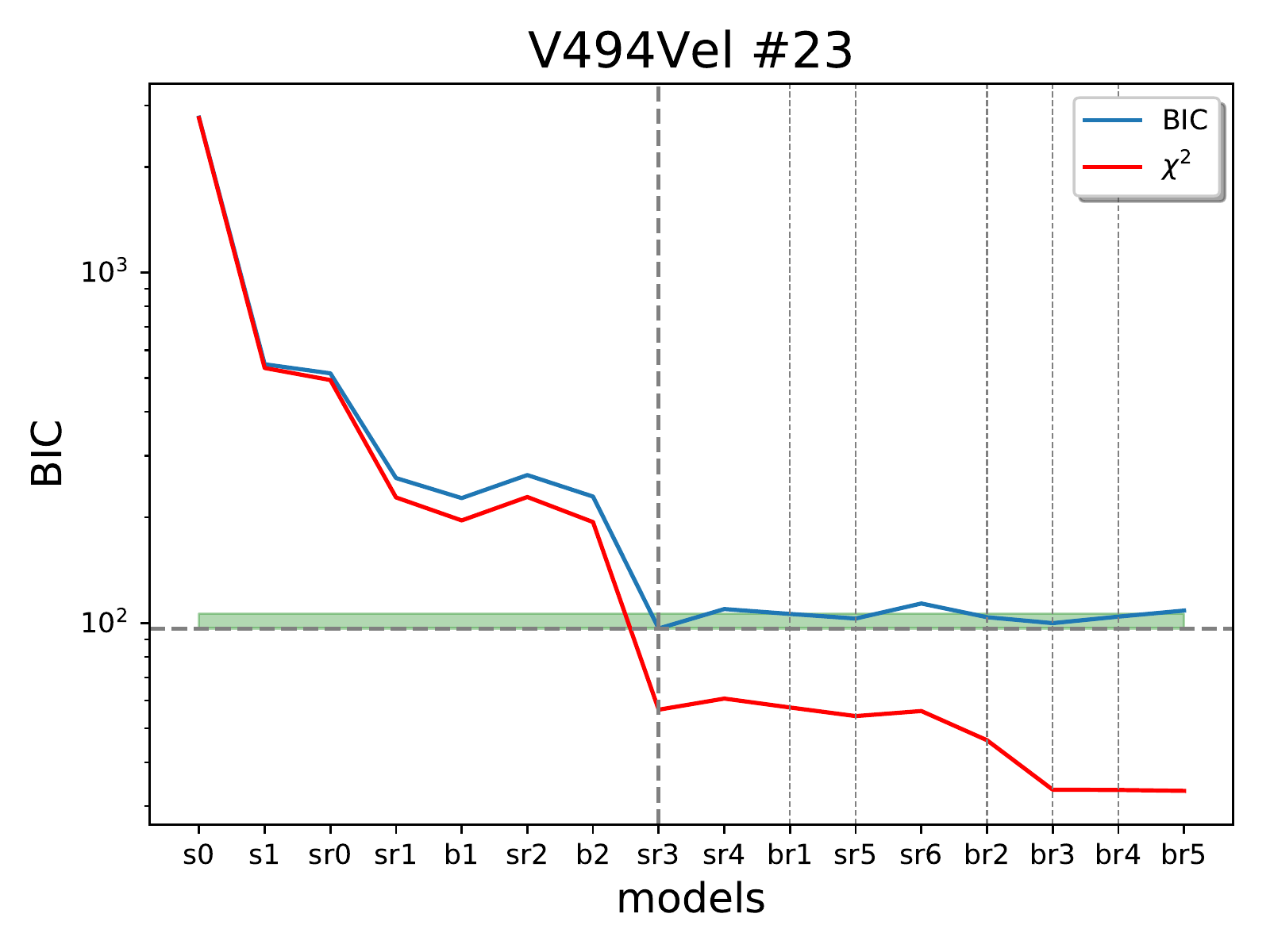}
\caption{As Fig\,\ref{fig:BIC2}. BIC and $\chi^2$ per model for each target.}
\label{fig:BIC2}
\end{figure*}

\section{\modif{Most likely models for each target as determined with the BIC.}}

\begin{sidewaystable}
\caption{Best-fit parameters of the selected models for AC Her (\#1). The best model with the lowest BIC (BIC$_0$) has the BIC value in black. For a BIC difference ($\Delta BIC$) smaller than 2 (weak evidence for the best model) the BIC value is in green. For 2<$\Delta$BIC<6 (positive evidence for the model with smallest BIC) the BIC value is in orange and for 6<$\Delta$BIC<10 the BIC value is red (moderetaly strong evidence for the model with the smallest BIC; see Sect.\,\ref{sec:strategy}). }             
\label{tab:resACHer}      
\centering                        
\begin{tabular}{c c c c c c c c c c c c c c c c c c c c c c c c }
 model & BIC & $\chi^2_\mathrm{red}$  & $f_{prim,0}$ & $f_{sec,0}$ & $f_{bg0}$ & $d_{sec}$ & $d_{back}$ & $T_\mathrm{ring}$ & $\theta$ & $\delta \theta$ & $inc$\\ 
 \hline 
  s0 & 36.0 & 1.0& $91.4^{0.3}_{-0.3}$ & - & - & - & -& $5810^{2710}_{-2240}$ & - & - & - \\ [1pt] 
  s1 & \textcolor{green}{37.6} & 1.0 & - & -& $8.1^{0.5}_{-0.5}$ & -& $-2.4^{1.0}_{-0.9}$ & - & - & - & - \\ [1pt] 
 sr0 & \textcolor{red}{45.0} & 1.1& $78.3^{12.4}_{-43.8}$ & -& $8.3^{0.5}_{-0.7}$ & - & -& $5920^{2750}_{-3066}$& $1.3^{5.8}_{-0.8}$& $2.0^{2.0}_{-1.4}$ & - \\ [1pt] 
\hline 
 $PA$ & $c_1$ & $s_1$ & $c2$ & $s2$ & $x_0$ & $y_0$ & rM & UD$_\mathrm{prim,0}$ & UD$_\mathrm{sec,0}$\\ 
 \hline 
  s0 &   - & - & - & - & - & - & - & - & - & - \\ [1pt] 
  s1 &   - & - & - & - & - & - & - & -& $1.1^{0.5}_{-0.6}$ & - \\ [1pt] 
 sr0 &   - & - & - & - & - & - & - & - & - & - \\ [1pt] 
  \hline
\end{tabular}
\end{sidewaystable}

\begin{sidewaystable}
\caption{Best-fit parameters of the selected models for AI Sco.}             
\label{tab:resAISco}      
\centering                        
\begin{tabular}{c c c c c c c c c c c c c c c c c c c c c c }
 model & BIC & $\chi^2_\mathrm{red}$  & $f_{prim,0}$ & $f_{sec,0}$ & $f_{bg0}$ & $d_{sec}$ & $d_{back}$ & $T_\mathrm{ring}$ & $\theta$ & $\delta \theta$ & $inc$\\ 
 \hline 
   br2 & 229.2 & 3.7& $41.1^{7.0}_{-6.7}$& $24.7^{6.5}_{-6.9}$& $1.1^{0.4}_{-0.4}$ & - & -& $1498^{63}_{-59}$& $4.6^{0.2}_{-0.2}$& $0.9^{0.1}_{-0.1}$& $47.3^{2.3}_{-2.6}$ \\ [1pt] 
 br3 & \textcolor{red}{237.5} & 3.9& $56.7^{2.8}_{-2.6}$& $8.4^{1.9}_{-1.6}$& $1.3^{0.5}_{-0.5}$ & - & -& $1449^{81}_{-82}$& $4.7^{0.3}_{-0.3}$& $0.9^{0.2}_{-0.3}$& $53.8^{3.4}_{-4.1}$ \\ [1pt]
 \hline \hline
  model  & $PA$ & $c_1$ & $s_1$ & $c2$ & $s2$ & $x_0$ & $y_0$ & rM & UD$_\mathrm{prim,0}$ & UD$_\mathrm{sec,0}$\\ 
 \hline 
 br2 & $179^{3}_{-3}$& $0.92^{0.04}_{-0.05}$& $-0.29^{0.12}_{-0.11}$ & - & -& $-0.13^{0.03}_{-0.03}$& $-0.09^{0.02}_{-0.03}$& $2.7^{1.1}_{-0.8}$ & - & - \\ [1pt] 
 br3 & $2^{3}_{-1}$& $-0.86^{0.14}_{-0.09}$& $-0.21^{0.20}_{-0.21}$& $-0.29^{0.14}_{-0.13}$& $0.22^{0.10}_{-0.10}$& $0.32^{0.08}_{-0.06}$& $0.24^{0.06}_{-0.06}$& $1.15^{0.54}_{-0.45}$ & - & - \\ [1pt]
  \hline
\end{tabular}
\end{sidewaystable}

\begin{sidewaystable}
\caption{Best-fit parameters of the selected models for EN TrA.}             
\label{tab:resENTrA}      
\centering                        
\begin{tabular}{c c c c c c c c c c c c c c c c c c c c c c }
 model & BIC & $\chi^2_\mathrm{red}$  & $f_{prim,0}$ & $f_{sec,0}$ & $f_{bg0}$ & $d_{sec}$ & $d_{back}$ & $T_\mathrm{ring}$ & $\theta$ & $\delta \theta$ & $inc$\\ 
 \hline 
 br2 & 88.7 & 1.0& $73.6^{1.5}_{-2.5}$& $7.8^{1.7}_{-1.2}$& $2.5^{0.9}_{-1.0}$ & - & -& $1445^{299}_{-210}$& $7.7^{1.7}_{-1.6}$& $1.8^{0.7}_{-0.6}$& $52.7^{7.8}_{-9.1}$ \\ [1pt] 
 sr3 & \textcolor{orange}{94.1} & 1.4& $68.9^{0.7}_{-0.8}$ & -& $4.0^{0.7}_{-0.8}$ & - & -& $1598^{274}_{-204}$& $2.9^{0.8}_{-0.5}$& $3.8^{0.8}_{-1.0}$& $62.5^{11.4}_{-9.6}$ \\ [1pt] 
 br3 & \textcolor{orange}{94.4} & 1.0& $69.8^{1.4}_{-1.1}$& $0.004^{0.005}_{-0.003}$& $4.0^{0.7}_{-0.7}$ & - & -& $1811^{477}_{-285}$& $6.7^{1.0}_{-0.8}$& $0.8^{0.3}_{-0.3}$& $60.6^{8.3}_{-11.2}$ \\ [1pt] 
 br4 & \textcolor{red}{98.1} & 1.0& $78.7^{2.7}_{-3.5}$& $0.003^{0.004}_{-0.002}$& $3.5^{0.8}_{-1.0}$ & - & -& $1376^{322}_{-222}$& $7.5^{1.5}_{-1.2}$& $1.1^{0.7}_{-0.4}$& $57.9^{10.8}_{-14.2}$ \\ [1pt] 
 \hline \hline
  model  & $PA$ & $c_1$ & $s_1$ & $c2$ & $s2$ & $x_0$ & $y_0$ & rM & UD$_\mathrm{prim,0}$ & UD$_\mathrm{sec,0}$\\ 
 \hline 
 br2 &  $13^{9}_{-7}$& $0.47^{0.13}_{-0.13}$& $0.71^{0.13}_{-0.16}$ & - & -& $-0.27^{0.09}_{-0.14}$& $-0.05^{0.05}_{-0.07}$& $3.2^{2.4}_{-1.4}$ & - & - \\ [1pt]
 sr3 &  $27^{6}_{-4}$& $0.49^{0.15}_{-0.14}$& $0.45^{0.25}_{-0.39}$ & - & - & - & - & - & - & - \\ [1pt] 
 br3 &  $29^{10}_{-11}$& $-0.81^{0.20}_{-0.12}$& $-0.01^{0.24}_{-0.27}$& $0.36^{0.35}_{-0.39}$& $0.27^{0.44}_{-0.53}$& $-0.11^{0.33}_{-0.37}$& $-1.46^{0.29}_{-0.29}$& $0.003^{0.004}_{-0.002}$ & - & - \\ [1pt] 
 br4 &  $27^{12}_{-12}$& $-0.60^{0.32}_{-0.22}$& $0.09^{0.42}_{-0.40}$& $0.32^{0.37}_{-0.47}$& $0.12^{0.50}_{-0.53}$& $-0.13^{0.41}_{-0.46}$& $-1.51^{0.37}_{-0.43}$& $0.009^{0.012}_{-0.006}$ & - & - \\ [1pt]
  \hline
\end{tabular}
\end{sidewaystable}

\begin{sidewaystable}
\caption{Best-fit parameters of the selected models for HD95767.}             
\label{tab:resHD95}      
\centering                        
\begin{tabular}{c c c c c c c c c c c c c c c c c c c c c c }
 model & BIC & $\chi^2_\mathrm{red}$  & $f_{prim,0}$ & $f_{sec,0}$ & $f_{bg0}$ & $d_{sec}$ & $d_{back}$ & $T_\mathrm{ring}$ & $\theta$ & $\delta \theta$ & $inc$\\ 
 \hline 
  br3 & 98.0 & 0.8& $59.9^{1.5}_{-2.5}$& $12.3^{2.3}_{-1.2}$& $1.1^{0.8}_{-0.7}$ & - & -& $986^{99}_{-83}$& $6.8^{0.4}_{-0.5}$& $0.9^{0.2}_{-0.2}$& $0.002^{0.003}_{-0.001}$ \\ [1pt] 
 br4 & \textcolor{orange}{103.7} & 0.9& $62.4^{3.4}_{-3.1}$& $11.1^{2.7}_{-1.8}$& $1.2^{0.8}_{-0.7}$ & - & -& $964^{101}_{-90}$& $7.0^{0.5}_{-0.5}$& $0.8^{0.2}_{-0.2}$& $0.01^{0.02}_{-0.01}$ \\ [1pt] 
 br5 & \textcolor{red}{106.7} & 0.9& $62.5^{4.1}_{-3.9}$& $11.2^{3.7}_{-2.2}$& $1.2^{0.8}_{-0.7}$& $-0.7^{2.3}_{-2.0}$ & -& $954^{106}_{-91}$& $7.0^{0.5}_{-0.5}$& $0.8^{0.2}_{-0.2}$& $0.01^{0.01}_{-0.01}$ \\ [1pt] 
 \hline \hline
 model  & $PA$ & $c_1$ & $s_1$ & $c2$ & $s2$ & $x_0$ & $y_0$ & rM & UD$_\mathrm{prim,0}$ & UD$_\mathrm{sec,0}$\\ 
 \hline 
 br3 & $165^{17}_{-18}$& $0.22^{0.21}_{-0.20}$& $0.34^{0.11}_{-0.09}$& $0.67^{0.17}_{-0.21}$& $0.11^{0.40}_{-0.46}$& $0.73^{0.13}_{-0.11}$& $-1.42^{0.16}_{-0.17}$& $0.24^{0.19}_{-0.15}$ & - & - \\ [1pt] 
 br4 & $168^{20}_{-21}$& $0.13^{0.23}_{-0.22}$& $0.34^{0.11}_{-0.09}$& $0.62^{0.19}_{-0.25}$& $0.15^{0.43}_{-0.52}$& $0.67^{0.14}_{-0.12}$& $-1.30^{0.21}_{-0.20}$& $0.4^{0.3}_{-0.2}$ & - & - \\ [1pt] 
 br5 & $166^{14.4}_{-13.8}$& $0.09^{0.21}_{-0.20}$& $0.32^{0.09}_{-0.07}$& $0.67^{0.17}_{-0.21}$& $0.05^{0.31}_{-0.32}$& $0.63^{0.13}_{-0.11}$& $-1.25^{0.19}_{-0.19}$& $0.34^{0.30}_{-0.23}$ & - & - \\ [1pt]
  \hline
\end{tabular}
\end{sidewaystable}

\begin{sidewaystable}
\caption{Best-fit parameters of the selected models for HD108015.}             
\label{tab:resHD108}      
\centering                        
\begin{tabular}{c c c c c c c c c c c c c c c c c c c c c c }
 model & BIC & $\chi^2_\mathrm{red}$  & $f_{prim,0}$ & $f_{sec,0}$ & $f_{bg0}$ & $d_{sec}$ & $d_{back}$ & $T_\mathrm{ring}$ & $\theta$ & $\delta \theta$ & $inc$\\ 
 \hline 
 sr6 & 229.2 & 3.7& $57.8^{0.2}_{-0.2}$ & -& $8.5^{0.4}_{-0.4}$ & - & -& $1155^{28}_{-27}$& $6.71^{0.06}_{-0.05}$& $0.4^{0.02}_{-0.02}$& $12.9^{4.7}_{-5.9}$ \\ [1pt]
 br3 & \textcolor{red}{235.9} & 3.9& $57.8^{0.2}_{-0.2}$& $0.003^{0.005}_{-0.002}$& $8.5^{0.4}_{-0.4}$ & - & -& $1154^{30}_{-28}$& $6.7^{0.1}_{-0.1}$& $0.43^{0.02}_{-0.02}$& $13.7^{4.7}_{-6.2}$ \\ [1pt] 
 \hline \hline
 model  & $PA$ & $c_1$ & $s_1$ & $c2$ & $s2$ & $x_0$ & $y_0$ & rM & UD$_\mathrm{prim,0}$ & UD$_\mathrm{sec,0}$\\ 
 \hline 
 sr6 &  $40^{16}_{-19}$& $-0.09^{0.09}_{-0.08}$& $0.21^{0.05}_{-0.05}$& $0.22^{0.05}_{-0.07}$& $-0.05^{0.14}_{-0.17}$& $0.09^{0.04}_{-0.04}$& $-0.75^{0.03}_{-0.03}$ & - & - & - \\ [1pt] 
 br3 &  $45^{14}_{-15}$& $-0.11^{0.08}_{-0.07}$& $0.20^{0.05}_{-0.05}$& $0.24^{0.04}_{-0.05}$& $-0.01^{0.11}_{-0.13}$& $0.09^{0.04}_{-0.04}$& $-0.76^{0.03}_{-0.03}$& $0.002^{0.004}_{-0.001}$ & - & - \\ [1pt] 
  \hline
\end{tabular}
\end{sidewaystable}

\begin{sidewaystable}
\caption{Best-fit parameters of the selected models for HD213985.}             
\label{tab:resHD213}      
\centering                        
\begin{tabular}{c c c c c c c c c c c c c c c c c c c c c c }
 model & BIC & $\chi^2_\mathrm{red}$  & $f_{prim,0}$ & $f_{sec,0}$ & $f_{bg0}$ & $d_{sec}$ & $d_{back}$ & $T_\mathrm{ring}$ & $\theta$ & $\delta \theta$ & $inc$\\ 
 \hline 
  sr3 & 133.5 & 1.9& $57.8^{0.6}_{-0.6}$ & -& $6.4^{0.6}_{-0.6}$ & - & -& $1203^{106}_{-93}$& $2.3^{0.4}_{-0.4}$& $3.3^{0.8}_{-0.6}$& $61.0^{2.5}_{-3.1}$ \\ [1pt] 
 sr5 & \textcolor{red}{140.5} & 1.9& $57.9^{0.6}_{-0.6}$ & -& $6.4^{0.6}_{-0.6}$ & - & -& $1206^{112}_{-93}$& $2.4^{0.4}_{-0.4}$& $3.1^{0.8}_{-0.6}$& $59.6^{3.1}_{-4.0}$ \\ [1pt] 
 sr4 & \textcolor{red}{140.6} & 2.0& $57.9^{0.6}_{-0.6}$ & -& $6.4^{0.6}_{-0.6}$ & - & -& $1200^{107}_{-90}$& $2.4^{0.7}_{-0.5}$& $3.2^{1.1}_{-1.0}$& $62.4^{2.6}_{-3.3}$ \\ [1pt] 
 br5 & \textcolor{red}{143.2} & 1.7& $63.5^{4.8}_{-6.1}$& $11.1^{5.6}_{-4.2}$& $5.0^{0.7}_{-0.8}$& $0.6^{2.3}_{-2.9}$ & -& $717^{110}_{-88}$& $4.4^{1.0}_{-0.8}$& $2.4^{0.7}_{-0.7}$& $64.7^{4.6}_{-5.7}$ \\ [1pt] 
 \hline \hline
 model  & $PA$ & $c_1$ & $s_1$ & $c2$ & $s2$ & $x_0$ & $y_0$ & rM & UD$_\mathrm{prim,0}$ & UD$_\mathrm{sec,0}$\\ 
 \hline 
  sr3 &  $93^{3}_{-4}$& $0.26^{0.18}_{-0.18}$& $0.83^{0.10}_{-0.14}$ & - & - & - & - & - & - & - \\ [1pt] 
 sr5 &  $92^{4}_{-5}$& $0.28^{0.17}_{-0.17}$& $0.81^{0.11}_{-0.15}$& $0.39^{0.38}_{-0.55}$& $-0.19^{0.59}_{-0.50}$ & - & - & - & - & - \\ [1pt] 
 sr4 &  $94^{3}_{-4}$& $0.000^{0.55}_{-0.49}$& $0.18^{0.49}_{-0.44}$ & - & -& $-0.20^{0.32}_{-0.31}$& $-0.22^{0.12}_{-0.13}$ & - & - & - \\ [1pt]
 br5 &  $81^{5}_{-5}$& $0.43^{0.19}_{-0.17}$& $0.64^{0.20}_{-0.26}$& $0.407^{0.38}_{-0.61}$& $-0.22^{0.46}_{-0.41}$& $0.59^{0.13}_{-0.10}$& $-0.20^{0.10}_{-0.08}$& $0.06^{0.07}_{-0.04}$ & - & - \\ [1pt] 
  \hline
\end{tabular}
\end{sidewaystable}

\begin{sidewaystable}
\caption{Best-fit parameters of the selected models for HR4049.}             
\label{tab:resHR4049}      
\centering                        
\begin{tabular}{c c c c c c c c c c c c c c c c c c c c c c }
 model & BIC & $\chi^2_\mathrm{red}$  & $f_{prim,0}$ & $f_{sec,0}$ & $f_{bg0}$ & $d_{sec}$ & $d_{back}$ & $T_\mathrm{ring}$ & $\theta$ & $\delta \theta$ & $inc$\\ 
 \hline 
 br5 & 253.9 & 1.3& $62.9^{1.0}_{-1.2}$& $2.5^{1.1}_{-0.8}$& $17.0^{0.7}_{-0.7}$& $-3.3^{1.0}_{-0.5}$ & -& $711^{24}_{-23}$& $16.4^{0.5}_{-0.5}$& $0.8^{0.1}_{-0.1}$& $49.3^{3.2}_{-3.3}$ \\ [1pt]
 br4 & \textcolor{orange}{259.6} & 1.4& $63.4^{0.9}_{-1.2}$& $2.3^{1.1}_{-0.7}$& $16.8^{0.7}_{-0.7}$ & - & -& $720^{25}_{-24}$& $16.7^{2.5}_{-1.3}$& $0.9^{0.1}_{-0.1}$& $53.2^{13.6}_{-5.8}$ \\ [1pt] 
 \hline \hline
 model  & $PA$ & $c_1$ & $s_1$ & $c2$ & $s2$ & $x_0$ & $y_0$ & rM & UD$_\mathrm{prim,0}$ & UD$_\mathrm{sec,0}$\\ 
 \hline 
 br5 &  $63^{7}_{-6}$& $0.04^{0.09}_{-0.07}$& $-0.29^{0.23}_{-0.24}$& $0.30^{0.18}_{-0.19}$& $-0.76^{0.31}_{-0.15}$& $-0.57^{0.14}_{-0.15}$& $-0.11^{0.04}_{-0.05}$& $0.7^{0.6}_{-0.4}$& $0.54^{0.08}_{-0.10}$ & - \\ [1pt] 
 br4 & $80.1^{10.4}_{-10.7}$& $-0.01^{0.17}_{-0.10}$& $-0.18^{0.22}_{-0.35}$& $0.36^{0.20}_{-0.25}$& $0.04^{0.27}_{-0.43}$& $-0.75^{0.15}_{-0.17}$& $-0.15^{0.05}_{-0.06}$& $0.3^{0.4}_{-0.2}$& $0.57^{0.07}_{-0.10}$ & - \\ [1pt]
  \hline
\end{tabular}
\end{sidewaystable}

\begin{sidewaystable}
\caption{Best-fit parameters of the selected models for IRAS05208-2035.}             
\label{tab:resIRAS05}      
\centering                        
\begin{tabular}{c c c c c c c c c c c c c c c c c c c c c c }
 model & BIC & $\chi^2_\mathrm{red}$  & $f_{prim,0}$ & $f_{sec,0}$ & $f_{bg0}$ & $d_{sec}$ & $d_{back}$ & $T_\mathrm{ring}$ & $\theta$ & $\delta \theta$ & $inc$\\ 
 \hline 
  sr3 & 267.8 & 1.1& $90.6^{0.3}_{-0.3}$ & -& $5.0^{0.4}_{-0.4}$ & - & -& $7871^{1484}_{-1918}$& $6.1^{0.6}_{-0.7}$& $0.9^{0.3}_{-0.3}$& $53.8^{3.7}_{-5.0}$ \\ [1pt] 
 br1 & \textcolor{red}{277.7} & 1.1& $90^{0.2}_{-0.2}$& $0.56^{0.35}_{-0.34}$& $5.40^{0.33}_{-0.34}$ & - & -& $7821^{1519}_{-1976}$& $5.4^{0.5}_{-0.5}$& $0.6^{0.3}_{-0.4}$& $57.0^{3.5}_{-4.6}$ \\ [1pt]  
 \hline \hline
 model  & $PA$ & $c_1$ & $s_1$ & $c2$ & $s2$ & $x_0$ & $y_0$ & rM & UD$_\mathrm{prim,0}$ & UD$_\mathrm{sec,0}$\\ 
 \hline 
  sr3 & $155^{4}_{-4}$& $-0.84^{0.11}_{-0.08}$& $0.32^{0.13}_{-0.13}$ & - & - & - & - & - & - & - \\ [1pt] 
 br1 & $153^{4}_{-5}$ & - & - & - & -& $0.31^{0.14}_{-0.14}$& $-1.31^{0.14}_{-0.14}$& $0.05^{0.08}_{-0.04}$ & - & - \\ [1pt]
  \hline
\end{tabular}
\end{sidewaystable}

\begin{sidewaystable}
\caption{Best-fit parameters of the selected models for IRAS08544-4431.}             
\label{tab:resIRAS08544}      
\centering                        
\begin{tabular}{c c c c c c c c c c c c c c c c c c c c c c }
 model & BIC & $\chi^2_\mathrm{red}$  & $f_{prim,0}$ & $f_{sec,0}$ & $f_{bg0}$ & $d_{sec}$ & $d_{back}$ & $T_\mathrm{ring}$ & $\theta$ & $\delta \theta$ & $inc$\\ 
 \hline 
 br5 & 3343.1 & 2.4& $58.5^{0.4}_{-0.4}$& $5.2^{0.4}_{-0.4}$& $15.3^{0.2}_{-0.2}$& $-0.4^{0.8}_{-0.8}$ & -& $875^{10}_{-9}$& $14.27^{0.04}_{-0.04}$& $0.48^{0.01}_{-0.01}$& $21.3^{0.8}_{-0.8}$ \\ [1pt]
 br4 & \textcolor{orange}{3346.9} & 2.5& $58.3^{0.5}_{-0.6}$& $5.5^{0.5}_{-0.5}$& $15.3^{0.2}_{-0.2}$ & - & -& $891^{7}_{-7}$& $14.28^{0.04}_{-0.04}$& $0.48^{0.01}_{-0.01}$& $21.4^{0.8}_{-0.8}$ \\ [1pt] 
 \hline \hline
 model  & $PA$ & $c_1$ & $s_1$ & $c2$ & $s2$ & $x_0$ & $y_0$ & rM & UD$_\mathrm{prim,0}$ & UD$_\mathrm{sec,0}$\\ 
 \hline 
 br5 & $12^{3}_{-3}$& $0.38^{0.01}_{-0.01}$& $-0.23^{0.02}_{-0.02}$& $0.02^{0.03}_{-0.03}$& $-0.27^{0.01}_{-0.01}$& $0.36^{0.01}_{-0.01}$& $0.61^{0.02}_{-0.02}$& $0.02^{0.03}_{-0.02}$& $0.46^{0.02}_{-0.02}$ & - \\ [1pt] 
   br4 & $11^{3}_{-3}$& $0.38^{0.01}_{-0.02}$& $-0.23^{0.02}_{-0.02}$& $0.01^{0.03}_{-0.03}$& $-0.27^{0.01}_{-0.01}$& $0.36^{0.01}_{-0.01}$& $0.56^{0.02}_{-0.02}$& $0.02^{0.03}_{-0.02}$& $0.46^{0.02}_{-0.03}$ & - \\ [1pt] 
  \hline
\end{tabular}
\end{sidewaystable}

\begin{sidewaystable}
\caption{Best-fit parameters of the selected models for IRAS10174-5704.}             
\label{tab:resIRAS10174}      
\centering                        
\begin{tabular}{c c c c c c c c c c c c c c c c c c c c c c }
 model & BIC & $\chi^2_\mathrm{red}$  & $f_{prim,0}$ & $f_{sec,0}$ & $f_{bg0}$ & $d_{sec}$ & $d_{back}$ & $T_\mathrm{ring}$ & $\theta$ & $\delta \theta$ & $inc$\\ 
 \hline 
 sr1 & 45.5 & 0.9& $56.8^{8.9}_{-10.7}$ & -& $0.4^{0.5}_{-0.3}$ & - & -& $7700^{1578}_{-1877}$& $1.5^{0.3}_{-0.4}$& $0.6^{0.9}_{-0.5}$& $33.6^{2.5}_{-2.4}$ \\ [1pt] 
 sr2 & \textcolor{orange}{48.6} & 1.0& $57.3^{8.6}_{-10.5}$ & -& $0.4^{0.5}_{-0.3}$ & -& $-0.6^{3.0}_{-2.4}$& $7780^{1550}_{-1867}$& $1.5^{0.3}_{-0.4}$& $0.6^{0.8}_{-0.4}$& $33.6^{2.5}_{-2.4}$ \\ [1pt] 
 sr3 & \textcolor{orange}{48.9} & 0.9& $48.8^{9.5}_{-9.2}$ & -& $0.6^{0.6}_{-0.4}$ & - & -& $7767^{1552}_{-1934}$& $1.4^{0.2}_{-0.3}$& $0.6^{0.8}_{-0.4}$& $32.8^{2.6}_{-2.7}$ \\ [1pt] 
 sr5 & \textcolor{red}{54.0} & 0.9& $48.7^{9.3}_{-9.6}$ & -& $0.6^{0.6}_{-0.4}$ & - & -& $7773^{1579}_{-2011}$& $1.3^{0.3}_{-0.3}$& $0.8^{0.9}_{-0.6}$& $27.5^{10.6}_{-13.0}$ \\ [1pt] 
 sr4 & \textcolor{red}{54.7} & 0.9& $49.4^{7.9}_{-8.4}$ & -& $0.6^{0.6}_{-0.4}$ & - & -& $6992^{1962}_{-1924}$& $1.2^{0.3}_{-0.3}$& $1.2^{0.8}_{-0.7}$& $33.5^{3.4}_{-3.4}$ \\ [1pt]
 \hline \hline
 model  & $PA$ & $c_1$ & $s_1$ & $c2$ & $s2$ & $x_0$ & $y_0$ & rM & UD$_\mathrm{prim,0}$ & UD$_\mathrm{sec,0}$\\ 
 \hline 
 sr1 &  $13^{5}_{-6}$ & - & - & - & - & - & - & - & - & - \\ [1pt] 
 sr2 &  $13^{5}_{-6}$ & - & - & - & - & - & - & - & - & - \\ [1pt] 
 sr3 &  $11^{6}_{-6}$& $0.09^{0.20}_{-0.19}$& $-0.10^{0.18}_{-0.22}$ & - & - & - & - & - & - & - \\ [1pt] 
 sr5 & $14^{12}_{-9}$& $0.13^{0.21}_{-0.18}$& $-0.02^{0.27}_{-0.24}$& $0.12^{0.24}_{-0.30}$& $0.03^{0.17}_{-0.14}$ & - & - & - & - & - \\ [1pt] 
 sr4 & $14^{7}_{-7}$& $-0.00^{0.36}_{-0.34}$& $-0.09^{0.43}_{-0.36}$ & - & -& $-0.01^{0.05}_{-0.05}$& $-0.01^{0.05}_{-0.05}$ & - & - & - \\ [1pt]
  \hline
\end{tabular}
\end{sidewaystable}

\begin{sidewaystable}
\caption{Best-fit parameters of the selected models for IRAS15469-5311.}             
\label{tab:resIRAS15}      
\centering                        
\begin{tabular}{c c c c c c c c c c c c c c c c c c c c c c }
 model & BIC & $\chi^2_\mathrm{red}$  & $f_{prim,0}$ & $f_{sec,0}$ & $f_{bg0}$ & $d_{sec}$ & $d_{back}$ & $T_\mathrm{ring}$ & $\theta$ & $\delta \theta$ & $inc$\\ 
 \hline 
 br4 & 190.0 & 1.6& $57.2^{0.4}_{-0.4}$& $0.1^{0.1}_{-0.1}$& $12.8^{0.5}_{-0.6}$ & - & -& $818^{17}_{-17}$& $10.4^{0.5}_{-0.3}$& $0.39^{0.03}_{-0.03}$& $53.5^{1.8}_{-2.2}$ \\ [1pt] 
 sr6 & \textcolor{orange}{192.3} & 1.7& $55.9^{0.1}_{-0.1}$ & -& $13.5^{0.5}_{-0.5}$ & - & -& $853^{17}_{-16}$& $9.6^{0.2}_{-0.2}$& $0.37^{0.02}_{-0.02}$& $48.5^{2.6}_{-2.7}$ \\ [1pt] 
 br5 & \textcolor{orange}{193.1} & 1.6& $57.2^{0.4}_{-0.4}$& $0.007^{0.011}_{-0.005}$& $12.8^{0.6}_{-0.6}$& $0.09^{2.68}_{-2.82}$ & -& $819^{18}_{-17}$& $10.4^{0.3}_{-0.3}$& $0.39^{0.03}_{-0.03}$& $53.9^{1.8}_{-2.1}$ \\ [1pt] 
 \hline \hline
 model  & $PA$ & $c_1$ & $s_1$ & $c2$ & $s2$ & $x_0$ & $y_0$ & rM & UD$_\mathrm{prim,0}$ & UD$_\mathrm{sec,0}$\\ 
 \hline 
 br4 &  $64.1^{2.4}_{-2.2}$& $0.08^{0.04}_{-0.04}$& $-0.34^{0.04}_{-0.05}$& $-0.19^{0.06}_{-0.05}$& $0.30^{0.05}_{-0.05}$& $-0.91^{0.06}_{-0.05}$& $0.72^{0.03}_{-0.03}$& $0.27^{0.34}_{-0.20}$& $0.47^{0.06}_{-0.06}$ & - \\ [1pt]
 sr6 &  $57.4^{2.0}_{-2.2}$& $0.01^{0.04}_{-0.04}$& $-0.32^{0.03}_{-0.04}$& $-0.03^{0.06}_{-0.06}$& $0.18^{0.04}_{-0.04}$& $-0.96^{0.04}_{-0.04}$& $0.72^{0.03}_{-0.03}$ & - & - & - \\ [1pt] 
 br5 &  $63.9^{2.3}_{-2.2}$& $0.08^{0.04}_{-0.04}$& $-0.34^{0.04}_{-0.05}$& $-0.20^{0.05}_{-0.05}$& $0.30^{0.05}_{-0.05}$& $-0.92^{0.06}_{-0.05}$& $0.73^{0.03}_{-0.03}$& $0.01^{0.02}_{-0.01}$& $0.47^{0.06}_{-0.06}$ & - \\ [1pt] 
 
  \hline
\end{tabular}
\end{sidewaystable}

\begin{sidewaystable}
\caption{Best-fit parameters of the selected models for IRAS17038-4815.}             
\label{tab:resIRAS17}      
\centering                        
\begin{tabular}{c c c c c c c c c c c c c c c c c c c c c c }
 model & BIC & $\chi^2_\mathrm{red}$  & $f_{prim,0}$ & $f_{sec,0}$ & $f_{bg0}$ & $d_{sec}$ & $d_{back}$ & $T_\mathrm{ring}$ & $\theta$ & $\delta \theta$ & $inc$\\ 
 \hline 
 sr6 & 148.6 & 1.2& $71.7^{0.2}_{-0.2}$ & -& $4.2^{0.4}_{-0.4}$ & - & -& $2132^{118}_{-104}$& $5.3^{0.3}_{-0.3}$& $1.2^{0.1}_{-0.1}$& $36.7^{5.1}_{-7.8}$ \\ [1pt]
 sr4 & \textcolor{red}{154.8} & 1.3& $71.8^{0.3}_{-0.3}$ & -& $4.5^{0.4}_{-0.4}$ & - & -& $2203^{130}_{-109}$& $4.6^{0.1}_{-0.1}$& $1.1^{0.1}_{-0.1}$& $19.7^{4.7}_{-9.4}$ \\ [1pt] 
 br3 & \textcolor{red}{157.3} & 1.2& $71.8^{0.2}_{-0.2}$& $0.004^{0.007}_{-0.003}$& $4.2^{0.4}_{-0.4}$ & - & -& $2140^{120}_{-109}$& $5.4^{0.3}_{-0.3}$& $1.1^{0.1}_{-0.1}$& $37.7^{4.5}_{-6.3}$ \\ [1pt] 
 \hline \hline
 model  & $PA$ & $c_1$ & $s_1$ & $c2$ & $s2$ & $x_0$ & $y_0$ & rM & UD$_\mathrm{prim,0}$ & UD$_\mathrm{sec,0}$\\ 
 \hline 
 sr6 &  $158^{7}_{-8}$& $-0.09^{0.07}_{-0.08}$& $0.27^{0.12}_{-0.10}$& $-0.43^{0.19}_{-0.17}$& $0.29^{0.16}_{-0.15}$& $0.24^{0.07}_{-0.07}$& $0.77^{0.13}_{-0.11}$ & - & - & - \\ [1pt] 
 sr4 &  $166^{16}_{-26}$& $-0.11^{0.07}_{-0.03}$& $0.13^{0.04}_{-0.06}$ & - & -& $0.08^{0.05}_{-0.03}$& $0.71^{0.04}_{-0.06}$ & - & - & - \\ [1pt] 
 br3 &  $158^{6}_{-7}$& $-0.08^{0.07}_{-0.07}$& $0.25^{0.11}_{-0.09}$& $-0.44^{0.17}_{-0.16}$& $0.29^{0.16}_{-0.15}$& $0.24^{0.07}_{-0.07}$& $0.73^{0.11}_{-0.10}$& $0.52^{0.42}_{-0.36}$ & - & - \\ [1pt] 
  \hline
\end{tabular}
\end{sidewaystable}

\begin{sidewaystable}
\caption{Best-fit parameters of the selected models for IRAS18123+0511.}             
\label{tab:resIRAS18}      
\centering                        
\begin{tabular}{c c c c c c c c c c c c c c c c c c c c c c }
 model & BIC & $\chi^2_\mathrm{red}$  & $f_{prim,0}$ & $f_{sec,0}$ & $f_{bg0}$ & $d_{sec}$ & $d_{back}$ & $T_\mathrm{ring}$ & $\theta$ & $\delta \theta$ & $inc$\\ 
 \hline 
 sr5 & 83.2 & 0.8& $61.^{1.9}_{-2.8}$ & -& $19.2^{1.2}_{-1.4}$ & - & -& $1410^{137}_{-125}$& $2.4^{0.2}_{-0.2}$& $0.4^{0.4}_{-0.3}$& $25.9^{9.1}_{-13.4}$ \\ [1pt] 
 br2 & \textcolor{orange}{88.9} & 0.8& $43.7^{5.2}_{-5.7}$& $24.3^{5.2}_{-4.4}$& $17.7^{1.7}_{-2.3}$ & - & -& $967^{132}_{-126}$& $2.7^{0.5}_{-0.5}$& $0.9^{0.6}_{-0.5}$& $39.7^{5.7}_{-7.2}$ \\ [1pt] 
 sr3 & \textcolor{red}{92.1} & 1.1& $57.7^{1.7}_{-2.8}$ & -& $18.6^{1.1}_{-1.3}$ & - & -& $1353^{124}_{-118}$& $2.5^{0.2}_{-0.3}$& $0.4^{0.4}_{-0.3}$& $47.9^{2.6}_{-2.5}$ \\ [1pt] 
 sr6 & \textcolor{red}{92.5} & 0.8& $60.4^{2.1}_{-2.9}$ & -& $19.3^{1.4}_{-1.6}$ & - & -& $1389^{146}_{-132}$& $2.3^{0.3}_{-0.3}$& $0.5^{0.4}_{-0.3}$& $26.8^{9.5}_{-12.5}$ \\ [1pt] 
 \hline \hline
 model  & $PA$ & $c_1$ & $s_1$ & $c2$ & $s2$ & $x_0$ & $y_0$ & rM & UD$_\mathrm{prim,0}$ & UD$_\mathrm{sec,0}$\\ 
 \hline 
 sr3 &  $91^{3}_{-3}$& $0.15^{0.09}_{-0.08}$& $0.92^{0.05}_{-0.10}$ & - & - & - & - & - & - & - \\ [1pt] 
 br2 &  $117^{12}_{-11}$& $-0.34^{0.24}_{-0.22}$& $0.74^{0.15}_{-0.20}$ & - & -& $-0.33^{0.06}_{-0.07}$& $-0.04^{0.04}_{-0.04}$& $0.82^{0.39}_{-0.34}$ & - & - \\ [1pt] 
 sr5 &  $96^{14}_{-17}$& $0.39^{0.23}_{-0.23}$& $0.74^{0.15}_{-0.21}$& $0.60^{0.21}_{-0.30}$& $0.34^{0.34}_{-0.52}$ & - & - & - & - & - \\ [1pt] 
 sr6 &  $113^{16}_{-17}$& $-0.09^{0.33}_{-0.31}$& $0.79^{0.12}_{-0.17}$& $0.33^{0.38}_{-0.45}$& $0.63^{0.19}_{-0.40}$& $-0.07^{0.06}_{-0.05}$& $-0.002^{0.06}_{-0.06}$ & - & - & - \\ [1pt]
  \hline
\end{tabular}
\end{sidewaystable}

\begin{sidewaystable}
\caption{Best-fit parameters of the selected models for LR Sco.}             
\label{tab:resLRSco}      
\centering                        
\begin{tabular}{c c c c c c c c c c c c c c c c c c c c c c }
 model & BIC & $\chi^2_\mathrm{red}$  & $f_{prim,0}$ & $f_{sec,0}$ & $f_{bg0}$ & $d_{sec}$ & $d_{back}$ & $T_\mathrm{ring}$ & $\theta$ & $\delta \theta$ & $inc$\\ 
 \hline 
  sr3 & 98.8 & 1.2& $71.2^{0.4}_{-0.4}$ & -& $2.4^{0.5}_{-0.5}$ & - & -& $1200^{54}_{-50}$& $4.2^{0.2}_{-0.3}$& $1.3^{0.2}_{-0.6}$& $40.2^{4.1}_{-4.9}$ \\ [1pt] 
  sr6 & \textcolor{green}{99.4} & 1.0& $71.2^{0.4}_{-0.4}$ & -& $2.3^{0.6}_{-0.6}$ & - & -& $1209^{55}_{-52}$& $5.3^{0.6}_{-0.6}$& $1.0^{0.3}_{-0.2}$& $51.8^{5.8}_{-6.4}$ \\ [1pt]
 sr4 & \textcolor{red}{106.5} & 1.3& $71.1^{0.5}_{-0.4}$ & -& $2.4^{0.5}_{-0.5}$ & - & -& $1201^{53}_{-50}$& $4.2^{0.3}_{-0.3}$& $1.3^{0.2}_{-0.2}$& $41.2^{4.9}_{-7.1}$ \\ [1pt] 
 \hline \hline
 model  & $PA$ & $c_1$ & $s_1$ & $c2$ & $s2$ & $x_0$ & $y_0$ & rM & UD$_\mathrm{prim,0}$ & UD$_\mathrm{sec,0}$\\ 
 \hline 
 sr3 & $48^{8}_{-8}$& $0.23^{0.11}_{-0.09}$& $-0.28^{0.06}_{-0.08}$ & - & - & - & - & - & - & - \\ [1pt] 
 sr6 & $71^{11}_{-9}$& $0.17^{0.14}_{-0.19}$& $0.31^{0.14}_{-0.14}$& $-0.76^{0.20}_{-0.12}$& $0.37^{0.12}_{-0.12}$& $-0.33^{0.23}_{-0.24}$& $0.66^{0.14}_{-0.15}$ & - & - & - \\ [1pt] 
 sr4 & $45^{11}_{-9}$& $0.27^{0.15}_{-0.11}$& $-0.21^{0.15}_{-0.20}$ & - & -& $-0.10^{0.23}_{-0.19}$& $0.09^{0.25}_{-0.30}$ & - & - & - \\ [1pt] 
  \hline
\end{tabular}
\end{sidewaystable}

\begin{sidewaystable}
\caption{Best-fit parameters of the selected models for PS Gem.}             
\label{tab:resPSGem}      
\centering                        
\begin{tabular}{c c c c c c c c c c c c c c c c c c c c c c }
 model & BIC & $\chi^2_\mathrm{red}$  & $f_{prim,0}$ & $f_{sec,0}$ & $f_{bg0}$ & $d_{sec}$ & $d_{back}$ & $T_\mathrm{ring}$ & $\theta$ & $\delta \theta$ & $inc$\\ 
 \hline 
 b1 & 38.7 & 1.0 & -& $2.4^{1.6}_{-0.3}$& $4.3^{0.7}_{-0.8}$& $-1.9^{1.4}_{-1.2}$& $1.2^{1.8}_{-2.0}$ & - & - & - & - \\ [1pt] 
 b2 & \textcolor{orange}{43.7} & 1.1 & -& $4.6^{2.5}_{-1.9}$& $3.9^{0.7}_{-0.7}$& $-2.0^{1.2}_{-1.0}$& $1.5^{1.6}_{-1.9}$ & - & - & - & - \\ [1pt]
 sr3 & \textcolor{red}{45.4} & 1.1& $88.0^{1.3}_{-1.2}$ & -& $3.0^{1.0}_{-1.0}$ & - & -& $2136^{584}_{-377}$& $11.0^{2.2}_{-2.2}$& $0.9^{0.7}_{-0.6}$& $1.8^{4.4}_{-1.4}$ \\ [1pt] 
 \hline \hline
 model  & $PA$ & $c_1$ & $s_1$ & $c2$ & $s2$ & $x_0$ & $y_0$ & rM & UD$_\mathrm{prim,0}$ & UD$_\mathrm{sec,0}$\\ 
 \hline 
 b1 &  - & - & - & - & -& $-3.3^{35.6}_{-32.5}$& $-5.0^{21.6}_{-23.7}$ & -& $2.1^{0.5}_{-1.1}$ & - \\ [1pt] 
 b2 &  - & - & - & - & -& $-0.08^{33.1}_{-34.4}$& $-5.1^{22.8}_{-21.9}$ & -& $1.9^{0.5}_{-0.8}$& $9.5^{4.7}_{-5.9}$ \\ [1pt] 
 sr3 & $0^{22}_{-23}$& $-0.66^{0.31}_{-0.21}$& $0.43^{0.32}_{-0.44}$ & - & - & - & - & - & - & - \\ [1pt] 
  \hline
\end{tabular}
\end{sidewaystable}

\begin{sidewaystable}
\caption{Best-fit parameters of the selected models for SX Cen.}             
\label{tab:resSXCen}      
\centering                        
\begin{tabular}{c c c c c c c c c c c c c c c c c c c c c c }
 model & BIC & $\chi^2_\mathrm{red}$  & $f_{prim,0}$ & $f_{sec,0}$ & $f_{bg0}$ & $d_{sec}$ & $d_{back}$ & $T_\mathrm{ring}$ & $\theta$ & $\delta \theta$ & $inc$\\ 
 \hline 
 br2 & 97.8 & 0.9& $72.8^{3.3}_{-4.2}$& $16.8^{4.1}_{-3.2}$& $2.5^{0.6}_{-0.6}$ & - & -& $1124^{171}_{-144}$& $5.1^{0.9}_{-0.9}$& $1.0^{0.5}_{-0.6}$& $64.5^{5.4}_{-8.1}$ \\ [1pt] 
 br4 & \textcolor{orange}{101.4} & 0.8& $81^{2}_{-2}$& $12.4^{1.7}_{-1.4}$& $0.9^{0.7}_{-0.6}$ & - & -& $859^{191}_{-149}$& $6.8^{1.7}_{-1.6}$& $1.2^{0.7}_{-0.4}$& $39.5^{18.1}_{-23.7}$ \\ [1pt] 
 br3 & \textcolor{orange}{102.8} & 0.9& $76.0^{2.3}_{-2.4}$& $13.0^{2.4}_{-2.3}$& $2.4^{0.6}_{-0.7}$ & - & -& $1189^{206}_{-160}$& $4.3^{0.9}_{-0.7}$& $1.1^{0.4}_{-0.4}$& $55.7^{7.4}_{-13.8}$ \\ [1pt] 
 br5 & \textcolor{orange}{103.6} & 0.8& $81.1^{1.8}_{-1.9}$& $12.1^{1.7}_{-1.5}$& $0.9^{0.8}_{-0.6}$& $-1.9^{1.4}_{-1.2}$ & -& $905^{206}_{-164}$& $7.3^{1.8}_{-1.6}$& $1.1^{0.6}_{-0.4}$& $45.4^{15.4}_{-24.5}$ \\ [1pt]
 \hline \hline
 model  & $PA$ & $c_1$ & $s_1$ & $c2$ & $s2$ & $x_0$ & $y_0$ & rM & UD$_\mathrm{prim,0}$ & UD$_\mathrm{sec,0}$\\ 
 \hline 
 br2 & $132^{4}_{-10}$& $-0.79^{0.19}_{-0.13}$& $-0.26^{0.40}_{-0.32}$ & - & -& $-0.14^{0.04}_{-0.05}$& $0.47^{0.12}_{-0.09}$& $0.59^{0.45}_{-0.38}$ & - & - \\ [1pt] 
 br3 & $112^{12}_{-11}$& $-0.71^{0.20}_{-0.16}$& $0.25^{0.25}_{-0.26}$& $-0.29^{0.36}_{-0.30}$& $-0.65^{0.26}_{-0.19}$& $-0.10^{0.03}_{-0.03}$& $0.29^{0.06}_{-0.06}$& $1.74^{0.70}_{-0.53}$ & - & - \\ [1pt] 
 br4 & $49^{53}_{-26}$& $0.02^{0.42}_{-0.49}$& $0.54^{0.28}_{-0.49}$& $-0.14^{0.55}_{-0.46}$& $0.19^{0.43}_{-0.50}$& $-0.19^{0.04}_{-0.04}$& $0.41^{0.09}_{-0.08}$& $1.14^{0.51}_{-0.38}$ & - & - \\ [1pt] 
 br5 & $47^{24}_{-22}$& $0.019^{0.112}_{-0.12}$& $0.63^{0.25}_{-0.44}$& $-0.15^{0.49}_{-0.44}$& $0.30^{0.41}_{-0.53}$& $-0.20^{0.04}_{-0.05}$& $0.45^{0.09}_{-0.09}$& $0.98^{0.46}_{-0.34}$ & - & - \\ [1pt] 
  \hline
\end{tabular}
\end{sidewaystable}

\begin{sidewaystable}
\caption{Best-fit parameters of the selected models for U Mon.}             
\label{tab:resUMon}      
\centering                        
\begin{tabular}{c c c c c c c c c c c c c c c c c c c c c c }
 model & BIC & $\chi^2_\mathrm{red}$  & $f_{prim,0}$ & $f_{sec,0}$ & $f_{bg0}$ & $d_{sec}$ & $d_{back}$ & $T_\mathrm{ring}$ & $\theta$ & $\delta \theta$ & $inc$\\ 
 \hline 
 br5 & 483.8 & 2.1& $68.8^{1.5}_{-1.2}$& $1.5^{0.6}_{-0.2}$& $9.7^{0.3}_{-0.3}$& $-3.7^{0.4}_{-0.2}$ & -& $2619^{138}_{-132}$& $5.49^{0.03}_{-0.03}$& $0.03^{0.03}_{-0.01}$& $57.9^{1.6}_{-1.5}$ \\ [1pt]
 br4 & \textcolor{orange}{487.3} & 2.1& $68.1^{1.6}_{-1.0}$& $1.6^{0.1}_{-0.2}$& $9.6^{0.3}_{-0.3}$ & - & -& $2637^{136}_{-125}$& $5.5^{0.03}_{-0.03}$& $0.02^{0.02}_{-0.01}$& $57.5^{1.6}_{-1.4}$ \\ [1pt] 
 \hline \hline
 model  & $PA$ & $c_1$ & $s_1$ & $c2$ & $s2$ & $x_0$ & $y_0$ & rM & UD$_\mathrm{prim,0}$ & UD$_\mathrm{sec,0}$\\ 
 \hline 
 br5 & $45^{1}_{-1}$& $-0.03^{0.10}_{-0.14}$& $0.04^{0.03}_{-0.03}$& $1.0^{0.02}_{-0.03}$& $-0.29^{0.08}_{-0.07}$& $1.42^{0.02}_{-0.02}$& $1.07^{0.04}_{-0.04}$& $0.001^{0.001}_{-0.000}$& $0.40^{0.14}_{-0.14}$ & - \\ [1pt]
 br4 & $45^{1}_{-1}$& $0.03^{0.08}_{-0.13}$& $0.03^{0.03}_{-0.03}$& $1.0^{0.02}_{-0.03}$& $-0.30^{0.07}_{-0.07}$& $1.41^{0.02}_{-0.02}$& $1.08^{0.04}_{-0.04}$& $0.001^{0.001}_{-0.001}$& $0.32^{0.17}_{-0.16}$ & - \\ [1pt] 
  \hline
\end{tabular}
\end{sidewaystable}

\begin{sidewaystable}
\caption{Best-fit parameters of the selected models for V494\,Vel.}             
\label{tab:resV494Vel}      
\centering                        
\begin{tabular}{c c c c c c c c c c c c c c c c c c c c c c }
 model & BIC & $\chi^2_\mathrm{red}$  & $f_{prim,0}$ & $f_{sec,0}$ & $f_{bg0}$ & $d_{sec}$ & $d_{back}$ & $T_\mathrm{ring}$ & $\theta$ & $\delta \theta$ & $inc$\\ 
 \hline 
 sr3 & 96.5 & 0.8& $74.9^{0.8}_{-0.7}$ & -& $3.2^{0.4}_{-0.4}$ & - & -& $8186^{1255}_{-1599}$& $2.6^{0.5}_{-0.4}$& $2.2^{0.6}_{-0.6}$& $53.0^{1.7}_{-1.7}$ \\ [1pt] 
 br3 & \textcolor{orange}{99.9} & 0.5& $49.1^{10.1}_{-10.6}$& $34.0^{10.5}_{-10.0}$& $2.5^{0.5}_{-0.5}$ & - & -& $8163^{1279}_{-1632}$& $4.1^{1.0}_{-0.7}$& $1.9^{0.5}_{-0.6}$& $54.1^{3.9}_{-4.4}$ \\ [1pt]
 sr5 & \textcolor{red}{103.0} & 0.7& $75.5^{1.8}_{-1.1}$ & -& $3.3^{0.4}_{-0.4}$ & - & -& $8269^{1182}_{-1580}$& $2.8^{0.7}_{-0.5}$& $1.9^{0.7}_{-0.6}$& $49.0^{4.2}_{-4.1}$ \\ [1pt] 
 br2 & \textcolor{red}{103.9} & 0.7& $79.5^{1.1}_{-1.4}$& $4.0^{2.5}_{-1.0}$& $2.4^{0.5}_{-0.5}$ & - & -& $8046^{1319}_{-1669}$& $7.0^{0.6}_{-0.6}$& $0.2^{0.2}_{-0.2}$& $61.9^{3.4}_{-6.2}$ \\ [1pt] 
 br4 & \textcolor{red}{104.3} & 0.5& $50.6^{10.6}_{-12.0}$& $33.3^{11.8}_{-10.4}$& $2.4^{0.5}_{-0.5}$ & - & -& $7862^{1432}_{-1656}$& $4.3^{1.0}_{-0.8}$& $1.8^{0.5}_{-0.5}$& $54.0^{4.6}_{-5.3}$ \\ [1pt]
 br1 & \textcolor{red}{106.2} & 0.8& $72.8^{1.6}_{-2.7}$& $1.8^{3.0}_{-1.4}$& $3.3^{0.4}_{-0.4}$ & - & -& $8008^{1361}_{-1581}$& $2.5^{0.7}_{-0.5}$& $2.4^{0.8}_{-0.8}$& $55.8^{1.6}_{-1.6}$ \\ [1pt] 
 \hline \hline
 model  & $PA$ & $c_1$ & $s_1$ & $c2$ & $s2$ & $x_0$ & $y_0$ & rM & UD$_\mathrm{prim,0}$ & UD$_\mathrm{sec,0}$\\ 
 \hline 
  sr3 & & $96^{2}_{-2}$& $-0.23^{0.06}_{-0.07}$& $0.87^{0.07}_{-0.10}$ & - & - & - & - & - & - & - \\ [1pt] 
  br3 & $101^{7}_{-5}$& $-0.57^{0.18}_{-0.16}$& $0.61^{0.16}_{-0.20}$& $-0.04^{0.37}_{-0.33}$& $-0.34^{0.58}_{-0.37}$& $-0.41^{0.08}_{-0.09}$& $-0.02^{0.04}_{-0.04}$& $0.490^{0.35}_{-0.28}$ & - & - \\ [1pt]
  sr5 & $106^{10}_{-6}$& $-0.37^{0.08}_{-0.07}$& $0.71^{0.17}_{-0.25}$& $0.14^{0.32}_{-0.33}$& $0.69^{0.20}_{-0.37}$ & - & - & - & - & - \\ [1pt] 
  br2 & $96^{5}_{-5}$& $0.73^{0.13}_{-0.15}$& $0.03^{0.45}_{-0.20}$ & - & -& $1.55^{0.28}_{-0.28}$& $-0.45^{0.36}_{-0.26}$& $0.05^{0.06}_{-0.04}$ & - & - \\ [1pt]
  br4 & $102^{7}_{-6}$& $-0.55^{0.18}_{-0.17}$& $0.60^{0.17}_{-0.21}$& $0.17^{0.42}_{-0.42}$& $-0.22^{0.59}_{-0.46}$& $-0.41^{0.09}_{-0.10}$& $-0.02^{0.04}_{-0.05}$& $0.43^{0.39}_{-0.26}$ & - & - \\ [1pt] 
  br1 & $94^{2}_{-2}$ & - & - & - & -& $0.12^{0.03}_{-0.03}$& $-0.33^{0.03}_{-0.03}$& $0.03^{0.06}_{-0.03}$ & - & - \\ [1pt] 
  \hline
\end{tabular}
\end{sidewaystable}

\begin{table}
\caption{SED best fit parameters.}             
\label{tab:SEDfit}      
\centering                          
\begin{tabular}{c l c c c}        
\hline\hline                 
\# & Target &  Teff & $E(B-V)$ & $\log g$  \\ 
   &        &    [K]  & &   \\
\hline                        
1	&AC\,Her	&	$ 5260$	&	$ 0.22^{+ 0.16}_{- 0.07}$	&	$  2.2$ \\[1pt]
2	&AI\,Sco	&	$ 5210$	&	$ 0.63^{+ 0.07}_{- 0.27}$	&	$  1.8$	\\[1pt]
3	&EN\,TrA	&	$ 5800$	&	$ 0.17^{+ 0.21}_{- 0.11}$	&	$  1.3$\\[1pt]
4	&HD\,93662 &	$ 4250$	&	$ 0.30^{+ 0.09}_{- 0.10}$	&	$  0.5$	\\[1pt]
5	&HD\,95767	&	$ 7370$	&	$ 0.54^{+ 0.09}_{- 0.04}$	&	$  2.5$	\\[1pt]
6	&HD\,108015	&	$ 6750$	&	$ 0.13^{+ 0.10}_{- 0.02}$	&	$  1.8$	\\[1pt]
7	&HD\,213985	&	$ 8040$	&	$ 0.14^{+ 0.07}_{- 0.04}$	&	$  2.0$	\\[1pt]
8	&HR\,4049		&	$ 7750$	&	$ 0.18^{+ 0.22}_{- 0.18}$	&	$  2.0$	\\[1pt]
9	&IRAS\,05208 -2035	&	$ 4150$	&	$ 0.03^{+ 0.16}_{- 0.02}$	&	$  1.8$	\\[1pt]
10	&IRAS\,08544-4431	&	$ 7020$	&	$ 1.30^{+ 0.12}_{- 0.03}$	&	$  2.4$	\\[1pt]
11	&IRAS\,10174-5704	&	$ 5770$	&	$ 2.00^{+ 0.00}_{- 0.21}$	&	$  2.1$	\\[1pt]
12	&IRAS\,15469-5311	&	$ 7410$	&	$ 1.36^{+ 0.22}_{- 0.18}$	&	$  2.2$	\\[1pt]
13	&IRAS\,17038-4815	&	$ 4620$	&	$ 0.74^{+ 0.32}_{- 0.28}$	&	$  2.2$	\\[1pt]
14	&IRAS\,18123+0511	&	$ 4760$	&	$ 0.29^{+ 0.34}_{- 0.16}$	&	$  1.0$	\\[1pt]
15	&IRAS\,19125+0343	&	$ 7590$	&	$ 0.83^{+ 0.15}_{- 0.11}$	&	$  2.2$ \\[1pt]
16	&IW\,Car&	$ 6520$	&	$ 0.72^{+ 0.21}_{- 0.07}$	&	$  2.6$	\\[1pt]
17	&LR\,Sco	&	$ 6050$	&	$ 0.61^{+ 0.17}_{- 0.11}$	&	$  0.6$	\\[1pt]
18	&PS\,Gem	&	$ 6090$	&	$ 0.00^{+ 0.08}_{- 0.00}$	&	$  2.3$	\\[1pt]
19	&R\,Sct		&	$ 4660$	&	$ 0.16^{+ 0.40}_{- 0.16}$	&	$  0.9$	\\[1pt]
20	&RU\,Cen	&	$ 5800$	&	$ 0.18^{+ 0.18}_{- 0.12}$	&	$  0.0$	\\[1pt]
21	&SX\,Cen	& 	$ 5760$	&	$ 0.11^{+ 0.25}_{- 0.11}$	&	$  0.1$	\\[1pt]
22	&U\,Mon		&	$ 4750$	&	$ 0.07^{+ 0.38}_{- 0.07}$	&	$  0.1$	\\[1pt]
\hline                                   
\end{tabular}
\end{table}

\begin{figure*}
\centering
\includegraphics[width=5cm]{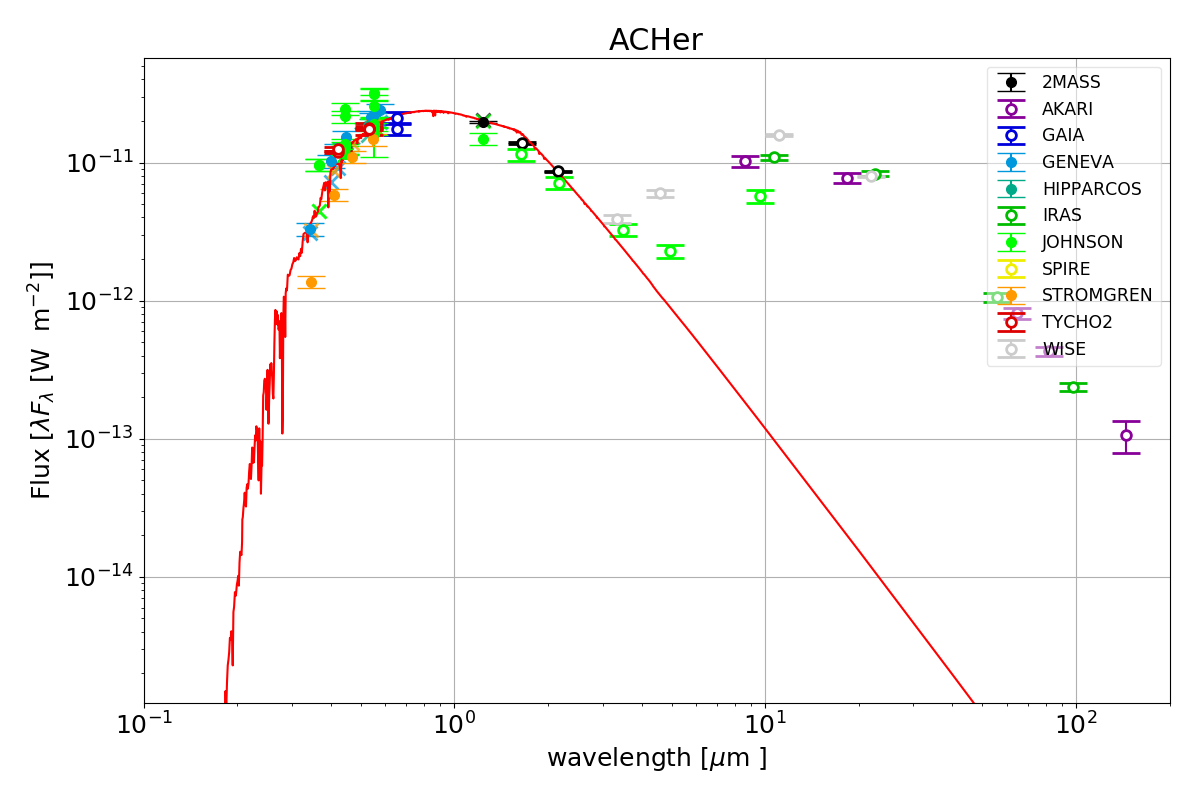}
\includegraphics[width=5cm]{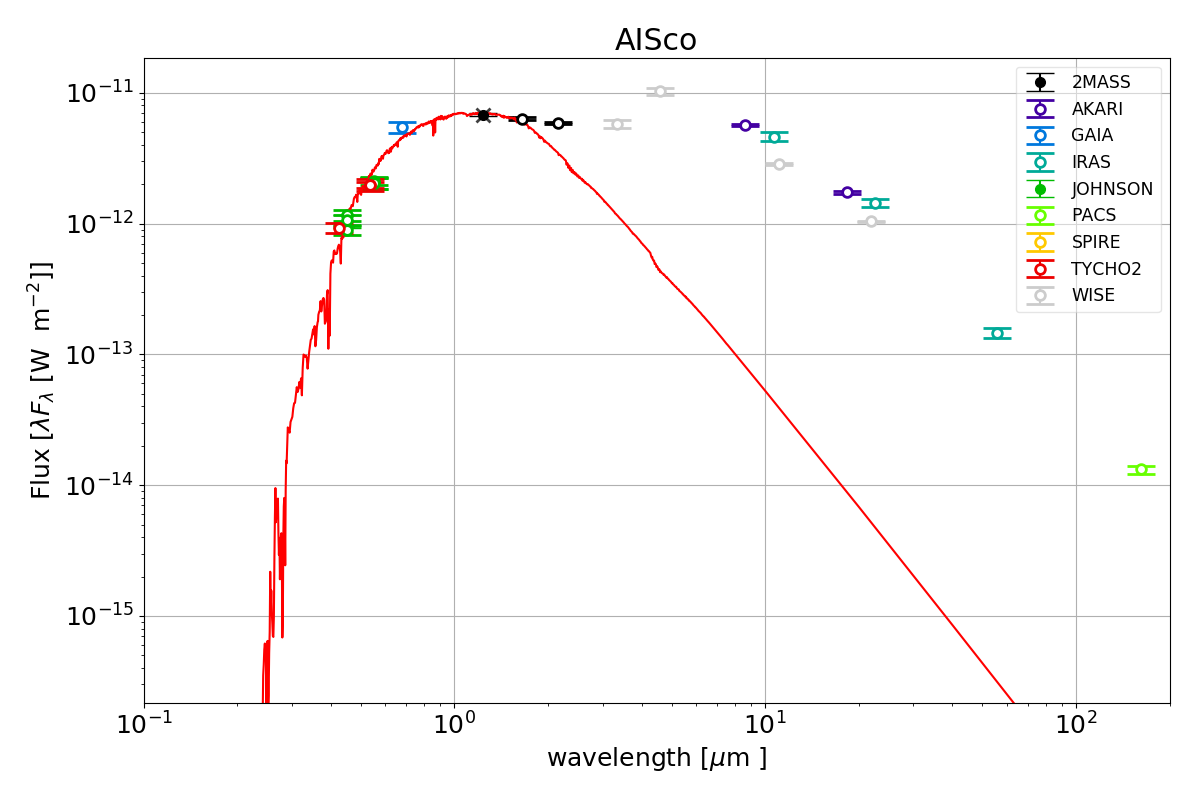}
\includegraphics[width=5cm]{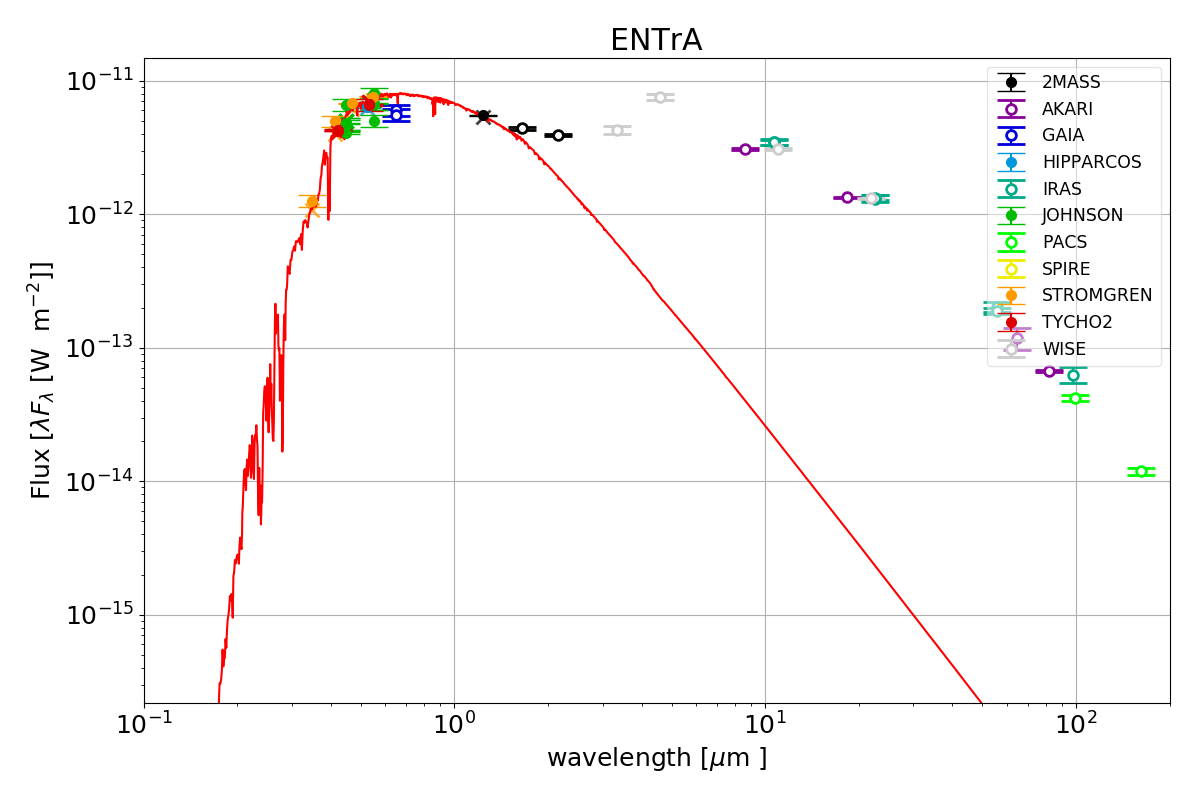}
\includegraphics[width=5cm]{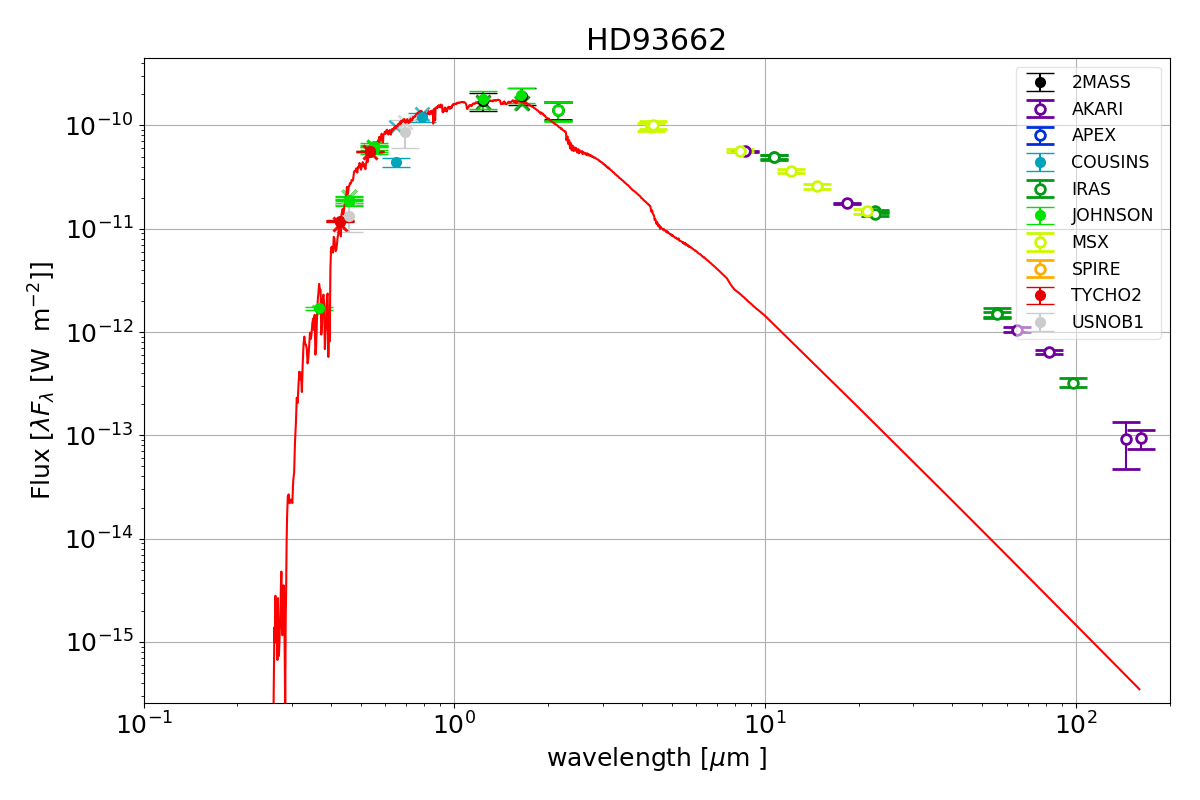}
\includegraphics[width=5cm]{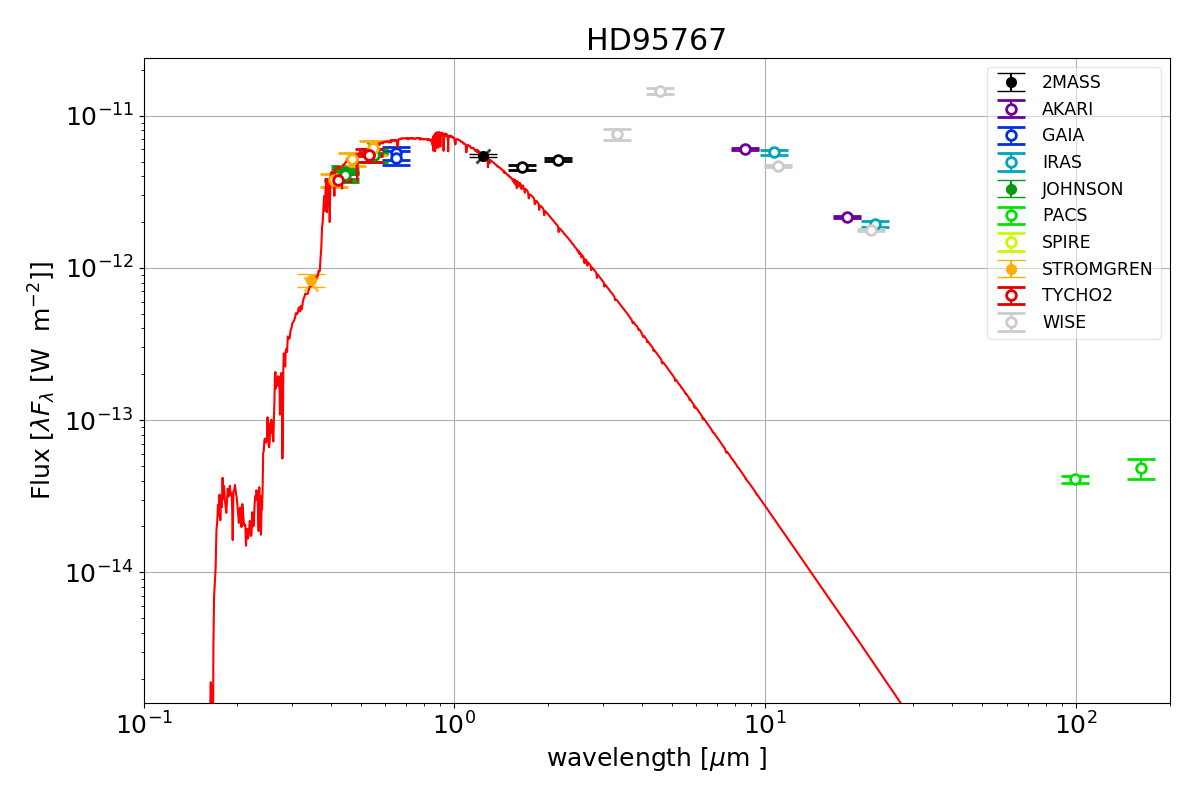}
\includegraphics[width=5cm]{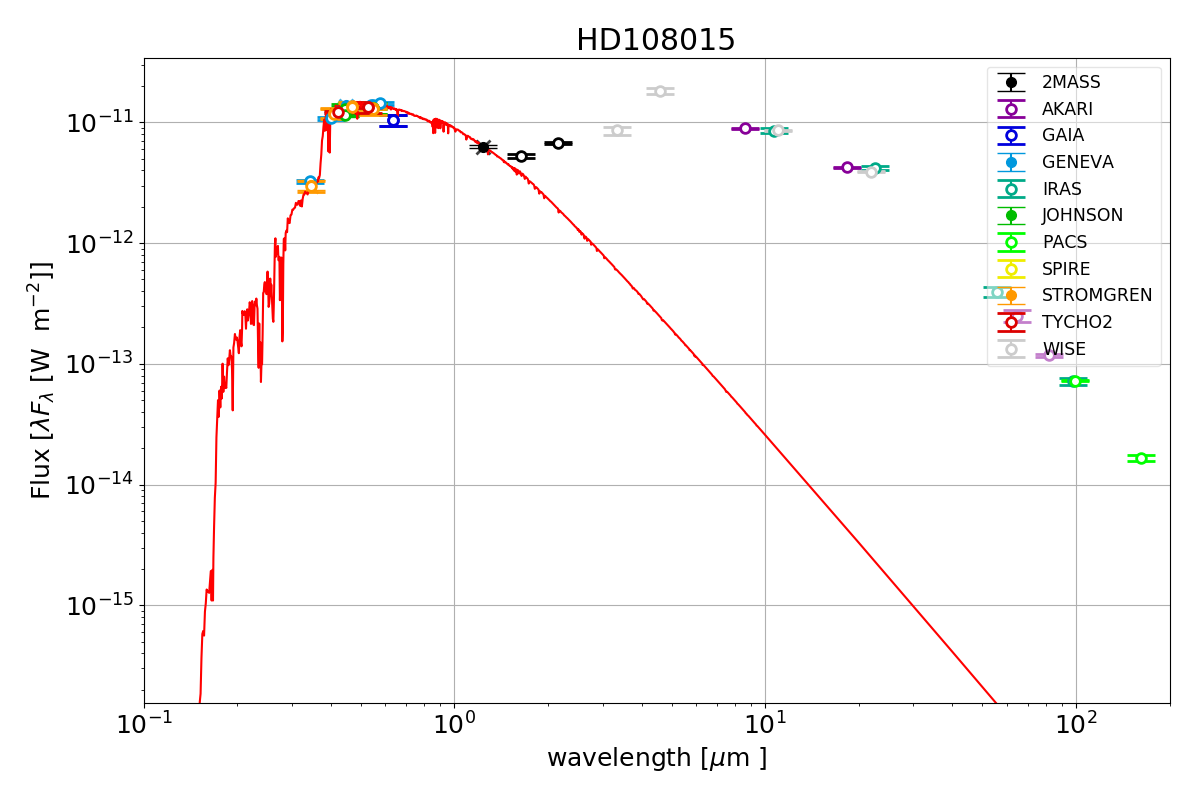}
\includegraphics[width=5cm]{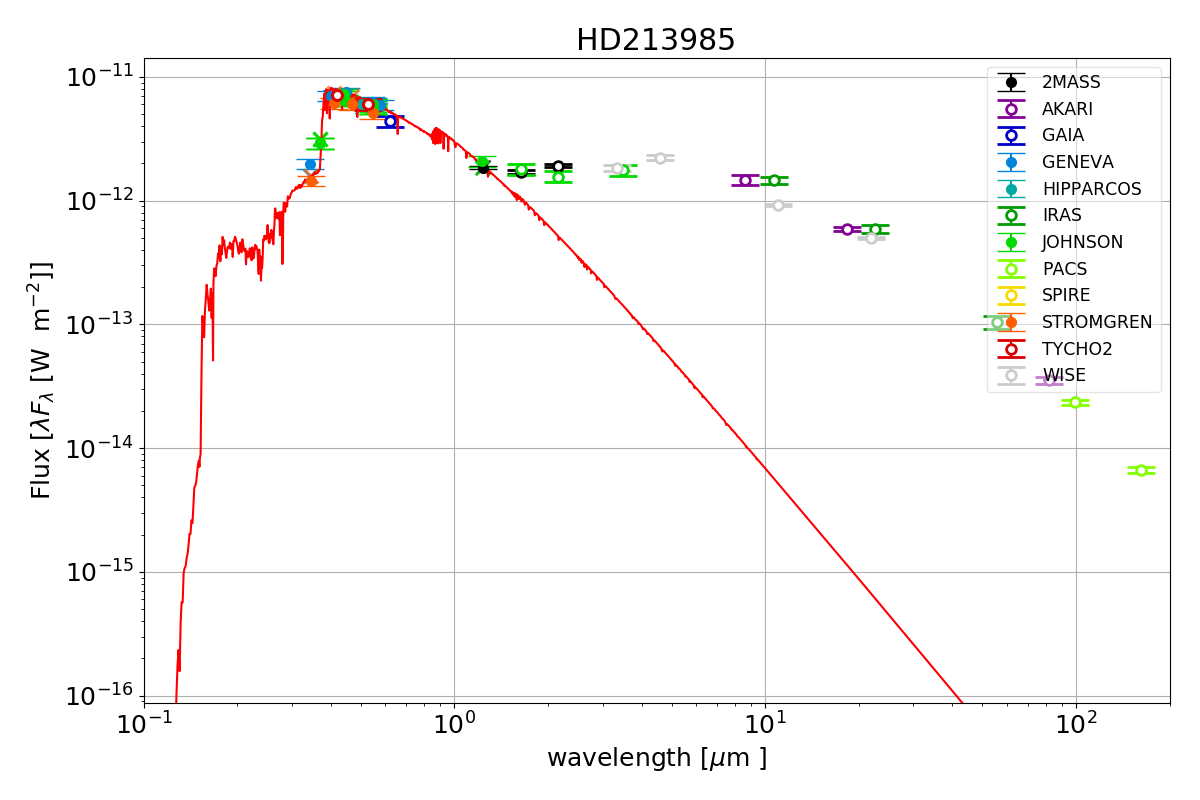}
\includegraphics[width=5cm]{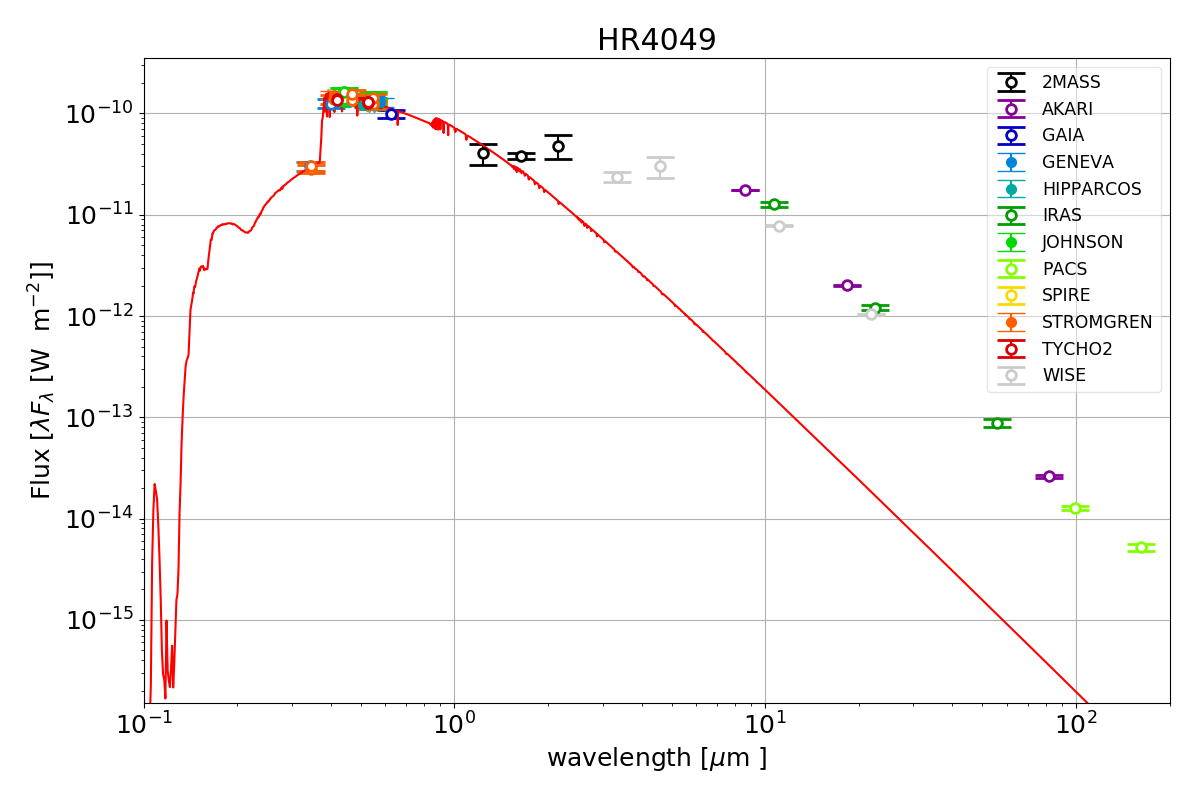}
\includegraphics[width=5cm]{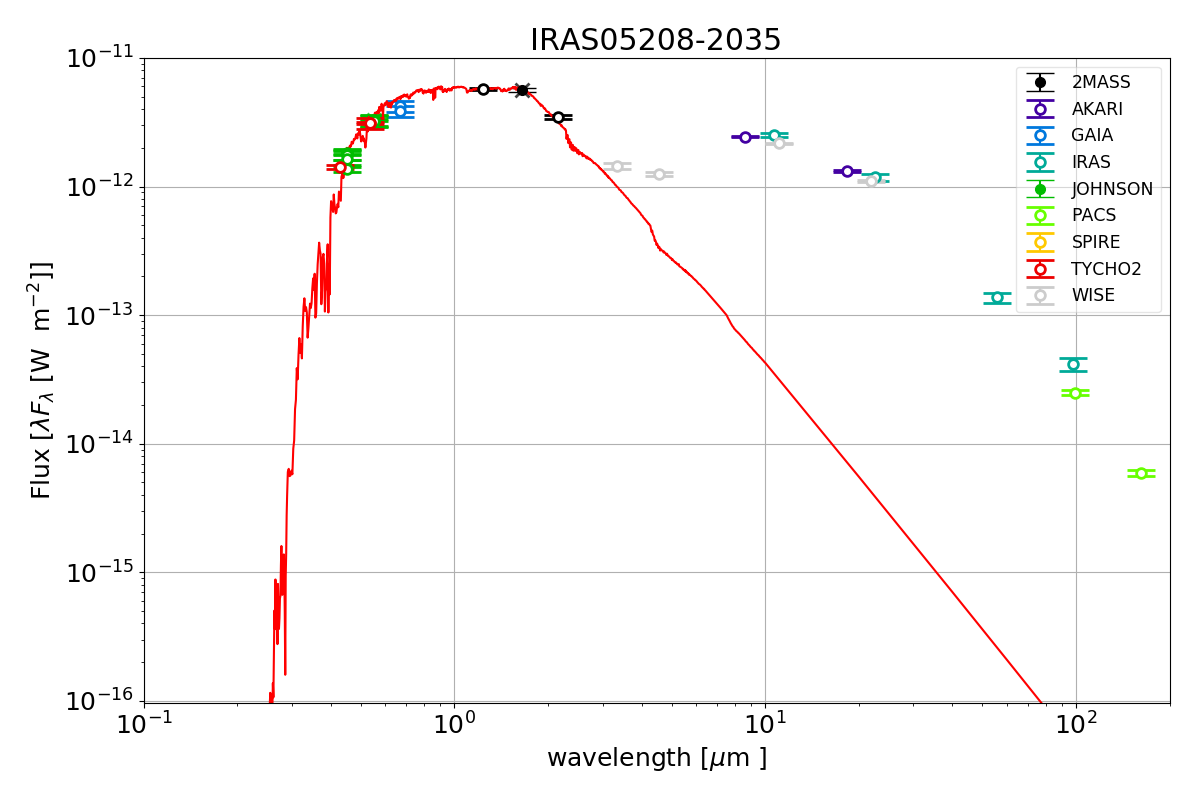}
\includegraphics[width=5cm]{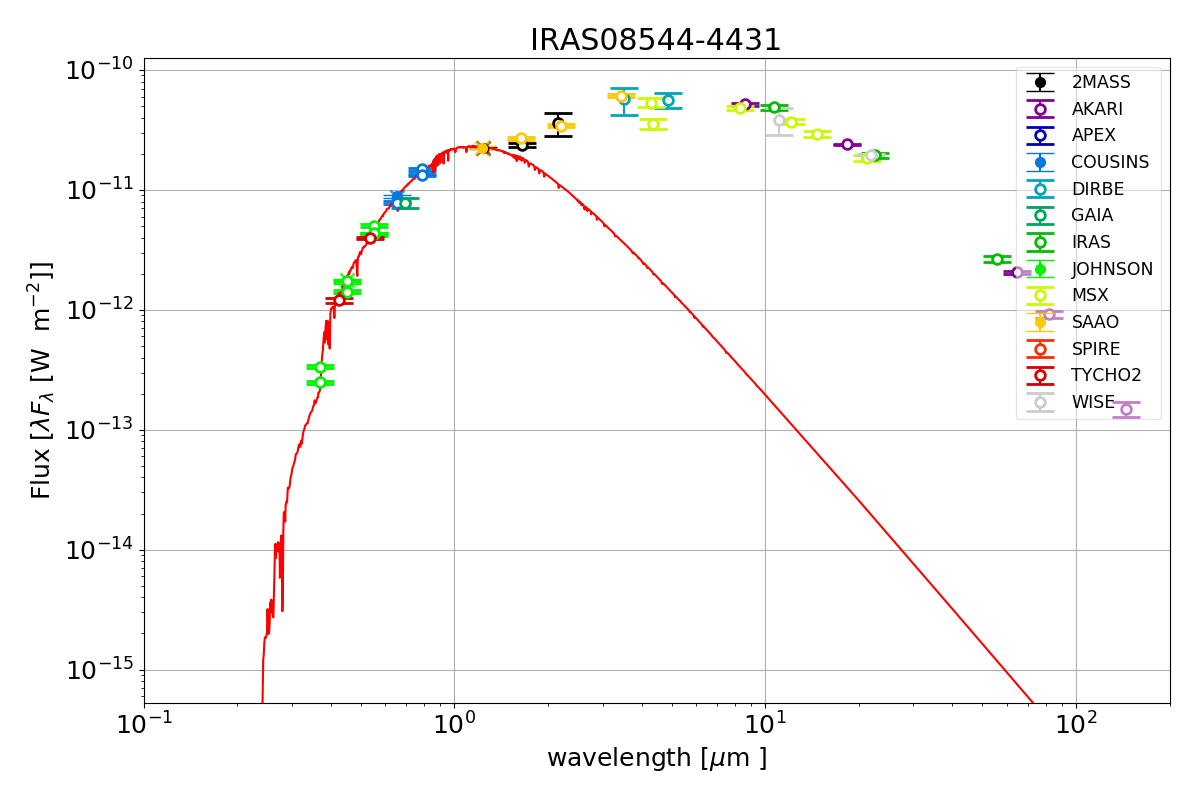}
\caption{Spectral energy distributions of the targets of our survey for which no SED was already published.}
\label{fig:SED1}
\end{figure*}

\begin{figure*}
\centering
\includegraphics[width=5cm]{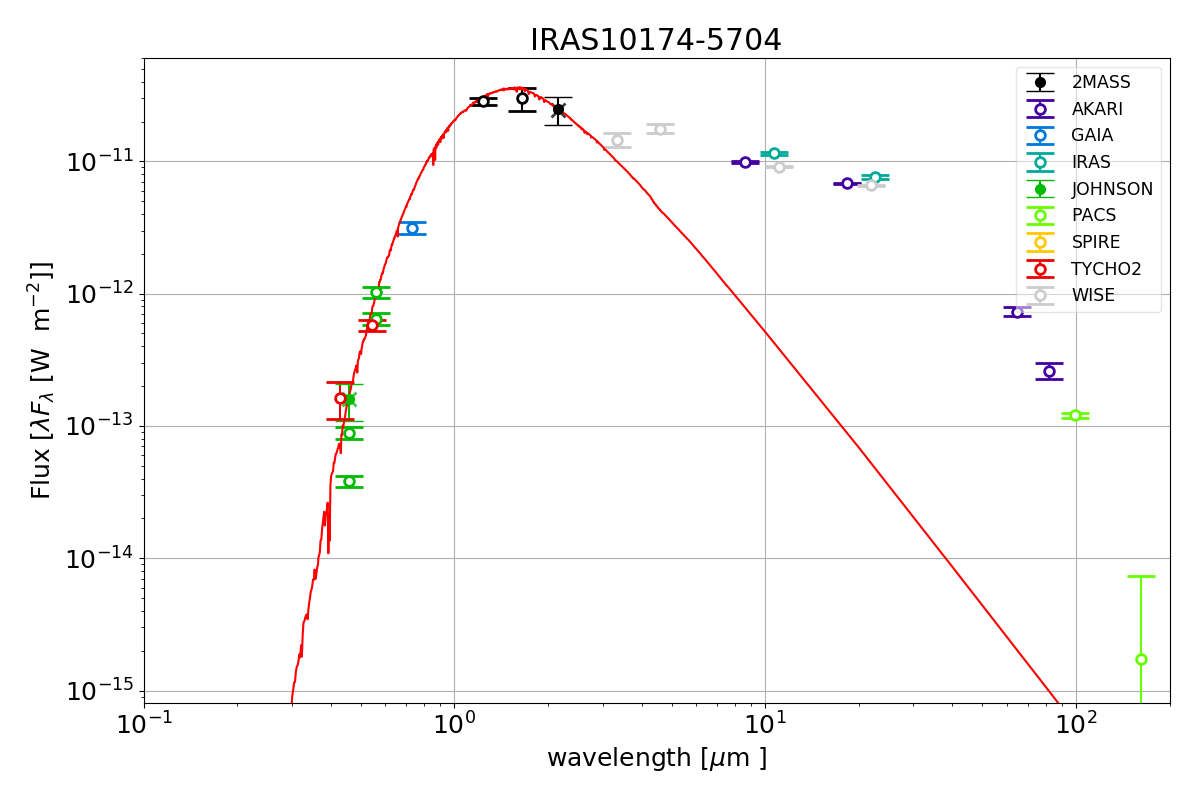}
\includegraphics[width=5cm]{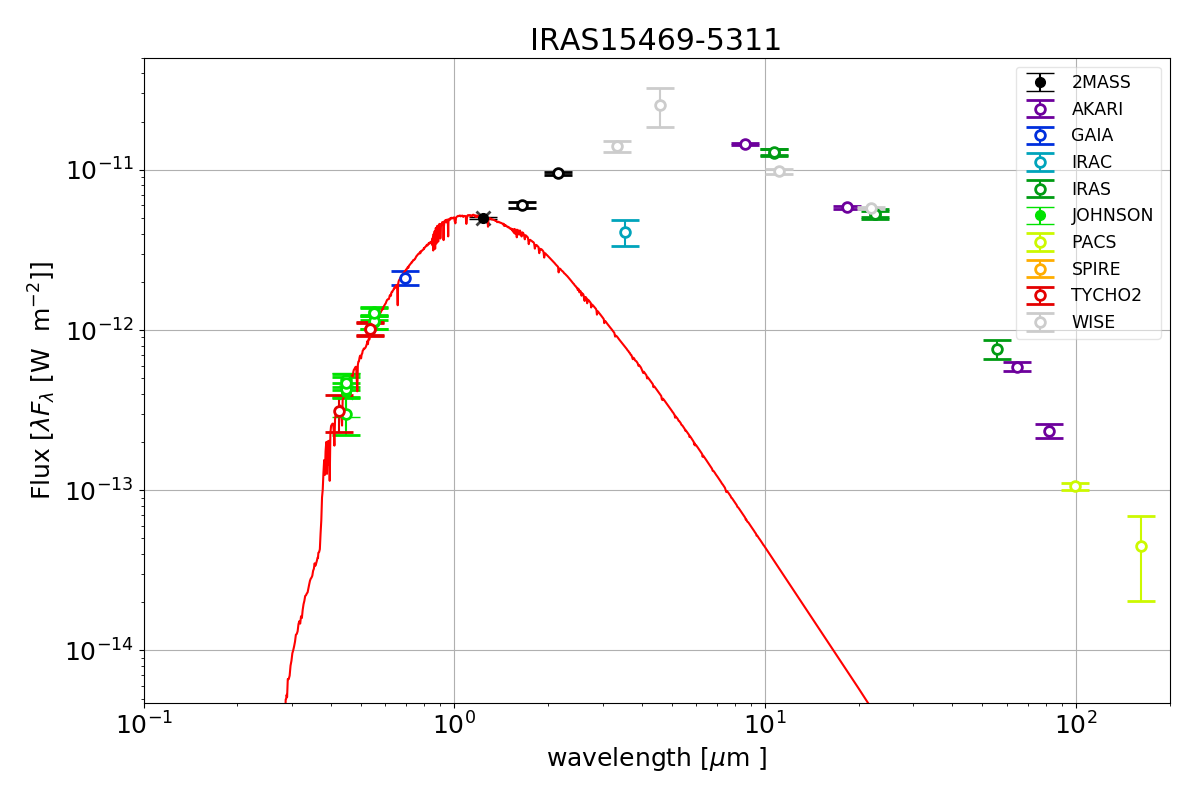}
\includegraphics[width=5cm]{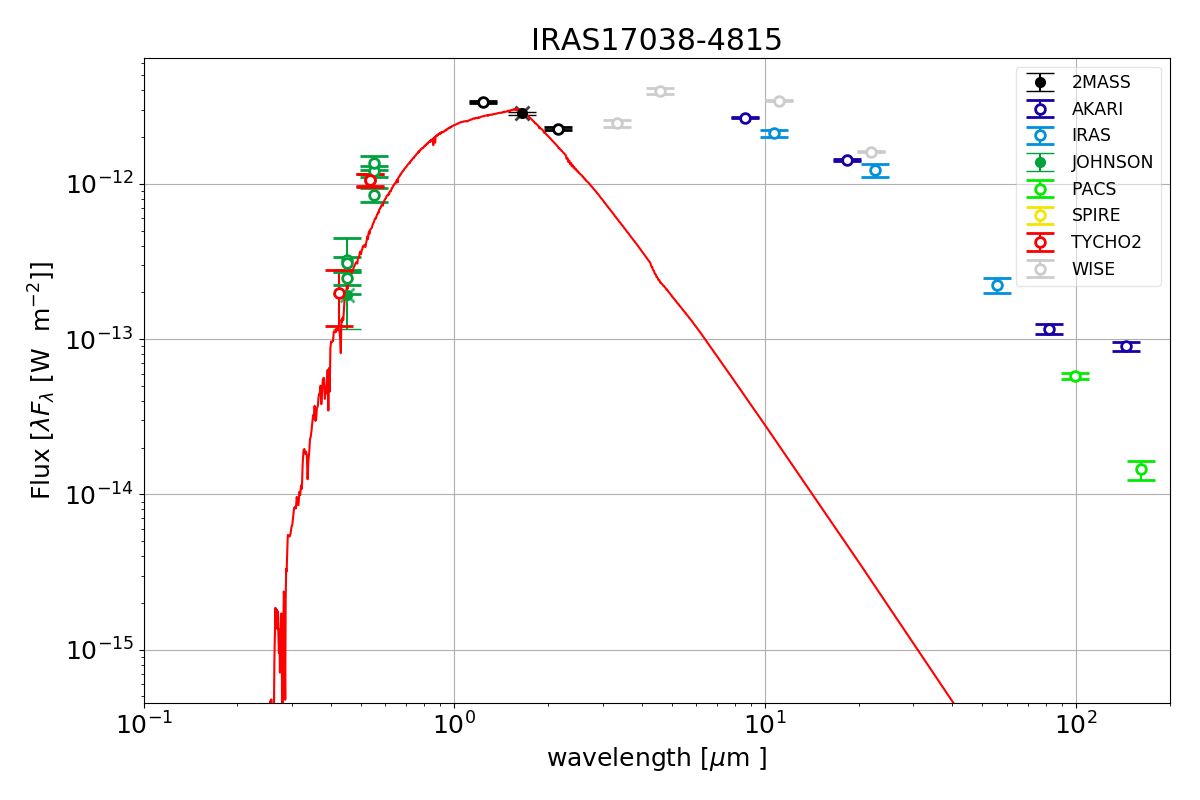}
\includegraphics[width=5cm]{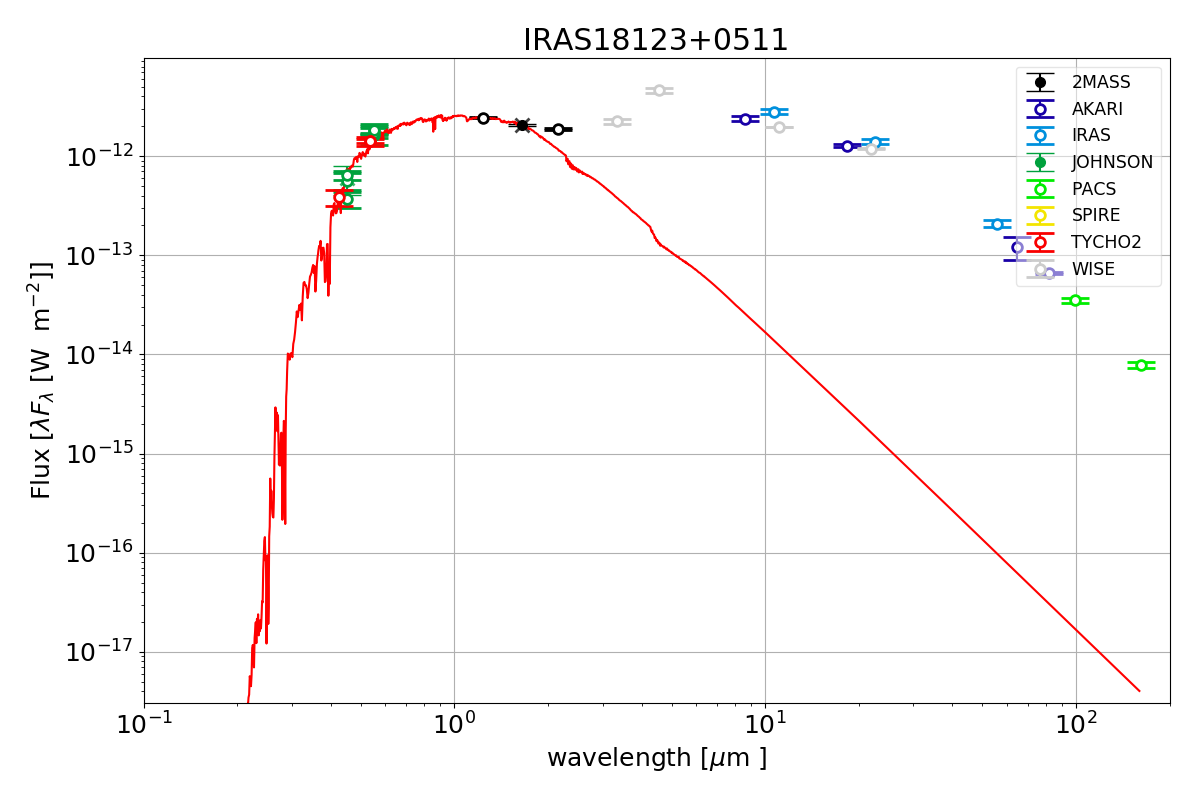}
\includegraphics[width=5cm]{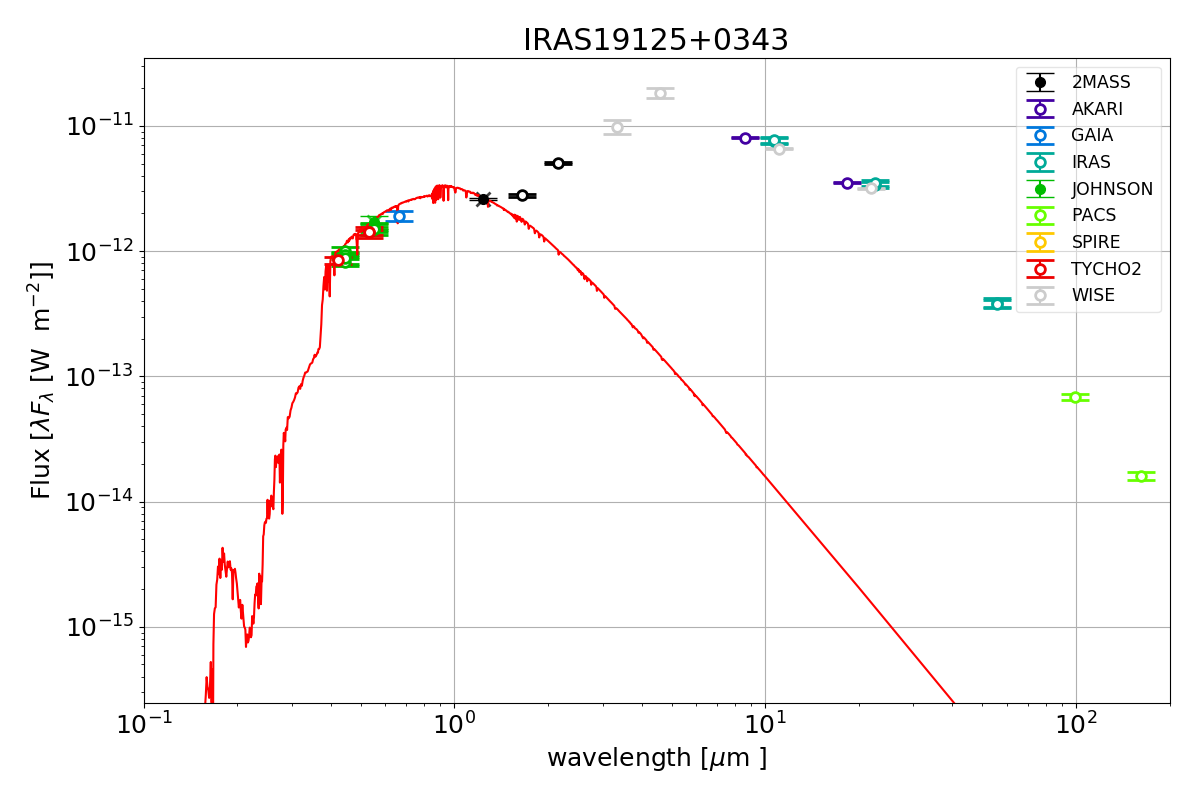}
\includegraphics[width=5cm]{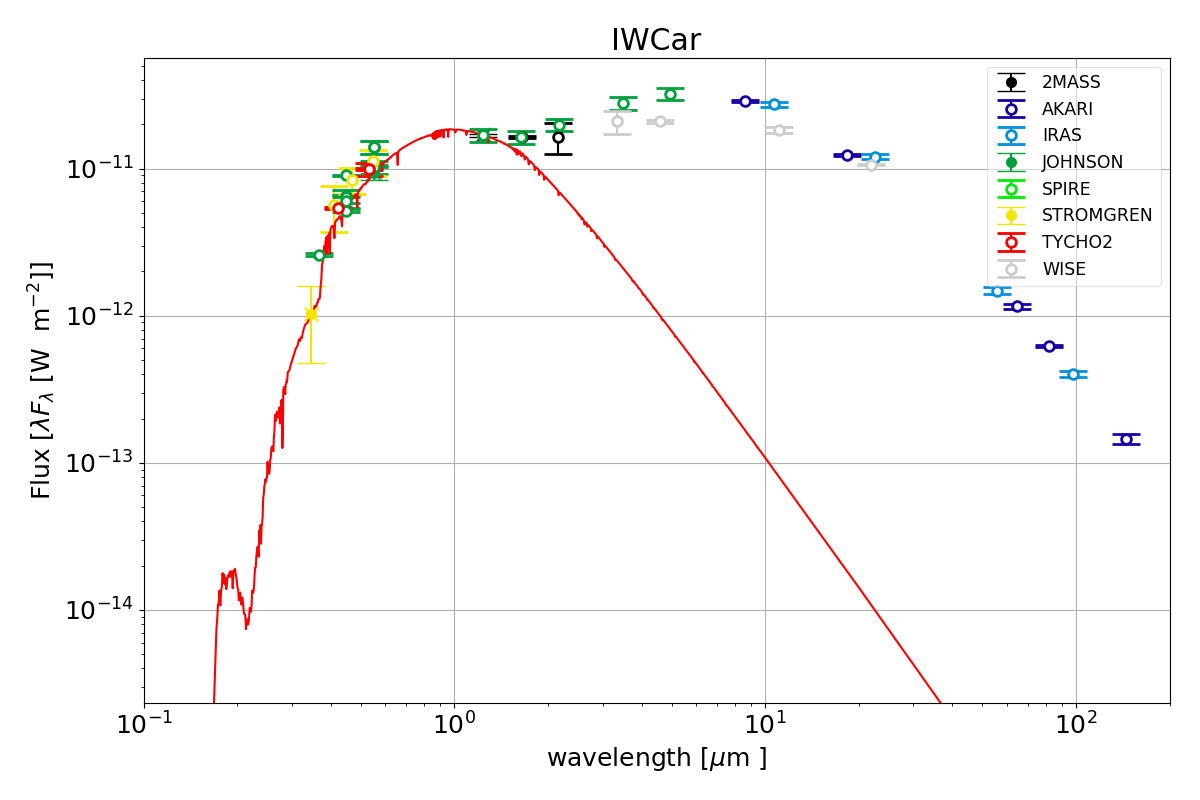}
\includegraphics[width=5cm]{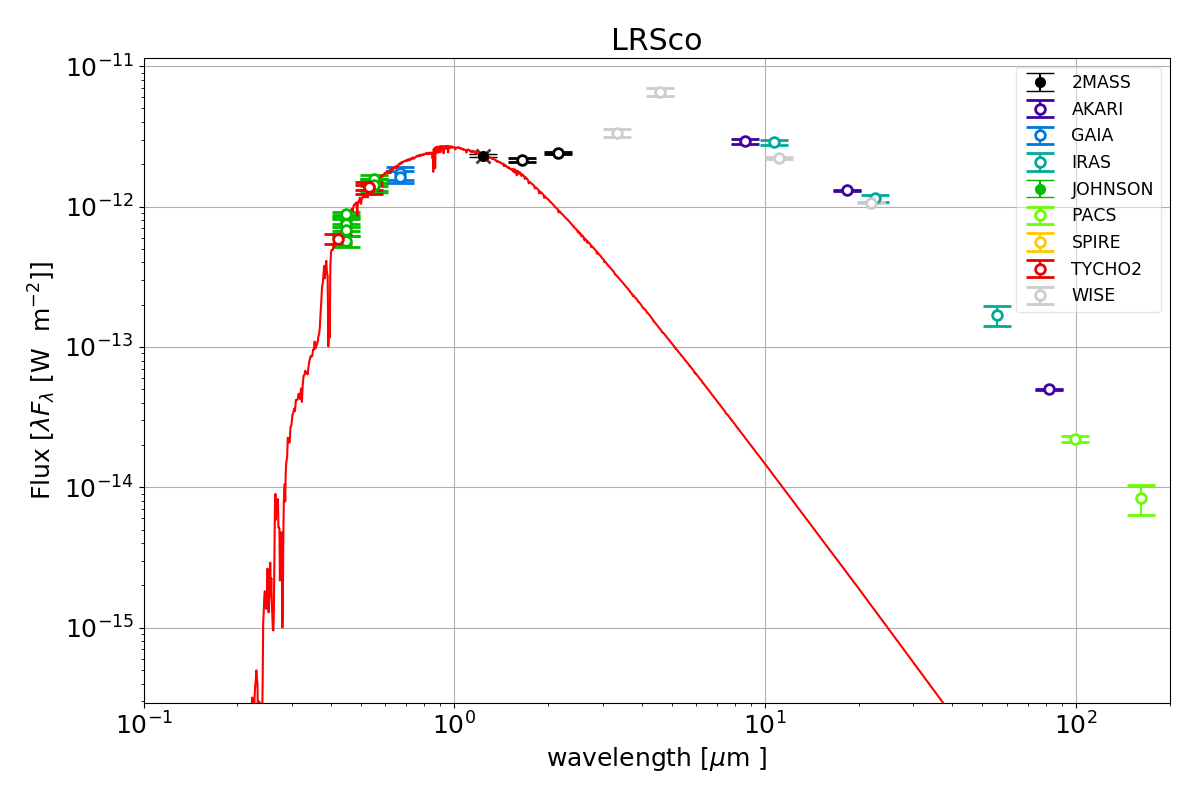}
\includegraphics[width=5cm]{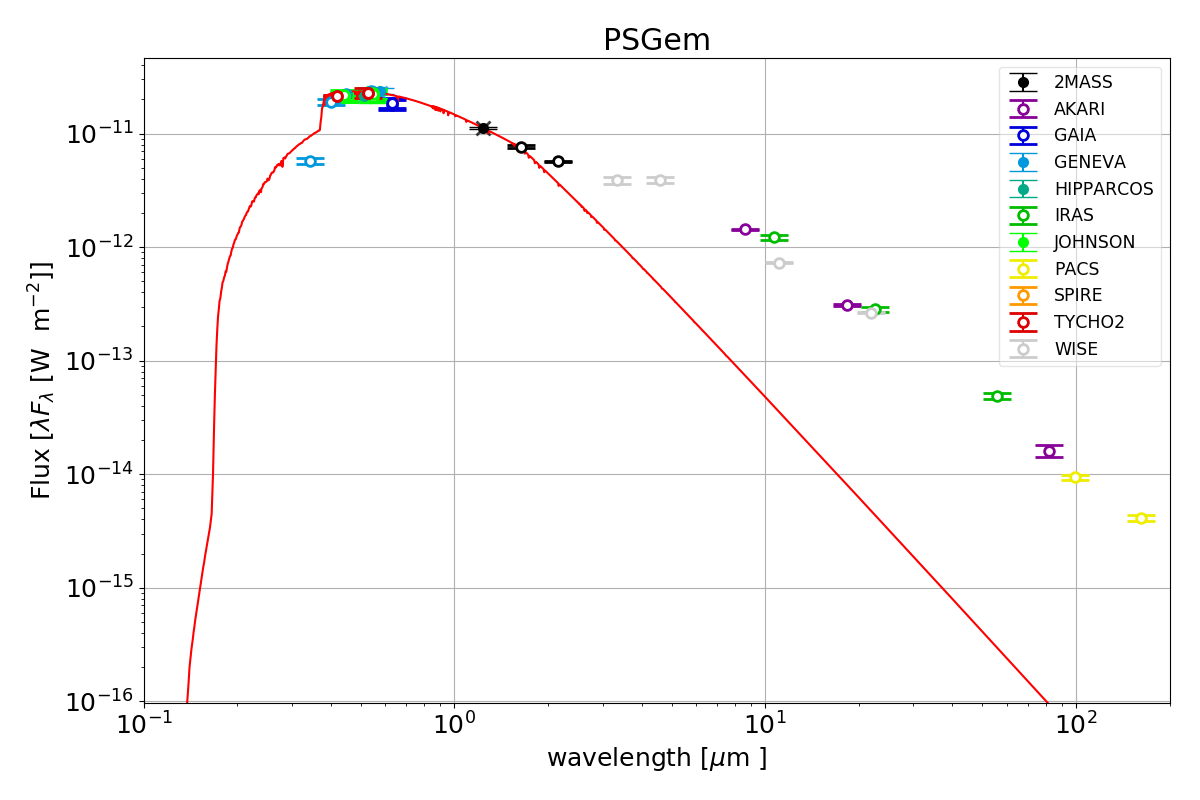}
\includegraphics[width=5cm]{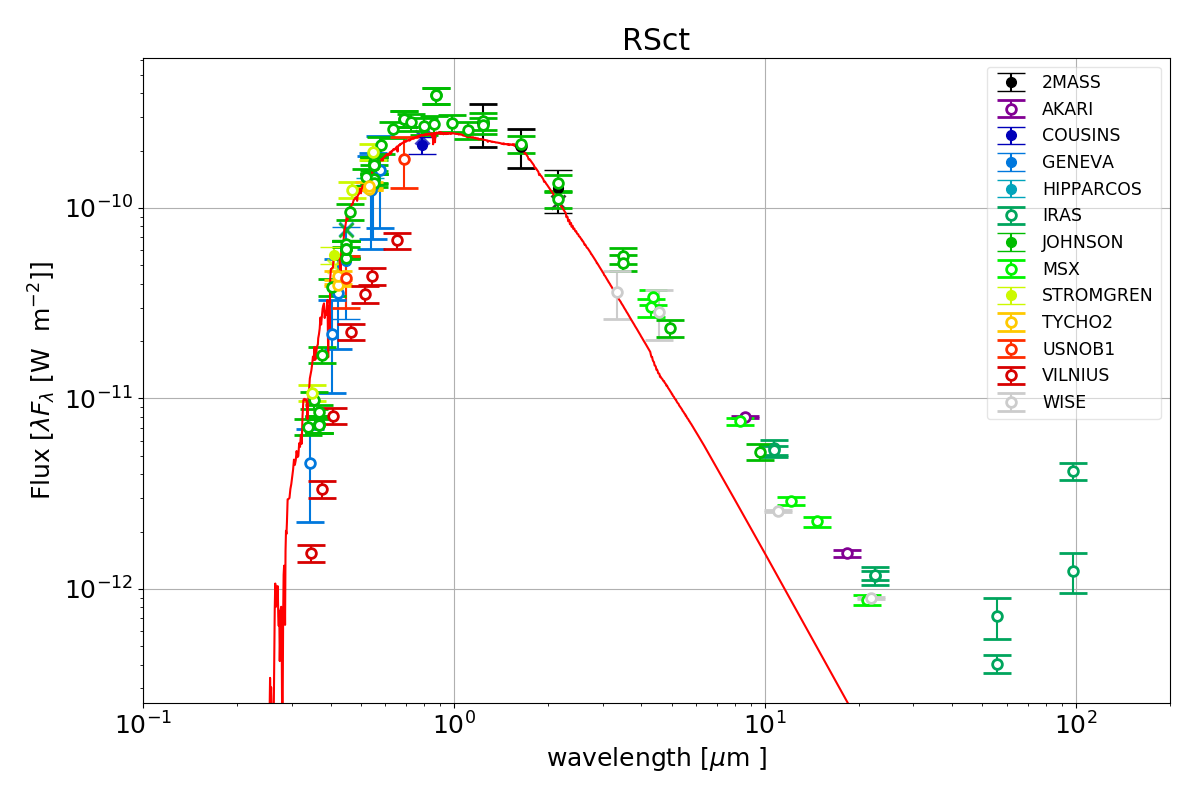}
\includegraphics[width=5cm]{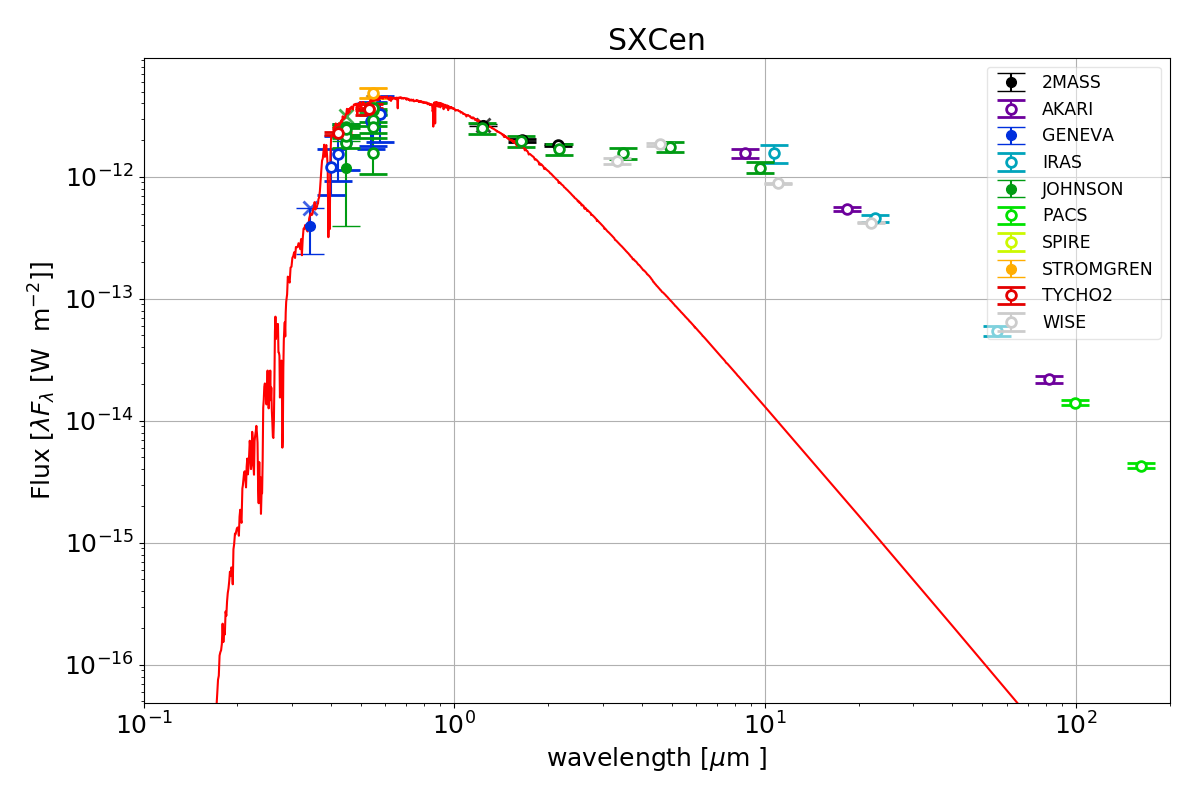}
\caption{Same as Fig.\,\ref{fig:SED1}.}
\label{fig:SED2}
\end{figure*}

\begin{figure*}
\centering
\includegraphics[width=5cm]{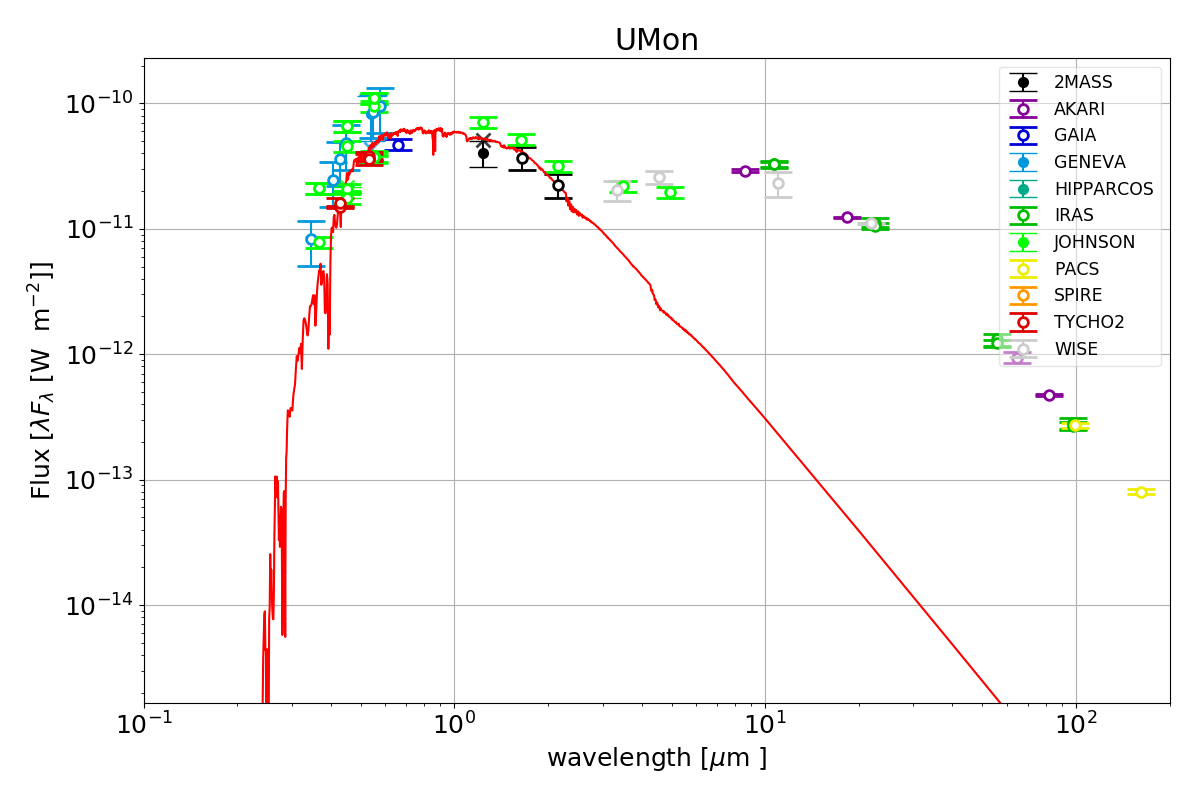}
\caption{Same as Fig.\,\ref{fig:SED1}.}
\label{fig:SED3}
\end{figure*}

\end{appendix}

\end{document}